\documentclass[12pt,a4paper,twoside]{book}
\renewcommand{\thechapter}{\Roman{chapter}}
\usepackage[utf8]{inputenc}
\usepackage[english]{babel}
\usepackage[T1]{fontenc}
\usepackage{amsmath}
\usepackage{amsfonts}
\usepackage{amssymb}
\usepackage{booktabs}
\usepackage{caption}
\usepackage[dvipsnames, table, x11names]{xcolor}
\usepackage{bbm}
\usepackage{pdfpages}
\usepackage{geometry}[bindingoffset=10mm]
\usepackage{graphicx}
\usepackage{tikz}
\usepackage{eso-pic}
\usepackage{lmodern}
\usepackage[numbers,sort&compress]{natbib}
\usepackage{diagbox}
\usepackage{color, colortbl}
\usepackage{hyperref}
\usepackage{lipsum} % for dummy text
\usepackage{etoolbox} % For conditional checks
\usepackage{array}  
\usepackage{emptypage} %no numbers from \cleardoublepage
\usepackage{booktabs}
\usepackage{fancyhdr}
\usepackage{mathtools}% http://ctan.org/pkg/mathtools für Feynman-Slash-Notation and underbracket
\usepackage{xfrac}
\usepackage{tabularx}
\usepackage[most]{tcolorbox}
\usepackage{xparse}
\usepackage{fontawesome}
\usepackage{varwidth}
\usepackage{pdfpages}
\usepackage{pax}
\usepackage{wrapfig} 
\usepackage{subcaption}
\usepackage{multicol}
\usepackage{placeins} %for FloatBarrier
\usepackage{sidecap}

\title{Analytic methods for perturbative Quantum Chromodynamics}
\author{Fabian Wunder}
\newcommand{\iu}{\mathrm{i}}
\newcommand{\ee}{\mathrm{e}}
\newcommand{\Qt}{Q_\text{T}}  			%Transversalimpuls
\newcommand{\dx}{\mathrm{d}}			%aufgerichtetes Differential
\newcommand{\as}{\alpha_\text{s}} 		%Kopplung starke WW
\newcommand{\NC}{N_\text{C}}		%number of colors
\newcommand{\CF}{C_\text{F}}		%CF
\newcommand{\CA}{C_\text{A}}
\newcommand{\eps}{\varepsilon}
\newcommand{\ghy}{{}_2\mathrm{F}_1}
\newcommand{\dilog}{\mathrm{Li}_2}

\newcommand{\goncharov}{\mathrm{G}}
\newcommand{\dPS}{\mathrm{dPS}}

\newcommand{\Lin}[1]{\mathrm{Li}_{#1}}
\newcommand{\Li}[2]{\mathrm{Li}_{#1}\left(#2\right)}

\newcommand{\Snp}[2]{\mathrm{S}_{#1}\left(#2\right)}

\newcommand{\MBint}[1]{\int_{-\iu\infty}^{\iu\infty}\frac{\dx #1}{2\pi \iu}}
\DeclareFontFamily{U}{wncy}{}
    \DeclareFontShape{U}{wncy}{m}{n}{<->wncyr10}{}
    \DeclareSymbolFont{mcy}{U}{wncy}{m}{n}
    \DeclareMathSymbol{\Sh}{\mathord}{mcy}{"58} 
\newcommand{\poha}[2]{\left(#1\right)_{#2}}
\newcommand{\BPV}{\mathcal{B}}
\newcommand{\APV}{\mathcal{A}}

\definecolor{burgundy}{RGB}{128, 0, 32}
\definecolor{dandelion}{rgb}{0.94, 0.88, 0.19}
\definecolor{dartmouthgreen}{rgb}{0.05, 0.5, 0.06}
\definecolor{cornflowerblue}{rgb}{0.39, 0.58, 0.93}
\definecolor{chromeyellow}{rgb}{1.0, 0.65, 0.0}
\definecolor{darkspringgreen}{rgb}{0.09, 0.45, 0.27}

\newcommand{\Red}[1]{\textcolor{red}{#1}}
\newcommand{\Blue}[1]{\textcolor{blue}{#1}}
\definecolor{accentcolor}{HTML}{800020} % Burgundy
\definecolor{quotebg}{gray}{0.95}

\newenvironment{fancychapter2}[4]{%
  \clearpage
  \thispagestyle{empty}% No header/footer on section title page
  \refstepcounter{chapter}
  \addcontentsline{toc}{chapter}{\protect\numberline{\thechapter}#1}%
  \markboth{\thechapter\ #1}{}% <-- Updates \headmark for subsequent pages
  \vspace*{-2cm}%
  \begin{tikzpicture}[remember picture,overlay]
    \fill[accentcolor] (current page.north west) rectangle ([yshift=-2.5cm]current page.north east);
    \node[anchor=north west, xshift=2cm, yshift=-1cm, text=white] 
      at (current page.north west) {\fontsize{36}{40}\selectfont \thechapter\quad #2};
  \end{tikzpicture}
  \vspace*{3cm}
  
  \begin{center}
  \ifstrempty{#1}{}{%
    {\centering{\Huge\bfseries #1}}\\[1cm]
  }
  \end{center}
  \ifboolexpr{ test {\ifstrempty{#3}} and test {\ifstrempty{#4}} }{}{%
    \begin{center}
      \begin{minipage}{0.8\textwidth}
        \colorbox{quotebg}{%
          \parbox{\dimexpr\linewidth-2\fboxsep}{
            \centering
            {\itshape #3}\\[0.5em]
            {\normalfont #4}
          }
        }
      \end{minipage}
    \end{center}
    \vspace{2cm}
    \FloatBarrier
  }
}{
  \ifdefined\ClearShipoutPicture
    \ClearShipoutPicture
  \fi
}

\pgfkeys{
  /block/.is family, /block,
  color/.store in=\blockcolor,
  title/.store in=\blocktitle,
  icon/.store in=\blockicon,
  type/.code={
    \def\blocktype{#1}%
    \ifstrequal{#1}{idea}{
      \def\blockcolor{blue}\def\blocktitle{Idea}\def\blockicon{\faLightbulbO}
    }{}%
    \ifstrequal{#1}{warning}{
      \def\blockcolor{red}\def\blocktitle{Warning}\def\blockicon{\faExclamationTriangle}
    }{}%
    \ifstrequal{#1}{note}{
      \def\blockcolor{green!50!black}\def\blocktitle{Note}\def\blockicon{\faStickyNoteO}
    }{}%
  },
  color=gray,
  title=Note,
  icon=,
  type=
}

\NewDocumentEnvironment{block}{O{}}
{
  \pgfkeys{/block,#1}%
  \begin{tcolorbox}[
    enhanced,
    breakable,
    width=\linewidth,
    colframe=\blockcolor!80!black,
    colback=\blockcolor!10!white,
    coltitle=\blockcolor!80!black,
    fonttitle=\bfseries,
    borderline west={14mm}{0pt}{\blockcolor!90!black},
    before skip=10pt,
    after skip=10pt,
    boxrule=0pt,
    left=17mm,
    right=5mm,
    notitle,                   % <-- stops any title rendering
    drop shadow,
    overlay={
      \ifx\blockicon\empty\else
        \node[anchor=center, xshift=7mm, text=white, font=\Large]
          at ([xshift=0pt]frame.west) {\blockicon};
      \fi
    }
  ]
}
{\end{tcolorbox}}
\NewDocumentEnvironment{summaryblock}{mmm}{
  \begin{block}[type=idea]
    #1
  \end{block}
  \begin{block}[color=purple,title=Custom,icon=\faPencilSquareO]
    #2
  \end{block}
  \begin{block}[color=Goldenrod,title=Result,icon=\faTrophy]
    #3
  \end{block}
}{}

\NewDocumentEnvironment{toolblock}{m}{
  \begin{block}[color=orange,title=Custom,icon=\faWrench]
    #1
  \end{block}
}{}

\NewDocumentEnvironment{readingblock}{m}{
  \begin{block}[color=brown,title=Custom,icon=\faBook]
    #1
  \end{block}
}{}
\begin{document}
%Mandatory titelpage
\includepdf[pages=1-2]{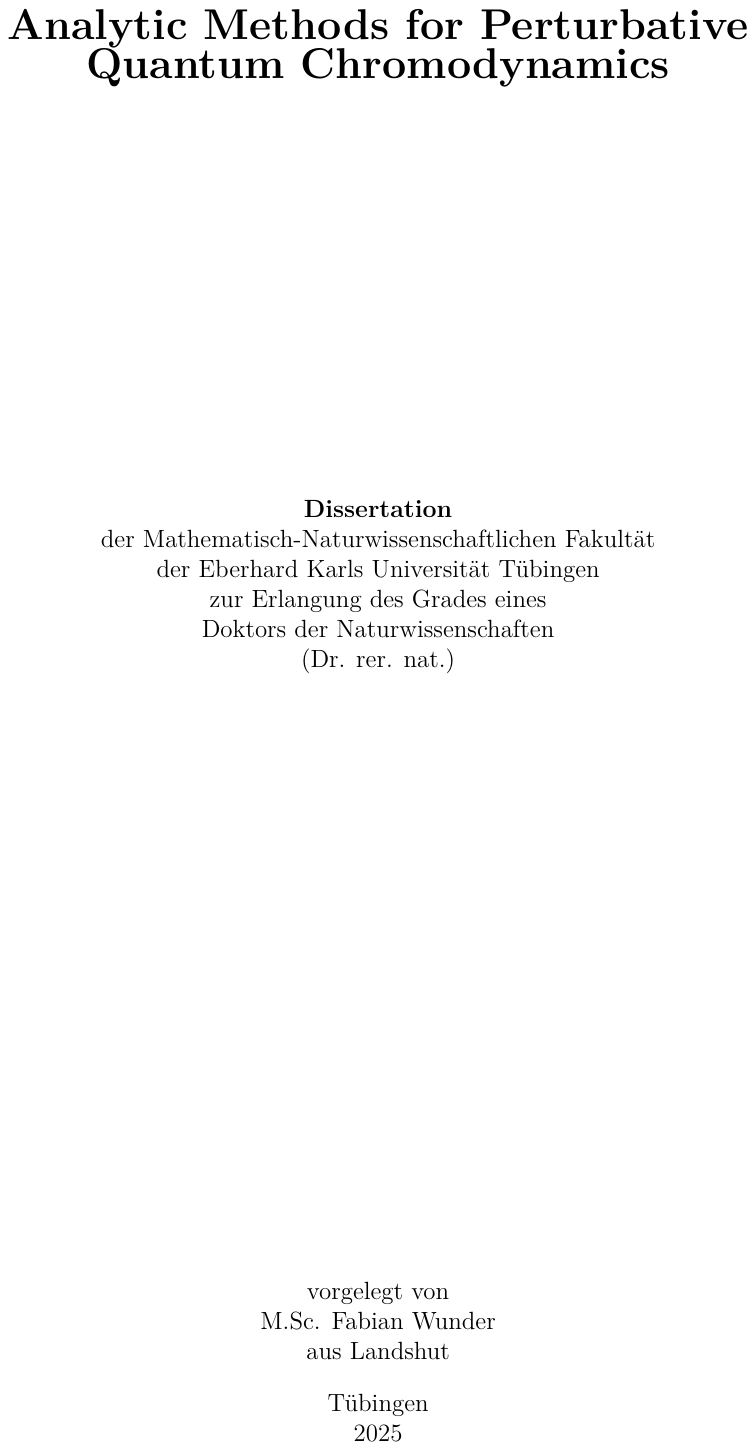}
%My custom titlepage
% =====================
% Title Page
% =====================
\begin{titlepage}
\pagecolor{burgundy} % Set full page background color
\thispagestyle{empty}

% --- Title and Information ---
\textcolor{white}{
\vspace*{5cm}
\begin{center}
{\Huge \bfseries Analytic Methods for Perturbative Quantum Chromodynamics}\\[1cm]
{\LARGE PhD Thesis}\\[2cm]
{\Large Institute for Theoretical Physics \\ University of Tübingen}\\[1cm]
{\large by Fabian Wunder \\ Advisor: Prof. Werner Vogelsang}\\[1cm]
December 2025
\end{center}
}

% --- Semi-transparent logo (drawn last so it's visible) ---
\begin{tikzpicture}[remember picture, overlay]
  \begin{scope}[opacity=0.2, blend mode=multiply] % blend mode forces effect
    \node[anchor=south, xshift=1cm, yshift=0cm]
      at (current page.south)
      {\includegraphics[width=10cm]{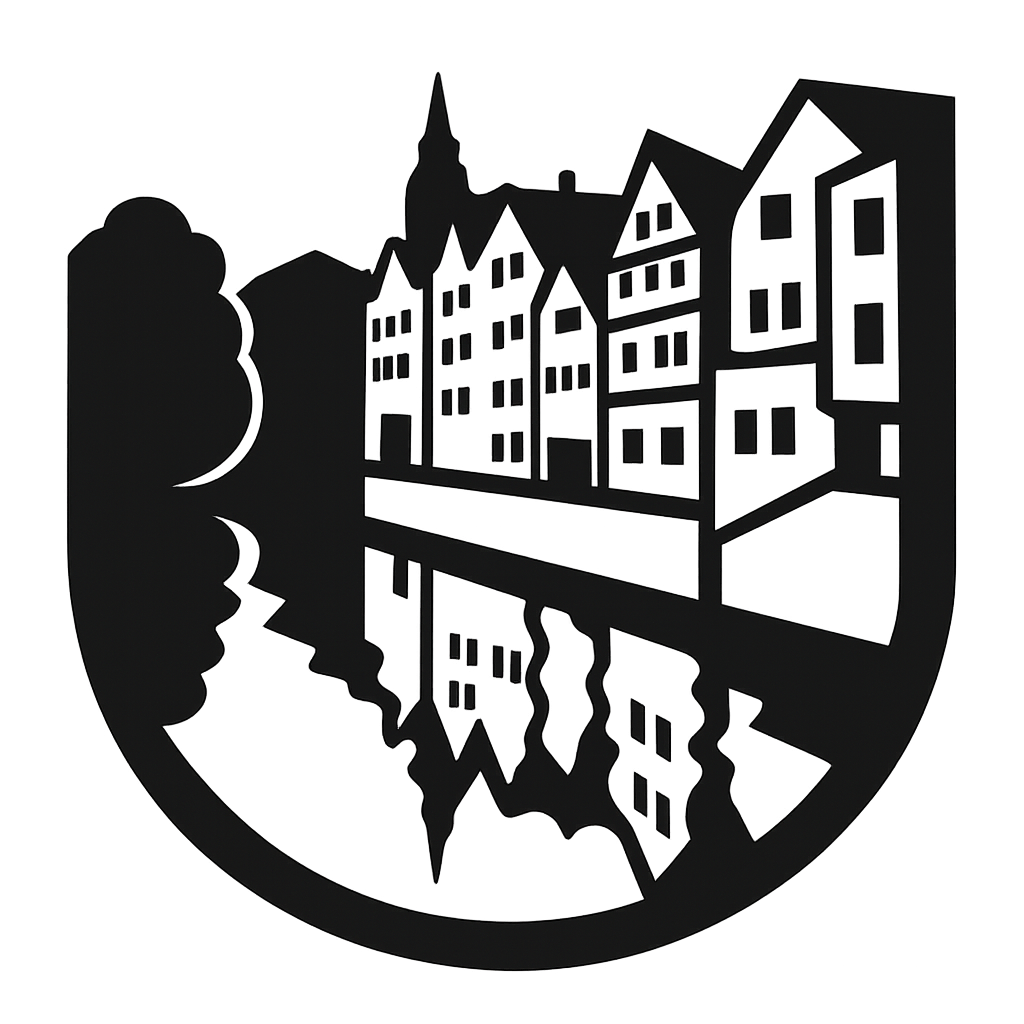}};
  \end{scope}
\end{tikzpicture}

\end{titlepage}

% Reset background color for the rest of the document
\nopagecolor
%%%
%%%
\newpage
\pagestyle{empty}
\phantom{.}
%%%
\newpage
%%%
\pagestyle{empty}
\phantom{.}\vspace{5cm}
\begin{center}
\textit{I do not mind if you think slowly,\\
 but I do object if you publish\\ more quickly than you think.}\\
\vspace{0.3cm}
 Wolfgang Pauli (1900\,-\,1958)
\end{center}
%%%
%%%
\newpage
\pagestyle{empty}
\phantom{.}
%%%
\newpage
%%% Inhaltsverzeichnis und Abstract
\pagestyle{plain}
\pagenumbering{roman}
\section*{Abstract}
\addcontentsline{toc}{section}{Abstract}
Nature, at the shortest distances accessible to contemporary colliders, is described by the Standard Model of particle physics with remarkable precision.
Since the discovery of the Higgs boson in 2012, progress has shifted from finding new resonances to exploring the Standard Model itself with unprecedented accuracy: advances now depend on extracting subtle consequences of established theory.
In this context, the theory of quantum chromodynamics (QCD) plays a central role.
Fundamentally, it describes the interactions of quarks and gluons, collectively referred to as partons. 
However, these are never observed in isolation but, due to the confining nature of the strong interaction, only in their bound hadronic states such as protons and neutrons, which make up most of the visible matter in the universe.

The central insight -- going back to Feynman -- that makes the dynamics of hadronic collisions accessible is factorization: the separation of observables into the universal long-distance dynamics of the hadrons encoded in parton distributions and process-specific short-distance coefficients that can be calculated in perturbation theory thanks to the asymptotic freedom of QCD at high energies.
This requires an interplay of different parts of the high-energy physics community:
(i) experiments supplying increasingly accurate data for observables, (ii) global analyses extracting parton distributions, (iii) precise perturbative calculations of hard scattering processes.

At present, a new collider, the Electron-Ion Collider (EIC), is under construction at Brookhaven National Laboratory in the USA. 
With polarized electron and proton beams and wide kinematic coverage, it will provide a new angle on the proton’s inner structure.
To make optimal use of the increased experimental precision, theory needs to keep pace.
This thesis contributes to this effort by enhancing analytic control over radiative corrections in hard scatterings, the kinematic dependence of structure functions in collider observables, and the dependence of parton densities on the energy scale at which they are probed.
 
A main part of this doctoral thesis is devoted to \textit{angular integrals}, key building blocks for phase-space integrals that sum over final-state radiation in a process.
Several powerful methods originally developed for Feynman integrals -- such as integration-by-parts identities, expansion by regions, differential equations, and dimensional shifts -- are adapted to this new setting to bring the knowledge about angular integrals to a modern standard, yielding several novel results.
Another topic is the analytic structure of certain one-loop integrals and the removal of spurious branch cuts by introducing single-valued polylogarithms.
This leads to a substantial compactification of the coefficient functions for semi-inclusive deep-inelastic scattering (SIDIS), a key process at the EIC.
Furthermore, a systematic expansion in higher powers of transverse momentum is developed and applied to the Drell-Yan process, allowing for a refined understanding of its factorization structure.
In addition, a new semi-analytical ansatz for the evolution of parton distributions is explored.
The unifying theme of these \textit{methods for perturbative QCD} is improving analytic understanding of processes relevant to collider phenomenology and the extraction of the proton structure.
\cleardoublepage
\section*{Zusammenfassung in deutscher Sprache}
\addcontentsline{toc}{section}{Zusammenfassung in deutscher Sprache (German version of the abstract)}
Phänomene auf den kleinsten Längenskalen, die moderne Teilchenbeschleuniger auf\-lösen können, lassen sich durch das Standardmodell der Teilchenphysik mit er\-staun\-licher Genauigkeit beschreiben.
Seit der Entdeckung des Higgs-Bosons im Jahr 2012 hat sich der Forschungsschwerpunkt von der Entdeckung neuer Resonanzen verlagert hin zu Präzisionstests des Standardmodells selbst: Weiterer Fortschritt hängt davon ab, subtile Effekte etablierter Theorie zu verstehen. In diesem Kontext spielt die Theorie der Quantenchromodynamik (QCD) eine zentrale Rolle.
Diese beschreibt auf fundamentaler Ebene die Wechselwirkung von Quarks und Gluonen, die kollektiv als Partonen bezeichnet werden.
Jedoch lassen sich diese nie isoliert beobachten, da die starke Wechselwirkung sie stets in Hadronen bindet, beispielsweise zu Protonen und Neutronen, die den Großteil der Materie im sichtbaren Universum ausmachen.

Die zentrale Idee, welche die Dynamik von Hadron-Kollisionen zugänglich macht, wird als Faktorisierung bezeichnet: eine Trennung von Observablen in einen uni\-ver\-sellen Bestandteil, der die niederenergetische interne Hadrondynamik beschreibt und einen prozessspezifischen Koeffizienten, der die hochenergetische Wechselwirkung cha\-rak\-te\-ri\-siert; letzterer lässt sich störungstheoretisch berechnen, da die QCD-Kopp\-lung bei hohen Energien klein ist.
Dies macht ein Zusammenspiel verschiedener Be\-rei\-che der Teilchenphysik notwendig: (i) Experimente, die immer genauere Daten lie\-fern, (ii) Bestimmung von Partonverteilungen unter Berücksichtigung aller ver\-füg\-ba\-rer Daten, (iii) präzise störungstheoretische Berechnungen von Hochenergie-Streu\-pro\-zessen.

Zur Zeit entsteht in den USA ein neuer Teilchenbeschleuniger, der Electron-Ion Collider (EIC).
Dieser wird die Kollision von polarisierten Elektronen- und Protonenstrahlen mit bisher unerreichter Genauigkeit vermessen, was neue Einblicke in die Protonstruktur ermöglichen wird.
Um diese Daten optimal zu nutzen, muss die Theorie schritthalten.
Diese Doktorarbeit leistet dazu einen Beitrag, indem sie die analytische Kontrolle über Strahlungskorrekturen in Streuprozessen, kinematische Abhängigkeiten von Observablen und die Energieabhängigkeit von Partonverteilungen erhöht.

Ein großer Teil der Arbeit ist sogenannten Winkelintegralen gewidmet, zentrale Bausteine von Phasenraumintegralen, die über die zusätzliche, im Streuprozess ent\-ste\-hen\-de, Strahlung summieren.
Verschiedene effiziente Methoden, die ursprünglich für so genannte Feynman-Integrale entwickelt worden sind -- zum Beispiel Partielle-Integrations-Identitäten, Entwicklung nach Bereichen, Differentialgleichungsmethoden und Dimensionsverschiebung -- werden für die neue Problemstellung adaptiert und auf Winkelintegrale angewendet.
Dies erlaubt eine Reihe neuer Resultate.
Ein weiterer Bereich der Arbeit befasst sich mit der analytischen Struktur spezieller Feynman Integrale und dem Beseitigen oberflächlicher Unstetigkeiten durch Verwendung einwertiger Polylogarithmen.
Dies führt zu einer deutlichen Vereinfachung der Strukturfunktionen in semi-inklusiver tief-inelastischer Streuung (SIDIS), einem wichtigen Prozess am EIC.
Des Weiteren wird eine systematische Entwicklung im Transversalimpuls untersucht und auf den Drell-Yan-Prozess angewendet um unser Verständnis der Faktorisierung zu verbessern.
Zusätzlich wird ein neuer semi-analytischer Ansatz zur Partonevolution vorgestellt.
Der gemeinsame Rahmen dieser analytischen Methoden für perturbative QCD ist es, unser Verständnis für Prozesse, die relevant für Kolliderphänomenologie und die Untersuchung der Protonstruktur sind, zu verbessern.
\cleardoublepage
\thispagestyle{empty}
\tableofcontents
%%%
%%%
%%%
\newpage
\pagestyle{plain}
\section*{Guide to read this thesis}
\addcontentsline{toc}{section}{Guide to read this thesis}
You hold in your hand -- or have on your screen -- a thesis about \textit{Analytical methods for perturbative Quantum Chromodynamics}.
Overall, this work amounts to well over four-hundred pages -- including three-hundred pages of peer-reviewed journal publications which contain the main scientific contributions of the doctoral work -- and only the most dedicated will enjoy reading it cover-to-cover.
This thesis was written with different readers in mind for which different parts of this thesis may be of interest.
This preface should help guide the reader to what they are looking for, see Table \ref{tab:SuggestedReading} for a quick overview.

\begin{table}[htb]
\centering
\begin{tabular}{p{4cm}|p{10cm}}
\hline
Reader & Suggested chapters\\
\hline
Non-physicist & Start with \ref{sec:Intro1} and \ref{sec:Intro2}, from there try to make it to \ref{sec:Intro7} (it is not necessary to get all the details). Also read \ref{sec:LogarithmKepler} to get a glimpse at the impact mathematical methods can have.\\
\hline
Non-particle physicist & Start with \ref{sec:Intro3} for an introduction to the broader field and read to \ref{sec:Intro7} for an overview of what this thesis is about specifically.\\
\hline
pQCD expert & Start with \ref{sec:Intro5} and \ref{sec:Intro6} for an intro and read the thesis summary of \ref{sec:Intro7}.
Skip to \ref{ch:Publications} and read only the publication summaries.\\
\hline 
Graduate student in pQCD & Work through the technical chapters \ref{ch:SpecialFunctions} and \ref{ch:Methods}, then you are well prepared to read the publications of chapter \ref{ch:Publications} that interest you.  \\
\hline
\end{tabular}
\caption[Reader guide for the thesis.]{Suggested minimal reading to get the relevant information from this thesis for different readers.}
\label{tab:SuggestedReading}
\end{table}

Chapter \ref{ch:Intro} gives the context for the content of this work.
It starts with a brief history of particle physics spanning a big arc from philosophical discussions in antiquity up to the completion of the Standard Model, explains where perturbative quantum chromodynamics (pQCD) fits into this picture, and summarizes the specific topics of this thesis.
However, to cater to different prior knowledge of the field and different enthusiasm regarding historical expositions, different entry points may be chosen.

Sections \ref{sec:Intro1} and \ref{sec:Intro2} are intended to be readable by non-physicists.
Along the historical journey through selected stories of the study of matter, we also introduce several of the concepts that are foundational to work in high energy physics.
It prepares for \ref{sec:Intro3}, where quantum field theory and collider physics are introduced.
A physicist not working on quantum field theory may start reading here.
Section \ref{sec:Intro4} gives a brief introduction into quantum chromodynamics (QCD), the main theoretical stage of the investigations of this thesis.
If you feel familiar with particle physics but not with QCD in particular, this is a good entry point.
Experts, that are not interested in the didactical exposition upfront, can start this thesis at \ref{sec:Intro5}, which briefly summarizes the state of contemporary collider physics, before going to \ref{sec:Intro6} that introduces the field of perturbative QCD as relevant to this thesis.
The novel contributions of this work are then summarized in \ref{sec:Intro7}.
 
Chapter \ref{ch:SpecialFunctions} introduces a main mathematical player in this thesis: special functions.
It comes with its own brief digression on the history of the logarithm to showcase the potential importance the right mathematical tools can have to science -- again this part should be accessible for a broader audience.
The remainder of this chapter one-by-one introduces the special functions that are relevant to the work in perturbative QCD.
This technical part of this chapter may be skipped at first reading and be treated as a reference when the relevant functions show up later in the publications.

Chapter \ref{ch:Methods} introduces the methods that form the foundation of the calculations performed in the publications.
These are more technical and intended for the reader who wants to learn these tools.
The exposition is centered around simple introductory examples rather than aiming for complete generality or a presentation of the state-of-the-art.
For this, there are dedicated references to the literature.

Chapter \ref{ch:Publications} contains the main scientific content of this thesis.
It goes through all the publications that were written in the course of this doctoral work.
Upfront each publication -- which are reprints of the journal versions -- is a short summary of the paper's content and how it connects to the overall story.
For a comprehensive overview over the main achievements of this thesis, reading these short summaries may suffice.
At the start of chapter \ref{ch:Publications}, there is a more detailed guide to reading the publications.

Finally, this thesis concludes in chapter \ref{ch:Conclusion} with a summary and a brief outlook on further research.

To make this thesis visually more appealing, several colored text blocks with icons are used to highlight different forms of additional material to the main text:
\begin{block}[type=note]
This block is used for clarifying examples and extra information that might be known to expert readers but not all.
\end{block}

\begin{block}[type=idea]
This block is used for key ideas and the scientific questions addressed in the publications.
\end{block}

\begin{toolblock}{
This block marks important tools for calculations.
}\end{toolblock}

\begin{readingblock}{
This block gives literature for further reading on a specific topic.
}\end{readingblock}

\begin{block}[color=purple,title=Custom,icon=\faPencilSquareO]
This block is used to refer to the methodology used in a publication.
\end{block}

\begin{block}[color=Goldenrod,title=Result,icon=\faTrophy]
This block is used to highlight key results of this thesis.
\end{block}

%\begin{block}[type=warning]
%This is a **Warning block** with the exclamation icon centered in the bar.
%\end{block}
\clearpage
\section*{Acknowledgments}
\addcontentsline{toc}{section}{Acknowledgments}
This doctoral thesis would not have been possible without the support of many wonderful people.
First and foremost, I want to thank my advisor Werner Vogelsang for his continued support during the entire course of this doctoral work -- even when it took slightly longer than initially anticipated. 
He created an environment that allowed me to explore the many interesting topics discussed in this work, allowed for much freedom to choose what to pursue, encouraged publication of interesting results, and gave plenty of opportunity to travel to schools and conferences.
These were a wonderful possibility to get to know the QCD community -- a thank you to all the nice people I had the pleasure to meet in the past years -- and keep the motivation afloat.
In this regard, I want to give special emphasis to the QCD Masterclass of 2023, outstandingly organized by François Arleo and Stéphane Munier in the lovely abbey of Saint-Jacut-de-la-Mer. 
While the first years of this doctoral work were under the shadow of covid, the mudflats of Brittany and the surrounding physics program helped ignite a lasting spark of inspiration.
Formative for the course of the doctoral journey was the participation in the DFG research group FOR2926 \textit{Next Generation Perturbative QCD for Hadron Structure: Preparing for the Electron-Ion Collider}.
The regular meetings gave ample room for presentation of results, exchange, and ideas to go forward.
At this point a warm thank you to all members of FOR2926 and those who made it possible!

Most of the time, research is more fun and feels more meaningful if you have co-authors, especially the right ones.
In this regard, I would like to thank Valery Lyubovitskij and Alexey Zhevlakov for their collaboration when working on my first publication, Juliane Haug for the many successful projects, Oliver Schüle for his contributions to \texttt{POMPOM}, Werner Vogelsang for his collaboration in the Drell-Yan project across countless Zoom sessions, and importantly Vladimir Smirnov for reaching out to a younger colleague to join forces in common projects that combined our expertise. 
Being great colleagues does not require common publications though.
There have been plenty of people around in the D7 during the past six years that made the workplace much more enjoyable.
For this my special thanks go to fellow bachelor/master/PhD students and postdocs Maurizio Abele, Dominik Bammert, Ignacio Borsa, Greta Bösinger, Lana Dambacher, Alexander Fürlinger, Juliane Haug, Colin Heckmeyer, Timothy Herl, Astrid Hiller-Blin, Felix Kunzelmann, Kevin Kurz (honorary member), Markus Löchner, Santiago Lopez, Daniel Rein, Simon Reinhardt, Jakob Rudoll, Timo Schreyer, Ozan Semin, Martin Vollmann, and David Weiler -- of course also for activities beyond the workplace.
The welcoming research atmosphere rests of course on the permanent members of the floor, Prof. Jan-Philipp Burde, Prof. Thomas Gutsche, Prof. Barbara Jäger, Dr. Valery Lyubovitskij, Prof. Hugo Reinhardt, Prof. Marc Schlegel, Dr. Marco Stratmann, and Prof. Werner Vogelsang, 
and of course also on our always supportive administrative staff, Ingrid Estiry and Sabine Werner -- thank you especially for the best department Christmas parties!

A successful PhD depends not only on the work-life, but also on the support around it.
I am grateful to my friends who were there to celebrate success and comfort during more difficult times.
Most importantly, Vivek Sehra and Kenny Fohmann for countless hours of talking about PhD life from mutual experience in the bars of Tübingen, and Kevin Kurz for extended coffee breaks at the university to discuss all important and less important aspects of life.
Of great importance to me is my family. I want to thank my parents Klaus and Martina for their continued support whenever I needed it and understanding that also doctoral students sometimes have work to do and cannot always come home, and of course my favorite brother and sister, Niklas and Isabella, respectively. It has always been a joy when you came to Tübingen in the past years.

Finally, I want to express my a\textcolor{blue}{b}iding gratitude to Jule for her love and s\textcolor{blue}{u}pport in a mult\textcolor{blue}{i}tude of roles in the past years, as \textcolor{blue}{b}eloved partner, co-a\textcolor{blue}{u}thor, proof-reader of this thes\textcolor{blue}{i}s, office-mate, and flatmate.

\vspace{2cm}
\noindent{}Enjoy reading,\\
{}\\
Fabian Wunder, Tübingen, December 2025.
\newpage

\section*{List of publications}
\addcontentsline{toc}{section}{List of publications}
This is a list of all publications that form part of this thesis.
The candidate made substantial contributions to all of them in all parts of the scientific process and contributed in approximately equal amounts to scientific ideas, programming, analysis, and paper writing.
\begin{itemize}
\item[\cite{Lyubovitskij:2021ges}] Valery E. Lyubovitskij, \textbf{Fabian Wunder}, Alexey S. Zhevlakov\,(2021): \textit{New ideas for handling of loop and angular integrals in D-dimensions in QCD}. In: \href{https://doi.org/10.1007/JHEP06(2021)066}{JHEP 06 (2021) 066}, e-print: \href{https://arxiv.org/abs/2102.08943}{arXiv:2102.08943}, re-print: sec.\,\ref{pub:1}.

\item[\cite{Haug:2022hkr}] Juliane Haug, \textbf{Fabian Wunder}\,(2023): \textit{The massless single off-shell scalar box integral — branch cut structure and all-order epsilon expansion}. In: JHEP 02 (2023) 177, e-print: \href{https://arxiv.org/abs/2211.14110 }{arXiv:2211.14110}, re-print: sec.\,\ref{pub:2}.

\item[\cite{Haug:2023eqg}] Juliane Haug, \textbf{Fabian Wunder}\,(2023): \textit{The massless non-adjacent double off-shell scalar box integral — branch cut structure and all-order epsilon expansion}. In: \href{https://doi.org/10.1007/JHEP05(2023)059}{JHEP 05 (2023) 059}, e-print: \href{https://arxiv.org/abs/2302.01956}{arXiv:2302.01956}, re-print: sec.\,\ref{pub:3}.

\item[\cite{Wunder:2024btq}] \textbf{Fabian Wunder}\,(2024): \textit{Asymptotic behavior of angular integrals in the massless limit}. In:  \href{https://doi.org/10.1103/PhysRevD.109.076022}{Phys.\,Rev.\,D 109 (2024) 7, 076022}, e-print: \href{https://arxiv.org/abs/2403.09773}{arXiv:2403.09773}, re-print: sec.\,\ref{pub:4}.

\item[\cite{Lyubovitskij:2024civ}] Valery E. Lyubovitskij, Werner Vogelsang, \textbf{Fabian Wunder}, Alexey S.\\ Zhevlakov\,(2024): \textit{Perturbative $T$-odd asymmetries in the Drell-Yan process revisited}. In: \href{https://doi.org/10.1103/PhysRevD.109.114023}{Phys.\,Rev.\,D 109 (2024) 11, 114023}, e-print: \href{https://arxiv.org/abs/2403.18741}{arXiv:2403.18741  }, re-print: sec.\,\ref{pub:5}.

\item[\cite{Haug:2024asl}] Juliane Haug, Oliver Schüle, \textbf{Fabian Wunder}\,(2024): \textit{A semi-analytical $x$-space solution for parton evolution — Application to non-singlet and singlet DGLAP equation}. In: \href{https://doi.org/10.1007/JHEP07(2024)072}{JHEP07 (2024) 072}, e-print: \href{https://arxiv.org/abs/2404.18667}{arXiv:2404.18667}, re-print: sec.\,\ref{pub:6}.

\item[\cite{Smirnov:2024pbj}] Vladimir A. Smirnov, \textbf{Fabian Wunder}\,(2024): \textit{Expansion by regions meets angular integrals}. In: \href{https://doi.org/10.1007/JHEP08(2024)138}{JHEP08 (2024) 138}, e-print: \href{https://arxiv.org/abs/2405.13120}{arXiv:2405.13120}, re-print: sec.\,\ref{pub:7}.

\item[\cite{Haug:2024yfi}] Juliane Haug, \textbf{Fabian Wunder}\,(2025): \textit{Angular integrals with three denominators via IBP, mass reduction, dimensional shift, and differential equations}. In: \href{https://doi.org/10.1007/JHEP03(2025)141}{JHEP 03 (2025) 141}, e-print: \href{https://arxiv.org/abs/2410.18177}{arXiv:2410.18177}, re-print: sec.\,\ref{pub:8}.

\item[\cite{Haug:2025sre}] Juliane Haug, \textbf{Fabian Wunder}\,(2025): \textit{Single-valued representation of unpolarized and polarized semi-inclusive deep inelastic scattering at next-to-next-to-leading order}. In: \href{https://link.aps.org/doi/10.1103/s8wt-hsxx}{Phys.\,Rev.\,D. 112 (2025) 114036}, e-print: \href{https://arxiv.org/abs/2505.18109}{arXiv: 2505.18109}, re-print: sec.\,\ref{pub:9}.

\item[\cite{Haug:2025ava}] Juliane Haug, Vladimir A. Smirnov, \textbf{Fabian Wunder}\,(2025):\\ \textit{On multi-propagator angular integrals}. In: \href{https://doi.org/10.1007/JHEP10(2025)001}{JHEP 10 (2025) 001}, e-print:\\ \href{https://arxiv.org/abs/2508.00693}{arXiv:2508.00693}, re-print: sec.\,\ref{pub:10}.
\end{itemize}
\noindent{(All author lists in alphabetical order by convention of the research field.)}

\cleardoublepage
%%% Hauptteil
%%%
\pagestyle{fancy}
\pagenumbering{arabic}
\setcounter{page}{1}
\begin{fancychapter2}{What is particle physics and what is this thesis about?}{A tale of particles and integrals}{Alles ist Wechselwirkung. (Everything is interaction.)}{Alexander von Humboldt}
\label{ch:Intro}

\begin{wrapfigure}[20]{r}{0.35\textwidth}
\centering
\setlength{\wrapoverhang}{2em} % pushes the entire wrapfigure into the margin
\includegraphics[width=0.3\textwidth]{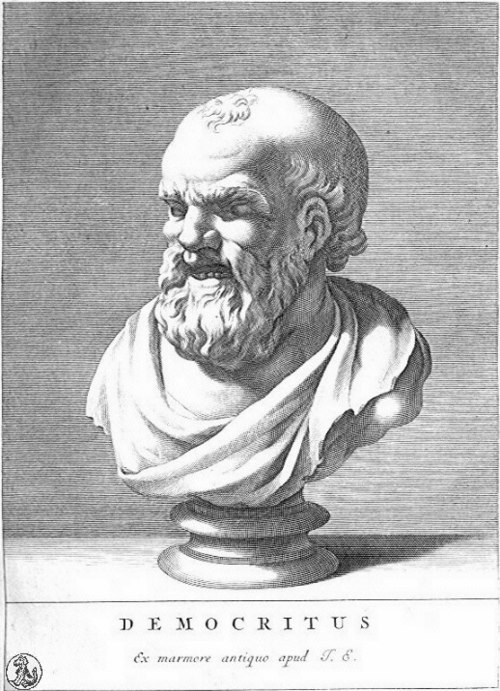}
\captionsetup{margin=5pt} 
\caption[Copper engraving after bust of Democritus (from \cite{HHU_Demokrit_Image}).]{\textit{Don't worry, it's just particles.}\\
 \footnotesize{Copper engraving after a bust of Democritus} \cite{HHU_Demokrit_Image}}
\label{fig: Democritius}
\end{wrapfigure}
\section{The idea of particles}
\label{sec:Intro1}
\textit{``What is the world made of?''} is a question that sits close to the heart of human curiosity.
Predating modern science by two millennia, the ancient Greeks speculated about what happens when we cut matter into smaller and smaller pieces.
Can we proceed indefinitely or will we reach some fundamental uncuttable layer?
The semi-mythical inventor of the latter hypothesis was Leucippus of Miletus in the mid-5th century BC, who coined the phrase \textit{atom}\footnote{In ancient Greek ``$\alpha\tau o\mu o\varsigma$'' (a-tomos) means ``un-cuttable/in-divisible''.} for these tiniest pieces \cite{Atomists:2010}.
This idea was popularized by his student Democritus of Abdera, who made it central to his philosophy \cite{Atomists:2010}. 
He envisioned a deterministic mechanistic world of tiny geometric shapes, differentiated only by form, order, and position, drifting through the void and interacting through collisions.

However, this concept was only popular with a minority of philosophers.
Most influentially, Aristotle -- arguing for a theory of four continuous ``elements'' of matter (earth, air, fire, water) and a distinct irreducible ``organizing principle'' required to form a substance from them -- objected to the concept of atoms \cite{Aristotle:1970}.
He criticized the impossibility of the void and the lack of purpose in deterministic motion.
Ironically, since only fragments of Democritus' writings were passed down through centuries, what we know today about the ancient atomistic theory mostly survived in critiques of it such as Aristotle's \textit{Physics} \cite{AristotlePhyiscs:1961} -- which was much more popular with thinkers in the following millennia throughout antiquity, the Middle Ages and at early European universities up until the 1700s.

One of the exceptions among the ancient Greeks was Epicurus, who took up Democritus' ideas and reworked them into a systematic philosophy of life -- allowing for free will by introducing the possibility of sudden, unpredictable deflections of the atoms \cite{Epikur:1968}.
In the 1st century BC, Lucretius brought these thoughts to the Latin world in his work \textit{De rerum Natura} \cite{Lucretius:1968}.
Transcribed in Carolingian times, it waited in a German monastery to be rediscovered in the Renaissance by Poggio Braciolini during the Council of Constance (1414-1418) \cite{Ricci:2020poggio}.

For the Greeks, the debate was a philosophical one.
There was no experimental evidence -- nor the intent to systematically search for one -- for the proposed existence of atoms.
The rediscovery of the ancient texts in the early modern period also started out philosophical.
Lucretius' writing put Giordano Bruno into the position to criticize the dominant Aristotelian view \cite{Knox:2018Bruno} and Pierre Gassendi streamlined the atom-in-void-physics to be acceptable in a Christian world \cite{GassendiBritannica}.
Still the concept was far from mainstream. 
Importantly, René Descartes -- preferring a world made of continua -- objected to the concept of a void and indivisible building blocks of matter \cite{Slowik2005DescartesPhysics}.
Nevertheless, his mechanistic approach strengthened the position that the behavior of matter is determined solely by its physical properties.

Building on the ideas of Robert Boyle \cite{BoyleBritannica}, a strong proponent of corpuscular theory -- atomism applied in the laboratory --, a milestone was set by Isaac Newton's \textit{Philosophiae Naturalis Principia Mathematica} (1687) \cite{Newton1833philosophiae}, which was written in explicit opposition to Descartes.
Describing the physical world in terms of ``hard and impenetrable particles'' had a lasting impact on the foundations of physics as an empirical science in the modern sense.
To this day, Newtonian point particles moving along trajectories will be among the first things a student of physics sees in university \cite{demtroder2003experimentalphysik}.
The central new ingredient, and what sets modern science apart from what came before, is the connection of theory with observation -- only an explanation which lives up to experimental testing can be considered valid.

\begin{wrapfigure}[14]{l}{0.35\textwidth}
\centering
\vspace{-0.2cm}
\setlength{\wrapoverhang}{2em} 
\includegraphics[width=0.3\textwidth]{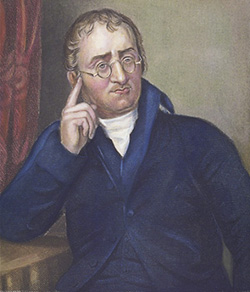}
\captionsetup{margin=5pt} 
\caption[Painting of John Dalton by Joseph Allen (from \cite{Dalton_JosephAllen_1814_Wikimedia}).]{\textit{It's Lego pieces playing billiards.}\\ \footnotesize{Painting of John Dalton} \cite{Dalton_JosephAllen_1814_Wikimedia}}
\label{fig: Dalton}
\end{wrapfigure}

The empirical ``cutting'' of matter was carried out by early chemists.
Keeping the Aristotelian term ``element'', but moving away from its original metaphysical meaning and its use in medieval alchemy, Boyle and later Antoine Laurent de Lavoisier gave an operational definition of elements as substances that cannot be decomposed by any known chemical means \cite{LavoisierBritannica}.

The first observation that directly tied it to the modern concept of atoms was John Dalton's ``law of multiple proportions'' presented in his \textit{New System of Chemical Philosophy} (1808) \cite{dalton2010new}: When two elements make different compounds, the amount of one element that combines with a fixed amount of the other always comes in simple whole-number ratios.

\vspace{0.2cm}
\begin{block}[type=note]
\textbf{Example of Dalton's law:}\\
One gram of carbon combines with either $1.3$ grams of oxygen (to form CO) or $2.6$ grams of oxygen (to form CO$_2$). The ratio of these oxygen masses is 1:2.
\end{block}
\vspace{0.2cm}

At this point, atoms could be considered ``useful fiction'' -- no direct observation was possible, but envisioning chemical compounds as consisting of tiny molecules which are made of atoms of different elements that recombine in chemical reactions was -- and is to this day -- a tremendously helpful bookkeeping device in chemistry, used to compile the periodic table by Dmitri Mendeleev in 1869 \cite{MendeleevBritannica}. 
Based on the experimental data at the time, William Prout hypothesized in 1815 that all atomic masses were multiples of hydrogen, proposing that it may build up other atoms \cite{ProutBritannica} -- foreshadowing the discovery of the proton a hundred years later.

\section{The development of particle physics}
\label{sec:Intro2}
Direct observation on length scales smaller than what the human eye can see came with the microscope.
The first observation directly linked to the existence of atoms came in 1827, when the Scottish botanist Robert Brown described the phenomenon that pollen dissolved in water shows random motion \cite{lavenda1985brownian}.
Subsequent experimentation with small inorganic particles ruled out life as a possible cause.
An explanation came in 1905, with Albert Einstein demonstrating that the probability distribution of Brownian particles follows from diffusion equations caused by random collisions with much smaller particles of the solvent \cite{einstein1906theory}.  
Three years later, Jean Baptiste Perrin experimentally verified Einstein's predictions \cite{perrin2013brownian} for which he received the Nobel prize in 1926 -- finally bringing atoms from mathematical devices in chemistry to accepted physical reality.
The award cites ``work on the discontinuous structure of matter'' \cite{NobelPrize1926} tying back to the ancient philosophical debate.
Curiously, already in \textit{De rerum natura} Lucretius attributed the motion of ``dust particles dancing in the sunlight'' \cite{Lucretius:1968} to underlying collisions with much smaller atoms.
While missing out on the fact that most of it is caused by air current, his wording comes remarkably close to a description of Brownian motion.

Even before the existence of atoms was settled, evidence appeared that they are not indivisible objects.
In 1897, Joseph Thomson discovered that the rays in cathode ray tubes\footnote{Cathode rays were later used to build television screens.} could be deflected in electric and magnetic fields, revealing a fixed charge-to-mass ratio that stayed constant across different cathode materials \cite{thomson1897xl}, earning him the 1906 Nobel prize.
This indicated that these were in fact particles, later named \textit{electrons} \cite{falconer1987corpuscles}.
Robert Millikan's oil drop experiment (1909) \cite{Milikan:1913} -- winning him the 1923 Nobel prize -- measured the electron mass to be two-thousand times lighter than any known atom.
Together this indicated that they most likely were part of atoms, opening the way to study atomic sub-structure.

Another discovery that pointed at a substructure came in 1896, when Henri Becquerel intended to use uranium salts in a fluorescence experiment \cite{becquerel1896radiations}.
He expected that after ``charging'' them in sunlight, they might emit radiation he could detect with photographic plates.
Initial experiments confirmed that expectation.
Then, due to cloudy weather, he had to postpone further experimentation and stored the uranium salts with the photographic plates in a dark drawer.
To his surprise, he discovered the photographic plates to be darkened even though the uranium was not exposed to the sunlight before.
Based on systematic studies, Marie and Pierre Curie inferred it could not be caused by fluorescence, but the radiation needed to come from within the uranium atoms -- but how exactly was a mystery at the time.
\textit{Radioactivity}, as this phenomenon was called, turned out to transmute elements into other elements -- against the established chemical wisdom.
For their work on radioactivity, Becquerel and the Curies shared the 1903 Nobel prize \cite{NobelCurieRadioactivity}.

Further progress in the understanding of the structure of matter came with scattering experiments.
To resolve a structure of a certain length scale, the wavelength of the radiation must be no longer than that magnitude.
In the scattering framework, atoms showed up in 1912, when Max von Laue used X-ray diffraction to resolve crystals \cite{FriedrichKnippingLaue1912}.
He observed a regular interference pattern with regularly spaced bright dots that can be well explained by regularly spaced, point-like scattering centers, for which he received the 1914 Nobel prize.
% for orientation, typical scales are given in Table \ref{tab: Scales}.
%\begin{table}
\vspace{-0.1cm}
{\begin{block}[type=note]
\begin{small}
\textbf{Typical length scales:}\vspace{0.2cm}
\\
\begin{tabular}{c | c c c}
\hline\hline
Structure & Length scale & Energy & Possible probe \\
\hline
Human hair & $10^{-4}$\,m & 0.01\,eV & Visible light (Naked eye)\\
Cells & $10^{-5}$\,m & 0.1\,eV & Visible light (Microscope)\\
Molecules & $10^{-9}$\,m & 1\,keV & X-rays (Laser)\\
Atoms & $10^{-10}$\,m & 10\,keV & X-rays (Laser)\\
Nuclei & $10^{-14}$\,m & 100\,MeV & $\alpha$-rays (Fixed Target) \\
Proton & $10^{-15}$\,m & 1\,GeV & Electrons (Fixed Target) \\
Quarks & $10^{-18}$\,m & 1\,TeV & Protons (Collider) \\
\hline\hline
\end{tabular}
\addcontentsline{lot}{table}{I.0 Table of typical length scales for orientation.}
\end{small}
%\caption{Typical length scales, the necessary radiation energy to access them, and typical experimental probes to reach them.}
\end{block}
}
%\label{tab: Scales}
%\end{table}

\begin{wrapfigure}[15]{l}{0.35\textwidth}
\centering
\vspace{-0.75cm}
\setlength{\wrapoverhang}{2em} 
\includegraphics[width=0.28\textwidth]{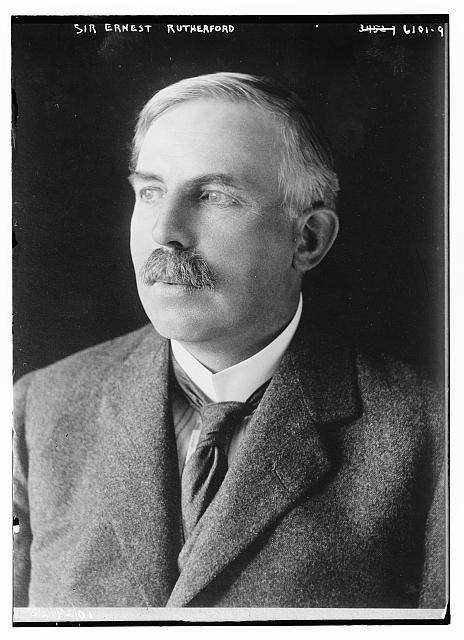}
\vspace{-0.25cm}
\captionsetup{margin=5pt} 
\caption[Photograph of Sir Ernest Rutherford (from \cite{Rutherford_LC_ggbain36570}).]{\textit{I got to the core of the matter.}\\
\footnotesize{Photograph of Sir Ernest Rutherford} \cite{Rutherford_LC_ggbain36570}}
\label{fig: Rutherford}
\end{wrapfigure}

In 1911, Ernest Rutherford, already a Nobel laureate at the time for his work on radioactivity and its classification into $\alpha$-, $\beta$-, and $\gamma$-radiation, performed his now famous gold foil experiment: shooting energetic $\alpha$-particles on a very thin gold foil he expected most of them to pass through unimpeded or only slightly deflected -- ``shooting a 15-inch naval shell on a tissue'' \cite{CERNCourier2011Rutherford}, as he phrased it.
To his surprise, he\footnote{Or more precisely the people actually running the experiment: Rutherford's research assistant Hans Geiger, who later became professor in Tübingen between 1929 and 1936 \cite{GeigerBritannica}, and his undergraduate student Ernest Marsden.} observed some of the $\alpha$-particles being deflected by large angles and even bouncing off the foil \cite{Rutherford1911Scattering}.
This could only be explained by nearly all of the atom's mass being centered in a region orders of magnitude smaller than the atom itself,
leading to a model of the atom with a positively charged, dense nucleus surrounded by orbiting electrons.

\begin{wrapfigure}{r}{0.575\textwidth}
\centering
\vspace{-0.2cm}
\setlength{\wrapoverhang}{2em} 
\includegraphics[width=0.5\textwidth]{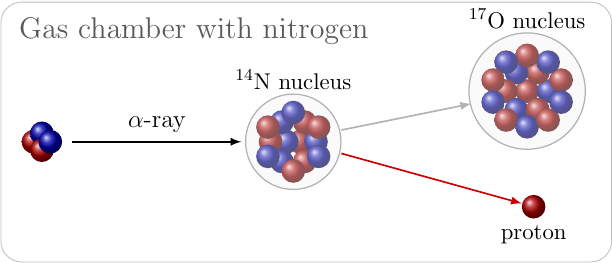}
\captionsetup{margin=5pt}
\caption[The discovery of the proton.]{The discovery of the proton: Rutherford shot $\alpha$-particles on nitrogen gas observing a \textit{proton} escaping from the nucleus while nitrogen turned to oxygen -- proving that larger nuclei are built up from protons and simultaneously living the alchemists' dream of turning one element into another.}
\label{fig: Proton discovery}
\end{wrapfigure}

The lightest of these nuclei, the core of the hydrogen atom, is the \textit{proton}\footnote{Greek for ``first''.}.
In 1919, Rutherford was shooting $\alpha$-particles on nitrogen and observed a proton being knocked out of it, as illustrated in Fig.\,\ref{fig: Proton discovery} \cite{Rutherford01061919}.
This was evidence that larger nuclei consist of them and is widely considered as the discovery of the proton\footnote{In honor of Prout's hypothesis a century earlier, Rutherford initially suggested the name ``prouton'' \cite{Romer1997ProtonOrProuton}.}.

Theoretical physics at the time knew of two fundamental interactions.
Gravity, introduced by Newton in his \textit{Principia}, and electromagnetism, described by James Clerk Maxwell's equations (1865) \cite{Maxwell:1865}.
The discovery of the nucleus made up of \textit{nucleons} -- in 1932 James Chadwick\footnote{Which won Chadwick the 1935 Nobel prize.} discovered the neutron as the second part of the nucleus beside the proton \cite{Chadwick:1932}-- made it evident that there needed to be another force keeping the nucleus together against the electromagnetic repulsion of the protons, and probably also responsible for the nuclear reactions observed in radioactive nuclei. This force was called the \textit{nuclear force}. 
On atomic length-scales, only the electromagnetic force is relevant.
Hence, it was Maxwell's equations Rutherford based the interpretation of his gold foil experiment on; electric repulsion grows with the inverse of the distance, so a smaller scattering center allows for a larger repulsive force.
For an accelerated electron, such as when it is orbiting a nucleus, Maxwell's equations predict the emission of radiation.
This radiation carries away energy resulting in an in-spiral of the electron into the nucleus \cite{LandauLifshitzCTF}.
Carrying out the calculation, it would be expected to happen on a timescale of a fraction of a nanosecond.
This contradicts the observation of stable atoms.
To resolve the problem, physics needed a revolution.

This came in the form of \textit{quantum mechanics}.
Postulating the existence of discrete stable orbits in 1913 -- inspired by earlier work of Max Planck on black body radiation \cite{Planck+1969+107+117}\footnote{For introducing the action quantum $\hbar$ he received the Nobel prize in 1918.} and Einstein's explanation of the photoelectric effect \cite{Einstein:1905photo}\footnote{Which won Einstein the 1921 Nobel prize.}--, Niels Bohr had early success in describing the spectrum of hydrogen, the simplest atom \cite{Bohr:1913}\footnote{Which won Bohr the 1922 Nobel prize and a villa in Copenhagen with lifelong free beer supply sponsored by Carlsberg \cite{BohrCarlsberg}.}.
However, trying to make it work on the next simplest case, the hydrogen molecule ion, H$_2^+$ -- one electron orbiting two proton nuclei --, failed, as Wolfgang Pauli learned in his PhD thesis \cite{PauliPhD}.
The breakthrough came with Werner Heisenberg \cite{Heisenberg1925Umdeutung}\footnote{At the time, Heisenberg was workationing on Helgoland, in the North Sea off the German coast \cite{heisenberg1969quantenmechanik}.} and Erwin Schrödinger \cite{Schrodinger:1926}\footnote{At the time, Schrödinger was workationing in a hut in Arosa, in the Swiss Alps \cite{Moore1989Schrodinger}.} in 1925 finding formulations to describe the dynamics on the atomic scale in an entirely new language.
Objects at these distances do not behave like small billiard balls any more but show wave-like properties such as interference -- as first pointed out by Louis de Broglie in his doctoral work for which he received the Nobel prize in 1929. 
Crucially however, the mathematical formulation of these waves are complex-valued amplitudes in an abstract configuration space and are ``real'' only insofar as they determine the probability of a measurement outcome -- though Epicurus might have been intrigued about this addition of randomness to particle motion.

\begin{wrapfigure}[14]{l}{0.43\textwidth}
\centering
\setlength{\wrapoverhang}{2em} 
\includegraphics[width=0.4\textwidth]{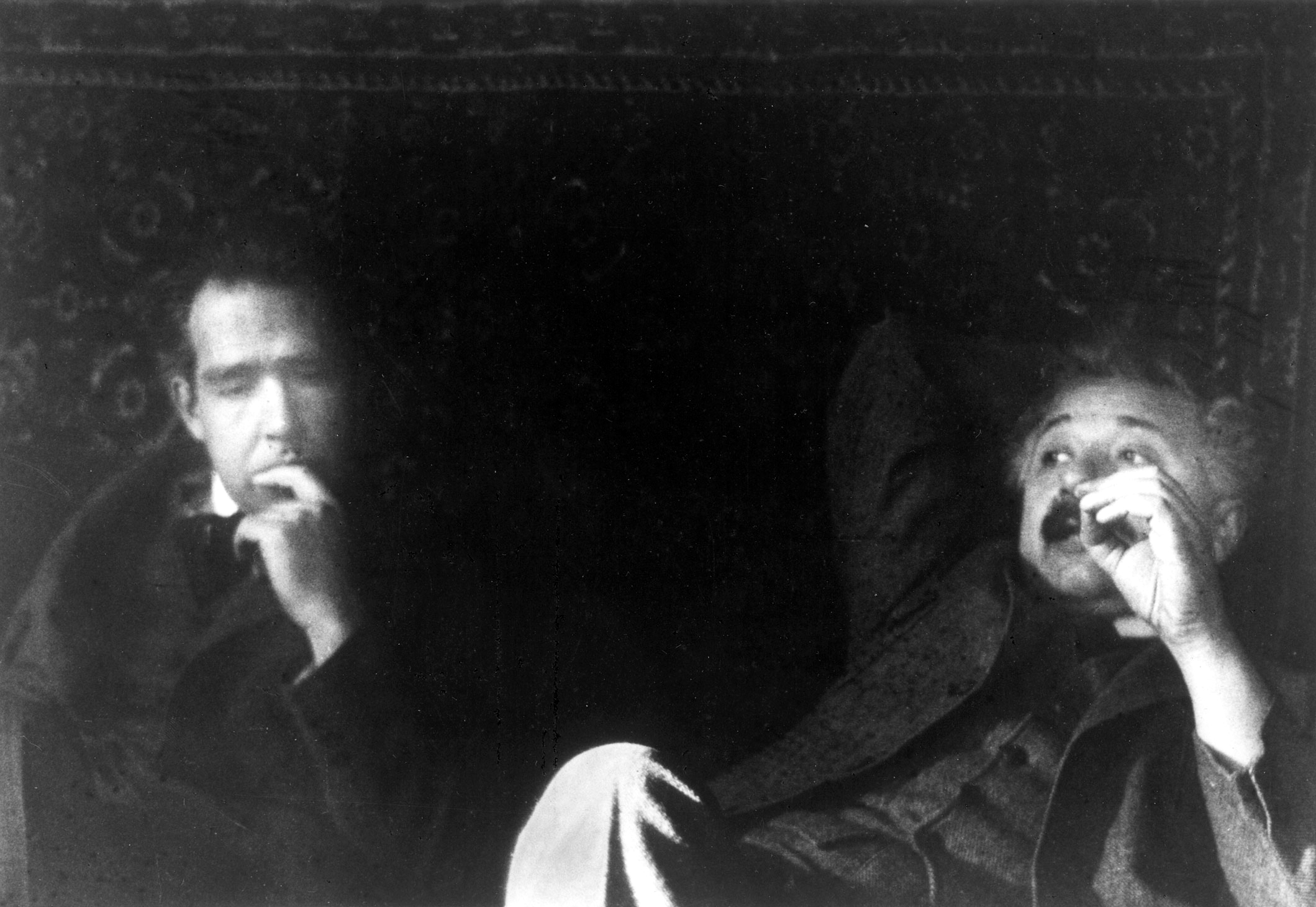}
\captionsetup{margin=5pt} 
\caption[Photograph of Niels Bohr and Albert Einstein taken by Paul Ehrenfest (from \cite{Bohr_Einstein_Ehrenfest_1925}).]{\textit{The world is quantum. We predict measurements. --- Let's imagine being a particle...}\\
\footnotesize{Niels Bohr and Albert Einstein} \cite{Bohr_Einstein_Ehrenfest_1925}}
\label{fig: Bohr and Einstein}
\end{wrapfigure}

The orthodox \textit{Copenhagen interpretation}, very much influenced by Bohr, bluntly states that physics can only make statements about measurements.
This disconnect from interpretability and determinism uneased many physicists, first and foremost Einstein, and is an active area of research to this day \cite{Bricmont2016MakingSenseQM}.
This debate on interpretation however did not stop the adoption of quantum physics in application -- it provided theoretical foundations for chemistry by explaining electron bonds, predicted atomic spectra, and opened the road to the investigation of matter on even shorter length scales.

From the get-go, it was clear that the Schrödinger equation was not the end of the story.
Already in 1905, Einstein had developed his special theory of relativity formulating physics in a way that is independent of the frame of reference, unifying the transformation properties of mechanics and electromagnetism \cite{Einstein:1905SRT}.
By the 1920s it was accepted that any fundamental theory should respect this principle.

A first generalization of the Schrödinger equation was presented by Pauli. 
He incorporated the behavior of particles in magnetic fields, introducing a ``classically non-describable two-valuedness'' to the wave function \cite{Pauli1927MagElectron}, later called \textit{spin} and shown to be associated with a form of intrinsic angular momentum.
Spin has an associated quantum number that takes either integer values $s=0,1,2,\dots$ (for particles called \textit{bosons}) or half-integer values $s=1/2,3/2,\dots$ (for particles called \textit{fermions}); electrons and protons both are spin-$1/2$ fermions.
Pauli's ``exclusion principle'' -- fermions may not agree in all their quantum numbers defining their state and position -- gave the first deep explanation why matter is stable: the atom's electron clouds need to ``stay apart'' \cite{Pauli1925Exclusion}.
This earned him the 1945 Nobel prize.
The spin is proportional to the magnetic moment\footnote{The property that makes magnets magnetic.} of the particles with a proportionality factor $g$, which will play an important role in the further story.

Only two years after the publication of Schrödinger's work, Paul Dirac managed to derive a relativistic quantum theory, showing that magnetism is a relativistic quantum phenomenon and predicting the existence of anti-particles from interpreting the emergent ``extra solutions'' to his equation \cite{Dirac:1928}.
Furthermore, his equation predicted a $g$-factor of two for the electron, which was in excellent agreement with available data.
The founding fathers of quantum mechanics, Heisenberg and Schrödinger jointly with Dirac, were awarded the 1932 and 1933 Nobel prizes, respectively. 

The Dirac equation describes the motion of single relativistic quantum particles.
Observing annihilation of electrons with their equal mass, opposite charge anti-particles (positrons), in cloud chambers called for a more general description that allows for varying particle number.
This is achieved by \textit{field theory}.

A \textit{field} is a quantity that takes a value at every point in space and time \cite{LandauLifshitzCTF} -- filling the void that disturbed Aristotle.
The equations of motion for a field are derived from a scalar function, the \textit{Lagrangian} $\mathcal{L}$, using the \textit{principle of least action} -- maybe \textit{the} central concept  to theoretical physics dating back to the 1600s and a fascinating story of itself \cite{Lanczos:1949}.
To follow the further story it will suffice to keep in mind that ``formulating a theory'' will mean ``finding a Lagrangian $\mathcal{L}$ for the field''.

\vspace{0.3cm}
\begin{block}[type=note]
\textbf{Toy example of a field theory -- Vibrating string $\phi(x,t)$:}
\begin{wrapfigure}{l}{0.45\textwidth}
\includegraphics[width=0.45\textwidth]{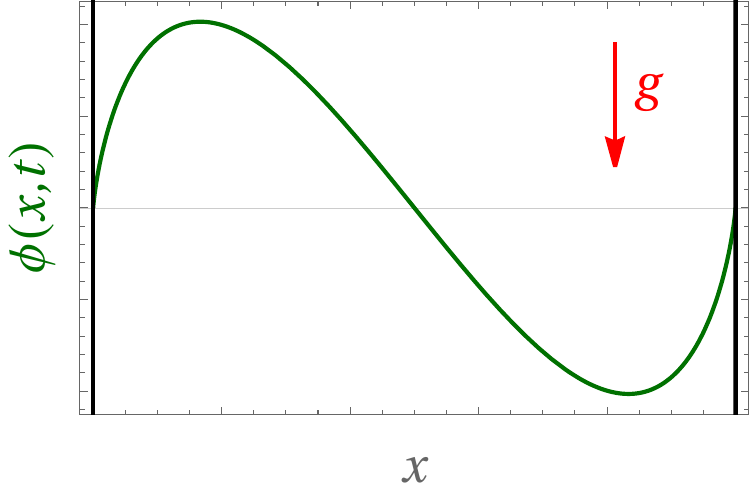}
\caption[Vibrating string fixed between two walls in the gravitational field. Created with \texttt{Mathematica} \cite{Mathematica13}.]{Vibrating string with mass density $\rho$ and string tension $\kappa$ fixed between two walls in the gravitational field.}
\end{wrapfigure}

Lagrangian: 
\begin{align}
\mathcal{L}(\dot{\phi},\phi^\prime,\phi)=\frac{\rho}{2}\left(\dot{\phi}\right)^2-\frac{\kappa}{2}\left(\phi^\prime\right)^2-\rho g\phi.
\end{align}
\vspace{0.4cm}
Equation of motion: 
\begin{align}
&0=\frac{\partial}{\partial t}\frac{\partial \mathcal{L}}{\partial \dot{\phi}}+\frac{\partial}{\partial x}\frac{\partial \mathcal{L}}{\partial \phi^\prime}-\frac{\partial \mathcal{L}}{\partial\phi}
\nonumber
\\
&\phantom{0}=\rho \ddot{\phi}-\kappa\phi^{\prime\prime}+\rho g\,.
\end{align}
\vspace{0.2cm}
\end{block}

\vspace{0.3cm}
Field theory has a long pedigree in classical physics.
The concept started out in electromagnetism, allowing for local interaction and the transmission of force.
For gravity, Newton's original theory was action-at-a-distance, a key critique at the time and disliked by Newton himself \cite{NewtonBentley1693}.
In 1915, Einstein\footnote{Yes, him again, for the fourth time. Later, Einstein also invented a refrigerator \cite{EinsteinSzilard1930}.} succeeded in a field theoretic formulation of gravity with his general theory of relativity \cite{Einstein1915Feldgleichungen,Einstein1916Grundlage}.
The predicted deviations from Newtonian theory stood up to all empirical tests.
With the fundamental classical theories described in terms of fields it seemed natural to bring the concept to the quantum world, introducing \textit{quantum field theory}.
\newpage
\section{Particles in quantum field theory}
\label{sec:Intro3}

Quantum field theory (QFT) \cite{peskin2018introduction} gives an entirely new language for the description of particles:
\begin{block}[type=idea]
\begin{itemize}
\item The fundamental degrees of freedom are quantum fields in space and time.
\item Particles are excitations of these fields.
\end{itemize}
\end{block}
\noindent{}This explains why for example two electrons have the exact same properties anywhere in the universe -- they follow from the underlying universal electron field.

However, it turned out to be anything but straightforward to consistently construct such a theory.
The fields allow for arbitrarily small and arbitrarily large distances -- corresponding to arbitrarily large and arbitrarily small energies, respectively. 
This manifested in all kinds of infinities showing up in the calculations of the pioneers of QFT, most notably Heisenberg and Pauli \cite{heisenberg1929quantendynamik}.
Taming these divergences was up to a new generation of physicists.

In the 1940s and 50s Richard Feynman \cite{Feynman:1949}, Julian Schwinger \cite{Schwinger:1948}, Sin-Itiro Tomonaga \cite{Tomonaga:1946} and Freeman Dyson \cite{Dyson:1949}\footnote{Not the inventor of the bagless vacuum cleaner.} succeeded in formulating \textit{quantum electrodynamics} (QED), a theory describing the interaction of light and matter in terms of spin-1 bosons, the \textit{photons}, and spin-1/2 fermions, respectively \cite{QEDhistory}.
The dimensionless coupling constant $\alpha$, called fine-structure constant has the approximate value of $1/137$.
This small number is well-suited for what is called \textit{perturbation theory}\footnote{Besides, it is also the inverse of the hospital room number Pauli died in \cite{EnzMeyenn1988}.}, a standard tool in all areas of physics.
Typically, equations in physics are such that an exact solution is impossible in practice; the best we can reasonably achieve is controlled approximation. 

In QFT, typical observables are \textit{cross-sections} $\sigma$.
They parameterize the probability for a scattering event with specified incoming and outgoing particles.
Then a perturbative series for $\sigma$ in the small parameter $\alpha$ looks like, e.\,g. to third order,
\begin{align}
\sigma=C_0+\alpha\,C_1+\alpha^2\,C_2+\alpha^3\,C_3+\mathcal{O}\!\left(\alpha^4\right),
\label{eq: perturbative series cross-section}
\end{align}
where $C_{0,1,2,3}$ are coefficients that can be calculated much more easily than the full $\sigma$\footnote{At least until someone figures out a method to do it directly \cite{Arkani-Hamed:2013jha}.}.
The higher orders, that are not calculated, are suppressed by a factor of $\alpha^4$ -- which is denoted by $\mathcal{O}\!\left(\alpha^4\right)$.
This sets the order of magnitude for the error of our approximation, provided the higher-order coefficients do not blow up in size.
The standard naming convention, when speaking of the order to which a process has been calculated is leading order (LO) for the first non-vanishing term, next-to-leading order (NLO) for the first correction, next-to-next-to-leading order (NNLO) for the second correction, and so on.\footnote{Because at some point the number of Ns gets large, it is customary to abbreviate them starting from the third order as N$^3$LO.}

\begin{wrapfigure}[17]{r}{0.35\textwidth}
\centering
\setlength{\wrapoverhang}{2em} 
\includegraphics[width=0.3\textwidth]{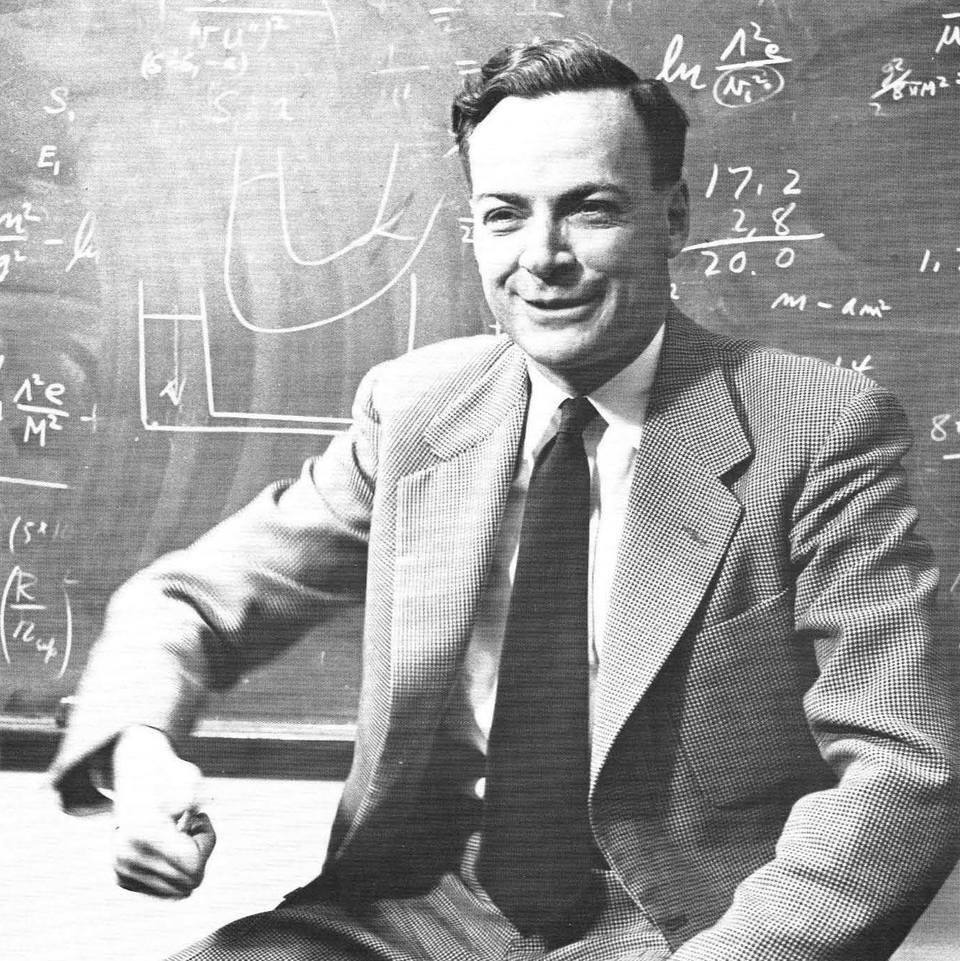}
\captionsetup{margin=5pt} 
\caption[Yearbook picture of Richard Feynman (from \cite{Feynman_1959_Wikimedia}).]{\textit{Enough of the kindergarten stuff, it's serious calculations now.  Somewhere, we need to draw the line.}\\ \footnotesize{Yearbook picture of Richard Feynman} \cite{Feynman_1959_Wikimedia}}
\label{fig: Feynman}
\end{wrapfigure}

Feynman developed a powerful tool for the perturbative calculations in QFT, so-called \textit{Feynman diagrams} \cite{Feynman:1949}, that graphically encode the calculations required for \textit{scattering amplitudes} $\mathcal{M}$.
These are complex-valued probability amplitudes for processes; from their absolute square one can extract the coefficients for the cross-section from eq.\,\eqref{eq: perturbative series cross-section}.
In Feynman diagrams, particles are drawn as lines of different type.
These lines meet at vertices, representing an interaction. 
Each vertex comes with a factor of the coupling constant, in QED the electric charge $e$ proportional to $\sqrt{\alpha}$.
Hence, diagrams with more vertices are relatively suppressed and only contribute to higher orders in the perturbative series.
With a given set of rules for allowed vertices and a given scattering process one can draw all possible diagrams which start out with the initial state particles and produce the final state.

\begin{block}[type=note]
\textbf{Feynman diagrams for electron-positron scattering:}
\begin{equation*}
%\iu\mathcal{M}_{e^+ e^-}=
\underbrace{\vcenter{\hbox{\includegraphics[scale=0.3]{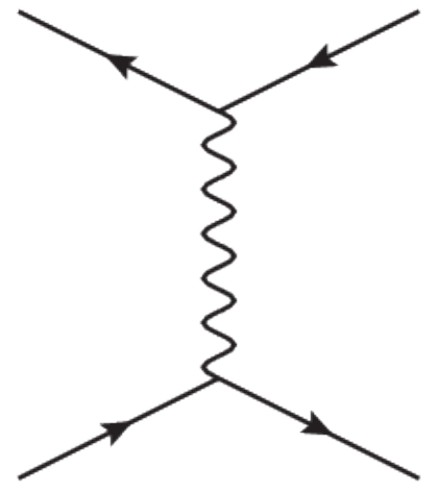}}}+\vcenter{\hbox{\includegraphics[scale=0.3]{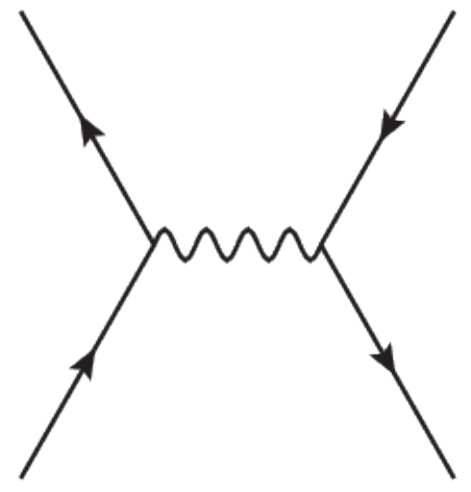}}}}_{\text{\normalsize{leading order contribution}}}+\underbrace{\vcenter{\hbox{\includegraphics[scale=0.3]{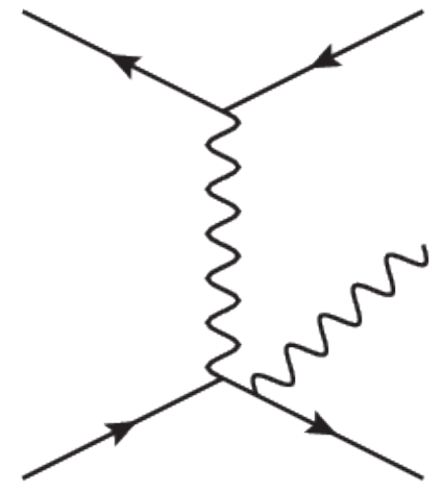}}}\quad+\dots}_{\text{\normalsize{real corrections}}}+\underbrace{\vcenter{\hbox{\includegraphics[scale=0.3]{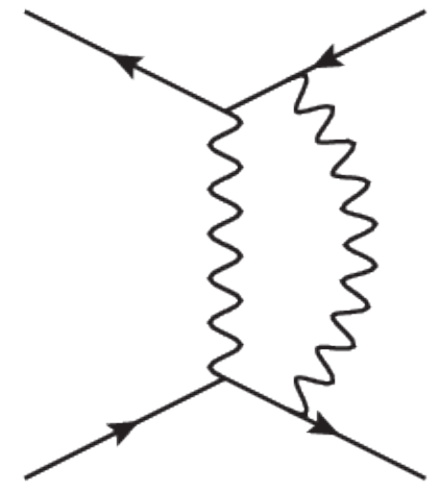}}}+\dots}_{\text{\normalsize{virtual corrections}}}
\end{equation*}
Each diagram translates into a contribution for the scattering amplitude. Incoming particles are on the left, outgoing on the right of the diagrams. Electrons and positrons are represented by straight lines, photons by squiggly lines (propagators); arrows indicate direction of charge flow, not momentum.
Every combination of propagators and electron-photon vertices that connects initial and final states is allowed.
\end{block}

As an example, let us consider the scattering of an electron with a positron.
They are both charged particles, so we expect them to interact via their electromagnetic field.
In the Feynman diagram language, this is reflected by the exchange of a photon.
Importantly, due to the nature of quantum mechanics, any intermediate state that leads to the same outcome can contribute.
Hence, there is a second contribution where the electron and positron annihilate into a photon which subsequently decays into a ``new'' electron-positron pair.
These processes are indistinguishable since we cannot tell the measurement outcome apart -- all electrons are the same, there is no way to ``mark'' them.

There can also be additional interaction.
For one, radiation in the form of extra photons can be emitted.
These are called \textit{real corrections}, since the photons in the final state are, in principle, measurable.
For another, radiation can be emitted and reabsorbed before reaching the final state; an example is the exchange of more than one photon between the electron and positron.
These are called \textit{virtual corrections}, since the extra ``particles'' only appear as (unobservable) intermediate states.
While observed particles always obey their relativistic energy-momentum relation $p^2=m^2$ (with four-momentum $p$ and mass $m$), and at each vertex four-momentum is conserved overall, virtual particles can be \textit{off-shell}, meaning $p^2\neq m^2$.

As stated above, the rules allow for any such correction representable by a valid Feynman diagram.
For real radiation, this means that extra particles can be produced in any way that is in accordance with overall momentum conservation.
To include all of them in a cross-section we need to sum over all these momenta.
This is done by a so-called \textit{phase-space} integral \cite{peskin2018introduction}, formally
\begin{equation}
\sigma_{2\rightarrow n}=\frac{1}{F}\!\int\dx\mathrm{PS}_{n}\,\overline{|\mathcal{M}_{2\rightarrow n}|^2}\,,
\label{eq: cross-section}
\end{equation}
where $\mathrm{PS}_{n}$ denotes the phase-space integral for $n$ outgoing particles, $F$ is the kinematic Møller flux factor, and the bar indicates an average and a summation over incoming and outgoing quantum numbers, in the case of QED only spin.

Also for virtual corrections, any momentum flow in the diagram that is consistent with the momentum conservation rules can contribute. 
For each closed loop, this leaves one momentum -- the one flowing through the loop -- unconstrained, hence it is to be integrated over in a so-called \textit{loop} or \textit{Feynman integral} \cite{peskin2018introduction}, schematically
\begin{equation}
\int\!\frac{\dx^4 l}{(2\pi)^4}\frac{\mathcal{N}(l)}{\mathcal{D}(l)}\,,
\end{equation}
where the numerator $\mathcal{N}$ and denominator $\mathcal{D}$ are determined by the Feynman rules for the loop diagram.
A large part of the thesis will be dedicated to the calculation of these loop and phase-space integrals.

The loop integrals present an immediate obstacle in the formulation of a QFT.
For diagrams where the integrand $\mathcal{N}/\mathcal{D}$ does not decay quickly enough as a function of the loop momentum $l$, the integral diverges due to contributions from large momenta, called an \textit{ultraviolet (UV)} divergence.
Naively, one might lean towards dismissing a theory producing such infinite predictions for correction terms.
But instead of giving up on QFT, physicists tried to find a way to extract meaningful predictions.

The first step is \textit{regularization} that means quantifying the divergences in a well-defined way.
The ad-hoc approach is to introduce a cut-off parameter $\Lambda$ on large loop momenta, resulting in an intermediate expression depending on $\Lambda$.
Then, in a second step, one hopes to find a way to make the final result independent of this arbitrary parameter.
Physically, one can hope that this is possible in a meaningful way, since an observed quantity should only depend on typical energy scales of the experiment, not arbitrarily large ones -- corresponding to arbitrarily small length scales that are expected to be ``blurred out''.
In the 1930s, Heisenberg, Pauli, and Dirac managed the first step, but could not make the second one in a well-defined way.
Dirac was \textit{very dissatisfied with the situation} \cite{Kragh1990Dirac}.

After World War II, a re-birth of particle physics came with the Shelter Island conference, organized by Robert Oppenheimer\footnote{Best known for his contribution to the atomic bomb as scientific head of the Manhattan Project and from the movie \textit{Oppenheimer (2023)}.}, June 2 to 4, 1947 at Ram's Head Inn\footnote{Still in business as of 2025.}, Shelter Island\footnote{A small island in a bay at the northern end of Long Island.}, New York.
There, two important new experimental discoveries were presented.
First, the \textit{Lamb shift}: Two energy levels of the hydrogen atom, which Dirac's theory predicted to be the exact same, turned out to be slightly different.
Second, the electron's $g$ factor was measured to be $g=2.0023\dots$, slightly higher than predicted by Dirac's theory \cite{QEDhistory}.

On the train ride home from the conference back to Cornell University in upstate New York, Hans Bethe managed to calculate the experimentally observed Lamb shift by realizing that it is caused by the influence the radiation field has on the electron mass, and crucially that such an effect is also present when measuring the mass of a free electron \cite{Bethe:1947}.
The regularized divergences in the calculations are the same for both cases, the final expression for the Lamb shift is in terms of the physically observed mass, a finite result, free of divergences.

Soon after, applying a similar procedure for the electron charge, Schwinger calculated\footnote{Not using Feynman diagrams, as Schwinger deemed them a gross oversimplification of the pretty algebra.} the anomalous magnetic moment of the electron to be
\begin{align}
g=2+\frac{\alpha}{\pi}=2.0023\dots
\end{align}
in agreement with the experimental value for $g$ reported on Shelter Island \cite{Schwinger:1948}.

This procedure of absorbing the divergences into the experimentally determined quantities became known as \textit{renormalization}, a method independently developed by Tomonaga in post-war Japan \cite{Tomonaga:1946}.
Using the diagrammatic method Feynman\footnote{Who loved playing the bongos and does so in the movie \textit{Oppenheimer (2023)}.} introduced in 1948, Dyson proved that this program to produce divergence-free predictions for observables in QED works order-by-order in perturbation theory \cite{Dyson:1949}.
This firmly established QED as a valid theory and resulted in the 1965 Nobel prize for Feynman, Schwinger, and Tomonaga.

The success of QED in predicting cross-sections made it a model for further theory development.
For the following story, the concept of \textit{gauge theories} -- for which QED is the foundational example -- is essential.
Already back in 1926, Vladimir Fock pointed out that the Schrödinger equation for an electron in an electromagnetic field,
\begin{equation}
\frac{1}{\iu}\frac{\partial \psi}{\partial t}-\frac{1}{2 m_e}(\vec{\nabla} +\iu e\vec{A})^2\,\psi-e\phi\psi=0\,,
\label{eq: Schrodinger emag}
\end{equation} 
is invariant under a group of local transformations, parameterized by the function $\omega(x)$ \cite{Fock1926InvariantForm}.
Concretely, transforming the electromagnetic potentials $\vec{A}$, $\phi$, and the electron spinor $\psi$ according to
\begin{align}
\vec{A}^\prime(x)=\vec{A}(x)+\vec{\nabla}\omega(x)\,,\,\,
\phi^\prime(x)=\phi(x)-\frac{\partial\omega(x)}{\partial t}\,,\,\,
\psi^\prime(x)=\ee^{-\iu e\omega(x)}\psi(x)\,,
\label{eq: U1 gauge symmetry}
\end{align}
the primed quantities describe the exact same physical situation, since when one of them solves eq.\,\eqref{eq: Schrodinger emag} so does the other.
For this property, Hermann Weyl coined the term \textit{gauge invariance} \cite{Weyl1929Elektron}.
As known from the work of Emmy Noether in 1918 \cite{Noether1918}, this abstract symmetry has directly measurable consequences: every continuous symmetry has an associated conserved quantity to it, in this particular case, the electric charge.

As mentioned earlier, field theories are formulated in terms of their Lagrangian.
Now turning the logic upside down and promoting the observed invariance of the equations of motions to a fundamental symmetry property the Lagrangian needs to satisfy, one finds that the Lagrangian for the electromagnetic field is uniquely determined by the transformation law eq.\,\eqref{eq: U1 gauge symmetry}.
Since for any $x$, the $\ee^{-\iu e\omega(x)}$ form the abelian group $\mathrm{U}(1)$, electromagnetism is termed an \textit{abelian gauge theory}.

Gauge theories play a special role in QFT due to their \textit{renormalizability}, meaning that the program of absorbing infinities into the parameters of the theory works with a finite number of parameters and at all energy scales.\footnote{This was however only figured out in 1971.}
All theories that are part of today's \textit{Standard Model} are such gauge theories, but with different gauge groups.
Essentially, the requirement of relativistically covariant, renormalizable QFTs served as a roadmap for the search for possible consistent theories we will look at in the following.

We left the storyline of the composition of matter with the discovery of the nucleus and its constituents, protons and neutrons.
When the neutron's $g$-factor was measured in 1939 by Luis Alvarez and Felix Bloch\footnote{Nobel laureates in 1968 and 1952, respectively.} \cite{Alvarez:1940}, it came out as
\begin{equation}
g_\text{Neutron}\approx-3.826\,,
\end{equation}
far away from the value expected for point-like Dirac particles.
This was an early hint at a substructure of nucleons.
In the 1930s and 40s, the nuclear force was studied in terms of non-relativistic potential models that allowed for the description of a broad range of nuclear reactions from $\alpha$-decay \cite{Gamow1928AlphaDecay} to the luminosity of the sun \cite{Bethe:1939}\footnote{For this work, Hans Bethe received the Nobel prize in 1967.}.

Regarding the understanding of the nuclear force in terms of QFT, two theories were developed in the mid 1930s, that captured different phenomena.
For one, Enrico Fermi introduced the QFT of \textit{weak nuclear force} to explain the $\beta$ decay \cite{Fermi1934Beta}.
For another, Hideki Yukawa proposed that the force that keeps the nucleus together, called the \textit{strong nuclear force}, is due to the exchange of \textit{mesons} \cite{Yukawa:1935}\footnote{From the Greek word for middle, since their mass $m_\pi$, which Yukawa predicted from the range of the nuclear force sits between the proton and electron mass.}.
In contrast to electromagnetism, the exchange boson of Yukawa's theory is massive.
This leads to an exponential decay of the force proportional to $\ee^{-m_\pi r}$, hence making it short ranged.

Both Fermi's and Yukawa's theories were successful in their respective domains but neither was a renormalizable QFT, hence their applicability was confined to restricted energy scales.
Especially, the presence of massive force carriers -- required to explain the short-rangedness of the nuclear forces -- was troublesome, since it was well established that the bosons of any gauge theory were necessarily massless.

In the 1920s, the list of known ``elementary'' particles was very concise, only the proton and the electron.
When Dirac predicted anti-particles with opposite charge \cite{Dirac:1931}, there was initial hope that one might be able to interpret the proton as the electron's anti-particle.
However, this idea needed to be abandoned quickly due to the huge mass difference between the particles.
In 1932, the actual anti-particle to the electron, the \textit{positron} was discovered in cosmic rays at the same mass as the electron as predicted by Dirac's equation \cite{Anderson:1933}.

\begin{wrapfigure}[28]{r}{0.5\textwidth}
\centering
\vspace{0.2cm}
\setlength{\wrapoverhang}{5em} 
\includegraphics[width=0.4\textwidth]{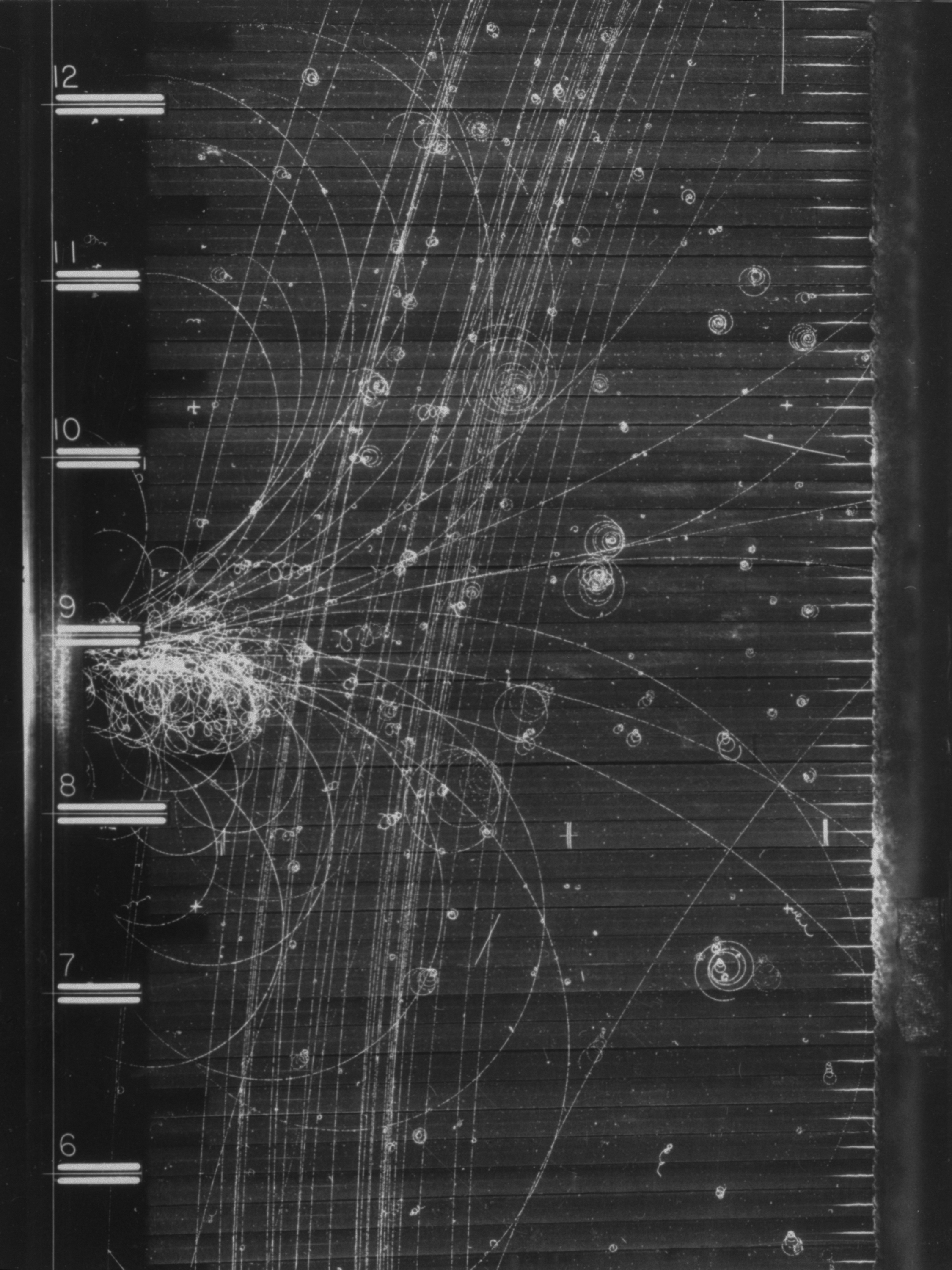}
\caption[Tracks in a bubble chamber (from \cite{BerkeleyLab1960CosmicRay}).]{Tracks of particles in a bubble chamber. Electromagnetic fields deflect the particles on characteristic tracks depending on the specific charge-to-mass ratio. 
Tracks can start, e.\,g., when high energy photons create electron-positron pairs.
Tracks can end when particles decay into uncharged particles. (from \cite{BerkeleyLab1960CosmicRay})}
\label{fig: Bubble chamber}
\end{wrapfigure}

When in the same year the neutron was discovered and found to have nearly the same mass as the proton \cite{Chadwick:1932}, Heisenberg immediately tried to stem the tide of an ever growing list of ``elementary'' particles:
Can proton and neutron be interpreted as being two different states of the same underlying particle, the nucleon?
Borrowing from the concept of spin, he introduced the \textit{isospin} \cite{Heisenberg1932Isospin},
an abstract quantum number that made the proton and neutron the isospin-up and isospin-down states of the nucleon, respectively.
In this picture, the strong force acts symmetrically on both components of the isospin vector, mathematically an $\mathrm{SU}(2)$-invariance. 

However, the list stayed clean only for so long.
In 1936, a new particle, the muon \cite{Muon:1937} was identified in cosmic rays.
Following observations of cosmic rays through cloud and later also bubble chambers and early experiments at accelerators in the 50s like at the Berkeley Bevatron \cite{lofgren1956bevatron}\footnote{The first machine that could reach energies of a billion electronvolts, abbreviated as BeV at the time. Today the same unit is commonly called GeV.}, produced an ever growing number of particles \cite{BevatronKmesons,BevatronStrangeMesons,BevatronTauMesons,Antiproton:1955,BevatronAntineutron}, including in 1947 the fitting candidate for the boson predicted by Yukawa, the pion \cite{Lattes1947Pion}.
While the situation is pretty straightforward with cosmic rays showing up as tracks in a detector like a bubble chamber (see Fig.\,\ref{fig: Bubble chamber}), at this point a short comment is in order to clarify what it means to ``find'' a particle at an accelerator.

As the most straightforward example, let us search for new particles interacting with the electromagnetic field.
If we collide electrons and positrons with sufficient energy, they can annihilate to form an intermediate virtual photon that can subsequently decay into a particle-antiparticle pair of new type.
For that to go along with energy conservation, the center-of-mass-energy of the  collision must be at least twice the mass of the new particles -- that's Einstein's $E=mc^2$ in collider practice.\footnote{You will not see much of $c$ or of the Planck action quantum $\hbar$ in the following, since we use so-called \textit{natural units} throughout that set $c=\hbar=1$.}
These new particles then show up as tracks in the detector from which we can deduce their properties.

\begin{block}[type=note]
\textbf{Example of particle creation at a collider:}

\flushleft
\begin{minipage}{0.49\textwidth}
If electrons and positrons are collided with sufficient center-of-mass-energy, they can produce a muon-antimuon ($\mu^-\mu^+$) pair that shows up in the detector as a 200 times heavier version of the electron.
\end{minipage}
\begin{minipage}{0.49\textwidth}
\centering
\includegraphics[width=\linewidth]{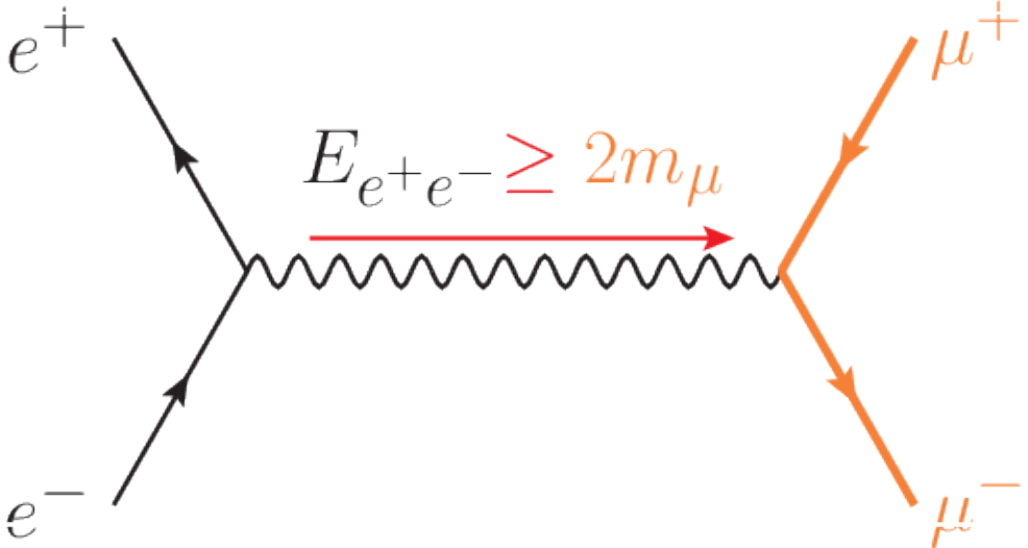}
\end{minipage}

\vspace{0.2cm}
The muon was first discovered in cosmic rays in 1936 \cite{Muon:1937}.
The honor of being the first ``new'' particle to be found at a collider goes to the antiproton produced in proton-proton collisions at the Bevatron in 1955 \cite{Antiproton:1955}.
\end{block}
However, there are also particles that decay too quickly to make it to the detector. 
In the following, we will discuss how we can nevertheless infer their existence.

In the language of Feynman diagrams each line between two vertices, called a \textit{propagator}, contributes to the scattering amplitude with a factor proportional to $1/(p^2-m^2)$, where $p$ is the relativistic four-momentum and $m$ the mass of the particle traveling along the line.
Now let us assume we have a new particle of mass $M$ in our theory which interacts with charged particles.
When we look at so-called $s$-channel contributions depicted in Fig.\,\ref{fig: s-channel scattering}, the first diagram gives a factor of
\begin{align}
\frac{1}{(p_1+p_2)^2-M^2}\,.
\end{align}
We see that whenever the sum of the incoming momenta is large enough and approaches the mass of the particle, the propagator becomes infinitely large.
This is, of course, unphysical.

\begin{figure}[h]
\vspace{-0.2cm}
\begin{equation}
\vcenter{\hbox{\includegraphics[width=0.25\textwidth]{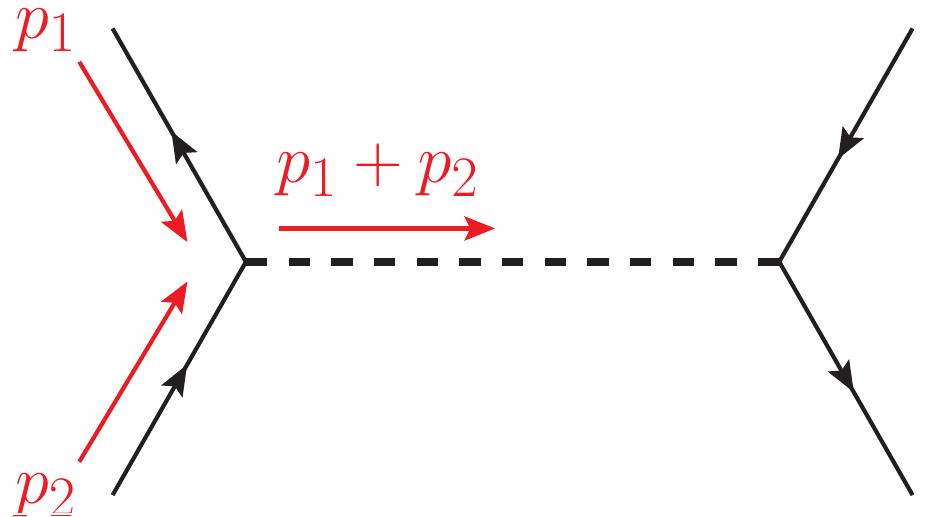}}}
+\vcenter{\hbox{\includegraphics[width=0.25\textwidth]{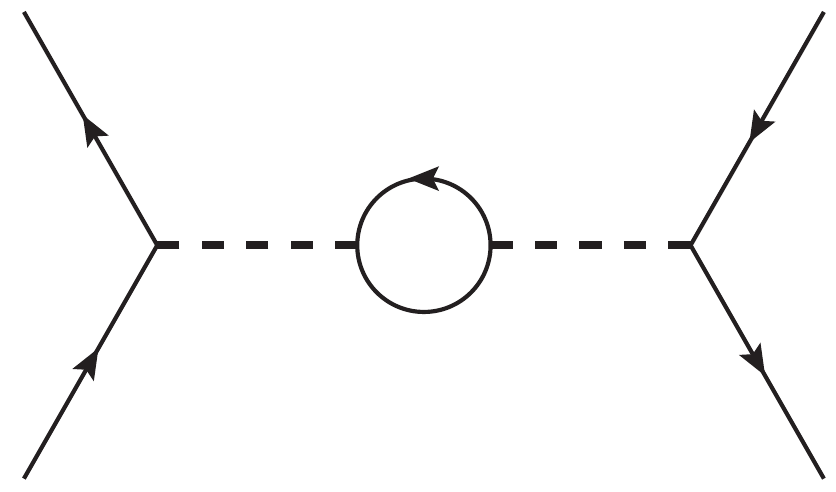}}}+
\vcenter{\hbox{\includegraphics[width=0.25\textwidth]{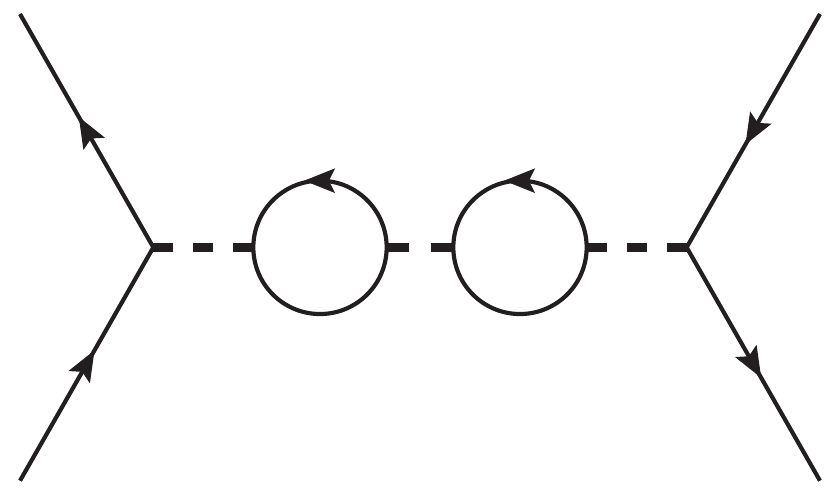}}}+\dots
\end{equation}
\caption[$s$-channel scattering Feynman diagrams. Created with \texttt{JaxoDraw} \cite{Binosi:2003}.]{$s$-channel scattering with a hypothetical new massive particle in the intermediate state represented by a dashed line. Loop corrections produce an imaginary part in the propagator that is responsible for the Breit-Wigner resonance peak at the mass of the dashed particle.}
\label{fig: s-channel scattering}
\end{figure}

However, taking into account the $s$-channel loop corrections, they add up in a way that they for one, shift the mass pole to a new value $M_\text{ren}$ (as was already mentioned when we talked about renormalization), and for another they introduce an imaginary part\footnote{Imaginary parts of topologically more involved loop integrals will be in the focus of publications \ref{pub:2} and \ref{pub:3}.} proportional to the physical \textit{decay width} $\Gamma$ of the particle\footnote{$1/\Gamma$ is the average lifetime of the particle before it decays.}.
So the total amplitude, taking all these infinitely many contributions into account, gets a factor
\begin{align}
\iu\mathcal{M}\sim\frac{1}{(p_1+p_2)^2-M_\text{ren}^2-\iu M_\text{ren}\Gamma}\,.
\end{align}
\begin{figure}
\centering
\begin{subfigure}[t]{0.45\textwidth}
\centering
\vspace{-0.2cm}
\includegraphics[width=0.95\linewidth]{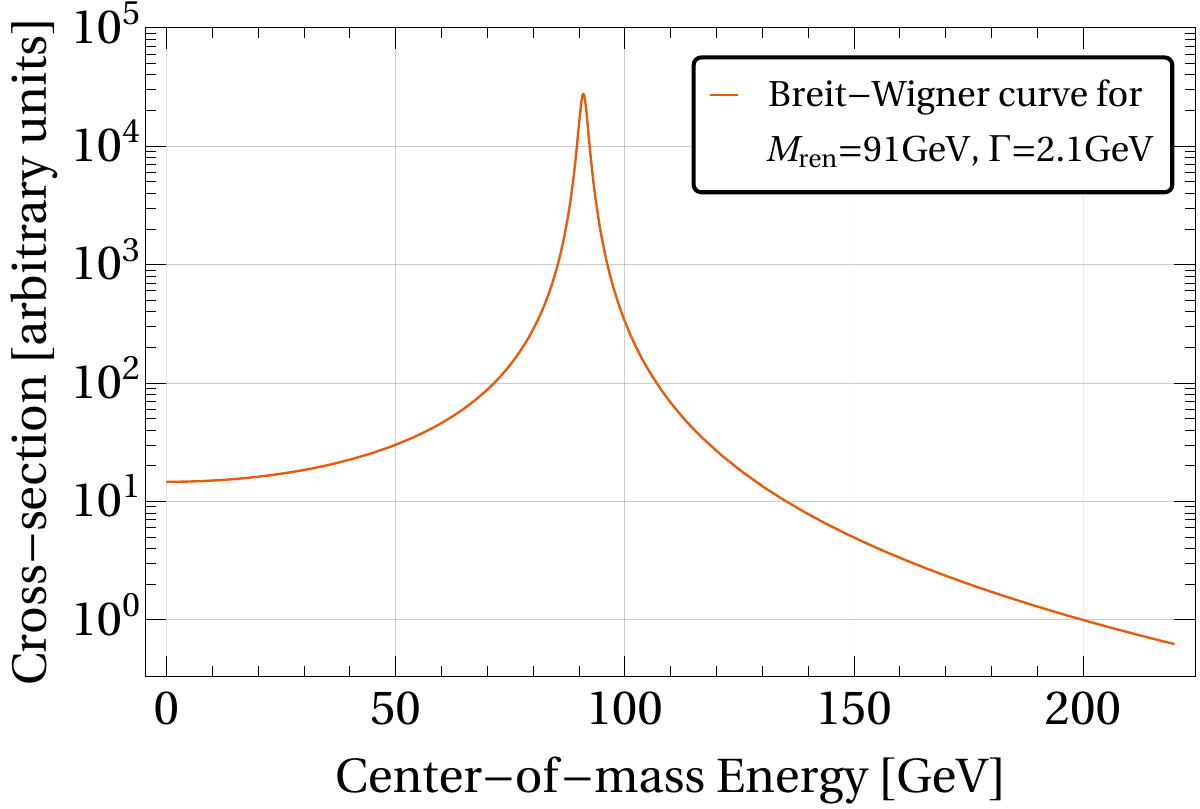}
\caption{Breit-Wigner curve from eq.\,\eqref{eq: Breit-Wigner} for a mass {$M_\text{ren}=91$\,GeV} and decay width {$\Gamma=2.1$\,GeV} plotted on a logarithmic scale.}
\label{fig: Breit-Wigner}
\end{subfigure}
\hspace{3em}
\centering
\begin{subfigure}[t]{0.45\textwidth}
\vspace{-0.4cm}
\includegraphics[width=0.95\linewidth]{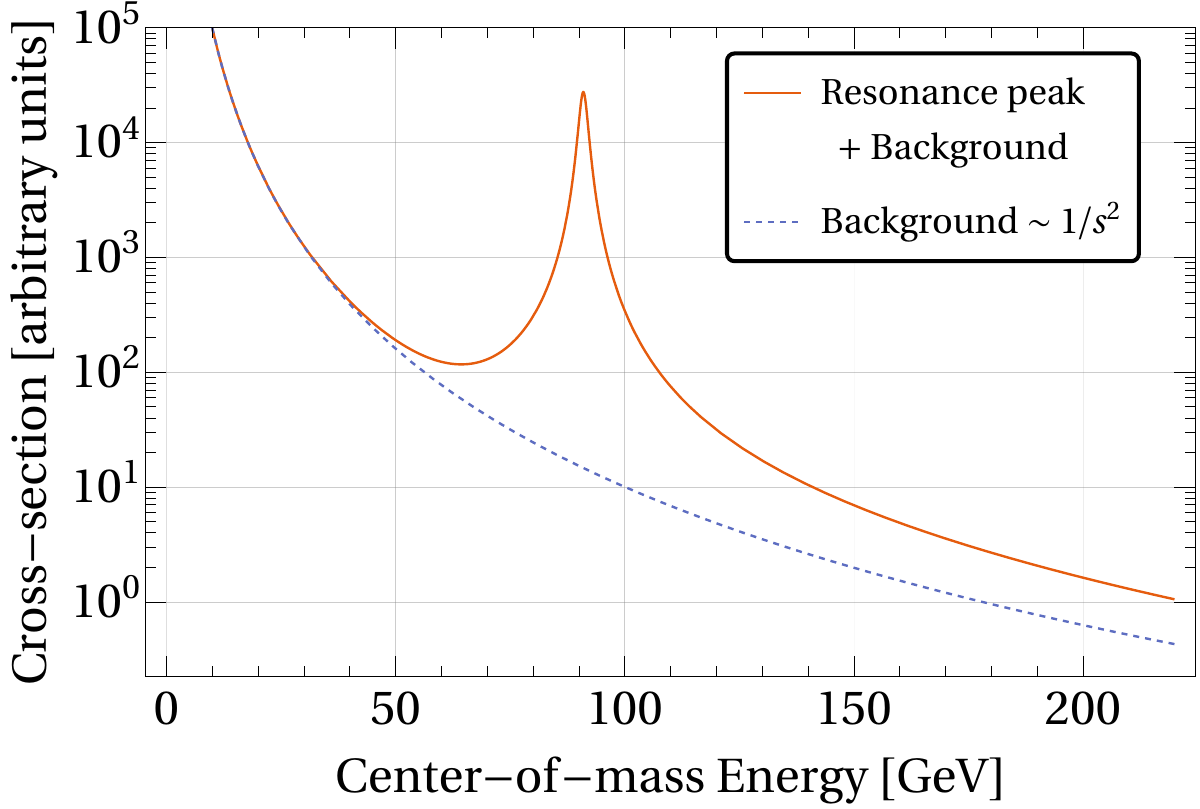}
\vspace{-0.4cm}
\caption{Resonance curve from the left panel with an added background from massless particles in the $s$-channel proportional to $1/(p_1+p_2)^4$.}
\label{fig: Resonance}
\end{subfigure}
\vspace{-0.2cm}
\caption[Plots of Breit-Wigner curves. Created with Mathematica \cite{Mathematica13}.]{The approximate expected cross-section for a hypothetical massive particle with a specific mass and decay width. The left panel shows the pure contribution due to the new massive particle, the right shows it on top of the background from massless (or much lighter) particles, e.g. from photons -- this signature is what we need to look for in our detector data.}
\label{fig: New Particle}
\end{figure}
To obtain the cross-section, per eq.\,\eqref{eq: cross-section}, we have to take the absolute square of the expression.
This results in a behavior of the cross-section as
\begin{align}
\sigma\sim\frac{1}{\left((p_1+p_2)^2-M_\text{ren}^2\right)^2+ (M_\text{ren}\Gamma)^2}\,,
\label{eq: Breit-Wigner}
\end{align}
known as the \textit{Breit-Wigner resonance curve} plotted in Fig.\,\eqref{fig: Breit-Wigner}.
If we combine this with the known background of the established theory with massless (or much lighter) exchange particles, like the photon in QED, the hypothesized massive particle will cause a characteristic resonance peak, depicted in Fig.\,\eqref{fig: Resonance} for some example values for mass and decay width.
In reality, the prominence of the peak will depend on relative coupling strengths, and the actual values for $M_\text{ren}$ and $\Gamma$.\footnote{Also, we ignored potential interference contributions of the massive particle with the ``background'' on matrix element level. These can alter the shape of resonance peaks.}

What we can see clearly from the resonance curve in Fig.\,\eqref{fig: Resonance} is that the peak can be orders of magnitude higher than the background, hence we have a clear signal, robust to experimental noise.
However, to observe a significant excess we need an accelerator capable of reaching center-of-mass energies $E_\text{CMS}=\sqrt{(p_1+p_2)^2}$ close to the mass of the new particle.
Furthermore, from Fig.\,\eqref{fig: Resonance} we see that the observed particle mass of the resonance is the renormalized mass $M_\text{ren}$ that includes the interaction of the particle with the field rather than the mass parameter $M$ we used in the original Lagrangian of the theory.

As already mentioned above, a lot of these particle resonances were identified by experimentalists in the 40s and 50s contributing to an ever growing \textit{particle zoo} \cite{Brown_Dresden_Hoddeson_1989}.
It was time for theoretical physics to restore order.
Most of the newly found particles fell into two categories.
For one \textit{baryons}, these showed properties similar to protons and neutrons, examples are the $\Lambda$, $\Sigma$, $\Xi$, $\Omega$, and $\Delta$ resonances \cite{Baryons:2010}.
For another \textit{mesons}, these showed properties similar to the pion $\pi$ of Yukawa theory, examples are the $K$, $\eta$, $\rho$, and $\omega$ resonances \cite{ParticleDataGroup:2024cfk}.

Experimentalists used cross-sections, decay products, and angular distributions to determine the quantum numbers associated with each particle\footnote{A ``quantum number'' in this context can be understood as a label that remains unchanged under the strong interaction because it is tied to a symmetry of the theory (remember the Noether theorem).}.
Extending the isospin concept of Heisenberg, a new quantum number, called \textit{strangeness}, was introduced by Murray Gell-Mann \cite{GellMann:1953}\footnote{At this point, we see that the tradition of using Greek terminology is grossly broken -- \textit{xenicity} (from the Greek xenos for stranger) would have been a dignified option. Strangely, Murray Gell-Mann had a deep interest in language and even published about linguistic evolution \cite{GellMann:2011}.}.
Isospin and strangeness allowed for a combination into a single symmetry group, called $\mathrm{SU}(3)$ flavor\footnote{$\mathrm{SU}(3)$ is the group formed by the three-by-three unitary complex matrices with unit determinant you can think of as rotations acting on abstract vector spaces of particle labels.}, with the strong force assumed to act the same on each representation vector of this symmetry group.
Each symmetry group allows only for specific dimensions for these representations, which physically correspond to a number of particles that have similar properties. For $\mathrm{SU}(3)$, the fundamental representation is three-dimensional.
Mathematically, this can be used to build up other so-called \textit{irreducible} representations with defined transformation properties. For $\mathrm{SU}(3)$, the first few have dimensions $1$ (singlet), $6$ (sextet), $8$ (octet), and $10$ (decuplet).
If all this group theory sounds Greek\footnote{Well actually, the naming of the multiplets derives from Latin.} to you, do not worry too much and go along with the vibe of smart physicists in the 60s trying to organize particles into lists with common features using abstract symmetry principles.
 
In what they called the \textit{eightfold way} (1961), Gell-Mann \cite{GellMann:1962} and Yuval Ne'eman \cite{Neeman:1961jhl} identified the spin-0 mesons and spin-1/2 baryons with two octets, the spin-3/2 baryons with a decuplet.
For one, this explained the similar masses within each multiplet, especially among the baryons.
For another, it gave a prediction for ``missing'' particles in each multiplet, i.e. resonances that were not observed yet but were required to fill up the multiplet.
For example, at the time, the baryon decuplet was missing a particle with strangeness $-3$.
And indeed, in 1964, the corresponding $\Omega^-$ was observed with the predicted properties \cite{OmegaMinus:1964}.

An immediate question that was raised by this symmetry scheme was the following: Why are there no observed resonances corresponding to the fundamental 3-dimensional representation?
These hypothetical particles would have non-integer electric charge of $-1/3$ and $2/3$.
In 1964, Gell-Mann \cite{Gell-Mann:1964ewy} and George Zweig \cite{Zweig:1964ruk}, a PhD student of Feynman, proposed that the baryons and mesons are bound states of the unobserved triplet ``particles'' and called them \textit{quarks}\footnote{A term Gell-Mann took from James Joyce's novel \textit{
Finnegans Wake}, where the phrase \textit{Three quarks for Muster Mark} occurs; apparently Joyce was thinking of the German dairy product. So in James Joyce's work, quark should rhyme with Mark, however Gell-Mann originally intended a ``quork'' pronunciation (rhyming with pork) by his own account \cite{MerriamWebsterQuark2025}. While reading, pronounce in your head as you please.}.
Following the names of the isospin and strangeness quantum numbers, the three quarks got the names \textit{up} (u), \textit{down} (d), and \textit{strange} (s).
This fit well into the group theoretical structure and a non-zero mass attributed to the strange quark gave a natural explanation why the flavor symmetry is only approximate.
One problem with the quark model was that apparently the $\Omega^-$ was a bound state of three strange quarks in the same state -- violating the Pauli principle.

At this point the quarks were more of an accounting device and their reality outside group theory was doubtful -- a situation reminiscent of the theoretical state of atoms in the 1900s.
There were strong doubts that the strong interaction could be described by a gauge theory since it was known already to Pauli that exchange bosons in gauge theories are necessarily massless and hence necessarily long-ranged, in sharp contrast to the properties of the strong force -- that keeps nucleons together against their electromagnetic repulsion but is irrelevant already on atomic scales \cite{Leutwyler:2012ax}.
In 1954, Chen Ning Yang\footnote{Nobel laureate in 1957 together with Tsung Dao Lee for the investigation of parity of elementary particles. Both lived to remarkable ages, 103 and 97, respectively.} and Robert Mills nevertheless showed that isospin can be turned into a local gauge symmetry and showcased how to formulate gauge theories for non-abelian groups \cite{Yang:1954}.
In their honor, such theories are called \textit{Yang-Mills theories}.
A lot of different theoretical alternatives were developed, going by names such as current algebra \cite{GellMann:1962,GellMann:1964curr}\footnote{Maybe it's currents that are fundamental and not particles?}, $S$-matrix theory \cite{Heisenberg1943_SMatrix,Dyson:1949Smat,Mandelstam:1958Smat,regge1959introduction,Chew:1962Smat}\footnote{Maybe we need to abandon the underlying Lagrangian?}, and string theory \cite{Veneziano1968_Amplitude,Koba:1969,Paton:1969,Viraso:1969,Goto:1971ce,Ramond:1971gb,Neveu:1971rx,Goddard:1972iy}\footnote{Maybe hadrons have string-like components, rather than point-like? Later modern string theory developed as a distant cousin in the immodest hunt for a \textit{theory of everything} (or nothing) \cite{Schwarz:2007yc,Ellis:1986mm}.}.

A breakthrough came with the discovery of the \textit{Higgs mechanism} by Robert Brout, François Englert and Peter Higgs \cite{Higgs:1964ia,Englert:1964}\footnote{Englert and Higgs received the Nobel prize in 2012, after the Higgs boson, predicted by their theory, was found at the LHC.}.
By so-called \textit{spontaneous symmetry breaking} -- an underlying symmetric potential can lead to a non-symmetric ground state\footnote{Imagine trying to balance a pencil on its tip on a table. All forces acting on the pencil are rotationally symmetric about the tip of the pencil but the tiniest perturbation will cause its state to spontaneously change into a non-symmetric one, ``tipping over'' as engineers call it.} -- massless gauge bosons in Yang-Mills theories can acquire a mass. 
Based on this idea, the \textit{electroweak} theory was developed \cite{Lee:1956,Wu:1957,Feynman:1958Fermi,Glashow:1961,Weinberg:1967,Salam:1968rm}, generalizing QED by unifying it with the weak interaction.
Where Fermi's theory had point-like interactions, the new theory predicted three new massive cousins of the photon, the $Z$ and the $W^\pm$ bosons.
This theory is a QFT based on an $\mathrm{SU}(2)\times\mathrm{U}(1)$ gauge group, where $\mathrm{SU}(2)$ is the group of two-by-two unitary complex matrices with unit determinant.
This group is non-abelian, meaning its members do not commute, in contrast to the abelian $\mathrm{U}(1)$ of electromagnetism -- hence there was the first successful Yang-Mills theory.
In 1971, during his PhD, Gerard 't Hooft with his supervisor Martinus Veltman \cite{Hooft:1971,tHooft:1972} showed that such theories are indeed renormalizable, for which they were awarded the 1999 Nobel prize.
The door was open for a QFT of the strong force.

\section{The world inside the proton -- Quantum chromodynamics}
\label{sec:Intro4}
From electron-proton scattering experiments by Robert Hofstadter at Stanford in the 1950s, it was  established that the proton is a finite-sized object with its charge distributed within a radius of the order of $1\,\text{fm}=10^{-15}\,\text{m}$ \cite{Hofstadter:1956}.
This program won Hofstadter the 1961 Nobel prize and demonstrated that high-energy electrons can serve as precision probes of the internal structure of the proton.

\begin{wrapfigure}[32]{l}{0.3\textwidth}
\centering
\includegraphics[width=0.2\textwidth]{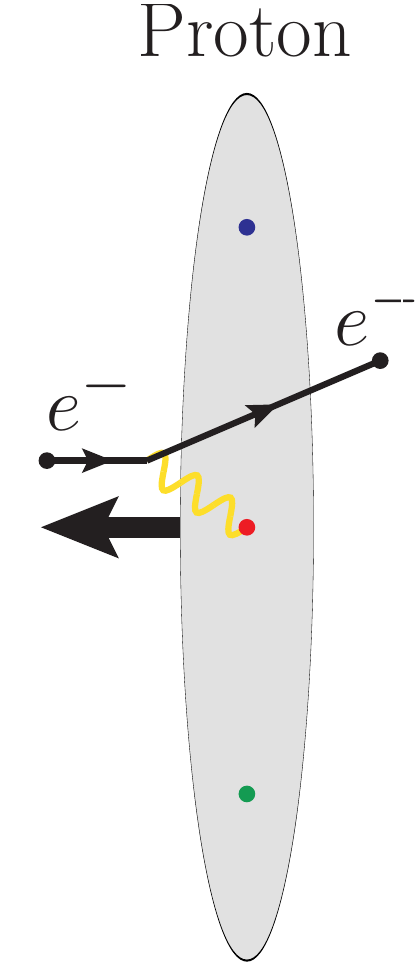}
\caption[Sketch of deep-inelastic scattering at SLAC in the parton model. Created with \texttt{JaxoDraw} \cite{Binosi:2003}.]{Deep-inelastic scattering at SLAC in the center-of-mass frame of the collision according to the parton model. The electron hits a single parton, the relativistic length-contraction makes the proton pancake-shaped (longitudinal length contraction is drawn approximately to scale for SLAC energies).}
\label{fig:SLAC_DIS}
\end{wrapfigure}
Starting in 1968, the MIT-SLAC collaboration led by Jerome Friedman, Henry Kendall, and Richard Taylor conducted \textit{deep inelastic scattering experiments} at the Stanford Linear Accelerator (SLAC)\footnote{For which they were awarded the Nobel prize in 1990.}.
With an electron beam hitting a fixed proton target, they could get to center-of-mass energies of around 6\,GeV, which corresponds to distance scales well smaller than the proton.
Hence, at this energy, the interior of the proton could be resolved for the first time.
The proton breaks up in the process resulting in hadronic debris that could not be detected at SLAC, but observing the scattered electron allows to reconstruct what happened in the collision process using momentum conservation.
When the results were published in 1969 \cite{Breidenbach:1969kd,MITSLAC:1969}, they were -- to the surprise of many -- compatible with scattering at point-like constituents within the proton.
This was inferred from a prediction made by James Bjorken that at high energies the structure functions parameterizing the cross-section did not depend separately on energy and four-momentum transfer but only their ratio, termed \textit{Bjorken}-$x$ or $x_\text{B}$ for short \cite{Bjorken:1969}.

Feynman gave a compelling physical picture to describe the observation \cite{Feynman:1969}.
Assuming the proton is a bound state of point-like constituents, called \textit{partons}, these interact on time scales corresponding to the proton mass in the proton's rest frame.
In the center-of-mass frame of the collision, the proton will move with highly relativistic velocity and exhibit time dilation.\footnote{A consequence of Einstein's special theory of relativity.}
Hence, the interaction between the partons happens on time-scales much longer than the time it takes the electron to traverse the proton.
Therefore, it will interact with constituents that behave, at least in first approximation, as free particles.
Additionally, length contraction\footnote{Another consequence of Einstein's special theory of relativity.} makes the proton pancake-shaped in this frame, depicted in figure \ref{fig:SLAC_DIS}, which makes multiple re-scatterings unlikely.
The probability for the scattering now depends on how likely it is for the electron to interact with a parton of specific momentum.
Since the partons are inside the fast-moving proton, the much slower internal transverse motion can be neglected, so only the \textit{collinear} momentum fraction $\xi$ is relevant.
Calling this probability to find a parton $a$ of momentum fraction $\xi$ of the parent proton the \textit{parton distribution function} $f_{a/p}(\xi)$ (PDF), the cross-section in the so-called \textit{parton model} becomes
\begin{align}
\dx\sigma_{e^- p\rightarrow e^- X}(P,q)=\sum_a \int_0^1\dx\xi\,f_{a/p}(\xi)\,\dx\sigma_{e^- a\rightarrow e^- X}(\xi P,q)\,.
\end{align}
Here, the sum runs over all parton species $a$, $q$ is the four-momentum of the virtual photon, and the $\dx\sigma$ indicates a cross-section is \textit{differential}, i.e. not fully integrated over all kinematically allowed phase-space configurations of the final state. 
In the case of DIS, by observing the scattered electron's momentum, $q^2$ and $x_\text{B}$ are measured and we need to compute the differential cross-section for fixed values of $q^2$ and $x_\text{B}$ to compare with data.
In this picture, the interaction of the electron with the parton is to leading order a pure QED process, hence Feynman could calculate a prediction without knowing about a QFT for the strong force and $x_\text{B}$ can be identified with the collinear momentum fraction $\xi$ by momentum conservation.

The SLAC experiment raised the question whether there was a correspondence between Feynman's partons and Gell-Mann's group-theoretical quarks. 
As already mentioned, a key problem with the quark model was the presence of hadrons where apparently all three constituent quarks were in the same state. 
To reconcile this with the Pauli principle, there needed to be additional degrees of freedom.
Moo-Young Han and Yoichiro Nambu were the first to try this idea in 1965, by splitting each quark into three states in their model \cite{Han:1965}.
In 1971, Gell-Mann and Harald Fritzsch refined the idea by introducing a new quantum number, called \textit{color} \cite{Fritzsch:1972jv}\footnote{The name is purely analogy based and is in no way related to the colors we see with our eyes. The latter are a matter of QED -- and biology.}.
They proposed it as the conserved quantity under a new $\mathrm{SU}(3)$ symmetry group, which is exact -- in contrast to the approximate flavor symmetry discussed earlier.
Equipping the hadrons with three identical quarks with antisymmetric color indices -- given the names red (r), blue (b), and green (g) -- solved the Pauli statistics problem.

A first test of this idea could be made by comparing the cross-sections of muon-production with hadron production in electron-positron collisions.
Before, we have already looked at the production of muon-pairs.
The same way, quark-antiquark pairs can be produced where only the charge factor is different.
By counting all processes where hadrons are produced to the cross-section, we ensure that we sum over all processes that involve quark-antiquark production, without knowing about the details of how the quark-to-hadron transformation\footnote{This process is called \textit{fragmentation}.} takes place.
If we now assume quarks come in $\NC$ ``color copies'' and in the three flavors known from hadron spectroscopy, we would predict
\begin{align}
\frac{\sigma(e^+e^-\rightarrow\text{hadrons})}{\sigma(e^+e^-\rightarrow\mu^+\mu^-)}=\NC\left[e_\text{u}^2+e_\text{d}^2+e_\text{s}^2\right]=\frac{2\NC}{3}\,,
\end{align}
and indeed, the experimental data was in agreement with a value of two for this ratio, matching Gell-Mann's and Fritzsch's prediction of $\NC=3$ \cite{Gross:2022hyw}\footnote{At higher energies, additional heavier quarks can be produced, adding to the ratio above their mass threshold.}.
Similarly, a color factor of three reconciled the predicted decay-rate of the neutral pion to photons with data \cite{Gross:2022hyw}.
Motivated by this success, Gell-Mann and Fritzsch interpreted the color symmetry as a gauge group.
As mentioned before, 't Hooft and Veltman had already established at that point that such a theory is indeed renormalizable.

This leads to a QFT, similar in spirit to QED, named \textit{Quantum Chromodynamics} or QCD for short \cite{Fritzsch:1973pi}.
The particle associated with the color force was termed \textit{gluon}\footnote{Fritzsch proposed the Greeky \textit{chromon} but Gell-Mann's proposal stuck \cite{Gross:2022hyw}.}.
They themselves carry color charge and come in eight color combinations.
In contrast to the uncharged photon, this makes the gluons self-interact.
In 1973, David Gross, David Politzer, and Frank Wilczek \cite{Gross:1973,Politzer:1973} discovered that this results in a reduction of coupling strength at increasing energies.\footnote{For which they were awarded the Nobel prize in 2004.}
This property of \textit{asymptotic freedom} explains, on the one hand, the quasi-free behavior of the partons in DIS, the Bjorken scaling of the cross-section, and predicts systematic deviations from exact scaling. 
On the other hand, the coupling increases at low energies.
In principle, this is consistent with the \textit{confinement} of quarks and gluons within bound hadronic states and the lack of observation of free quarks and gluons.
However, the large value of the coupling constant poses a profound technical challenge, not allowing for a perturbative treatment.
In a non-relativistic model with only a single heavy quark, one can show that the potential in a quark-antiquark bound state is proportional to $\frac{1}{r}$ for short distances, just as the Coulomb potential, but rises linearly with separation of the pair \cite{Gross:2022hyw}.
This can be understood from the gluon field's self-interactions that make the field lines lump together to strings between the color charges instead of spreading out uniformly.
The increase in potential makes it such that at some separation the field energy crosses the threshold where it is energetically more favorable to break the string and create another quark-antiquark pair from the gluon field.
Hence, no free quarks are ever observed at long distances.
However, in full QCD, with three generations of quarks, how confinement works still remains to be understood.\footnote{Overall, it is very interesting that QCD is a quantum theory not formulated in terms of observable particles.}

Exciting news hit the physics world in November 1974, when at Brookhaven \cite{JPsi2:1974} and nearly simultaneously at SLAC \cite{JPsi1:1974}, a new sharp resonance at around 3\,GeV was found.
The associated particle was named $J/\psi$ and further studies, revealing an entire spectrum of states, showed that it was the first instance of heavy quarkonia -- a bound state of a new type of quark-antiquark pairs.
The new heavy quark with mass of about $1.3$\,GeV, already theorized in 1970 \cite{Glashow:1970Charm}, received the name \textit{charm} quark and naturally added to the existing quark flavors.
Finding quarks in resonances, even though as pairs, helped elevating quarks from group theoretic obscurities to proper particles in the broader physics community.
The discovery did not only fit with QCD but also electroweak theory required a partner to the strange quark as the up is to the down.
So when in 1977 the bottom quark was discovered with a mass of about 4\,GeV \cite{Bquark:1977}, it was expected to also have a partner, the top quark, finally discovered in 1995 at Fermilab's Tevatron with a mass above 150\,GeV \cite{Topquark:1995a,Topquark:1995b}.

\begin{figure}[t]
\centering
\includegraphics[width=0.45\textwidth]{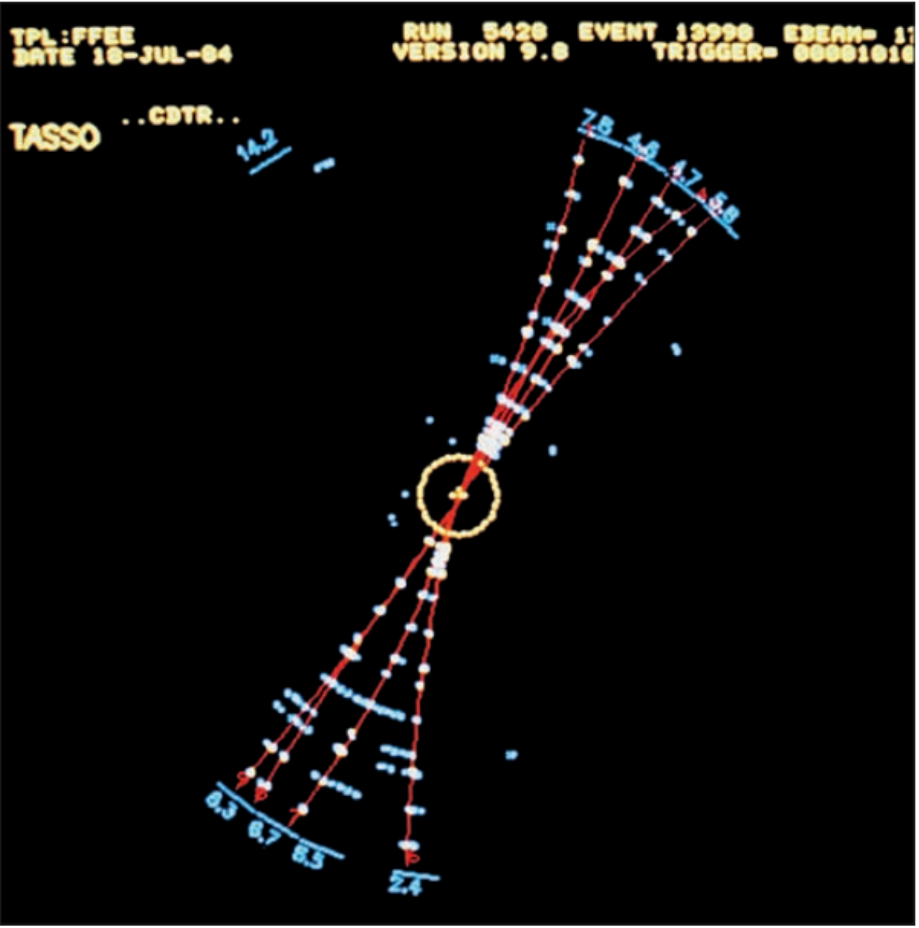}
\includegraphics[width=0.45\textwidth]{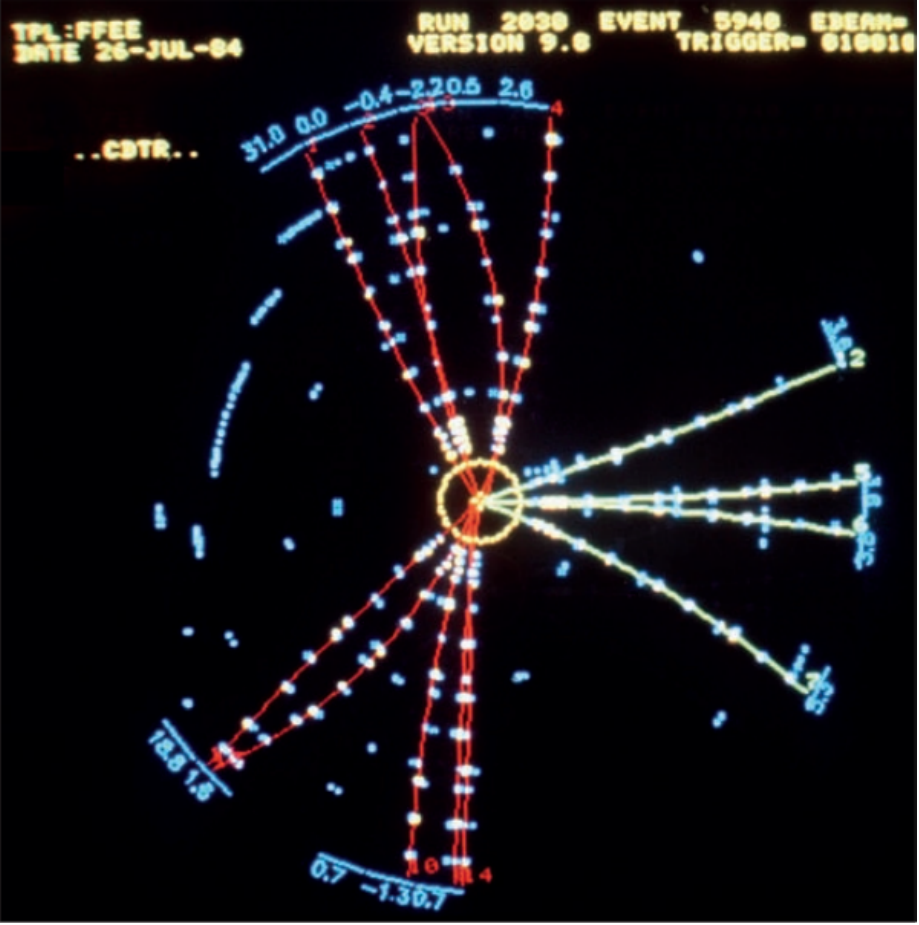}
\caption[Image of jet events in the PETRA collider at DESY (from \cite{Fritzsch2012_historyQCD}; Image credit: Oxford University PPU.).]{Jet events in the PETRA collider at DESY. The observed tracks in the detector are the hadrons (and their decay products) resulting from the initially created partons. On the left, a two-jet event initiated by a quark-antiquark pair. On the right a three-jet event attributed to the additional radiation of a gluon at high energies -- the gluon's discovery signature. (from \cite{Fritzsch2012_historyQCD}; Image credit: Oxford University PPU.)}
\label{fig:JetEventsDESY}
\end{figure}

Crucial for the success of QCD, however, was establishing the existence of the gluon.
As already discussed, in a high-energy electron-positron collision quark-antiquark pairs can be produced.
By energy-momentum conservation, these move apart at high velocities in opposite directions.
Therefore, it is to be expected that we will observe two \textit{jets} -- a spray of hadrons with approximately collinear momenta to the partons they originate from -- in our detector, in a nearly back-to-back configuration.
Indeed, this was observed with the TASSO detector at the PETRA collider at DESY, Hamburg in 1978 \cite{TASSO:1979zyf}, see the left panel of figure \ref{fig:JetEventsDESY}.
At the high-energy collision, due to asymptotic freedom, QCD is perturbative.
Hence, from the Feynman rules, one expects that in some cases an extra gluon with large momentum is radiated, resulting in its own third jet.
In 1979, such events were detected at DESY \cite{TASSO:1979zyf}, marking the ``discovery'' of the gluon, and establishing QCD as \textit{the} theory of the strong interaction, see the right panel of figure \ref{fig:JetEventsDESY}.

\begin{block}[type=idea]
\textbf{QCD as a field theory}\\
QCD consists of two kinds of fields: fermionic spin-1/2 quark fields $\psi_f^i(x)$ in $\NC=3$ colors $i$ that come in six flavors $f=u,d,s,c,b,t$ and bosonic spin-1 gluon fields $A_\mu^a$ in $\NC^2-1=8$ adjoint colors $a$.

From local $\mathrm{SU}(3)$ invariance follows the Lagrangian
\begin{equation}
\mathcal{L}_\text{QCD}(\psi_f,A_\mu)=-\frac{1}{4}G_{a,\mu\nu} G^{a,\mu\nu}+\sum_f \bar{\psi}_f(\iu D_\mu\gamma^\mu-m_f)\psi_f\,,
\end{equation}
with the \textit{gluon field strength tensor} $G^{\mu\nu}_a=\partial^\mu A^{\nu}_a-\partial^\nu A^\mu_a-g_s f_{abc} A^\mu_b A^\nu_c$ and the covariant derivative $D^\mu=\partial^\mu+\iu g_s A^\mu_a t_a$. The \textit{generators} $t_a$ and the \textit{structure constants} $f_{abc}$ form the \textit{color algebra} $[t_a,t_b]=\iu f_{abc} t_c$. Color factors in QCD are commonly expressed through $\CA=\NC$, associated with the emission of a gluon from a gluon, and $\CF=(\NC^2-1)/(2\NC)$.
$\gamma^\mu$ is a Dirac-matrix that forms the Dirac algebra $\lbrace\gamma^\mu,\gamma^\nu\rbrace=2g^{\mu\nu}$.
The QCD Lagrangian is commonly quantized in the path integral formalism and requires gauge fixing and the introduction of Faddeev-Popov ghost terms to preserve unitarity.
$g_s$ is the strong coupling constant connected with the QCD expansion parameter as $\as(\mu)=\frac{g_s^2(\mu)}{4\pi}$.
The QCD  running coupling is at one-loop given by
\begin{align}
\as(\mu)=\frac{4\pi}{\beta_0\,\log\frac{\mu^2}{\Lambda_\text{QCD}^2}}\,,\quad\text{with } \beta_0=\frac{11}{3}\CA-\frac{4}{3}T_\text{F} n_f\,,
\end{align}
where $T_\text{F}=\frac{1}{2}$, $n_f$ is the number of active quarks at the scale $\mu$, and $\Lambda_\text{QCD}\approx 300\,\text{MeV}$ is the intrinsic QCD scale that governs when QCD becomes non-perturbative. It is not present on the level of the classical Lagrangian and enters the theory via the renormalization procedure.
The currently accepted numerical value for the coupling strength is $\as(M_Z)\approx 0.1180\pm 0.0009$ \cite{ParticleDataGroup:2024cfk}.
\end{block}

\begin{figure}[h]
\includegraphics[width=\textwidth]{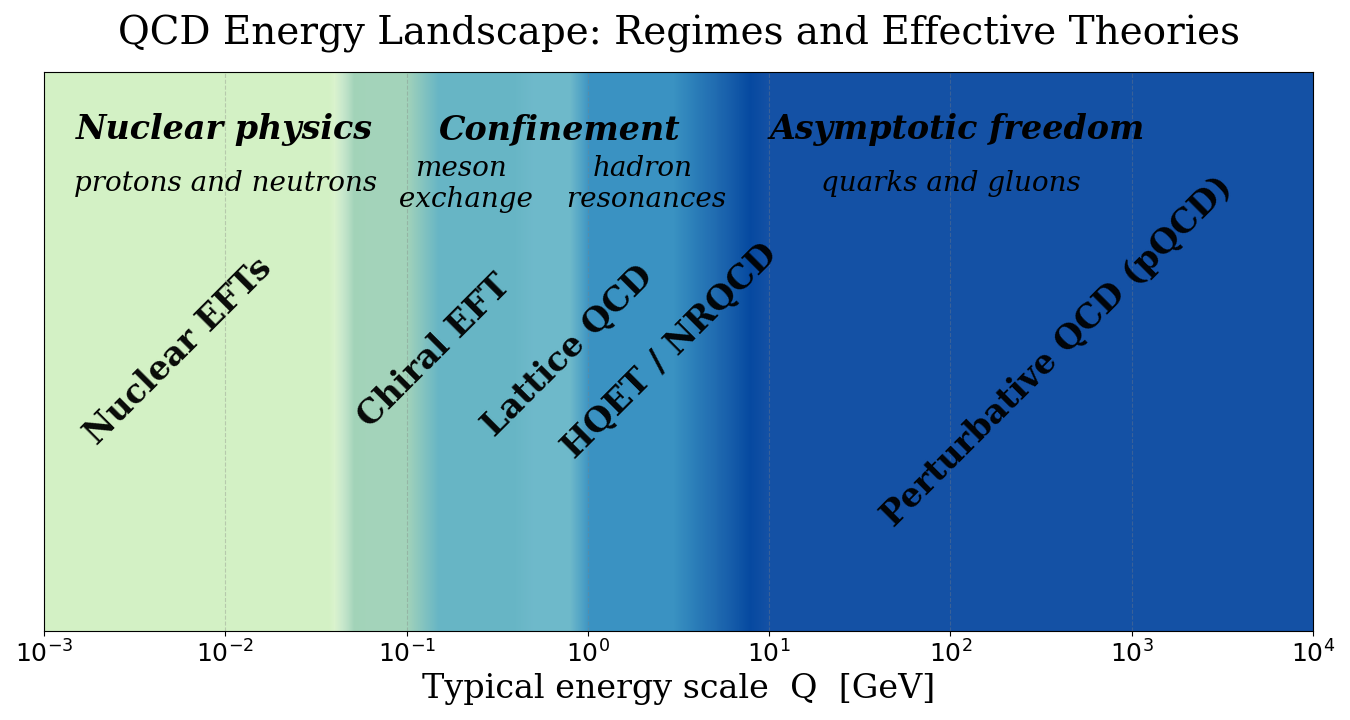}
\caption[Overview of the QCD energy landscape. Created with \texttt{NumPy} \cite{harris2020array} and \texttt{Matplotlib} \cite{Matplotlib}.]{Approximate guide to which approach to QCD is used at different energy scales. $Q$ refers to the energy scale at which the strong interaction is probed.}
\label{fig:QCD_energy_landscape}
\end{figure}

Despite its deceptively simple looking Lagrangian which, except for the quark masses, contains no explicit energy scales, the physics described by QCD is very different, depending on the scale $Q$ the strong interaction is probed at, see Fig.\,\ref{fig:QCD_energy_landscape}.
Only at high enough energies above a few $\text{GeV}$ the interactions can be described in terms of the fundamental degrees of freedom of the theory, the quarks and gluons. This is the domain of \textit{perturbative QCD} \cite{Collins:2011zzd}, which set the stage for the studies conducted in this work.
Going down with the energy, the strong force binds the quarks together in short-lived bound states of quarks and hadron resonances set in. This domain can be described in an \textit{effective field theory (EFT)} approach \cite{Weinberg:1978kz}, meaning that the high-energy modes of the field are integrated out, examples of such approaches are non-relativistic QCD (NRQCD) \cite{Caswell:1986,Bodwin:1995} and heavy-quark effective theory (HQET) \cite{Eichten:1990,Grinstein:1990}. Alternatively, lattice QCD \cite{Davies:2004,FlavourLatticeAveragingGroupFLAG:2024oxs} -- a first principles approach that takes the QCD Lagrangian and calculates expectation values of observables on a discretized space-time lattice -- is successful in this domain in predicting, for example, the hadronic mass spectrum.
At still lower energies, only a special set of very light mesons remains relevant.
These are light because they are the remnants of a broken symmetry of QCD \cite{Nambu:1961} -- whenever a continuous symmetry breaks, the theory produces nearly massless \textit{Goldstone bosons} \cite{Goldstone:1961eq}.
In QCD these appear as the pions (and, with three flavors, kaons  and one eta), and they form the dynamical degrees of freedom of \textit{chiral} perturbation theory \cite{Gasser:1983yg,Gasser:1984gg}.
This framework can be extended to also incorporate baryons, and when applied to multi-nucleon systems leads to the modern theory of chiral EFT \cite{Weinberg:1990,Weinberg:1991}.
At even lower scales, the momentum transfer becomes smaller than the pion mass, so the pions themselves can be integrated out and nuclear interactions reduce to short-range effective forces between protons and neutrons \cite{Epelbaum:2009}.
This is the domain of nuclear physics.

After this short digression into the richness of QCD down the energy line -- that led us, from a different perspective, back the timeline that culminated in the discovery of quarks and gluons -- we will return to the scales most relevant to modern collider phenomenology for the remainder of this work: the turf of perturbative QCD.
However, we need to keep in mind that in the detector, which is far away from where the hard QCD interaction happens, only hadrons are observed, thus there is always the need to make contact with the non-perturbative part of QCD.
This happens by either defining variables in a way that is insensitive to the low-energy dynamics, an example being the total production cross-section of hadrons at electron-positron colliders, or by parameterizing the non-perturbative structure in terms of universal functions that can be extracted from data, similar to what we have seen with Feynman's parton model for DIS.

Also, the perturbative part of the calculation that can be performed on the level of quarks and gluons poses several technical challenges for higher-order calculations due to the structure of QCD as a non-abelian gauge theory.
As typical for a QFT, loop integrals in QCD contain UV divergences.
However, regularization with a cut-off is not compatible with gauge and Lorentz symmetry.
Hence, to regularize them in a way that is compatible with the symmetries of QCD, \textit{dimensional regularization} was developed by Juan Giambiagi and Carlos Bollini \cite{Bollini:1972} and independently by 't Hooft and Veltman \cite{tHooft:1972}.
The idea is to treat the space-time dimension $d$ as a variable, chosen such that all integrals are well defined and then use analytic continuation -- a method from complex analysis that uniquely continues functions beyond their original domain of definition -- to connect back with the physical theory in $d=4$ dimensions. 
The method is mathematically elegant but admittedly abstract and we will look at it in some detail in section \ref{ch:Methods}.

Besides the UV divergences, there is also trouble in the infrared, i.\,e. for small momenta.
Here, the presence of massless particles -- gluons and light quarks that are commonly treated as massless -- makes it such that calculations on parton level exhibit so-called soft and collinear divergences.
Considering the propagator of two momenta $p$ and $k$ associated with two massless particles,
\begin{equation}
\frac{1}{(p+k)^2}=\frac{1}{2p\cdot k}=\frac{1}{2E_p E_k(1-\cos\theta_{pk})},
\label{eq:IRprop}
\end{equation}
we see that it becomes singular if the energies $E_{p,k}$ tend to zero, meaning soft particles, or the momenta become parallel, i.e. $\theta_{pk}\rightarrow 0$, meaning collinear particles.
When calculating a (differential) cross-section, we are faced with the problem of integrating over additional radiation in higher orders of perturbation theory.
For example, integrating eq.\,\eqref{eq:IRprop} over all angles $\theta_{pk}$ results in
\begin{align}
\frac{1}{2E_p E_k} \int_0^\pi\dx\theta_{pk}\,\underbrace{\frac{\sin\theta_{pk}}{1-\cos\theta_{pk}}}_{\approx 2/\theta_{pk} \text{ for }\theta_{pk}\rightarrow 0}=\infty\,.
\end{align}

Therefore, also the phase-space integrals need to be calculated with a regulator, most conveniently and conventionally also using dimensional regularization.
Here, the basic integral to occur in a wide range of QCD processes is the (two-denominator) \textit{angular integral in $d$ dimensions}
\begin{align}
\int\dx\Omega_{d-1}(k)\frac{1}{(p_1\cdot k)^{j_1}(p_2\cdot k)^{j_2}}.
\label{eq:Angular_integral_first}
\end{align}
The earliest occurrence in the literature of such an integral can be traced to Richard Ellis in 1980, calculating hadron production in hadron-hadron collisions \cite{Ellis:1980}.
Since then, angular integrals have appeared in all sorts of pQCD calculations in the past 45 years 
\cite{Schellekens:1981,Duke:1982,vanNeerven:1985,Beenakker:1988,Matsuura:1989, Matsuura:1990,Hamberg:1991,Mirkes:1992,Gordon:1993,Bolzoni:2010,Schlegel:2012,Anastasiou:2013,Ringer:2015,Anderle:2016,Campbell:2017,SIDISQTNLO,Lionetti:2018,Specchia:2018,Bahjat-Abbas:2018,Hekhorn:2019,Baranowski:2020,Alioli:2022,Assi:2023,Pal:2023,Catani:2023,Ahmed:2024owh,Devoto:2024,Agarwal:2024gws,Baranowski:2024ysi,Rein:2024,Rein:2025qhe,Rein:2025pwu,Baranowski:2025}.

In this thesis we will thoroughly study this class of integrals, give the first ``full'' result of \eqref{eq:Angular_integral_first} for arbitrary masses and with complete control over the dependence on dimensionality and generalize to kinematics with more external particles.

Before we will finally narrow down on the specific topics within perturbative QCD treated in this thesis, it is however in order to briefly return back to one of the triumphs of modern physics:
A unified, consistent description of all known elementary particles and their interactions, except for gravity\footnote{Which is orders of magnitude smaller than all the other forces at the energy scales accessible to contemporary particle physics.}.

\section{The completion of the Standard Model of particle physics and the precision era}
\label{sec:Intro5}
The combination of the electroweak theory with QCD constitutes the current \textit{Standard Model of particle physics} (SM).
From its theoretical formulation in the 1970s to today, the SM has been a success story, even more so than its inventors might have anticipated.
One after another, all predicted particles were found (see e.g. Fig.\,\ref{fig:ZandHiggsResonances}) and with the observation of the Higgs boson in 2012 at the Large Hadron Collider (LHC) at CERN the Standard Model was finally completed \cite{ATLAS:2012yve,CMS:2012qbp}.

\begin{figure}
\centering
\includegraphics[width=0.5\textwidth]{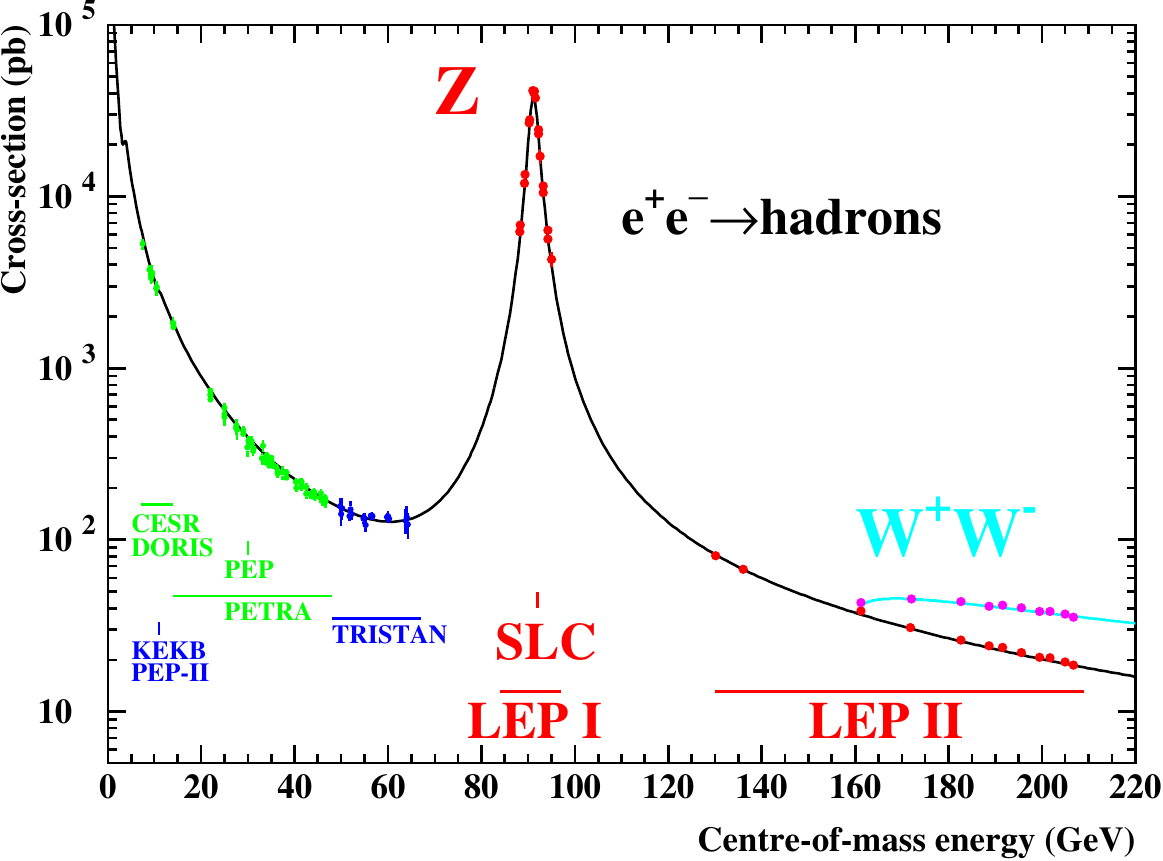}
\hspace{3em}
\includegraphics[width=0.4\textwidth]{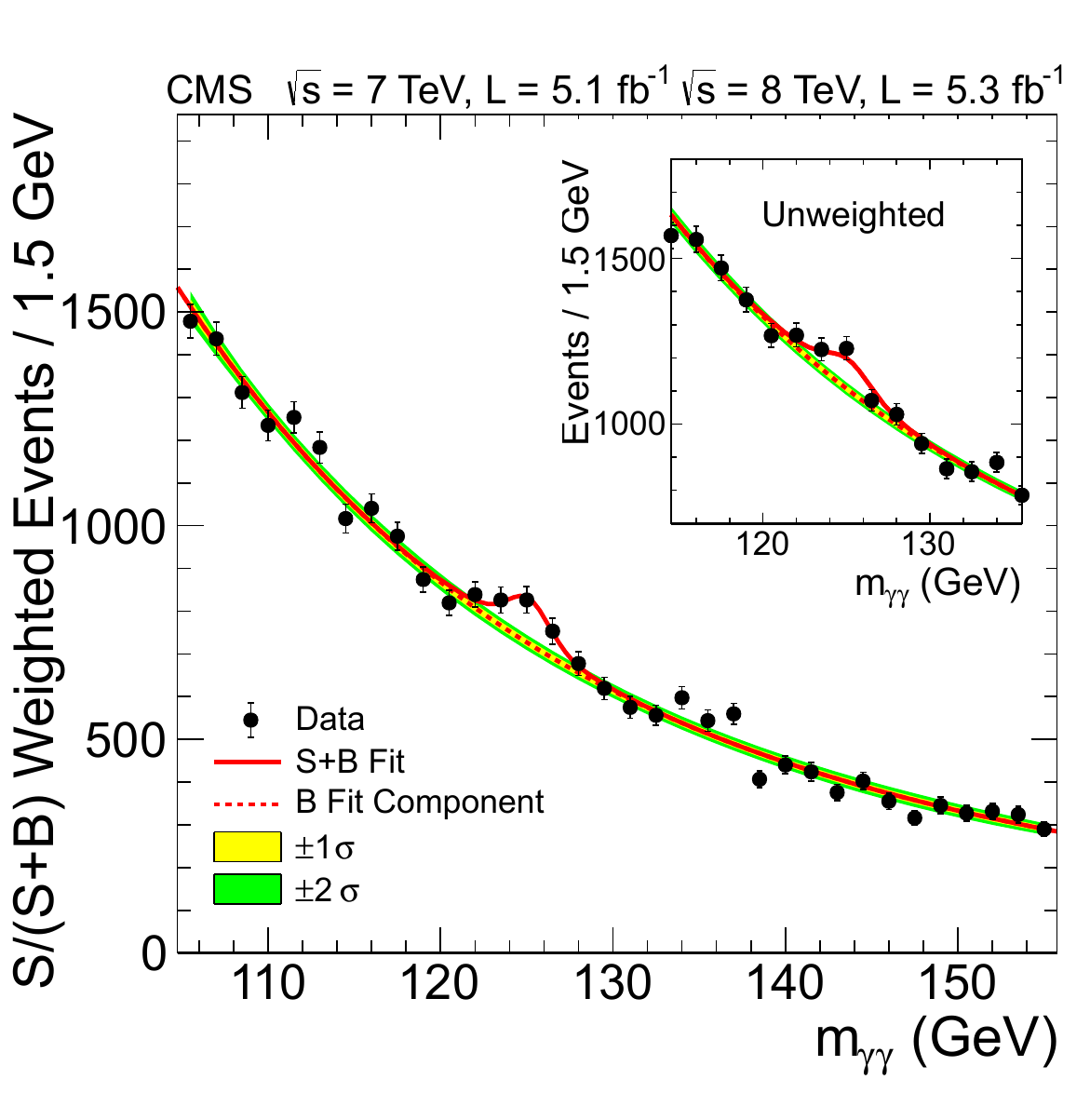}
\caption[$Z$ and Higgs boson resonance peaks (from \cite{ZBoson:2006} and \cite{CMS:2012qbp}, respectively).]{The $Z$ boson (left, from \cite{ZBoson:2006}) and Higgs boson (right, \cite{CMS:2012qbp}) resonances. 
Note the characteristic Breit-Wigner-like shape of the $Z$ peak and the comparatively more subtle signature of the Higgs boson, necessitating a higher degree of precision to disentangle from the background.}
\label{fig:ZandHiggsResonances}
\end{figure}

Since then, it has stood up to all experimental tests and no direct signs of beyond the Standard Model (BSM) physics have been observed \cite{ParticleDataGroup:2024cfk}: No supersymmetry \cite{Martin:1997ns}, no string theory \cite{Dienes:1996du}, no earth devouring black holes at the LHC \cite{CMS:2012yf}.
This led to a shift on the discovery frontier from hunting for new particles and resonances to gathering precision information about the Standard Model itself.

A prime example for the theoretical sophistication that has been reached is the anomalous magnetic moment of the electron which has been calculated at the parts-per-billion level \cite{Laporta:1992pa,Aoyama:2012,Aoyama:2018}, matching the experimental value to the same accuracy \cite{Hanneke:2006,Hanneke:2008,Hanneke:2011}\footnote{The closely related anomalous magnetic moment of the muon has been a strong contender for a sign of BSM physics. However the latest studies put it back into agreement with SM predictions within uncertainties \cite{Aliberti_2025}.}.
In the context of this thesis, it is interesting to note that a close cousin of the angular integrals played a role in the pioneering multi-loop calculations of the electron $g-2$ \cite{Laporta:1994mb}.

In general, particle physics has entered a precision era \cite{Heinrich:2020ybq}: The age of spectacular discoveries gave way to the age of spectacularly difficult calculations.
Making use of the high statistics of measurements provided by modern colliders for a wide range of processes requires theoretical predictions to match the same accuracy.

This required the development of an advanced technology for multi-loop integrals: Starting from direct one-loop calculations in the early days of QCD by 't Hooft \cite{tHooft:1978}, the introduction of \textit{integration-by-parts (IBP)} identities by Fyodor Tkachov and Konstantin Chetyrkin in 1981 reduced the number of integrals to be calculated to a manageable set of master integrals and made higher-loop calculations feasible \cite{Tkachov:1981}.
This method was turned into an algorithm by Stefano Laporta in 2000, opening up the multi-loop frontier \cite{Laporta:2000}.
Subsequently, in 2010, Vladimir Smirnov proved that the number of master integrals is always finite \cite{Smirnov:2010hn}.
Already in 1997, Smirnov together with Martin Beneke, introduced \textit{expansion by regions}, a versatile tool for the calculation of Feynman integrals in kinematic limits.
To calculate master integrals exactly, the most powerful method on the market uses differential equations.
It was introduced by Anatoli Kotikov \cite{Kotikov:1991} in 1991 and algorithmized by Johannes Henn in 2013 \cite{Henn:2013pwa}. 
This allows to understand the solution to Feynman integrals in terms of \textit{polylogarithms} and their generalizations \cite{Panzer:2014caa}.

The calculation of amplitudes is routinely done at the two-, three-, and sometimes four-loop level (e.g. \cite{Agarwal:2024twoloop,Caola:2021rqz,vonManteuffel:2016xki}, for a review see \cite{Heinrich:2020ybq}) and the formal study has led to a much deeper understanding of their structure \cite{Parke:1986gb,Goncharov:2010,Lee:2013hzt}.
As an example, in 1996, Oleg Tarasov found that Feynman integrals in different space-time dimensions are connected by algebraic relations \cite{Tarasov:1996br}. 
Applications and adaptions of these techniques will be central to this thesis.

\section{The role of perturbative QCD in today's particle physics}
\label{sec:Intro6}
Even though its main objective is not to directly discover new particles -- it deals with a known theory of particles known to exist --, there are several reasons to care about perturbative QCD:
\begin{itemize}
\item[(i)] Due to the rather large value of the strong coupling even in the perturbative regime, perturbative corrections are sizeable, necessitating the calculation of higher orders in QCD perturbation theory.
\item[(ii)] Protons are the projectiles of choice as initial states for colliders, so knowing their structure is important to precision collider phenomenology.
\item[(iii)] Additionally, studying the proton structure in its own right provides us with a refined image of a fundamental building block of matter that makes up most of the visible matter in the universe.
\item[(iv)] This allows to understand how QCD leads to fundamental properties of matter, for example how the quarks and gluons contribute to the spin of the proton.
\end{itemize}
Overall, it is a dual role between precision tests of the SM and continuing Rutherford's work of mapping out what the world is made of.

To reach the high energies for precision tests of the Standard Model, using protons as beam particles has several advantages:
They are stable, abundantly available\footnote{The protons used at CERN originate from a simple bottle of hydrogen gas.}, easy to accelerate due to their charge, and nearly two-thousand times more massive than electrons.
This poses an additional challenge for calculating cross-sections, since now the initial state is not a fundamental particle but a complicated bound state of QCD.
However, as we already know from the success of the original parton model, this problem can be addressed by separating long- and short-distance dynamics.
In QCD, Feynman's idea lives on under the name of \textit{factorization}, foundational to perturbative QCD \cite{Collins1989,Collins:2011zzd}.

In its basic form it states that suitable collider observables, for example the cross-section of deep inelastic scattering of electrons off protons, can be decomposed into two parts: 
one universal, which only depends on the internal composition of the proton in terms of its partons -- now a collective term for quarks and gluons -- and another, process-specific, which describes the short-distance interaction of the electron with the partons.
Structurally, factorization formulas look like 
\begin{align}
\underbrace{\tcboxmath[colback=dandelion!75!white,colframe=black]{\mathrm{d}\sigma_{p\rightarrow X}(x,Q)}}_\text{\normalsize{(i): data}}=\underset{a}{\boldsymbol{\sum}} \underbrace{\tcboxmath[colback=dartmouthgreen!50!white,colframe=black]{f_{a/p}(x,\mu_\text{F})}}_\text{\normalsize{(ii): fit}}\boldsymbol{\otimes} \underbrace{\tcboxmath[colback=cornflowerblue!50!white,colframe=black]{C_{a\rightarrow X}(x,Q,\mu_\text{F})}}_\text{\normalsize{(iii): calculation}}
\underbrace{\,+\,\mathcal{O}\!\left(\frac{m_p}{Q}\right)}_\text{\normalsize{correction}}
\,.
\label{eq:FactorizationSchematic}
\end{align}
The three parts of this equation are:
\begin{itemize}
\item[(i)] The (differential) cross-section \textcolor{orange}{$\dx\sigma_{p\rightarrow X}(x,Q)$} on hadron level that can be observed at the collider.
\item[(ii)] The universal parton distribution \textcolor{dartmouthgreen}{$f_{a/p}(x,\mu_\text{F})$} that parametrizes the hadron structure and encodes the probability for a parton to participate in the hard scattering.
\item[(iii)] The process-specific hard scattering coefficient function \textcolor{blue}{$C_{a\rightarrow X}$} on parton level that can be calculated in perturbative QCD due to asymptotic freedom.
\end{itemize}
The $\otimes$ symbol denotes a (Mellin-)convolution integral that couples the PDF with the scattering coefficient. Broadly, it sums over all kinematically allowed momentum fractions. By \textit{Mellin transformation} it can be converted into an ordinary product.
Corrections $\mathcal{O}(\dots)$ to this picture are power suppressed by the hard scale $Q$, i.e. the four-momentum transfer from the electron to the proton, which is typically much larger than the proton mass $m_p\approx 1\,\text{GeV}$ .
Physically, such a factorization is possible precisely since there is this separation of scales $Q\gg m_p$.
The interactions between quarks and gluons within the proton happen at timescales governed by the proton mass while the hard interaction occurs at a much shorter timescale $1/Q$ at which the partons appear quasi-free to an external high-energy probe such as a scattering electron.  
The scale separation can be formalized within the framework of renormalization which introduces a dependence of both PDF and the hard coefficient on the factorization scale $\mu_\text{F}$, which controls which interactions are viewed as part of the proton structure and what is considered as part of the partonic scattering. 
Varying the scale leads to so-called \textit{evolution equations} for the parton distributions, known as \textit{Dokshitzer-Gribov-Lipatov-Altarelli-Parisi\footnote{Giorgio Parisi was awarded the 2021 Nobel prize for his work on disorder and fluctuations in complex systems.} (DGLAP) equation} for PDFs \cite{Dokshitzer1977,Dokshitzer1980,Gribov1972,Altarelli1977}.
To give intuition to the reader unfamiliar with factorization, figure \ref{fig:SIDIS_schematic} illustrates how the parts of equation \eqref{eq:FactorizationSchematic} relate to a collider process.
\begin{figure}[t]
\centering
\includegraphics[width=0.95\textwidth]{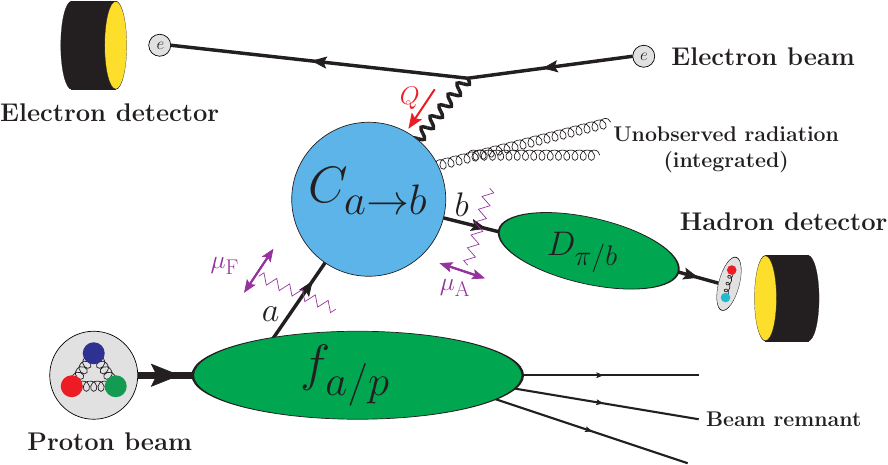}
\caption[Schematic sketch of the factorization for semi-inclusive deep inelastic scattering. Created with \texttt{JaxoDraw} \cite{Binosi:2003}.]{Schematic sketch of the factorization for semi-inclusive deep inelastic scattering, i.e. electron-proton scattering where a hadron gets produced and measured in the detector.
Here, the non-perturbative parts that link hadronic and partonic degrees of freedom are the PDF $f_{a/p}$ and the \textit{fragmentation function} $D_{\pi/b}$ (this parametrizes the probability that parton $b$ hadronizes into a hadron, here a pion $\pi$. The corresponding factorization formula thus has two ``green blocks'' that get convoluted). The hard part of the scattering depends only on the partons $a$ and $b$ and can be calculated perturbatively. The energy scales used to separate the two parts, called factorization and fragmentation scales, are denoted by $\mu_\text{F}$ and $\mu_\text{A}$, respectively. 
Unobserved radiation is to be summed over all kinematically allowed configurations in a phase-space integral.
This picture is for illustrative purposes only; more formally, one would need to take the absolute square of the depicted ``amplitude'' to get to a cross-section (see eq.\,\eqref{eq: cross-section}) and a proper Feynman-diagrammatic depiction of the quantities of eq.\,\eqref{eq:FactorizationSchematic}.}
\label{fig:SIDIS_schematic}
\end{figure}

By the structure of eq.\,\eqref{eq:FactorizationSchematic}, we see that accurate predictions are only possible when all three parts -- experiment, PDF extraction, and perturbative calculation -- work together.
Only precision data from experiment and high-order theory computations of coefficient functions across several observables allow for the precise extraction of the parton distribution in a \textit{global analysis}, meaning the best fit of all available channels.
Precise PDFs in turn are required to make predictions for collider observables.

To match the precise data collected by experiments, for a large part by the LHC in recent years, perturbative calculations need to be performed to higher orders in $\as$.
For processes with only very few observed particles, the usual observables are structure functions that parameterize differential cross-sections.
The calculation of the hard coefficients is done analytically, becoming increasingly more difficult with every order and each kinematic scale involved.

\begin{figure}[htb]
\includegraphics[width=\textwidth]{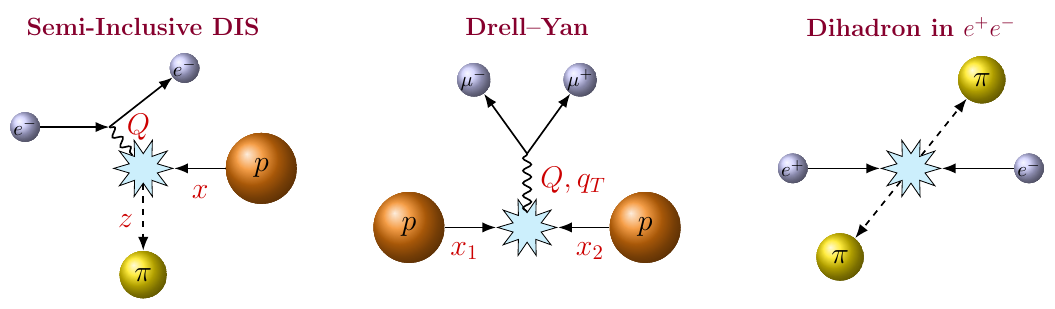}
\caption[Schematic sketch of key processes for perturbative QCD. Adapted from \cite{Boussarie:2023izj}.]{Schematic sketch of key processes for perturbative QCD.}
\label{fig:QCD_processes}
\end{figure}
For perturbative QCD, relevant observables besides DIS, which we looked at when discussing the parton model, include among others lepton production in proton-proton collisions, called Drell-Yan process \cite{Drell1970,Peng:2016ebs}, heavy quark production \cite{Beenakker:1988,ParticleDataGroup:2024cfk}, jet observables \cite{Sapeta:2015gee}, and energy-energy correlators \cite{Neill:2022lqx}.
The kinematics of selected key processes are depicted in figure \ref{fig:QCD_processes}.
As a representative example, we may look at the development of accuracy in perturbative predictions for deep-inelastic scattering:
The LO was put on a field theoretic foundation in the early 1970s \cite{DISLOa,DISLOb}, the NLO has been calculated in 1978 \cite{DISNLO}, the NNLO in 1991 \cite{DISNNLOa}, the N$^3$LO in 2005 \cite{DISN3LOa,DISN3LOb}.
These were the results for the coefficient functions that only depend on the single momentum fraction variable $x$, for unpolarized beams, photon exchange, and massless quarks.
All these aspects can and have been generalized, adding complexity.
Coming a few years after the unpolarized results, polarized DIS has been calculated to NLO in 1979 \cite{DISpolNLO}, to NNLO in 1994 \cite{DISpolNNLO}, and to N$^3$LO in 2022 \cite{DISpolN3LO}, also $W/Z$ exchange has been incorporated \cite{DISN3LOc}.
Both these calculations are more complicated in comparison to standard DIS, since they need to deal with the $\gamma_5$-matrix which is notoriously badly suited for an extension to $d$ dimension as required for dimensional regularization.
Adding heavy quarks introduces a second scale $\frac{m_q^2}{Q^2}$ that makes the computation much more involved.
Hence, fully analytic results are currently only known to NNLO \cite{Blumlein:2016xcy}, while using an expansion in the heavy quark mass allowed to push to N$^3$LO recently \cite{Ablinger:2025awb}.

All these results are for inclusive DIS, where only the scattered electron is observed and the hadronic part of the final state is fully integrated over.
When instead observing one of the hadrons produced in the collision, the process is called \textit{semi-inclusive} deep inelastic scattering (SIDIS).
This is interesting for QCD since it allows to study how the partons of the hard scattering fragment into hadrons.
In SIDIS, one has an additional variable $z$ that gives the longitudinal momentum fraction of the outgoing parton carried by the identified hadron.
This problem with two invariant momentum fractions poses additional computational challenges compared to DIS, so for a long time only the NLO had been known \cite{Altarelli:1979kv,Baier:1979sp,Furmanski:1981,deFlorian:1997} and after preliminary work in \cite{Anderle:2016,Haug:2021}, only recently the full NNLO has been calculated \cite{Goyal:2023unpol,Bonino:2024unpol,Bonino:2024pol,Goyal:2024pol,Goyal:2024emo}.
Besides its collinear momentum, one can also observe the transverse momentum of the hadron $P_{h,\text{T}}$.
Besides the higher number of invariants that complicate the analytic calculation, the extra scale is also an additional challenge for consistent treatment by factorization.
Here, one has to distinguish two cases.
For $P_{h,\text{T}}^2\sim Q^2$, both scales are equally hard and factorization works in terms of collinear parton distribution functions.
For this case, the structure functions are only known to NLO \cite{SIDISQTNLO}, which corresponds to $\mathcal{O}(\as^2)$ since at least one hard radiation need to take place for non-zero transverse momentum.
Here, it is instructive to look at the behavior when $P_{h,\text{T}}^2$ becomes much smaller than $Q^2$.
Typically, when there are multiple scales involved in a perturbative calculation, logarithms of these scales appear in the result as logarithms $L=\log(P_{h,\text{T}}^2/Q^2)$ in each order of perturbation theory.
Now if $P_{h,\text{T}}^2\ll Q^2$, the combination $\as L$ is no longer a small quantity and a perturbative expansion in powers of $\as L$ is no longer meaningful.
This calls for a reorganization of the perturbative expansions by \textit{resummation} that takes the offending logarithms to all orders in the coupling and sums them in an exponential series \cite{Sterman:1986aj,Catani:1989ne}. 
This works due to the universality of these logarithms that are governed by renormalization group equations \cite{Collins:1984xc}.
For $P_{h,\text{T}}^2\ll Q^2$, the process also becomes sensitive to the transverse motion of partons within the initial state hadron and the transverse dynamics in the fragmentation process.
This changes the factorization picture in that region into a hard function at scale $Q^2$ and transverse-momentum dependent (TMD) parton distributions \cite{Boussarie:2023izj},
a foundational study for SIDIS in the TMD regime is \cite{Bacchetta:2006tn}.
While PDFs are a one-dimensional scan through the hadron, TMDs provide a more detailed, three-dimensional picture, since they are sensitive to the transverse dynamics in the hadron.
There is also an ongoing effort to systematically include higher powers in $P_{h,\text{T}}\ll Q$ into the small transverse momentum picture and to better understand the matching in the intermediate region between the collinear and TMD regimes \cite{Collins:2016SIDIS}.

Another extension of DIS deals with the power suppressed corrections to the factorization formula.
This involves a zoo of new so-called \textit{higher-twist} parton distributions \cite{Jaffe:1982Twist4,Ellis:1982wd,Harindranath:1997kk,Dressler:1999zi,Eguchi:2006mc,Bacchetta:2006tn,Mukherjee:2009uy,PhysRevD.95.074017} that becomes increasingly relevant once these corrections are the limiting factor to increase theoretical precision \cite{PhysRevD.107.014013,Abir:2023fpo}.

All these studies of multi-scale processes are of particular importance due to the upcoming Electron-Ion Collider (EIC) \cite{Accardi:2012,AbdulKhalek:2021gbh}, which is currently under construction at Brookhaven National Laboratory (BNL) on Long Island, New York in the USA.\footnote{Not too far from where the Shelter Island conference was held.};
start of operation is planned for 2035.
This will provide an entirely new window on the structure of QCD matter, since the EIC will be equipped with polarized beams which allow to study the spin structure and the full set of TMD PDFs.
Thanks to older experiments, for example HERA at DESY \cite{Abramowicz:1998ii} or RHIC at BNL \cite{Muller:2006ee}, there is already data for various observables in polarized electron proton scattering, which help settle fundamental questions -- for example that a significant portion of the proton spin is carried by the gluons \cite{deFlorian:2014yva}.
However at the EIC, the high luminosity, the multitude of detectable final states, and the often multi-differential cross-sections offer unprecedented data that require theory to keep pace.
For polarized PDFs, percent-level accuracy is achievable -- similar to the state-of-the-art for unpolarized PDFs reached today \cite{NNPDF:2024dpb} --, and the first determination of antiquark and gluon TMDs will be possible \cite{Accardi:2012}.

%\begin{block}[type=note]
%\textbf{Contemporary Colliders: }\\
%  \begin{tabular}{llllll}
%    \hline
%    Collider & Laboratory & Type & Dates &
%    $\sqrt{s}$ [GeV] & Luminosity [$\mathrm{cm}^{-2}\mathrm{s}^{-1}$] \\
%    \hline
%    Tevatron & Fermilab & $p\bar p$ & 1987--2012 &
%    1960 & $4\times10^{32}$ \\
%    LEP      & CERN     & $e^{+}e^{-}$ & 1989--2000 &
%    90--209 & $10^{32}$ \\
%    HERA     & DESY     & $e^{-}p / e^{+}p$ & 1992--2007 &
%    320 & $8\times10^{31}$ \\
%    PEP-II   & SLAC     & $e^{+}e^{-}$ & 1999--2008 &
%    10.5 & $1.2\times10^{34}$ \\
%    KEKB     & KEK      & $e^{+}e^{-}$ & 1999--2010 &
%    10.6 & $2.1\times10^{34}$ \\
%    RHIC     & BNL      & $pp,\ AA$ & 2000--2025 &
%    200--500 & $\sim10^{32}$ \\
%    LHC      & CERN     & $pp$ & 2009-- &
%    14000 & $10^{34}$ \\
%    EIC      & BNL/JLab & $ep,\ eA$ & 2035(?)-- &
%    20--140 & $10^{33}$--$10^{34}$ \\
%    \hline
%  \end{tabular}
%\end{block}
\section{What is this thesis about?}
\label{sec:Intro7}
For processes with complicated final states, for example SIDIS with observed transverse momentum but also jet or heavy quark production, the differential cross-section is often not calculated fully analytically but with parton-level event generators \cite{Biedermann:2017yoi,Helenius:2019gbd,Bierlich:2022pfr,Bewick:2023tfi,Banfi:2023mhz,NNLOJET:2025rno}.
Using appropriate subtraction schemes that deal with all divergent configurations at a specified perturbative order \cite{Frixione:1995ms,Catani:1996vz,Anastasiou:2003gr,Binoth:2004jv,Gehrmann-DeRidder:2005btv,Catani:2007vq,DelDuca:2016ily,Czakon:2017wor,Caola:2017dug,Bertolotti:2022aih}, the remaining phase space integration can be performed numerically.
This is a highly successful program, NLO calculations in this setup have been automatized by the demand of the LHC, and are currently pushed to NNLO.
For usage with the EIC, the consistent inclusion of polarized initial states at NNLO is a frontier.

For the determinations of parton densities in a fit, it is especially important to have fast and reliable numerical implementations. This is one case where analytically known coefficient functions shine, since they are expressed through special functions that can be evaluated by dedicated efficient implementations to any desired precision. 
Besides, knowing coefficient functions analytically helps understanding the mathematical structure of perturbation theory -- for example which functions and constants appear -- and reveal patterns between different processes.
This motivates pushing the boundary for which processes fully analytic coefficient functions can be calculated and a main focus of the presented work is dedicated to that project\footnote{This program may also be useful to the calculation of analytic subtraction terms within subtraction schemes, though.}. 

\begin{figure}
\centering
\includegraphics[width=0.8\textwidth]{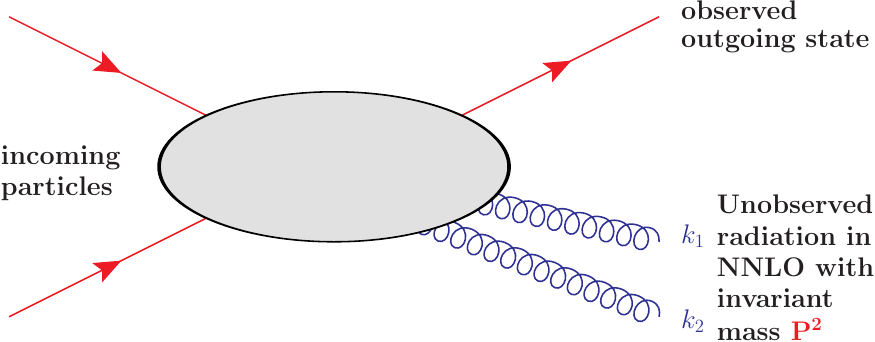}
\caption[Kinematics for the two-denominator angular integral. Created with \texttt{JaxoDraw} \cite{Binosi:2003}.]{Kinematics for the two-denominator angular integral. The grey blob symbolizes any way the external lines can be connected to a valid Feynman diagram.}
\label{fig:Kinematics_for_angular_integral_2denom}
\end{figure}

\subsection{Phase-space integrals: Angular integrals}
Historically, analytic treatment of phase-space integrals has for a large part revolved around the angular integral with two denominators.\footnote{A comprehensive account of its usage in the literature is provided in publication \ref{pub:1}.}
This is sufficient for the calculation of double real corrections in any process with $2\rightarrow 1+X$ kinematics, meaning that we observe one final-state particle, see Figure \ref{fig:Kinematics_for_angular_integral_2denom}.
Examples include SIDIS, the Drell-Yan process, semi-inclusive hadron-hadron scattering, heavy-quark production, and prompt-photon production.

The phase-space integral associated with the additional radiation of the particles $k_1$ and $k_2$ in figure \ref{fig:Kinematics_for_angular_integral_2denom} in $d=4-2\eps$ dimensions is
\begin{align}
\int\dx\mathrm{PS}_{2,\Red{P}}=\int\frac{\dx^{d-1}\Blue{k_1}}{(2\pi)^{d-1} 2\Blue{k_{1}^0}}
\int\frac{\dx^{d-1} \Blue{k_2}}{(2\pi)^{d-1}2\Blue{k_{2}^0}}\,(2\pi)^d \delta^d\left(\Red{P}-\Blue{k_1}-\Blue{k_2}\right).
\end{align}
Using the momentum-conserving delta function, this can be simplified to the spherically symmetric form
\begin{align}
\int\dx\mathrm{PS}_{2,\Red{P}}=\frac{\left(\Red{P^2}\right)^{-\eps}}{2 (4\pi)^{2-2\eps}}
\int\dx\Omega_{3-2\eps}(\Blue{k})\,,
\end{align}
where $\int\dx\Omega_{3-2\eps}(\Blue{k})$ integrates over all angles of $k_1$ or equivalently $k_2$ in the center-of-mass system of the observed particles, where $k_1$ and $k_2$ are back-to-back.

For a given process, we need to calculate the angular integral over a squared matrix element,
\begin{align}
\int\dx\Omega_{3-2\eps}(\Blue{k})|\mathcal{M}|^2(\Red{p_1},\Red{p_2},\Red{q},\Blue{k_1},\Blue{k_2})\,,
\end{align}
where $|M|^2$ depends on the incoming momenta $p_{1,2}$, the final state observed momentum $q$ and the momenta $k_1$ and $k_2$ of the additional radiation unobserved; by momentum conservation it is $P=p_1+p_2-q=k_1+k_2$.
Here and in the following, ``observed'' means that the momentum is fixed for the angular phase-space integration over $k_{1,2}$.
In the center-of-mass-frame of $P$ by definition $\mathbf{P}=0$, so the particles $\mathbf{p_1}$, $\mathbf{p_2}$, and $\mathbf{q}$ are kinematically confined to a plane.\footnote{Bold letters indicate the spatial part of a vector in Minkowski space, $p=(E_p,\mathbf{p})$.}
Therefore, two angles suffice to parameterize any scalar product involving $k_1$ and $k_2$ with these vectors.
These two angles, $\theta_1$ and $\theta_2$, need to be made explicit in the integration measure $\dx\Omega_{3-2\eps}(k)$ by
\begin{equation}
\dx\Omega_{3-2\eps}(\Blue{k})=\dx\Blue{\theta_1}\sin^{1-2\eps}\Blue{\theta_1}\dx\Blue{\theta_2}\sin^{-2\eps}\Blue{\theta_2}\,\dx\Omega_{d-3}
\end{equation}
and require non-trivial integration over $|M|^2(\theta_1,\theta_2)$.
By the kinematic constraint, $|M|^2$ does not depend on any of the $d-3$ other spatial components of $k_{1,2}$ thus we can integrate them out which simply gives a constant factor $\Omega_{d-3}$.

For a tree-level matrix element -- the case we are concerned with for fully real corrections -- the complicated angle dependence lies in the denominators.
Any dependence in the numerator can be canceled against these or can be viewed as an inverse denominator.
Due to the restricted kinematics, products of angular dependent denominators are often reducible by \textit{partial fractioning} identities such as
\begin{align}
\frac{1}{p_1\cdot k_1\,p_1\cdot k_2}=\frac{1}{p_1\cdot P}\frac{p_1\cdot(k_1+k_2)}{p_1\cdot k_1\,p_1\cdot k_2}=\frac{1}{p_1\cdot P}\left[\frac{1}{p_1\cdot k_1}+\frac{1}{p_1\cdot k_2}\right]
\end{align}
or for massless $k_{1,2}$
\begin{align}
\frac{1}{p_1\cdot k_1\,p_2\cdot k_1\,q\cdot k_1}&=\frac{1}{P\cdot k_1}\frac{(p_1+p_2-q)\cdot k_1}{p_1\cdot k_1\,p_2\cdot k_1\,q\cdot k_1}\nonumber\\
&=\frac{2}{P^2}
\left[
-\frac{1}{p_1\cdot k_1\,p_2\cdot k_1}+\frac{1}{p_1\cdot k_1\,q\cdot k_1}+\frac{1}{p_2\cdot k_1\,q\cdot k_1}
\right],
\end{align}
where we used $0=k_2^2=(P-k_1)^2=P^2-2P\cdot k_1$.

For $2\rightarrow 1$ kinematics, performing the complete partial fractioning procedure until we hit irreducible denominator products, we end up with at most two denominators, raised to integer powers $j_{1,2}$. 
For most of the literature, these integrals were parameterized in an explicit coordinate system in the form
\begin{align}
\int_0^\pi\dx\Blue{\theta_1}\int_0^\pi\dx\Blue{\theta_2}\frac{\sin^{1-2\eps}\Blue{\theta_1}\sin^{-2\eps}\Blue{\theta_2}}{(\Red{a}+\Red{b}\cos\Blue{\theta_1})^{j_1}(\Red{A}+\Red{B}\sin\Blue{\theta_1}\cos\Blue{\theta_2}+\Red{C}\cos\Blue{\theta_1})^{j_2}}\,.
\label{eq:VanNeervenIntegral}
\end{align}
Due to their universality, these integrals, called \textit{van Neerven integrals} in this work, were tabulated for fixed integer powers and to a few orders in $\eps$ in the late 1980s and early 1990s and used ever since for a multitude of calculations in pQCD.

\begin{figure}
\centering
\includegraphics[width=0.8\textwidth]{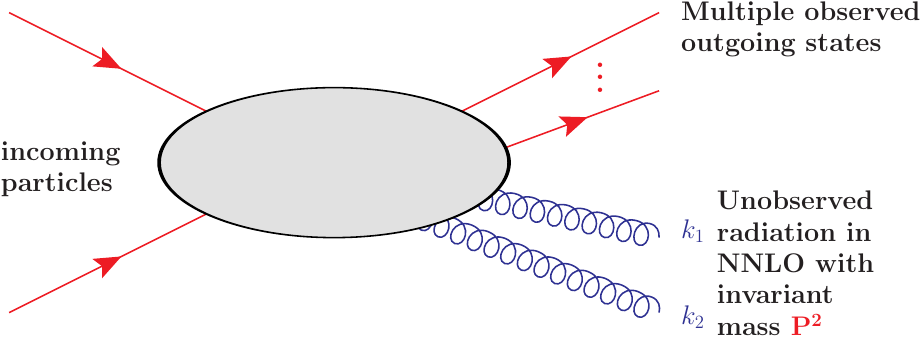}
\caption[Kinematics for processes requiring more than two denominators in the angular integral. Created with \texttt{JaxoDraw} \cite{Binosi:2003}.]{Kinematics for processes requiring more than two denominators in the angular integral. The gray blob symbolizes any way the external lines can be connected to a valid Feynman diagram.}
\label{fig:Kinematics_for_angular_integral_multidenom}
\end{figure}

Considering a kinematics with more than one observed particle, i.e., $2\rightarrow n+X$ with $n>1$ (see Figure \ref{fig:Kinematics_for_angular_integral_multidenom}), the van Neerven integral is no longer sufficient.
Starting from two observed outgoing momenta $q_1$ and $q_2$, the vectors $\mathbf{p_1}$, $\mathbf{p_2}$, $\mathbf{q_1}$, and $\mathbf{q_2}$ are no longer confined to a plane.
Hence, an additional angle $\theta_3$ appears in the scalar products with $k_1$ and $k_2$ that needs to be made explicit in the integration measure.
In general, if we have $n$ observed particles, we need to write 
\begin{align}
\int\dx\mathrm{PS}_{2,\Red{P}}=\frac{\left(\Red{P^2}\right)^{-\eps}}{2 (4\pi)^{2-2\eps}}
\int_0^\pi&\dx\Blue{\theta_1}\sin^{1-2\eps}\Blue{\theta_1}\dx\Blue{\theta_2}\sin^{-2\eps}\Blue{\theta_2}\dots\nonumber\\
\times&\dots\underbrace{\dx\Blue{\theta_{n+1}}\sin^{1-n-2\eps}\Blue{\theta_{n+1}}}_{\text{explicit }\theta \text{ dependence}}\,\,\underbrace{\dx\Omega_{2-n-2\eps}}_{\text{spher. sym.}}
\end{align}
and integrate a matrix element which depends on angles $\theta_{1,\dots,n+1}$.
Also, more independent momentum vectors mean that partial fractioning down to two denominators no longer works in general. 

The first systematic study of these multi-denominator angular integrals, using so-called \textit{Mellin-Barnes} integrals, was conducted by Gabor Somogyi in 2011 \cite{Somogyi:2011}.
He defined the general angular integral in $d$ dimensions in a parameterization independent and explicitly rotationally symmetric way as
\begin{align}
\Omega_{j_1,j_2,\dots,j_n}(\Red{v_1},\dots,\Red{v_n};d)\equiv\int\dx\Omega_{d-1}(\Blue{k}) \prod_{i=1}^n\frac{1}{(\Red{v_i}\cdot \Blue{k})^{j_i}}\,,
\end{align}
with $d$-vectors that are normalized by their energy component,
\begin{align}
&\Red{v_i}=(1,\Red{\mathbf{v}_i})\,,\\
&\Blue{k}=(1,\Blue{\mathbf{k}})
=(1,\dots,\cos\Blue{\theta_n}\prod_{i=1}^{n-1}\sin\Blue{\theta_i},\dots,\cos\Blue{\theta_2}\sin\Blue{\theta_1},\cos\Blue{\theta_1})
\,.
\end{align}
The common integral normalization, that avoids constants associated with the spherically symmetric part of the integration and is chosen such that the two-denominator integral has the same normalization as the van Neerven integrals, is
\begin{align*}
{I_{j_1,\dots,j_n}^{(\#\text{non-zero masses})}(v_{ij};\eps)}\equiv\frac{\Omega_{j_1,j_2,\dots,j_n}(v_1,\dots,v_n;4-2\eps)}{\Omega_{1-2\eps}}\,,
\end{align*} 
where the arguments, by the rotational symmetry of the angular integral, can only depend on the scalar products $v_{ij}=v_i\cdot v_j$ with $i\leq j$.
A ``mass'' in this context means $v_{ii}\neq 0$.

To use the integrals in a practical calculation, what we need is an expansion in the dimensional regularization parameter $\eps$ in a series of the form
\begin{align}
I_{j_1,\dots,j_n}^{(\#\text{non-zero masses})}(v_{ij};\eps)=\frac{c_{-1}(v_{ij})}{\eps}+c_0(v_{ij})+\eps\,c_1(v_{ij})+\eps^2\,c_2(v_{ij})+\dots\,.
\end{align}
The massless one-denominator integrals $I^{(0)}_{j_1}(\eps)$ are scaleless and can thus be calculated easily.
These were known since the formulation of angular integrals.
Starting from the massive one-denominator integral $I^{(1)}_{j_1}(v_{11},\eps)$ and massless two-denominator integral $I^{(0)}_{j_1,j_2}(v_{12},\eps)$, which are single-scale, the calculations become non-trivial.
Each further mass and further denominator adds more scales, quickly increasing the difficulty.
\begin{table}
\centering
\begin{tabular}{c|ccccc}
\backslashbox{\# denom.}{\# masses} & 0 & 1 & 2 & 3 & 4 \\
\hline
0 &  $\infty^+$ & & & & \\
1 &  $\infty^+$ & $\infty$ & & & \\
2 &  $\infty^+$ & $1$ & $0$ & & \\
3 &  $0$ & ? & ? & ? & \\
4 &  ? & ? & ? &  ? & ? \\
%\dots & \dots & \dots & \dots & \dots & \dots \\
%$n$ &  ? & ? & ? & ? & ?\\
\end{tabular}
\caption[Orders to which the $\eps$-expansion had been known before the work on this thesis.]{Orders to which the $\eps$-expansion had been known before the work on this thesis for different numbers of denominators and masses. $\infty$ denotes a known expansion to all orders. For the scaleless cases $(0,0)$ and $(1,0)$ this is a trivial result, for $(2,0)$ and $(1,1)$ it involves the expansion of a \textit{Gauss hypergeometric function}. The $(2,1)$ integral was known in $d$ dimensions in terms of an \textit{Appell function} but expanded only to order $\eps^1$. The $+$ denotes a form that is suited to deal with soft limits in which kinematic variables vanish.}
\label{tab:Status_of_eps_before}
\end{table}
The status of known $\eps$-expansion as of 2020 \cite{vanNeerven:1985,Beenakker:1988,Somogyi:2011}, before the work on this thesis, is shown in table \ref{tab:Status_of_eps_before}.
The apparent lack of interest in the case of more than two denominators can be explained by the restricted kinematics that was considered throughout the literature that allowed for partial fractioning down to two denominators.
\begin{block}[type=idea]
\textbf{Partial fraction decomposition:}\\
When denominators are linearly dependent, a reduction in the number of denominators is possible. The maximum number of linearly independent denominators is
\begin{align}
&\# \text{independent denominators in rest frame}
\nonumber\\
=&\,\,\# \text{incoming}  + \# \text{observed outgoing} - 1\,,
\end{align}
where the $-1$ is due to the rest frame condition. For example for $2\rightarrow 2+X$, this means we need to deal with three-denominator angular integrals.
\end{block}
Also, the double-massive two-denominator integral is finite in $d=4$ so in many cases higher orders in the expansion were not needed. Nevertheless, there were unsuccessful attempts to calculate it in the literature (see e.\,g. \cite{Somogyi:2011}).

In this thesis, we systematically extend the body of knowledge about angular integrals.
This starts with a systematic formulation of the partial fraction decomposition in terms of linear dependence that naturally leads to an elementary yet powerful identity to reorganize the products of denominators, which we called \textit{two-point splitting lemma}.

\begin{toolblock}{
\textbf{Two-point splitting lemma:}\\
Constructing a new energy-normalized vector $v_{(12)}$ as a linear combination of $v_1$ and $v_2$ as ${
v_{(12)}\equiv(1-\lambda_{(12)})\,v_1+\lambda_{(12)}\,v_2
}$
yields
	\begin{align}
		\frac{1}{v_1\cdot k\,\,v_2\cdot k}=\frac{\lambda_{(12)}}{v_1\cdot k\,\,v_{(12)}\cdot k}+\frac{1-\lambda_{(12)}}{v_2\cdot k\,\,v_{(12)}\cdot k}\,,
		\label{eq:twopointsplit}
	\end{align}
which holds for any value of the constant $\lambda_{(12)}$.
Now, we can impose the condition $v_{(12)}^2=0$ and by solving a quadratic equation, we can choose $\lambda_{(12)}$ such that $v_{(12)}$ is a massless vector. 
In this case, eq.\,\eqref{eq:twopointsplit} expresses the product of two massive denominators in terms of a sum of products with only a single mass each.
Integrating both sides results in a \textbf{mass reduction} for angular integrals. This identity also has an intuitive geometric representation that is introduced in publication \ref{pub:1} and extensively used in \ref{pub:4}.
}\end{toolblock}

The mass reduction formula directly leads to the first genuinely new result of this work: By expressing the double-massive two-denominator integral in terms of single-massive integrals, we are in the position to extend the result from $d=4$ to all orders in $\eps$ in terms of \textit{generalized polylogarithms} \cite{Goncharov:2001} -- using the corresponding, also novel, all-order single-mass result based on the $d$-dimensional result established by Somogyi \cite{Somogyi:2011}.
This constitutes one of the key outcomes of publication \ref{pub:1} and showcased the potential for improvement motivating further work.
Besides, in \ref{pub:1}, first attempts are made to transfer techniques from loop integration to angular integrals, such as \textit{integration-by-parts} (IBP) for the reduction to a small set of master integrals where $j_{1,2}=0,1$.
Such a transfer of powerful calculational technology, battle-tested in highly-complex multi-loop calculations, was a driving factor for much of the further progress in the modest study of angular integrals.

Despite publication \ref{pub:1} giving all-order $\eps$-expansions for all two-denominator integrals, that turned out not to be all that is needed.
In special cases, for example when one needs to integrate over one of the ``masses'' starting from the massless case, one needs a form that smoothly turns the massive version of the integral into the massless one.
This is a non-trivial problem since $\eps$-expansion and kinematic limits cannot be interchanged in general.
For the two-denominator case, that problem was solved in publication \ref{pub:4}, mainly by clever application of the two-point splitting lemma.

Continuing the transfer of methods developed for loop integration, the study of kinematic limits was systematized in publication \ref{pub:7}, by using \textit{expansion by regions} on angular integrals with multiple denominators.
Generalizing these results beyond the small-mass limit to the full kinematics was achieved for the three-denominator case in publication \ref{pub:8} using a good part of the arsenal of modern integral calculations: IBP relations, dimensional shifts, and differential equations.
The investigation of angular integrals concludes with publication \ref{pub:10}, where the results of publication \ref{pub:8} are pushed to four denominators and several aspects of the angular integral family for the general case of $n$ denominators are discussed, including a reduction to master integrals which works without the \textit{Laporta algorithm} \cite{Laporta:2000}, the standard tool for this task.
Furthermore, publication \ref{pub:10} uses the method of dimensional recurrence, known from the study of loop integrals, and extends it to a form that decomposes angular integrals into \textit{branch integrals} -- a novel concept -- that have fewer scales than the original integral.
Conceptually, this is the same kind of simplification that allowed the initial success with the double-massive two-denominator integral and indeed, branch integrals allowed to calculate the massless three-denominator integral to all orders in $\eps$ including the treatment of soft limits.

Overall, the work presented in this thesis pushed the status of angular integrals from the situation of table \ref{tab:Status_of_eps_before} to what is presented in table \ref{tab:Status_of_eps_after}.
Equipped with these new tools, the door is now open to apply angular integrals to kinematics beyond $2\rightarrow 1+X$.
\begin{table}
\centering
\begin{tabular}{c|ccccc}
\backslashbox{\# denom.}{\# masses} & 0 & 1 & 2 & 3 & 4 \\
\hline
0 &  $\infty^+$ & & & & \\
1 &  $\infty^+$ & $\infty^+$ & & & \\
2 &  $\infty^+$ & $\infty^+$ & $\infty^+$ & & \\
3 &  $\infty^+$ & $1$ & $1$ & $1$ & \\
4 &  $0$ & $0$ & $0$ & $0$ & $0$ \\
\dots & \dots & \dots & \dots & \dots & \dots \\
$n$ & $-1^\ast$ & $-1^\ast$ & $-1^\ast$ & $-1^\ast$ & $-1^\ast$\\
\end{tabular}
\caption[Orders to which the $\eps$-expansion is known after the work on this thesis.]{Orders to which the $\eps$-expansion is known after the work on this thesis for different numbers of denominators and masses. $\infty$ denotes a known expansion to all orders. The $+$ denotes a form that is suited to deal with soft limits in which kinematic variables vanish. The $\ast$ indicates conjectured results based on the structure observed for explicit values of $n$.}
\label{tab:Status_of_eps_after}
\end{table}

\subsection{Loop integrals: Branch cuts}
In between the work on phase-space integrals, several other projects were carried out in the course of the doctoral work presented here that are unified by the common objective to increase the analytic control over calculations in perturbative QCD.
Closest to the study of phase space integrals was the study of loop integrals, explicitly the investigation of the \textit{branch cut} structure of the massless one-loop box integral with a single off-shell particle.
A branch cut is associated with a discontinuous imaginary part in the complex plane. These imaginary parts in amplitudes can have very real consequences as we have seen with the Breit-Wigner resonance peak above.
While the box integral has been well-known in the literature for a long time, making the imaginary parts explicit has only been discussed to order $\eps$, but higher orders are relevant for NNLO calculations where additional poles are present.
In particular, the box integral suffers from spurious branch cuts that are present in parts of the result and only cancel in the sum. 
We were able to remove them by expressing the integral through a class of special functions called \textit{single-valued polylogarithms}.
This allowed us to present a unified result for all kinematic regions -- those relevant to DIS, Drell-Yan, and electron-positron annihilation -- in publication \ref{pub:2}.
This work was generalized shortly after to two diagonally opposite off-shell particles in publication \ref{pub:3}.
While not immediately phenomenologically relevant in its own right, since the result does not encompass all double off-shell boxes, this constitutes an interesting case of simplification by generalization:
Considering the double off-shell integral, which is more complicated on the surface in contrast to the single off-shell case, a hidden symmetry between the kinematic variables in the result is uncovered.

\subsection{Phenomenology: SIDIS and Drell-Yan}
The results from \ref{pub:2} turned out to be directly useful to phenomenology.
In publication \ref{pub:9}, we applied them to the recently published NNLO results for SIDIS.
These results included several case distinctions that could be traced back to the spurious branch cuts in the box integral.
Using this insight, we re-expressed the SIDIS coefficient functions through single-valued polylogarithms, which removed the case distinctions, and provided a unified and significantly more compact result.
This improved analytic form is both useful for more efficient numerical implementations as well as for further analytic handling, for example in context of a Mellin transform. 
Both these aspects are of immediate interest for using NNLO SIDIS results in future fits of polarized parton distribution functions that tell us about the spin content of the proton.

Another application to phenomenology, where imaginary parts of loop integrals played a role, was the study of so-called $T$-odd observables in the Drell-Yan process in publication \ref{pub:5}.
Here, a focus was on the transverse momentum spectrum, especially in the matching region between collinear and TMD physics.
In this context, we developed a power expansion in transverse momentum in the collinear framework that systematically extends to arbitrary higher powers where previous work stopped at leading power.

\subsection{Parton distribution functions: Evolution}
While the majority of publications deal with methods applicable to the hard part of the QCD interaction, in a side project we engaged with parton distributions directly.
Specifically, we introduced a novel semi-analytic method to solve the evolution equation. It uses a basis of analytic functions in the momentum fraction $x$ that is motivated by the analytic structure of the evolution equation and translates the integro-differential equation into a system of ordinary differential equations that is subsequently solved by a numeric matrix exponential.
For a proof-of-principle, we used this method in publication \ref{pub:6} to solve the DGLAP equation to leading order and showed its potential to build it into a practically useful evolution tool in the future -- either by incorporating higher orders into the evolution or by applying it to other evolution equations.
An interesting use case comes in combination with the power expansion of publication \ref{pub:5} that introduces derivatives of PDFs.
Here, analytic control over the $x$-space representation allows for analytic differentiation.
 
\subsection{Summary}
After this introduction has provided the necessary physical context for the themes of this thesis and briefly introduced the different lines of work, the next two chapters are dedicated to more technical aspects that are the foundations of the analytic methods developed in this thesis.
In chapter \ref{ch:SpecialFunctions}, we will put a key player for analytic methods into the spotlight, special functions, starting with a historical digression on how knowing about the right functions can shape science.
Afterwards, chapter \ref{ch:Methods} discusses selected analytic methods that are relevant for the central part of this thesis.
This comes with chapter \ref{ch:Publications}, where we go one by one through the publications that were produced during the doctoral work -- providing a summary of the problem they treat, the methods used, the results, and their implications.
The publications span a variety of new contributions to analytic methods in perturbative QCD and are, as already described above, concerned with phase-space integrals, loop integrals, application to phenomenology, and parton distribution functions.

To summarize, five of the ten publications are dedicated to elevating the state-of-the-art for angular integrals.
Transferring methods known from loop integrals, we extend the formerly limited knowledge to more masses, more external particles, and more control over the behavior in $d$ dimensions.

Two other publications are concerned with the imaginary part of loop integrals using a special class of functions called \textit{single-valued polylogarithms}.
In another publication, these results are then applied to simplify the expressions for the SIDIS coefficient functions at NNLO. 
Besides, methods for the systematic expansion of the Drell-Yan process for small transverse momentum and the semi-analytic evolution of parton distributions are discussed in a publication each.

The common theme is increased analytic control over processes relevant for collider phenomenology and the extraction of the proton structure.
To paraphrase the EIC proposal \cite{Accardi:2012} and put it in dramatic words:
\begin{block}[type=idea]
The aim of this thesis is to study analytic methods for perturbative QCD to understand the glue that binds us all.
\end{block}
\end{fancychapter2}
\cleardoublepage
\begin{fancychapter2}{Math and physics: Special functions}{Logarithms and friends}{Es ist nicht das Wissen, sondern das Lernen, was den größten Genuss gewährt. (It is not knowledge, but the act of learning, which grants the greatest enjoyment.)}{Carl Friedrich Gauss}
\label{ch:SpecialFunctions}
\vspace{-0.5cm}
\noindent{}The developments of mathematics and physics are deeply intertwined and there has always been a fruitful back and forth between new mathematics being invented driven by the motivation to solve a physics problem or new mathematical techniques being applied to better understand physics.
A list of examples includes calculus \cite{Leibniz1684NovaMethodus,Newton1711Analysis} and Newtonian mechanics \cite{Newton1833philosophiae}, differential geometry \cite{Gauss1828Disquisitiones,Riemann1868Hypothesen} and general relativity \cite{Einstein1915Feldgleichungen,Weyl1919RaumZeitMaterie} as well as functional analysis \cite{Hilbert1906Spektrum}\footnote{In a stroke of genius foresight, Hilbert coined the term \textit{spectrum} for the set of eigenvalues of an operator in analogy to the spectrum of light before it was later realized that indeed the spectrum of light emitted by atoms can be described by eigenvalues of an operator.} and quantum mechanics \cite{Schrodinger:1926}.

In this chapter, we want to put a spotlight on one important actor in this interplay of maths and physics that often plays a significant role: \textit{special functions}.
They are analogous to atoms in an analytic calculation.
Most of the time, they do not get the attention of elegant mathematical theories, and more often than not get only a mention in the appendix of a physics paper.
Due to their special importance in this thesis, here they get their own place in the story.

Special functions are what separates ``analytic'' and ``numerical'' calculations.\footnote{In the usual terminology of the field.}
If we succeed in expressing the results of a complicated calculation in terms of members of this exclusive club, we can understand the result by using the full body of knowledge we have for these functions -- including analytic behavior and efficient algorithms for the numeric calculation at any point.
There is no clear-cut definition of what is and what is not a special function.
The terminology is more of an operational one: If they are useful, appear in multiple types of calculations, or exhibit an interesting structure in their own right it is worth considering them as ``special'' enough to give them a name and study their properties.

We want to begin our story with the \textit{logarithm}, which for one is a prime showcase of the influence special functions can have on scientific discovery, and for another is the ancestor of a whole family of functions that play a central role in the calculation of scattering processes.

\section{The logarithm and the scientific revolution}
\label{sec:LogarithmKepler}
It is the year 1601 and Johannes Kepler \cite{Caspar1993KeplerOnline} -- who had studied at the University of Tübingen, where he learned about the heliocentric worldview -- gets appointed imperial astronomer in Prague by Rudolph II, emperor of the Holy Roman empire.
His task is to compile the most accurate astronomical tables of the time.
Kepler's predecessor on the position, Tycho Brahe\footnote{Who had just died, presumably because of a bladder disease induced by keeping etiquette and not going to the toilet when dining with the emperor.}, had left him with a huge body of observational data.
To convert these into orbits for the planets, a lot of calculation work was ahead.
At the time, the main bottleneck to scientific calculation was the multiplication and division of numbers with many digits.
This task was laborious and error prone \cite{Gingerich2009KeplerRudolphine}.

Crunching numbers for nine years, Kepler realized that the planetary orbits match ellipses with one focal point at the sun which they traverse in a way that sweeps equal areas of the ellipse in equal times.
These findings were later called his first and second law.
However, his work on the Rudolphine tables was far from over.
So tirelessly, Kepler kept on multiplying large numbers until 1617.

In this year, Kepler became aware of a recent invention from Scotland, made by John Napier in 1614: the \textit{logarithm} \cite{Napier1614Descriptio}.
It has the property that for any positive numbers $x$ and $y$, the logarithm of their product and quotient is the sum and difference of their logarithms, respectively, i.e.
\begin{equation}
\log(x\cdot y)=\log(x)+\log(y)\quad\text{and}\quad\log\!\left(\frac{x}{y}\right)=\log(x)-\log(y).
\label{eq:LogFunctionalEquation}
\end{equation}
In modern language, these are \textit{functional equations}, for Kepler it was a divine aid for calculation:
Kepler was in possession of a large list of numbers and their corresponding logarithms -- compiled by Napier or Henry Briggs \cite{Briggs1617ChiliasPrima} and later made his own.
Now, instead of calculating products and quotients of the numbers $x$ and $y$, he could look up the values of $\log(x)$ and $\log(y)$ in the logarithm tables and add or subtract those values -- which is a lot easier than multiplication or division\footnote{Recall your experience from elementary school.}.
Then he needed to go through the logarithmic column of the table to find the value that produced this particular $\log$-value.
By Eq.\,\eqref{eq:LogFunctionalEquation}, this number is the product $x\cdot y$ or quotient $x/y$, that he wanted to calculate.

Analyzing astronomical data just got a lot easier and more reliable.
Every complicated multiplication could be replaced by an addition and two table look-ups.
Thanks to this invention, Kepler finally managed to complete his monumental work and could publish the Rudolphine tables in 1627 \cite{Kepler1627RudolphineTables}, something that he most probably would not have been able to achieve during his lifetime if it had not been for the logarithm.\footnote{A remark commonly attributed to Pierre-Simon Laplace has it that logarithms doubled the lifetime of an astronomer \cite{ClarkMontelle2011Logarithms}.}

Besides its computational power, the logarithm also offered conceptual insight.
For example, after Kepler had figured out planetary orbits were ellipses around the sun, he had data for their distance to the sun and their orbital periods given in table \ref{tab:Kepler_Planet_data}.

\begin{table}[ht]
\centering
\begin{tabular}{c|cc}
\toprule
\toprule
Planet & Mean distance to sun in AU & Orbital period in years \\
\midrule
Mercury & 0.388 & 0.2403 \\
Venus & 0.724 & 0.6152 \\
Earth & 1.000 & 1.0000 \\
Mars & 1.524 & 1.8808 \\
Jupiter & 5.200 & 11.8621 \\
Saturn & 9.510 & 29.4571 \\
\bottomrule
\bottomrule
\end{tabular}
\caption[Kepler's planetary data extracted from Tycho Brahe's observations (based on \cite{Kepler1619Harmonices}).]{Kepler's planetary data extracted from Tycho Brahe's observations \cite{Kepler1619Harmonices}.}
\label{tab:Kepler_Planet_data}
\end{table}

At first glance, there is no obvious connection between these numbers and also when plotting them in a graph as in Fig.\,\ref{fig:Kepler_data} there is nothing that directly catches the eye.
However, something astonishing happens when one takes Kepler's data and converts both entries of Tab.\,\ref{tab:Kepler_Planet_data} with his logarithm tables and graphs these new numbers:
As depicted in Fig.\,\ref{fig:3rd_law_loglog}, the ``stars'' literally align -- with a slope of $3/2$!

\begin{figure}[h]
\centering
\begin{subfigure}[t]{0.4\textwidth}
\centering
\includegraphics[width=\textwidth]{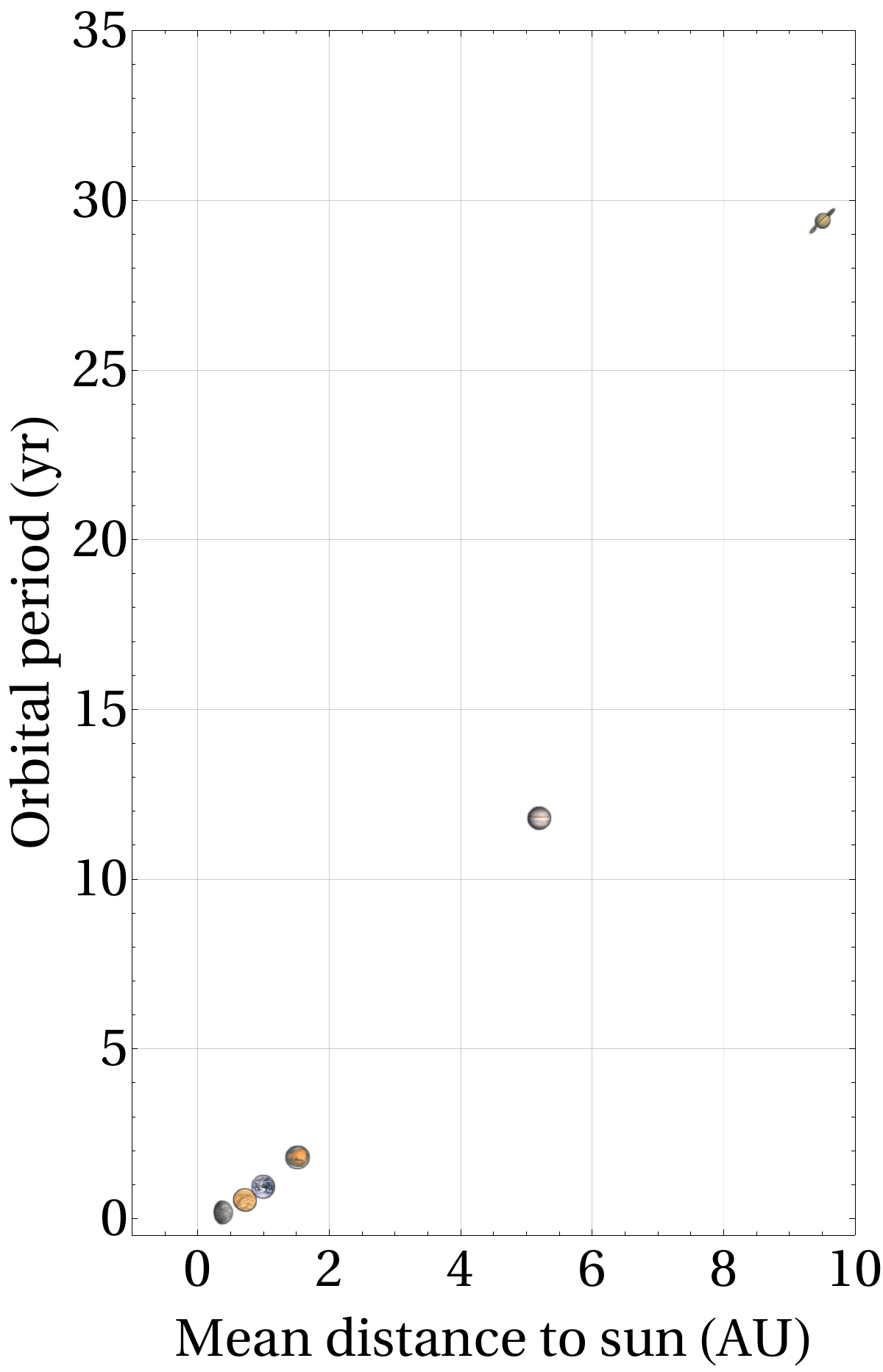}
\caption{Graphical representation of Kepler's planetary data. Besides a general trend, there is no obvious relation between distance and orbital period that catches the eye.}
\label{fig:Kepler_data}
\end{subfigure}
\hspace{3em}
\begin{subfigure}[t]{0.4\textwidth}
\centering
\includegraphics[width=\textwidth]{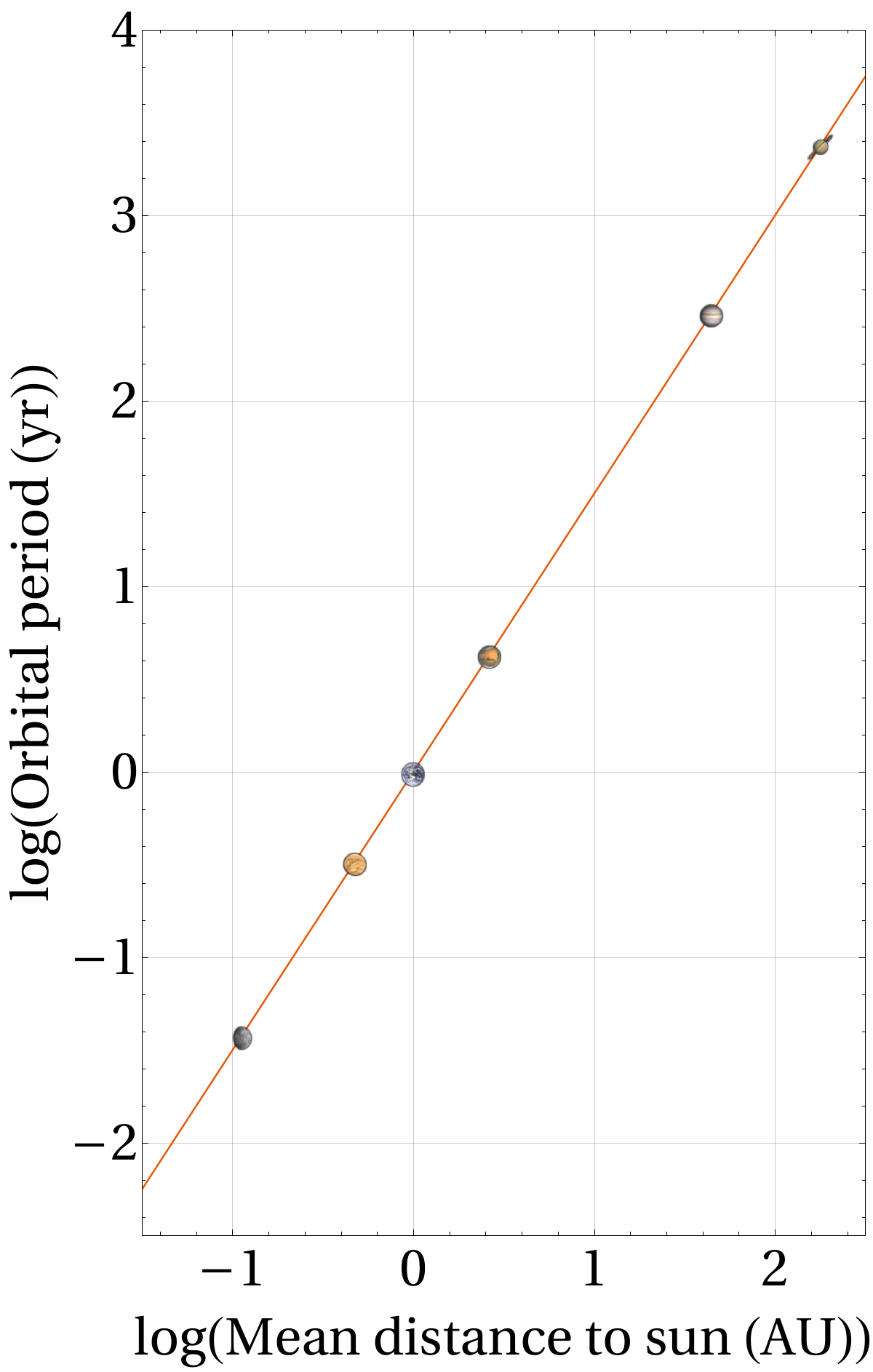}
\caption{Kepler's planetary data represented by taking the logarithm for mean distances to the sun and orbital periods. A straight line with slope of $3/2$ fits the data to extremely good precision. Only later it was found that instead of the mean distance it is more accurate to take the large semi-major axis of the ellipse -- the modern form of Kepler's third law.}
\label{fig:3rd_law_loglog}
\end{subfigure}
\caption[Graphical representation of Kepler's planetary data on linear and logarithmic scale. Created with \texttt{Mathematica} \cite{Mathematica13}.]{Laws of nature from logarithms: What looks complicated in the left plot becomes apparent in the right one.}
\end{figure}

It is not known whether Kepler really did this exercise, but it is quite plausible.
He discovered this relation -- which later became known as Kepler's third law\footnote{After it was understood that ``mean distance'' is to be replaced by the orbit's semi-major axis, which however is nearly the same for the planets in the solar system due to the low eccentricity of their orbits.} -- in 1619 \cite{Kepler1619Harmonices}, ten years after the other laws and only after logarithms had become part of his routine toolset.

The importance of logarithms to his work is showcased by the frontispiece artwork in the Rudolphine tables\footnote{In the artwork there is also a depiction of Kepler himself, working tirelessly in the basement of the science temple while high above the imperial eagle is performing the most important task -- spreading funding.}. Atop the temple of astronomy stands a muse with the number 6931472 in her halo, in modern notation the first digits after the decimal point\footnote{Decimal points were not a thing in Kepler's times, instead all numbers were multiplied by a sufficiently large multiple of 10 to perform the calculation with integers.} for the logarithm of $2$,
\begin{align}
\log(2)=0.6931472\dots
\end{align}
To Kepler it was the number he needed to add or subtract when doubling or halving values with Napier's tables, respectively.

The discoveries of Kepler paved the way for Newton to develop his theory of gravity in his \textit{Principia}, giving an explanation for Kepler's empirical laws.
In the meantime the logarithm spread in science and beyond.
The Royal Navy was sailing the world on logarithm tables for fast position determination \cite{Waters1958ArtOfNavigation}, the slide-rule, translating the table look-up into a mechanical device, became a core accessory for engineers up until the electronic computer took over -- when landing on the moon, Apollo 11 still had one on board as a backup \cite{OughtredSocietySlideRuleHistory}.

Now that we are hopefully convinced about the importance of the logarithm, we are sufficiently motivated to look at its mathematical properties.
This will lead naturally to systematic generalizations, first to the dilogarithm, then the polylogarithm, and finally generalized polylogarithms. 
All these functions are ubiquitous in perturbative calculations in QFT, where they arise in the calculation of both loop- and phase-space integrals.
In the context of dimensional regularization they appear as the coefficient functions in an $\eps$-expansion of those integrals.
Besides their use in physics, they are also subjects of study in pure mathematics. Their rich structure allows for deep insights in fields ranging from non-euclidean geometry to analytic number theory. 
For the phenomenologist it is worth studying the properties of these functions, since they often allow for considerable reduction in the length of analytic results for scattering amplitudes.

In modern mathematical language, the logarithm can be defined\footnote{Of course, it is also the inverse of the exponential function, $\ee^{\log(x)}=x$.} as
\begin{align}
\log x=\int_1^x\frac{\dx t}{t}\,.
\label{eq:log_def_Integral}
\end{align}
Using the geometric series
\begin{align}
\sum_{n=0}^\infty t^n=\frac{1}{1-t}
\end{align}
that holds for $|t|<1$ it is quickly established that the logarithm can also be represented as an infinite series in the form
\begin{align}
\log(1+x)=\sum_{n=1}^\infty\frac{(-1)^{n+1}x^n}{n}\,.
\label{eq:log_Series}
\end{align}

\begin{wrapfigure}{r}{0.5\textwidth}
\includegraphics[width=0.5\textwidth]{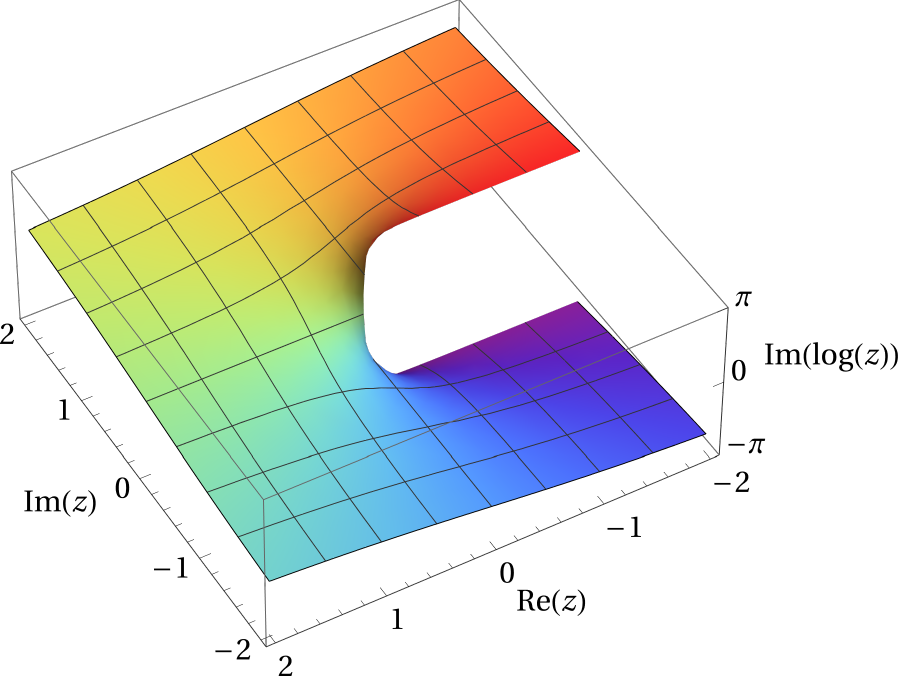}
\caption[Branch cut of the logarithm in the complex plane. Created with \texttt{Mathematica} \cite{Mathematica13}.]{The branch cut of the logarithm in the complex plane. At the negative real axis the imaginary part has a discontinuity.}
\label{fig:branchcut}
\end{wrapfigure}
For a deeper understanding of the analytical behavior of the logarithm, as for many special functions, it is highly beneficial to view it as a function of a complex variable.
Here, the integral definition in the form of Eq.\,\eqref{eq:log_def_Integral} is especially easy to generalize:
For complex argument $x$ we can integrate from $1$ to $x$, for definiteness on a straight line -- even though, by the rules of complex analysis, the integration path does not matter as long as we do not circle around the singularity at $t=0$.
The singularity at $t=0$ makes it such that we cannot reach the real negative axis from $1$ by a straight line.
Hence, the integration path must circle around it either above or below the singularity. 
This leads to a different value of the integral depending on the choice of path and results in a discontinuity of $\log(x)$ along the negative real axis.
Comparing values for $x<0$ with small positive and negative imaginary parts, denoted as $\pm\iu 0$, respectively, we have
\begin{equation}
\log(x+\iu 0)-\log(x-\iu 0)=2\pi \iu\,.
\end{equation}
Such a discontinuity, depicted in figure \ref{fig:branchcut}, is called \textit{branch cut} and will be front and center in publications\,\ref{pub:2} and \ref{pub:3}.

The logarithm also allows to generalize powers beyond rational exponents.
Taking the exponential on both sides of another functional equation of the logarithm\footnote{Kepler could make use of this equation to simplify the calculation of powers and roots.}
\begin{align}
\log(x^a)=a\log x
\end{align}
leads to the definition
\begin{equation}
x^a\equiv\ee^{a\log(x)}\,
\end{equation}
which is well defined also for non-rational $a$.
\section{The dilogarithm: Not revolutionary, yet delightful}
When the logarithm changed the world, how bad can it be to look for generalizations?
From the Taylor series of the logarithm in eq.\,\eqref{eq:log_Series}, we see that the function
\begin{align}
\mathrm{Li}_1(x)\equiv-\log(1-x)=\sum_{n=1}^\infty \frac{x^n}{n}
\label{eq:Li1 series}
\end{align}
has a particularly simple series representation.
Integrating this function and its series representation termwise with $\int\dx x/x$, Gottfried Wilhelm Leibniz, in letters to Johann Bernoulli in 1696 \cite{Bogner:Habilitation2017}, defined the \textit{dilogarithm} -- not yet under this name or with the $\dilog$ notation -- as
\begin{equation}
\mathrm{Li}_2(x)\equiv-\int_0^x\frac{\dx t}{t}\,\log(1-t)=\sum_{n=1}^\infty\frac{x^n}{n^2}\,.
\label{eq:dilog_def}
\end{equation}

Even though, in the following centuries, this function was studied by several famous mathematicians including Leonhard Euler (1776) \cite{Euler1776_Meditationes}, William Spence (1809) \cite{Spence1809}, Carl Johann Hill (1828) \cite{Hill1828}, Thomas Clausen (1832) \cite{Clausen:1832}, Ernst Kummer (1837) \cite{Kummer1837}, Niels Abel (1881) \cite{Abel1881}, Leonard Rogers (1907) \cite{Rogers1907}, and Niels Nielsen (1909) \cite{Nielsen1909}, the dilogarithm did not have the impact of the logarithm\footnote{It has not made it to school curricula in over 300 years.}, and was virtually unheard of in physics before the advent of QFT.
Since then, it immensely grew in popularity -- still confined to the circles of mathematicians and theoretical physicists, of course.
While mathematicians recognized its utility in hyperbolic geometry and number theory \cite{Zagier1991,Lewin1991}, it entered particle physics when it showed up in the calculation of loop integrals in \cite{Eriksson1961,nikishov1961radiative,tHooft:1978}.
While often referred to as \textit{Spence function} in the older literature, thanks to the efforts of Leonard Lewin and his classic book \cite{Lewin:1981}, \textit{dilogarithm} became the standard name and $\dilog$ the standard notation.

While the series definition only converges for $|x|<1$, the integral is an analytic continuation to the complex plane except for the branch cut inherited from $\log(1-x)$, which sits on the positive real axis starting at $z=1$.
Don Zagier, once the youngest professor in Germany, called the dilogarithm \textit{the only function with humor} due to its delightful mix of elementary definition and rich non-trivial structure \cite{Zagier1991}.
Most prominently, there are the functional equations of the dilogarithm.
For one, the inversion identity
\begin{align}
\dilog\!\left(\frac{1}{x}\right)=-\dilog(x)-\frac{\pi^2}{6}-\frac{1}{2}\log^2(-x)\,,
\end{align}
and for another the reflection identity
\begin{align}
\dilog(1-x)=-\dilog(x)+\frac{\pi^2}{6}-\log x \log(1-x)\,.
\end{align}
Putting these together, they connect six arguments of the dilogarithm, and up to constants and logarithmic terms,
\begin{align}
\dilog(x),\;\dilog\!\left(\frac{1}{1-x}\right),\;\dilog\!\left(\frac{x-1}{x}\right),\;-\dilog\!\left(\frac{1}{x}\right),\;-\dilog(1-x),\text{ and }-\dilog\!\left(\frac{x}{x-1}\right)
\end{align}
are identical.
The most interesting of the dilogarithm identities is \textit{Abel's five term relation}
\begin{align}
&\Li{2}{x}+\Li{2}{y}+\Li{2}{\frac{1-x}{1-x y}}+\Li{2}{1-x y}+\Li{2}{\frac{1-y}{1-x y}}\nonumber\\
=&\frac{\pi^2}{2}-\log x \log(1-x)-\log y\log(1-y)-\log\!\left(\frac{1-x}{1-x y}\right)\log\!\left(\frac{1-y}{1-x y}\right).
\label{eq:Abel5Term}
\end{align}
All single variable identities are valid in the entire complex plane as long as no argument is directly on the branch cut; analytic continuation of the 5-term relation may require additional terms.

Constructing functions that have the same functional equations as the dilogarithm -- without the logarithmic product terms -- but no branch cut leads to \textit{single-valued} versions of the dilogarithm.
These are discussed in publication \ref{pub:2}, and used again in publications \ref{pub:3} and \ref{pub:9}.
The imaginary part of the dilogarithm gives rise to the \textit{Clausen function} which is reviewed and used in publication \ref{pub:8}.

\begin{readingblock}{
\textbf{Further reading: } To start with the polylogarithm literature and fall in love with special functions, we warmly recommend Lewin's delightful book on the subject \cite{Lewin:1981}.
}\end{readingblock}
\section{Polylogarithms and generalizations}
\noindent{}\textbf{Classical Polylogarithms: }A natural generalization of the dilogarithm leads to the \textit{polylogarithms} \cite{Lewin:1981}
\begin{align}
\Li{s}{x}\equiv\sum_{n=1}^\infty\frac{x^n}{n^s}\,.
\label{eq:Lis series}
\end{align}
For the purpose of perturbative quantum field theory the case $s\in\mathbb{Z}$ is of interest, where the maximum value of $s$ that appears, corresponds to the order of perturbation theory one is working with.
At $x=1$ they take the value of the famous \textit{Riemann Zeta function}
\begin{align}
\Li{s}{1}=\sum_{n=1}^\infty\frac{1}{n^s}=\zeta(s)\,.
\end{align}
Those zeta values at positive integers appear as numerical constants in perturbative calculations and are commonly written as $\zeta(n)=\zeta_n$.
For the zeta values at even integers the well-known identity due to Euler connects them to rational multiples of $\pi^{2n}$,
\begin{align}
\zeta_{2n}=\frac{(2\pi)^{2 n}(-1)^{n+1}\mathrm{B}_{2n}}{2(2n)!}\,,
\label{eq:even_zeta_values}
\end{align}
where $B_n$ is the $n$-th \textit{Bernoulli number} defined as the $n$-th Taylor coefficient of $\frac{t}{\ee^t-1}$.
Remarkably, there is no such identity for $\zeta_{2n+1}$.\footnote{A transparent proof of Eq.\,\eqref{eq:even_zeta_values} relates the $\sum_{k=1}^\infty1/k^{2n}$ by symmetry to a sum over all integers and subsequently in a series of poles of a complex contour integral. A similar construction for odd integer values fails since the sum over positive and negative integers cancels.}
In fact $\zeta_3$, $\zeta_5$, $\zeta_7$, $\zeta_9$,\dots are conjectured to be algebraically independent of each other and $\pi$.
However, only for Apery's constant $\zeta_3$ has irrationality been proven by Roger Apéry in 1979\footnote{Granting him naming rights to $\zeta_3$.}.

The number of functional identities is much sparser among higher polylogarithms $\Li{n}{x}$.
Importantly, there is always the inversion relation
\begin{align}
		\Li{n}{z} \,+\, (-1)^n\,\Li{n}{\frac{1}{z}} \,=\, -\frac{\ln^n(-z) }{n!} \,+\, 2\sum_{k=1}^{\lfloor n/2\rfloor} \frac{\ln^{n-2k}(-z)}{(n-2k)!}\,\Li{2k}{-1}\,,
	\end{align}
which holds for $z\in\mathbb{C}\setminus[0,\infty)$ and finds explicit application in publication \ref{pub:2}.
However, as we will see shortly, there are plenty of relations for classical polylogarithms if we view them as members of a more general class of functions.

Starting from \eqref{eq:Li1 series} one can iteratively define the polylogarithm via the integral
\begin{align}
\Li{n+1}{x}=\int_0^x\frac{\dx t}{t}\,\Li{n}{t}\,.
\label{eq:PolyLogIntIterative}
\end{align}
This will open up a systematic path to generalization by allowing for integrating singularities other than $1/t$.
Alternatively, there is also the integral representation
\begin{align}
\Li{n}{x}=\frac{(-1)^{n-1}}{(n-1)!}\int_0^1 \frac{\dx t}{t}\,\log^{n-1}(t)\log(1-x t)\,.
\label{eq:PolyLogIntNielsen}
\end{align}

\noindent{}\textbf{Nielsen Polylogarithms: }
The natural generalization of the integral eq.\,\eqref{eq:PolyLogIntNielsen} is to allow for higher powers of $\log(1-x t)$. This gives rise to introducing \textit{Nielsen-polylogarithms} \cite{Nielsen1909,Kolbig:1983} defined as
\begin{align}
\Snp{n,p}{x}=\frac{(-1)^{n+p-1}}{(n-1)!p!}\int_0^1 \frac{\dx t}{t}\,\log^{n-1}(t)\log^p(1-x t)\,.
\end{align}
The choice of prefactors is explained by the identity $\int_0^1\dx t (-\log t)^n=n!$\,.

\noindent{}\textbf{Harmonic Polylogarithms: }
A way to generalize the iterative integration eq.\,\eqref{eq:PolyLogIntIterative} is to also allow for the functions $f(1;t)=1/(1-t)$ and $f(-1;t)=1/(1+t)$ in replacement of $f(0;t)=1/t$.
This idea leads to the definition of \textit{Harmonic polylogarithms} \cite{Remiddi:1999ew} by setting
\begin{align}
\mathrm{H}(\vec{0}_w;x)=\frac{1}{w!}\log^w(x)
\end{align}
together with the recursion 
\begin{align}
\mathrm{H}(\vec{m}_w;x)=\int_0^x\dx t f(a; t)\,\mathrm{H}(\vec{m}_{w-1};t)
\end{align}
for $\vec{m}_w\neq 0$, where $a$ is the leftmost index of $\mathrm{H}(\vec{m}_w;x)$.

\noindent{}\textbf{Multiple Polylogarithms: }
Starting from the series representation \eqref{eq:Li1 series} yet another direction of generalization is to allow for two arguments $x$, $y$, or possibly more.
The double-sum
\begin{align}
\sum_{n1,n_2=1}^\infty \frac{x^{n_1}y^{n_2}}{n_1 n_2}=\log(1-x)\log(1-y)\,,
\label{eq:logproduct}
\end{align}
does not give anything new. Hence, one restricts the second summation and defines
\begin{align}
\Li{1,1}{x,y}=\sum_{n_1>n_2>0}\frac{x^{n_1}y^{n_2}}{n_1 n_2}
\end{align}
as the simplest example of a \textit{multiple polylogarithm}\, \cite{Goncharov:2001}. Studying functions of this kind provides a setting that encompasses all the other logarithms we mentioned so far. From the above examples we already see that there is an interplay between series and integral representations.
This results in a rich structure among the polylogarithms.
As a first simple example of this, we notice that we can decompose the summation from eq.\eqref{eq:logproduct} in the following form
\begin{align}
\Li{1}{x}\Li{1}{y}&=\sum_{n1,n_2=1}^\infty \frac{x^{n_1}y^{n_2}}{n_1 n_2}=\sum_{n_1>n_2>0} \frac{x^{n_1}y^{n_2}}{n_1 n_2}+\sum_{n2>n1>0} \frac{x^{n_1}y^{n_2}}{n_1 n_2}+\sum_{n=1}^\infty \frac{(x y)^n}{n^2}
\nonumber\\
&=\Li{1,1}{x,y}+\Li{1,1}{y,x}+\Li{2}{xy}\,.
\label{eq:Li11 stuffle}
\end{align}
This \textit{stuffle relation} eq.\eqref{eq:Li11 stuffle}, together with the \textit{depth reduction} of $\Lin{1,1}$ in terms of $\Lin{2}$,\footnote{Eq.\,\eqref{eq:Li11reduction} can be established by a short calculation conveniently starting from the sum representation of $\Li{1,1}{x,y}+\Li{2}{x y}$ and turning it into an integral via $y^{n_2}/n_2=\int_0^y\dx t\,t^{n_2-1}$.} 
\begin{align}
\Li{1,1}{x,y}=-\Li{2}{x y}-\Li{2}{\frac{x}{x-1}}+\Li{2}{\frac{x(1-y)}{x-1}}\,,
\label{eq:Li11reduction}
\end{align}
directly gives a five term relation for $\Lin{2}$ equivalent to eq.\,\eqref{eq:Abel5Term}.

Generally, multiple polylogarithms are defined for $|x_i|<1$ by \textit{multiple nested sums} 
\begin{align}
\Li{m_1,\dots,m_k}{x_1,\dots,x_k}=\sum_{n_1>n_2>\dots>n_k>0}\frac{x_1^{n_1}\dots x_k^{n_k}}{n_1^{m_1}\dots n_k^{m_k}}\,.
\end{align}
Note that in the literature there is also the definition with the opposite convention $0<n_1<n_2<\dots<n_k$, which results in a reversed notation of indices and arguments.

The number of indices is called the \textit{depth} of the multiple polylogarithm, the sum of indices $\sum_{i=1}^k m_i$ its \textit{weight}.
The weight is an invariant preserved in identities between multiple polylogarithms, where one sets the weight of a product of two terms as the sum of their individual weights.
As an example note that in eqs.\,\eqref{eq:Abel5Term} and \eqref{eq:Li11 stuffle} all terms on both sides are of weight two.\footnote{Note that since $\dilog(1)=\zeta_2$ we assign weight two also to $\zeta_2$. Such an extension of the concept of weight to constants gives predictive power over which constants may appear in calculations from the structure of the associated integrals alone.}

These sums directly generalize the regular polylogarithm and Nielsen polylogarithms via
\begin{align}
\Snp{n,p}{x}=\Li{n+1,\scriptsize{\underbrace{1,\dots,1}_{p-1}}}{x,1,\dots,1}\,.
\end{align}
Harmonic polylogarithms with letters $\sigma_i=\pm1$ and $N_-$ cases of $-1$ can be represented as
\begin{align}
\mathrm{H}(\underbrace{0,\dots,0}_{m_1-1},\sigma_1,\dots,\underbrace{0,\dots,0}_{m_k-1},\sigma_k;x)=(-1)^{N_-}\Li{m_1,\dots,m_k}{\frac{x}{\sigma_1},\frac{\sigma_1}{\sigma_2}\dots,\frac{\sigma_{k-1}}{\sigma_k}}\,.
\end{align}

A further generalization of multiple polylogarithms are the $\mathrm{Z}$-sums \cite{Moch2002}. They are defined as
\begin{align}
\mathrm{Z}_{m_1,\dots,m_k}(n;x_1,\dots x_k)=\sum_{n\geq n_1>n_2>\dots>n_k>0}\frac{x_1^{n_1}\dots x_k^{n_k}}{n_1^{m_1}\dots n_k^{m_k}}\,
\end{align}
and provide a unifying setting to study multiple polylogarithms $\Li{m_1,\dots,m_k}{x_1,\dots,x_k}=\mathrm{Z}_{m_1,\dots,m_k}(\infty;x_1,\dots x_k)$, \textit{Euler-Zagier sums} $\mathrm{Z}_{m_1,\dots,m_k}(n)=\mathrm{Z}_{m_1,\dots,m_k}(n;1,\dots 1)$, and \textit{multiple zeta values} $\zeta(m_1,\dots,m_k)=\mathrm{Z}_{m_1,\dots,m_k}(\infty;1,\dots 1)$.

\noindent{}\textbf{Generalized Polylogarithms:}
To define multiple polylogarithms outside the domain where the defining sum converges,  an integral representation is useful. This is achieved by so-called \textit{iterated integrals} -- also known as \textit{generalized} or \textit{Goncharov polylogarithms (GPLs)} \cite{Borwein1998,Goncharov:2001} -- which are a natural generalization of the integral definition of harmonic polylogarithms.
We define recursively \cite{Duhr2012}
\begin{align}
G(a_1,\dots a_n;x)=\int_0^x\frac{\dx t}{t-a_1}\,G(a_2,\dots,a_n;t)\,,
\end{align}
with $G(x)=G(;x)=1$. The special case where all the $a_i$ are zero is defined by
\begin{align}
G(\vec{0}_n;x)=\frac{1}{n!}\log^n x\,.
\end{align}
Using the notation \cite{Duhr2012}
\begin{align}
\mathrm{G}_{m_1,\dots,m_k}(t_1,\dots,t_k)=\mathrm{G}(\underbrace{0,\dots,0}_{m_1-1},t_1,\dots,\underbrace{0,\dots,0}_{m_k-1},t_k;1)
\end{align}
we can express multiple polylogarithms in terms of iterated integrals via \cite{Frellesvig2016}
\begin{align}
\Li{m_1,\dots,m_k}{x_1,\dots,x_k}=(-1)^k\,\mathrm{G}_{m_1,\dots,m_k}\left(\frac{1}{x_1},\dots,\frac{1}{x_1\cdots x_k}\right)\,.
\end{align}
The inverted relation reads \cite{Frellesvig2016}
\begin{align}
G_{m_1,\dots,m_n}(a_1,\dots,a_n;x)=
(-1)^n \Li{m_1,\dots,m_n}{\frac{x}{a_1},\frac{a_1}{a_2},\dots,\frac{a_{n-1}}{a_n}}\,.
\end{align}

\noindent{}\textbf{Structural relations:}
Multiple polylogarithms are not independent of each other but satisfy a plethora of algebraic relations. The most important ones include
\begin{itemize}
\item[(a)] \textit{Stuffle relation} \cite{Frellesvig2016}:
\begin{align}
\Li{\vec{m}_1}{\vec{x}_1}\Li{\vec{m}_2}{\vec{x}_2}
=\sum_{\vec{m}\in\vec{m}_1\star\vec{m}_2}\Li{\vec{m}}{\vec{x}}\,,
\end{align}
where $\vec{m}_1\star\vec{m}_2$ is the \textit{stuffle product} of $\vec{m}_1$ and $\vec{m}_2$. It is defined by the relations $1\star\vec{m}=\vec{m}\star 1=\vec{m}$, for any set of indices $\vec{m}$, and inductively for indices $a$, $b$ and sets of indices $\vec{m}_1$, $\vec{m}_2$
\begin{align}
(a,\vec{m}_1)\star (b,\vec{m}_2)=(a,\vec{m}_1\star (b,\vec{m}_2))+(b,(a,\vec{m}_1)\star \vec{m}_2)+(a+b,\vec{m}_1\star \vec{m}_2)\,.
\end{align}
The corresponding argument $\vec{x}$ is constructed by multiplying $x_i$ and $x_j$ if the indices $m_i$ and $m_j$ are added in the stuffle. For example
\begin{align}
\Li{1}{x}\Li{1}{y}=&\Li{1,1}{x,y}+\Li{1,1}{y,x}+\Li{2}{x y}\,,\\
\Li{2}{x}\Li{1}{y}=&\Li{2,1}{x,y}+\Li{1,2}{y,x}+\Li{3}{x y}\,,\\
\text{or}\quad
\Li{1,1}{x,y}\Li{1}{z}=&\Li{1,1,1}{x,y,z}+\Li{1,1,1}{x,z,y}+\Li{1,1,1}{z,x,y}
\nonumber\\
&+\Li{2,1}{xz,y}+\Li{1,2}{x,yz}\,.
\end{align}
\item[(b)] \textit{Shuffle relation} \cite{Duhr2012}:
\begin{align}
\mathrm{G}(\vec{a};z)\mathrm{G}(\vec{b};z)=\sum_{\vec{c}\in\vec{a}\Sh\vec{b}}\mathrm{G}(\vec{c};z)\,,
\end{align}
where $\vec{a}\Sh\vec{b}$ denotes the set of all \textit{shuffles} of the vectors $\vec{a}$ and $\vec{b}$, i{.}e{.} all the ways of interlacing them while keeping the order within $\vec{a}$ and $\vec{b}$. For example
\begin{align}
\mathrm{G}(a,b;z)\mathrm{G}(c,d;z)=&\mathrm{G}(a,b,c,d;z)+\mathrm{G}(a,c,b,d;z)+\mathrm{G}(a,c,d,b;z)+\mathrm{G}(c,a,b,d;z)\nonumber\\
+&\mathrm{G}(c,a,d,b;z)+\mathrm{G}(c,d,a,b;z)\,.
\end{align}
\item[(c)] \textit{Rescaling invariance} \cite{Duhr2012}: If $a_n\neq 0$ it holds for any $k\in\mathbb{C}^*$
\begin{align}
\mathrm{G}(k\vec{a};k x)=\mathrm{G}(\vec{a};x)
\end{align}
\item[(d)] \textit{H\"older convolution} \cite{Duhr2012}: If $a_1\neq 1$ and $a_n\neq 0$ it holds
\begin{align}
\mathrm{G}(a_1,\dots,a_n;1)=\sum_{k=0}^n(-1)^k\mathrm{G}\left(1-a_k,\dots,1-a_1;1-\frac{1}{p}\right)\mathrm{G}\left(a_{k+1},\dots,a_n;\frac{1}{p}\right)\,.
\end{align}
\end{itemize}

It is conjectured that every multiple polylogarithm with at least one index equal to one can be reduced to multiple polylogarithms of lower depth.\footnote{According to Rudenko, who recently proved a version of the conjecture \cite{Rudenko2020}, it was first formulated by Goncharov in \cite{Goncharov:2001}. The form stated here is taken from Gangl, who attributes it to Goncharov \cite{Duhr2012}.}
As a special case, this conjecture implies that up to weight 3 all multiple polylogarithms can be reduced to regular polylogarithms.
Above, in eq.\,\eqref{eq:Li11reduction}, we have seen the example of $\Li{1,1}{x,y}$ reducing to $\Li{2}{x}$.
Indeed, $\Li{2,2}{x,y}$ is the first polylogarithm that does not reduce to the classical version \cite{Frellesvig2016}. 
This reduction property of low weight polylogarithms is often helpful to simplify analytic results.

The main usage of GPLs comes with integral calculation.
They are a completion of the rational functions with respect to integration, meaning that any function
\begin{align}
p(x)/q(x)\,G(a_1,\dots,a_n;x)
\label{eq:GintGeneral}
\end{align}
with polynomials $p(x),\,q(x)$ has a primitive again expressible in terms of Goncharov polylogarithms.
This makes them highly valuable for algorithmic integration \cite{Panzer:2014caa}.
Here, a direction of further generalizing GPLs is to allow for more general integration kernels than $1/(x-a)$.
While a lot of singularities can be reduced to that case -- $1/p(x)$ for any polynomial $p(x)$ can be reduced by partial fractioning and even $1/\sqrt{p(x)}$ can often be rationalized by a change of variables \cite{Besier:2018} -- there are Feynman integrals which have non-rationalizable singularities of the form $1/\sqrt{p(x)}$, where $p(x)$ is a degree four polynomial that describes an elliptic curve.
This gives rise to \textit{elliptic polylogarithms} \cite{Broedel:2019hyg}.
Recently, even more complex curves were introduced into the business \cite{Duhr:2024uid}.
However, these will not appear in this thesis and the worst we will have to deal with in angular integrals are rationalizable square roots.
\begin{toolblock}{
For practical work with GPLs in the computer algebra system \texttt{Mathematica} \cite{Mathematica}, there is the powerful package \texttt{PolyLogTools} \cite{Duhr:2019}.
For the integration of GPLs, this can be combined with the algorithm from Erik Panzer \cite{Panzer:2014caa}, unfortunately not yet part of \texttt{PolyLogTools}. 
To learn how to handle roots in the integration procedure, the paper \cite{Besier:2018} is highly recommended.
}\end{toolblock}
\begin{readingblock}
{
A good introduction to GPLs and their properties is given in the 2014 TASI lectures by Claude Duhr \cite{Duhr:2014woa}.
For a review of the state-of-the-art of special functions beyond GPLs, see the Snowmass white paper from 2022 \cite{Bourjaily:2022bwx}.
}
\end{readingblock}
\section{Gamma, Beta, and hypergeometric functions}
Besides polylogarithms, there are other classes of functions that appear in Feynman integrals.
Generally speaking, the arguments of polylogarithms are kinematic variables, i.e. combinations of momenta and masses.
The dependence of Feynman integrals on ``exponent-like'' quantities and also the space-time dimension appears in exponents of integral representations and is captured by the \textit{Gamma function} and its generalization.

The Gamma function is defined by
\begin{equation}
\Gamma(a)=\int_0^\infty\dx t\,t^{a-1}\,\ee^{-t}\,.
\end{equation}
Its fundamental property is
\begin{align}
\Gamma(a+1)=a\,\Gamma(a)\,,
\end{align}
hence it generalizes the factorial to non-integer arguments\footnote{This will become essential, once we discuss dimensional regularization.}.
The most relevant special values are $\Gamma(1)=1$ and $\Gamma(1/2)=\sqrt{\pi}$ from which all integer and half-integer values follow.

Important functional identities for the Gamma function are the Euler reflection identity
\begin{align}
\Gamma(a)\Gamma(1-a)=\frac{\pi}{\sin(\pi a)}
\end{align}
and the Legendre duplication formula
\begin{align}
\Gamma(a)\Gamma\!\left(a+\frac{1}{2}\right)=2^{1-2a}\sqrt{\pi}\,\Gamma(2a)\,.
\label{eq:LegendreDuplication}
\end{align}
To approximate the super-exponential growth of the Gamma function for large values, something that is often useful in the context of Mellin integrals (see sec.\,\ref{sec:Mellin}), one can use the Stirling formula
\begin{align}
\Gamma(a)\sim \sqrt{\frac{2\pi}{a}}\left(\frac{a}{e}\right)^a
\label{eq:Stirling}
\end{align}
or the useful simple corollary
\begin{align}
\Gamma(a+n)\sim a^n\Gamma(a)
\label{eq:GammaApprox}
\end{align}
where in this context $\sim$ denotes asymptotic equality, e.\,g. the ratio of both sides tends to one for $a\rightarrow\infty$.

To write down a series expansion, it is simplest to look at the logarithmic Gamma function, which can be expanded as
\begin{equation}
\log\Gamma(1+a)=-\gamma_\text{E}a+\sum_{n=2}^\infty \frac{(-1)^n \zeta_n}{n}\,a^n\,,
\end{equation}
where $\gamma_\text{E}=0.57721\dots $ is the Euler-Mascheroni constant and $\zeta_n$ are the integer values of the zeta function.
The derivative of the logarithmic gamma function is called the digamma function,
\begin{equation}
\psi(a)=\frac{\dx\log\Gamma(a)}{\dx a}=-\gamma_\text{E}+\sum_{k=0}^\infty \left(\frac{1}{1+k}-\frac{1}{a+k}\right)\,;
\end{equation}
higher derivatives define the polygamma family
\begin{align}
\psi^{(n)}(a)=\frac{\dx^n \psi(a)}{\dx a^n}=(-1)^{n+1}n!\sum_{k=0}^\infty\frac{1}{(a+k)^{n+1}}\,.
\end{align}

Going from one variable to two, the \textit{Beta function} is defined by the integral
\begin{align}
\mathrm{B}(a,b)=\int_0^1\dx t\, t^{a-1}(1-t)^{b-1}\,.
\label{eq:BetaFctInt1}
\end{align}
Importantly, this integral evaluates in terms of Gamma functions,
\begin{align}
\mathrm{B}(a,b)=\frac{\Gamma(a)\Gamma(b)}{\Gamma(a+b)}\,.
\label{eq:BetaAsGamma}
\end{align}
An alternative integral representation of the Beta function, which is useful occasionally, is
\begin{align}
\mathrm{B}(a,b)=\int_0^\infty\frac{\dx t\,t^{a-1}}{(1+t)^{a+b}}\,.
\end{align}
The special relevance of the Beta function is that we can use either of its integral forms to express a combination of Gamma functions as a parametric integral -- which is an important technique in evaluating Mellin-Barnes integrals.
As a direct generalization of eq.\,\eqref{eq:BetaFctInt1}, the Beta function can be generalized to $n$ variables as
\begin{align}
\mathrm{B}(a_1,\dots,a_n)=\int_0^1\dx x_1 x_1^{a_1-1}\cdots \int_0^1\dx x_n x_n^{a_n-1}\delta\!\left(1-\sum_{i=1}^n x_i\right)=\frac{\Gamma(a_1)\cdots\Gamma(a_n)}{\Gamma(a_1+\cdots+a_n)}\,.
\end{align}

Another important direction of generalization of eq.\,\eqref{eq:BetaFctInt1} is the combination with a ``kinematic'' variable $x$.
This connection is made via the integral
\begin{align}
\ghy(a,b,c;x)=\frac{\Gamma(c)}{\Gamma(b)\Gamma(c-b)}\int_0^1\dx t\,t^{b-1}(1-t)^{c-b-1}(1- x t)^{-a}\,,
\end{align}
which is a representation of the \textit{Gauss hypergeometric function}, defined by the series representation
\begin{align}
\ghy(a,b,c;x)=\sum_{n=0}^\infty\frac{\poha{a}{n}\poha{b}{n}}{\poha{c}{n}}\frac{x^n}{n!}
\end{align}
where the $\poha{a}{n}$ are \textit{Pochhammer symbols}
\begin{equation}
\poha{a}{n}=\frac{\Gamma(a+n)}{\Gamma(a)}=a(a+1)\dots(a+n-1)\,.
\end{equation}
The Gauss hypergeometric function $\ghy$ generalizes a lot of elementary functions, for example
$\ghy(a,b,b;x)=(1-x)^{-a}$ or $\ghy(1,1,2;x)=-\log(1-x)/x$.

What makes $\ghy$ valuable is the large body of known identities, which allow to perform non-trivial transformations.
From the definition, the symmetry property
\begin{equation}
\ghy(a,b,c;x)=\ghy(b,a,c;x)
\end{equation}
is evident.
The Gauss hypergeometric function allows for several changes of variables.
Most importantly, the Pfaff-Euler transformations
\begin{align}
\ghy(a,b,c;x)&=(1-x)^{-a}\ghy\!\left(a,c-b,c;\frac{x}{x-1}\right)
\\
&=(1-x)^{-b}\ghy\!\left(c-a,b,c;\frac{x}{x-1}\right)
\\
&=(1-x)^{c-a-b}\ghy(c-a,c-b,c;x)\,.
\label{eq:PfaffEuler3}
\end{align}
Besides, there are reflection  and inversion identities relating $z$ to $1-z$ and $1/z$, respectively \cite{Ananthanarayan:2025nsr} which are useful for analytic continuation; like the polylogarithms, $\ghy(a,b,c;z)$ has a branch cut on the positive real axis starting at $x=1$. 
For specific parameter values, there are also \textit{quadratic} transformations, for example
\begin{align}
\ghy(a,b,2b;x)=\left(1-\frac{x}{2}\right)^{-a}\ghy\!\left(\frac{a}{2},\frac{a+1}{2},b+\frac{1}{2};\left(\frac{x}{2-x}\right)^2\right)\,,
\end{align}
which is of particular relevance in publication \ref{pub:1} to establish a representation for single-massive one-denominator angular integrals.
For further identities, see the classic reference \cite{bateman_1953_cnd32-h9x80} or the Digital Library of Mathematical Functions (DLMF) \cite{NIST:DLMF}.

The series definition of hypergeometric functions naturally generalizes to the multi-variable case, however not uniquely.
In total, there are $14$ distinct hypergeometric functions in two variables, known as Appell \cite{Appell1882} and Horn functions \cite{Horn1931}.
In the context of this thesis, the most important of these is the Appell function $\mathrm{F}_1$, which is defined by the two-fold series
\begin{align}
\mathrm{F}_1(a,b_1,b_2,c;x_1,x_2)=\sum_{n_1,n_2=0}^\infty\frac{\poha{a}{n_1+n_2}\poha{b_1}{n_1}\poha{b_2}{n_2}}{\poha{c}{n_1+n_2}}\frac{x_1^{n_1}x_2^{n_2}}{n_1! n_2!}\,.
\end{align}
It is special among the two-variable hypergeometric functions, since it admits the one-fold, rather than two-fold, integral representation
\begin{align}
\mathrm{F}_1(a,b_1,b_2,c;x_1,x_2)=\frac{\Gamma(c)}{\Gamma(c-a)\Gamma(a)}\int_0^1\dx\,t\,t^{a-1}(1-t)^{c-a-1}(1- x_1 t)^{-b_1}(1- x_2 t)^{-b_2}\,.
\end{align}
The Appell function appears in the solution to the single-massive two-denominator angular integral.
Going to two masses, we will encounter the three-variable hypergeometric Lauricella function $\mathrm{F}_\text{B}^{(3)}$
\begin{align}
&\mathrm{F}_\text{B}^{(3)}(a_1,a_2,a_3,b_1,b_2,b_3,c;x_1,x_2,x_3)\nonumber\\
&=\sum_{n_1,n_2,n_3=0}^\infty \frac{\poha{a_1}{n_1}\poha{a_2}{n_2}\poha{a_3}{n_3}\poha{b_1}{n_1}\poha{b_2}{n_2}\poha{b_3}{n_3}}{\poha{c}{n_1+n_2+n_3}}\frac{x_1^{n_1}}{n_1!}\frac{x_2^{n_2}}{n_2!}\frac{x_3^{n_3}}{n_3!}\,.
\end{align}
\begin{readingblock}{
\textbf{Further reading: } Good references for hypergeometric functions are the classic book by Harry Bateman \cite{bateman_1953_cnd32-h9x80}, written in a purely mathematical context, and the recent review article on their use in the context of Feynman integrals \cite{Ananthanarayan:2025nsr}.
}\end{readingblock}
\end{fancychapter2}

\cleardoublepage
\begin{fancychapter2}{Analytic methods in perturbative QCD}{A tour through the armory}{You showed me colors you know I can't see with anyone else.}{Taylor Swift}
\label{ch:Methods}
This chapter gives a brief introduction to important methods in perturbative QCD calculations.
They are not meant as a replacement for a systematic textbook discussion (those are referenced after each section) or a summary of the  technical state-of-the-art, but are an invitation to get acquainted with the powerful methods developed for loop integrals in the past decades.
They form the basis for the developments in the publications, where they were developed further and applied in new contexts.

We start by discussing dimensional regularization and the related concept of $\eps$-expansion, a main form in which results are presented in the publications.
While this is of course well-known to practitioners of QFT, this exposition is given some room to make results of the form ``$\eps$-expansion of angular integrals'' meaningful for someone who is not familiar with the concept.
We continue in sec.\,\ref{sec:loop} by introducing loop integrals more generally, including tensor reduction and common representations as parametric integrals.
Their cousins, phase-space integrals, are introduced in \ref{sec:PSint} in the form that leads to angular integrals.
The subsequent sections are devoted to actual calculational methods: IBPs to reduce the number of integrals that need to be actually calculated in sec.\,\ref{sec:IBP}, expansion-by-regions to calculate integrals in kinematic limits in sec.\,\ref{sec:ExpByReg}, differential equations to calculate integrals in general kinematics in sec.\,\ref{sec:DiffEq}, and Mellin integral techniques in sec.\,\ref{sec:Mellin}.
Throughout the technical sections, we will mainly follow simple examples to illustrate the main ideas rather than striving for the highest degree of generality.

\section{Dimensional regularization and epsilon expansion}
\label{sec:DimReg}
Dimensional regularization is a central technique to make sense of calculations in perturbative QCD.
To explain the rationale behind this rather abstract idea, we will work through a simple example that illustrates the process of how we arrive at a prescription to obtain numbers from formally infinite expressions.

From the Feynman rules in QFT, we encounter integrals such as
\begin{align}
I=\int\frac{\dx^4 k}{(2\pi)^4}\frac{1}{k^2-m^2+\iu 0}
=\int\frac{\dx^3 k}{(2\pi)^4}\int_{-\infty}^\infty\dx k^0\frac{1}{(k^0-E_k+\iu 0)(k^0+E_k-\iu 0)}
\end{align}
where $E_k=\sqrt{\mathbf{k}^2+m^2}$ denotes the energy and the $+\iu 0$ \textit{Feynman prescription} ensures that we do not directly integrate over poles at $k^2=m^2$.

\begin{wrapfigure}[17]{r}{0.5\textwidth}
\includegraphics[width=0.45 \textwidth]{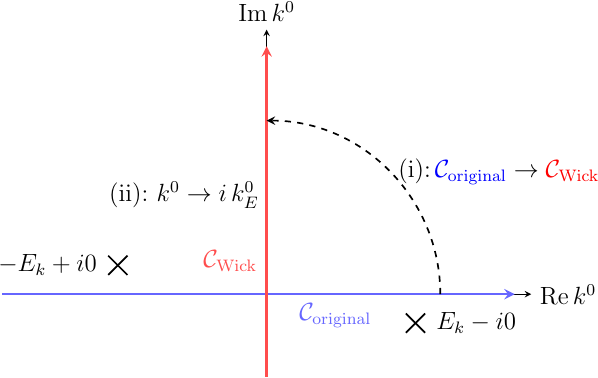}
\caption[Graphical depiction of Wick rotation.]{Graphical depiction of Wick rotating the original integration contour along the real axis onto the imaginary axis without crossing a pole.}
\label{fig:WickRot}
\end{wrapfigure}
Considering $k^0$ as a complex variable and using that the poles of the $k^0$ integration are only in the second and fourth quadrant in the complex $k^0$-plane, we can rotate the integration path counterclockwise from the real to the imaginary axis without changing the value of the integral.
With the new variable $k_\text{E}=(-\iu k^0,\mathbf{k})$ this leads to the euclidean integral
\begin{align}
I=-\iu\int\frac{\dx^4 k_\text{E}}{(2\pi)^4}\frac{1}{k_\text{E}^2+m^2}
\end{align}
where the $\iu 0$-prescription could be dropped since the integrand is never at the pole.
This procedure goes under the name \textit{Wick-rotation} and is depicted in figure \ref{fig:WickRot}.

Next, we can introduce spherical coordinates in four dimensions for the integral as $\dx^4 k_\text{E}=K^3\dx K\dx\Omega_4$ and integrate over the spherically symmetric part, resulting in
\begin{align}
I=-\frac{\iu\Omega_4}{(2\pi)^4}\int_0^\infty\dx K\frac{K^3}{K^2+m^2}\,.
\end{align}
Upon inspection, we see that due to the growth of the integrand for $K\rightarrow\infty$ the radial $K$ integral diverges at the upper boundary!
At this point one might be inclined to quit on the idea of QFT and go back to the drawing board.

However, before we give up so easily, we may recall what Bohr taught about quantum mechanics:
The job of the theoretical physicist is to predict measurement outcomes and the infinite integral we are looking at does not, on its own, correspond to an observable, but is only one contribution from a particular Feynman diagram.
So this infinity does not render our theory meaningless if we can formulate a prescription of how to organize the calculation such that we end up with finite predictions for measurable quantities.

Physically, we can assume that for one, our theory might not be meaningful up to arbitrarily large energy scales, and for another, a sensible observable should not be sensitive to energy scales much larger (or much smaller) than the typical scale of the experiment.
Our example integral diverged because of the $K\rightarrow\infty$ region -- in a well-defined observable this region, corresponding to arbitrarily large energies, should cancel between different contributions.
For this to be possible in a meaningful way, we however first need a consistent way to parameterize the divergences.
This process is called \textit{regularization}.
The regulator needs to be such that the expression becomes mathematically well-defined and that we recover the original expression when removing the regulator, at least for finite quantities where this step is meaningful.

For a regulator to be useful in practice, it needs to respect the symmetries of the theory.
For QCD, those are Lorentz and gauge invariance.
The most widely used regularization prescription,  precisely because it respects these symmetries, is called \textit{dimensional regularization}.
Its idea is to take the space-time dimension, which is four in physical space, and promote it to a new variable $d$ which can then be chosen such that the integrals become convergent.
A more exact formulation will be developed in the following.

Going back to our example integral $I$, and considering it in $d$ instead of $4$ dimensions we have
\begin{equation}
I(d)=\int\frac{\dx^d k}{(2\pi)^d}\frac{1}{k^2-m^2+\iu 0}\,.
\end{equation}
To give meaning to this expression, we may first think of $d$ as a positive integer, after all, the number of integrations that will be performed explicitly needs to always be a positive integer.
However, there is a way to extend the notion of ``integration'' beyond the integers.
Going through the same steps of Wick-rotation, introducing spherical coordinates, and integrating the spherically symmetric part as
\begin{equation}
\Omega_d=\frac{2\pi^{\frac{d}{2}}}{\Gamma\!\left(\frac{d}{2}\right)}
\end{equation} 
we get to
\begin{equation}
I(d)=-\frac{\iu\Omega_d}{(2\pi)^d}\int_0^\infty\dx K\,\frac{K^{d-1}}{K^2+m^2}\,.
\label{eq:IntInDdim}
\end{equation}
Now, since $\Gamma(d/2)$ and the exponential are not restricted to integer values, this expression is also meaningful for non-integer $d$.
Hence, from here on, we can consider $d$ as a complex valued variable, allowing us to use the powerful tools of complex analysis.

For the  $K$-integral to converge, we require $0<\mathrm{Re}(d)<2$.
Then it holds
\begin{align}
\int_0^\infty\dx K\,\frac{K^{d-1}}{K^2+m^2}=\frac{\pi}{2}\,\frac{m^{d-2}}{\sin\left(\frac{\pi d}{2}\right)}\qquad(\text{for }0<\mathrm{Re}(d)<2)\,.
\end{align}
While the integral on the left side is only defined on a strip in the complex $d$-plane, the right side is an analytic function for all complex $d$ with simple poles at the even integers.
This makes
\begin{align}
I(d)=-\frac{\iu\Omega_d}{(2\pi)^d}\,\frac{\pi}{2}\,\frac{m^{d-2}}{\sin\left(\frac{\pi d}{2}\right)}
\end{align}
the unique analytic continuation of the integral $I(d)$.

To recover a theory in 4 dimensions, we are interested in the analytic continuation at $d=4$.\footnote{For a convergent integral, this would just be the value of the original integral.}
Since typically, we encounter expressions containing $d/2$, it is customary to introduce a new variable $\eps$ by $d=4-2\eps$. 
The continuation to $d=4$ is then a Laurent expansion about $\eps=0$.
This is called $\eps$\textit{-expansion}, a central theme in a number of publications in this thesis.

For the example integral we have
\begin{align}
\mathcal{I}(\eps)\equiv\mu^{2\eps}I(4-2\eps)=-\frac{\iu S_\eps \,m^2}{(4\pi)^2 (1-\eps)}\left[-\frac{1}{\eps}+\log\frac{m^2}{\mu^2}-\eps\left(\frac{1}{2}\log^2\frac{m^2}{\mu^2}+\frac{\pi^2}{6}\right)+\mathcal{O}\!\left(\eps^2\right)\right]\,,
\label{eq:ExampleEpsExpansion}
\end{align}
where the scale $\mu$ has been introduced to avoid dimensionful logarithms\footnote{This scale $\mu$ naturally appears from the Feynman rules in dimensional regularization from the dimensionality of the coupling in $d$ dimensions.}
and
\begin{align}
S_\eps=\frac{(4\pi)^\eps}{\Gamma(1-\eps)}
\end{align}
is a conventional factor that is kept unexpanded.
The factor $S_\eps$ is a part of the rotationally-symmetric angular part, which is, in some form, shared by all integrals in dimensional regularization.\footnote{How to deal with integrals, which are not fully rotationally symmetric, in dimensional regularization is discussed in appendix A of publication \ref{pub:1}.}
Keeping a factor of $S_\eps$ per loop makes it such that no factors of $\gamma_\text{E}$ or $\log\pi$ ever appear as artifacts of the regularization.\footnote{Note that there are different normalization conventions; however, the $S_\eps$ here is popular in phenomenology to organize cancellations between real and virtual corrections.
Later in this chapter, for pure loop examples, we will often drop the $\mu^{2\eps}$ and normalize by $\iu \pi^{d/2}$ per loop. The exact convention does not matter as long as one remembers to get the constants right when comparing results with the literature.}

Overall, in dimensional regularization, we replace $I\rightarrow\mathcal{I}(\eps)$ and use the $\eps$-expansion about $\eps=0$ as its assigned ``value''.
When combining all contributions to an observable, we are now in the position to add the divergences in a well-defined way and hope that they cancel out if we organize everything correctly. 
Specifically, the example integral eq.\,\eqref{eq:ExampleEpsExpansion} has a $\frac{1}{\eps}$-pole which parameterizes the UV-divergent behavior of the original integral.
By renormalization, i.e., organizing the perturbative series in terms of finite physical quantities instead of the ``bare'' parameters of the Lagrangian, these poles will drop out from observables.

A few features of  eq.\,\eqref{eq:ExampleEpsExpansion}, representative for $\eps$-expansions in general, are worth pointing out:
\begin{itemize}
\item[(a)] In a perturbative calculation it can happen that $\mathcal{I}(\eps)$ gets multiplied by pole terms from other contributions. 
This makes it possible that higher order terms in the $\eps$-expansion can contribute and we cannot truncate the series at $\mathcal{O}(\eps^0)$ right away.
~
\item[(b)] We kept a factor of $1/(1-\eps)$ unexpanded for clarity (this pole at $\eps=1$ is a remnant of the divergence at $d=2$ of the integral in eq.\,\eqref{eq:IntInDdim}).
The resulting remainder is of \textit{uniform weight}:
Counting factors of $\eps$ as $-1$\footnote{This counting is motivated since $1/\eps$ corresponds to $\log\Lambda$ in a cut-off regularization.}, $\log$ as $1$, and $\pi$ as $1$, all terms in the expansion have weight $1$, or equivalently the coefficient of $\eps^{n}$ has weight $n+1$.
When setting up an $\eps$-expansion, this property, also called \textit{uniform transcendentality}, is a very useful guiding principle, showing that one has ``factored out'' all ``trivial'' $\eps$-dependence.
It should be noted that the $\pi$s in the expansion are of ``polylogarithmic'' origin and not $\pi$s from the spherical volume.
\item[(c)] The analytic structure in $\eps$ of integrals in dimensional regularization is very simple.
They are always analytic functions with isolated poles, never essential singularities or branch cuts.
This is very much in contrast to the analytic structure in kinematic variables, which may have a complex branch structure, e.g. in our example there is a branch cut for $m^2/\mu^2<0$.
\item[(d)] \begin{block}[type=idea]Generally, to obtain correct results in dimensional regularization, kinematic limits need to be taken first, only then expand in $\eps$.
\end{block}
We see that for example our result for $\mathcal{I}(\eps)$ in eq.\,\eqref{eq:ExampleEpsExpansion} has a $\log\frac{m^2}{\mu^2}$ which is not well defined for $m^2=0$.\footnote{In this particular case there is a suppression by  a factor of $m^2$, but that is ``accidental''.}
To properly treat the $m^2=0$ case, we have to keep the $\left(\frac{m^2}{\mu^2}\right)^{-\eps}$, from which it originated, unexpanded, or equivalently if we are handed the expansion, ``resum'' the $\log^n\frac{m^2}{\mu^2}$ terms in each order of $\eps$.\footnote{This resummation in $\eps$ is not to be confused with the resummation of logarithms in the perturbative series in $\alpha_s$.}

Then we have
\begin{align}
\mathcal{I}(\eps)=-\frac{\iu S_\eps \,\mu^2}{(4\pi)^2 (1-\eps)}\left(\frac{m^2}{\mu^2}\right)^{1-\eps}\left[-\frac{1}{\eps}-\frac{\pi^2}{6}\,\eps-\frac{7\pi^4}{360}\,\eps^3+\mathcal{O}(\eps^5)\right]\,.
\label{eq:ExampleEpsExpansionResummed}
\end{align}
Now, what is going on for $m^2\rightarrow 0$?
Depending on the value of $\eps$, we have
\begin{align}
\left(\frac{m^2}{\mu^2}\right)^{1-\eps}\overset{m^2\rightarrow 0}{\longrightarrow}\left\lbrace
\begin{array}{cc}
0 &\text{for } \mathrm{Re(\eps)}<1\\
\infty &\text{for } \mathrm{Re(\eps)}>1
\end{array}
\right..
\label{eq:m2casedistinctions}
\end{align}
To have a definite value, we better have the first case.
The $m\rightarrow 0$ limit probes the infrared behavior of the original integral from eq.\,\eqref{eq:IntInDdim}.
So even if we previously had to assume $2-2\eps<0$, which is the bad case of eq.\,\eqref{eq:m2casedistinctions}, for the integral to converge in the UV, we now need to use the opposite case in the dimensionally continued result to assign a meaningful value, namely $\mathcal{I}(\eps,m=0)=0$.
\end{itemize}

In general, treatment of UV divergence requires $\eps$ to be sufficiently large, while that of IR divergences requires it to be sufficiently small.
At this point, the treatment of both types of singularities with the same regulator, subject to mutually exclusive conditions, seems to be on somewhat shaky foundations.
For another way to argue that $0$ is indeed the correct value to assign to the integral, we go back to the integral form of eq.\,\eqref{eq:IntInDdim} and set $m^2=0$ there.
This means we have the integral
\begin{equation}
\int_0^\infty\dx K\, K^{1-2\eps}
\end{equation}
which converges for no value of $\eps$.
However, if we tell all mathematicians in the room to close their eyes for a moment, we can split the integral at some intermediate scale $\Lambda$ and calculate
\begin{align}
\int_0^\infty\dx K\, K^{1-2\eps}&=\underbrace{\int_0^\Lambda\dx K\, K^{1-2\eps}}_{\eps<1}+\underbrace{\int_\Lambda^\infty\dx K\, K^{1-2\eps}}_{\eps>1}\\
&=\frac{\Lambda^{2-2\eps}}{2-2\eps}-\frac{\Lambda^{2-2\eps}}{2-2\eps}=0.
\end{align}
Admittedly, this feels like cheating.
A way to make this logic more rigorous would be to split all integrals into UV and IR regions, do the analytic continuation separately for both, one with $\eps_\text{UV}$, the other with $\eps_\text{IR}$.
In a sensible observable, poles in $\eps_\text{UV}$ and $\eps_\text{IR}$ need to separately cancel.
However, it is more practical to embrace the magic of dimensional regularization that UV and IR poles can cancel each other.
This goes well with the argument that
\begin{align}
\int\frac{\dx^d k}{(2\pi)^d}\frac{1}{k^2}
\end{align}
needs to be zero anyways, since by dimensional analysis the result must be of dimension $(\text{mass})^{d-2}$ but there is no dimensionful scale the integral can depend on.
In short:
\begin{block}[type=idea]
Scaleless integrals vanish in dimensional regularization.
\end{block}

A key reason why it is possible to get away without distinguishing between $\eps_\text{UV}$ and $\eps_\text{IR}$ in the calculation of cross-sections is that the prescriptions of the respective analytic continuations happen sequentially.
First, the loop integral is evaluated with $\eps$ in the $\eps_\text{UV}$ range, all remaining divergences in the Feynman parameter integrals are of IR type, so, after the loop integration is done and analytic continuation in $\eps$ is performed, we can switch to $\eps$ in the $\eps_\text{IR}$ range and have again integrals that are convergent in some domain of $\eps$.
Also in phase-space integrals, all divergences are IR, so we have a domain of convergence from which we can analytically continue to $\eps=0$.

One interesting aspect of dimensional regularization is that it automatically removes all polynomial divergences and only creates poles for logarithmic singularities.
For example, the IR integral
\begin{align}
\int_0^\Lambda\dx K\,K^{-1-\eps}=-\frac{\Lambda^{-\eps}}{\eps}=-\frac{1}{\eps}+\log\Lambda+\mathcal{O}(\eps)
\end{align}
has a pole, but the more strongly divergent
\begin{align}
\int_0^\Lambda\dx K\,K^{-2-\eps}=-\frac{\Lambda^{-1-\eps}}{1+\eps}=-\frac{1}{\Lambda}+\mathcal{O}(\eps)
\end{align}
retains, maybe somewhat surprisingly at first sight, no pole but a finite value.
At this point, it is essential to stress that dimensional regularization assigns values by analytic continuation, not a limiting process.
The property of being sensitive to logarithmic divergences only is extremely convenient, since those show up in the renormalization procedure and carry physical information about the theory, e.g. the running of the coupling.

Staying with our example, we can discuss another aspect of dimensional regularization that becomes important later, especially in publication \ref{pub:4}.
We already mentioned that kinematic limits need to be taken before the $\eps$-expansion.
Now that we have $\mathcal{I}(\eps,m^2)$, let us assume that we want to use it to calculate an integral over $m^2$.\footnote{If one strictly thinks about $m$ as the mass of a particle we certainly will not do exactly that in practice, but structurally similar integrals appear in Feynman parameter or phase-space integrals.}
To make everything scaleless, we take $x\equiv \frac{m^2}{\mu^2}$ and look at 
\begin{align}
\mathcal{J}=\frac{1}{\mu^2}\int_0^1\frac{\dx x}{x^2}\mathcal{I}(\eps,\mu^2\,x)\,.
\end{align}

Using the $\eps$-expansion of eq.\,\eqref{eq:ExampleEpsExpansion}, this is ill defined, because in
\begin{align}
\mathcal{J}\overset{?}{=}-\frac{\iu S_\eps}{(4\pi)^2(1-\eps)}\int_0^1\frac{\dx x}{x}\left[-\frac{1}{\eps}+\log x+\dots\right]
\end{align}
the $x$-integrals diverge in every order of $\eps$.
We clearly see that $\eps$-expansion, or more precisely, the truncation of the expansion, does not always commute with integration.

Hence, providing $\eps$-expansions of integrals that are re-usable in different applications, requires a careful treatment of kinematic limits.
Here, the ``resummed'' form of the expansion -- which we already discussed in the context of the $m^2\rightarrow 0$ limit --, where the $m^{2-2\eps}$ is kept together, solves the issue.
With eq.\,\eqref{eq:ExampleEpsExpansionResummed} it is
\begin{align}
\mathcal{J}=-\frac{\iu S_\eps}{(4\pi)^2(1-\eps)}\int_0^1\dx x\,x^{-1-\eps}\left(-\frac{1}{\eps}-\frac{\pi^2}{6}\eps-\dots\right)\,.
\end{align}
Now, the $x$-integral is properly regulated for $\eps<0$.
This matches the earlier statement about ``small $\eps$ for IR divergences'' and we get
\begin{align}
\mathcal{J}=-\frac{\iu S_\eps}{(4\pi)^2(1-\eps)}\left[\frac{1}{\eps^2}+\frac{\pi^2}{6}+\mathcal{O}(\eps)\right]
\end{align}
where the $1/\eps^2$-pole comes from an overlap of the UV and IR poles.
Importantly, this calculation showcased how a term that was suppressed by a factor of $\eps$ transformed into a finite contribution.\footnote{Indeed, factors such as $\frac{\pi^2}{6}$ are quite typical to appear in physical cross-sections in higher orders of perturbation theory.}

We can summarize our findings as follows and, in passing, introduce some classifications regarding the $\eps$-expansion that will be helpful when discussing results of the publications, especially in the context of angular integrals:
\begin{itemize}
\item[\textbf{(1)}]\textbf{Defining integral representation in $d$-dimensions}\\
\textbf{Example: }
\begin{equation}
I(d)=\int\frac{\dx^d k}{(2\pi)^d}\frac{1}{k^2-m^2+\iu 0}
\end{equation}
\\
\textbf{What is it?: }Original integral with integration over $d$ instead of $4$ dimensions.
\\
\textbf{Usage: }Variable space-time dimension serves as regulator for divergences, structural properties are retained, starting point of calculations in dimensional regularization, useful e.\,g. for finding IBP relations.
\item[\textbf{(2)}]\textbf{Integral representation in $d$-dimensions}\\
\textbf{Example: }
\begin{equation}
I(d)=-\frac{\iu\Omega_d}{(2\pi)^d}\int_0^\infty\dx K\frac{K^{d-1}}{K^2+m^2}\,.
\end{equation}
\\
\textbf{What is it?: }
Representation of a function with $d$ as a complex variable, convergent for some domain of $d$ and a certain range of kinematic variables.
\\
\textbf{Usage: } Step towards calculation in terms of special functions, analytic continuation, and $\eps$-expansion. Typically, the fewer remaining integrations the more useful.
\item[\textbf{(3)}]\textbf{(Closed form) analytic result in $d$-dimensions}\\
\textbf{Example: }
\begin{equation}
I(d)=-\frac{\iu \Omega_d}{(2\pi)^d}\,\frac{\pi}{2}\,\frac{m^{d-2}}{\sin\left(\frac{\pi d}{2}\right)}\,.
\label{eq:ClosedFormTadpole}
\end{equation}
\textbf{What is it?: }The $d$-dimensional integral representation evaluated in terms of special functions, typically this involves Gamma and hypergeometric functions.\\
\textbf{Usage: }Reveals the analytic continuation, identities for special functions can be applied, can be used as starting point for the $\eps$-expansion.
If the integral representation can be evaluated in terms of known special functions, we can use all that is known about those functions e.g. alternative integral representations, variable transformations, or series expansions.
For complicated integrals, requiring multi-variable hypergeometric functions, this form is often less useful since the ``special'' functions become more ``arbitrary'' and useful reduction identities are not known.
\item[\textbf{(4)}]\textbf{$\eps$-expansion}\\
\textbf{Example: }
\begin{align}
\mathcal{I}(\eps)=-\frac{\iu S_\eps \,m^2}{(4\pi)^2 (1-\eps)}\left[-\frac{1}{\eps}+\log\frac{m^2}{\mu^2}-\eps\left(\frac{1}{2}\log^2\frac{m^2}{\mu^2}+\frac{\pi^2}{6}\right)+\mathcal{O}\!\left(\eps^2\right)\right],
\end{align}
\textbf{What is it?: }Laurent expansion around $\eps=0$ in the variable $\eps=(4-d)/2$.\\
\textbf{Usage: }Required to subtract poles between contributions and connect to the theory in $d=4$.
\item[\textbf{(5)}]\textbf{Resummed or asymptotic $\eps$-expansion}\\
\textbf{Example: }
\begin{equation}
\mathcal{I}(\eps)=-\frac{\iu S_\eps \,\mu^2}{(4\pi)^2 (1-\eps)}\left(\frac{m^2}{\mu^2}\right)^{1-\eps}\left[-\frac{1}{\eps}-\frac{\pi^2}{6}\,\eps-\frac{7\pi^4}{360}\,\eps^3+\mathcal{O}(\eps^5)\right]\,.
\end{equation}
\\
\textbf{What is it?: } $\eps$-expansion where some parts of the form $x^{-n\eps}$ are kept unexpanded to regularize kinematic limits.\\
\textbf{Usage: }Needed in situations where kinematic limits are relevant, e.\,g. for subsequent integration.
\item[\textbf{(6)}]\textbf{All-order $\eps$-expansion}\\
\textbf{Example: }
\begin{equation}
\mathcal{I}(\eps)=-\frac{\iu S_\eps \,\mu^2}{(4\pi)^2 (1-\eps)}\left(\frac{m^2}{\mu^2}\right)^{1-\eps}\sum_{k=0}^\infty
\frac{(-1)^k(2^{2k}-2)B_{2k}\pi^{2k}}{(2k)!}\eps^{2k-1}\,.
\end{equation}
\textbf{What is it?: }$\eps$-expansion (resummed or not) where all orders are explicitly known and which has not been truncated.\\
\textbf{Usage: }This is the ultimate form of $\eps$-expansion, no information is lost and all kinematic limits can be recovered.
Importantly an all-order expansion is not the same as a general ``analytic result in $d$ dimensions'' but a much more useful form.
\end{itemize}

\begin{toolblock}{
\textbf{Tadpole integral: }
The integral discussed in this section, commonly called \textit{tadpole integral}\footnote{Because the associated Feynman diagram looks just like a tadpole.}, will appear in several of the following examples.
With the same elementary methods discussed here, also the general case of a denominator raised to a power can be calculated.
For later reference, the result in $d$ dimensions is
\begin{align}
\int\frac{\dx^d k}{\iu \pi^{d/2}}\frac{1}{(-k^2+m^2-\iu 0)^j}=\frac{\Gamma\!\left(j-\frac{d}{2}\right)}{\Gamma(j)}\,(m^2-\iu 0)^{d/2-j},
\label{eq:Tadpole}
\end{align}
where we adapted the normalization of the loop integral to what we will use in later examples.
}\end{toolblock}

Dimensional regularization is suited for phase space integrals just as well as for loop integrals.
Here, divergences are always IR, since momentum conservation serves as a UV cutoff.
Phase space integrals can exhibit two kinds of IR divergences.
For one, \textit{soft divergences} associated with a momentum where all components become small, and for another, \textit{collinear divergences} that are associated with two massless particles becoming parallel on the light-cone such that their scalar product vanishes.

For angular integrals, both are relevant but appear in well separated form.
The angular integral itself has only collinear poles in its $\eps$-expansion that occur when the integration momentum becomes collinear to an external light-like vector.
Soft poles only arise in kinematic limits upon additional integration. 
They require a resummed form of the $\eps$-expansion.

\begin{readingblock}{\textbf{Further reading: }
An old but still useful review article discussing the basic principles shortly after dimensional regularization was introduced is \cite{Leibbrandt:1975}.
For a modern discussion of the formal definition of dimensional regularization, see section 2.4.2 of \cite{Weinzierl:2022eaz}.
For a treatment of subtleties arising in dimensional regularization and different ways to organize the calculation -- especially with algebraic objects that do not admit a natural continuation to $d$ dimensions -- see \textit{To $d$ or not to $d$} \cite{Gnendiger:2017pys}.
}\end{readingblock}

\section{Loop integrals, tensor reduction, and parametric representations}
\label{sec:loop}
In the last section, we have seen the first example of a loop integral, which had one loop momentum $k$ and a single denominator $1/(k^2-m^2)$.
Now, we want to generalize the discussion to loop integrals with $L$ loop momenta and a product of $n$ propagators.

In momentum space, such a scalar $L$-loop integral is defined as
\begin{align}
I(p_1,\dots,p_E;m_1^2,\dots,m_n^2;j_1,\dots,j_n,d)=
\int\left(\prod_{l=1}^L \frac{\dx^d k_l}{\iu\pi^{d/2}}\right)\frac{1}{D_1^{j_1}D_2^{j_2}\cdots D_n^{j_n}}
\label{eq:FeynmanIntegralMomSpace}
\end{align}
where the denominators $D_i$ have the form
\begin{equation}
D_i=-q_i^2+m_i^2-\iu 0
\end{equation}
with the $q_i$ given by
\begin{align}
q_i=\sum_{j=1}^L\alpha_{ij}k_j+\sum_{j=1}^E\beta
_{ij}p_j\quad (\alpha_{ij},\beta_{ij}\in\lbrace\pm 1,0\rbrace)
\end{align}
and $\lbrace p_1,\dots,p_E\rbrace$ is the set of external momenta.
Note that for convenience in the following discussion of pure loop integrals, we adopted a normalization different from the previous section, including a sign in the denominator.

The integral depends on two types of variables, kinematic and exponent-like.
By Lorentz invariance, the kinematic dependence of the integral is only on the invariant scalar products and masses, $\lbrace p_i\cdot p_j;m_i\rbrace$.
As we will see in the following, to systematically study Feynman integrals it is crucial to keep the dependence on the propagator powers $j_i$ and the dimension $d$ symbolic, even though in practice we will be interested in specific values for these. 
In most cases of interest, the $j_i$ will be integers and the dimensionality is to be expanded about an integer value, as we have seen in the previous section.

Feynman integrals arise from associated Feynman graphs by Feynman rules.
While each edge contributes with a denominator factor $1/(-q^2+m^2-\iu 0)$, where $q$ is the momentum flowing through the line, there can also appear numerator factors $p_i\cdot k_j$ or $k_i\cdot k_j$.
However, these can be removed by expressing them through inverse propagators -- in the multi-loop case this can give rise to additional propagators that correspond to larger graphs than the original one.

Also, there can be a tensorial structure formed by loop momenta $k_i^\mu$ in the numerator.
These can be systematically removed by first expressing the scalar part of the integrand such that its only dependence is on the squares $k_i^2$.
Then, the only tensor structure available is formed by the metric tensor $g^{\mu\nu}$ for even tensor-rank, while all odd-rank tensors vanish by anti-symmetry upon integration.
The additional factors of $k_i^2$, which arise by projecting onto the $g^{\mu\nu}$ structure, relate the integral to scalar integrals with shifted dimensions, since they appear in the radial part of the angular measure after Wick rotation.

While the dimensional shift method for tensor reduction works for arbitrary loop integrals, in the special case of $1$-loop integrals, the classical method for tensor reduction is the \textit{Passarino-Veltman reduction} \cite{Passarino:1978jh}.
As an example of this method, we consider the vector and tensor bubble integrals
\begin{align}
\BPV^\mu(p,m)=&\mu^{2\eps}\int\frac{\dx^d k}{\iu \pi^{d/2}}\frac{k^\mu}{(-k^2+m^2)(-(k+p)^2+m^2)}\\
\BPV^{\mu\nu}(p,m)&= \mu^{2\eps}\int\frac{\dx^d k}{\iu \pi^{d/2}}\frac{k^\mu k^\nu}{(-k^2+m^2)(-(k+p)^2+m^2)}\,.
\end{align}
These integrals can only depend on tensor structure built from $g^{\mu\nu}$ and $p^\mu$.
Therefore, they can be parameterized as
\begin{align}
\BPV^\mu(p,m)&=p^\mu \BPV_1(p^2,m)\\
\BPV^{\mu\nu}(p,m)&=p^\mu p^\nu \BPV_{21}(p^2,m)+g^{\mu\nu} \BPV_{22}(p^2,m)
\end{align}
where $\BPV_{1}$ and $\BPV_{21}$, $\BPV_{22}$ are scalar functions\footnote{These integrals are known as Passarino-Veltman functions. We use $\mathcal{B}$ instead of the standard $B$ to avoid confusion with the notation we use in the IBP reduction example of the bubble integral for integrals with different denominator powers.} that can be determined by contracting the equations with $p_\mu$ and $p_\mu p_\nu$, $g_{\mu\nu}$, respectively.
Using $g_{\mu\nu}g^{\mu\nu}=d$ and expressing the scalar product $2p\cdot k$ via inverse propagators as
\begin{align}
-2p\cdot k=[-(k+p)^2+m^2]-[-k^2+m^2]+p^2
\end{align}
we get the system of equations
\begin{align}
\left(
\begin{array}{ccc}
p^2 & 0 & 0\\
\frac{p^2}{2} & p^2 & 1\\
0 & p^2 & d
\end{array}
\right)
\left(\begin{array}{c}
\BPV_1\\
\BPV_{21}\\
\BPV_{22}
\end{array}\right)=
\left(\begin{array}{c}
-\frac{p^2}{2} \BPV_0\\ -\frac{1}{2}\APV_0\\
m^2 \BPV_0-\APV_0
\end{array}\right)
\end{align}
with the scalar tadpole integral
\begin{align}
\APV_0=\mu^{2\eps}\int\frac{\dx^d k}{\iu \pi^{d/2}}\frac{1}{-k^2+m^2}\,.
\end{align}
This system of equations can be solved for $\BPV_1$, $\BPV_{21}$ and $\BPV_{22}$ and we find
\begin{align}
\BPV_1&=-\frac{1}{2}\,\BPV_0\,,\\
\BPV_{21}&=\frac{1}{(d-1) p^2}\left[-\frac{d}{2}p^2 \BPV_1-m^2 \BPV_0-\frac{d-2}{2}\,\APV_0\right]\,,\\
\BPV_{22}&=\frac{1}{2(d-1)}\left[p^2 \BPV_1+2 m^2 \BPV_0-\APV_0\right].
\end{align}
In publication \ref{pub:1}, we will present a slight modification of the Passarino-Veltman reduction that uses an orthogonal basis constructed from the external momenta that allows for a direct projection on the scalar factors without having to solve a linear system of equations.

Overall, the strategies of numerator and tensor reductions allow us to focus on scalar integrals of the type of eq.\,\eqref{eq:FeynmanIntegralMomSpace}.
To be able to carry out the loop integral over $n$-propagators, a key trick is combining the propagators by \textit{Feynman parameterization} -- this method will be used analogously for angular integrals in publication \ref{pub:1}.
Starting from the partial fraction identity
\begin{align}
\frac{1}{a b}=\frac{1}{a-b}\left[\frac{1}{b}-\frac{1}{a}\right]
\end{align}
and recognizing the bracket as an integral evaluated at its boundaries, we can write
\begin{align}
\frac{1}{a b}=\frac{1}{a-b}\int_b^a\frac{\dx t}{t^2}=\int_0^1\frac{\dx x}{[x a+(1-x) b]^2},
\end{align}
where, in the last step, we made the substitution $t=x a+(1-x) b$.
This identity was first given by Feynman in \cite{Feynman:1949}, attributing the original idea to Schwinger. 
We can also symmetrize the identity in $a$ and $b$ and write the Feynman parameterization in the form
\begin{align}
\frac{1}{a b}=\int_0^1\dx x_1 \dx x_2\,\frac{\delta(1-x_1-x_2)}{(x_1 a+x_2 b)^2}\,.
\end{align}
This identity can be used to combine two denominators $D_1$ and $D_2$, 
\begin{align}
\int\frac{\dx^d k}{\iu \pi^{d/2}}\frac{1}{D_1 D_2}=
\int_0^1\dx x_1 \dx x_2\,\delta(1-x_1-x_2)\int\frac{\dx^d k}{\iu \pi^{d/2}}\frac{1}{(x_1 D_1+x_2 D_2)^2}\,,
\end{align}
where we can -- by completing the square -- write the new denominator again in the form of a propagator with a shifted loop momentum and a mass $\Delta^2$ which depends on the Feynman parameters, concretely 
\begin{align}
x_1 D_1+x_2 D_2=-(k+x_2 p)^2\underbrace{-x_1 x_2 p^2+m^2}_{\Delta^2}.
\end{align}
Using the translational invariance of the loop integral measure\footnote{At this point, it is important that dimensional regularization respects translational invariance.} to make the shift $k\rightarrow k-x_2 p$, we find
\begin{align}
\int\frac{\dx^d k}{\iu \pi^{d/2}}\frac{1}{D_1 D_2}&=
\int_0^1\dx x_1 \dx x_2\delta(1-x_1-x_2)\int\frac{\dx^d k}{\iu \pi^{d/2}}\frac{1}{(-k^2+\Delta^2)^2}\nonumber\\
&=\Gamma\!\left(2-\frac{d}{2}\right)\int_0^1\dx x\, [m^2-x(1-x)p^2]^{d/2-2}
\label{eq:FeynmanParamB11}
\end{align}
where we used the tadpole result from eq.\,\eqref{eq:Tadpole} to evaluate the loop integral.
We see that Feynman parameterization allowed us to exchange the integration over loop momenta for a parametric integration.

\begin{toolblock}{\textbf{Feynman parameterization: }
Feynman parameterization can be generalized to the case of $n$ denominators raised to arbitrary powers $j_i$.
The general formula reads
\begin{align}
&\frac{1}{D_1^{j_1} D_2^{j_2}\dots D_n^{j_n}}\nonumber\\
=&\frac{\Gamma(j_1+\dots+j_n)}{\Gamma(j_1)\dots\Gamma(j_n)}\int_0^\infty\frac{\dx x_1\dx x_2\dots \dx x_n\,x_1^{j_1-1}\dots x_n^{j_n-1}\delta\left(1-\sum_{i} x_i\right)}{(x_1 D_1+x_2 D_2+\dots+x_n D_n)^{j_1+\dots+j_n}}\,,
\label{eq:FeynmanParametrization}
\end{align}
where the sum in the delta function can run over any non-empty subset of Feynman parameters according to the Cheng-Wu theorem.
Depending on the choice, the condition in the delta function will restrict the upper integration bound.
Note that the prefactor of the integral is the inverse of the multivariate Beta function of the $j_i$.
}
\end{toolblock}
Essentially, eq.\,\eqref{eq:FeynmanParametrization} finds application whenever one knows an integration formula over one denominator $D_i^{j_i}$ and wants to generalize to an arbitrary number of denominators.
It will for example be used in \ref{pub:1} to establish representations for the angular integral with $n$ denominators from the one-denominator case.

Using Feynman parameterization on the general $L$-loop integral of eq.\,\eqref{eq:FeynmanIntegralMomSpace}, it can be brought into the form
\begin{align}
I=\Gamma\!\left(j-\frac{L d}{2}\right)\prod_{i=1}^n\int_0^\infty\left(\prod_{i=1}^n\dx x_i\frac{x_i^{j_i-1}}{\Gamma(j_i)}\right)\delta(1-\sum_i x_i)\frac{\mathcal{U}^{j-(L+1)d/2}}{\mathcal{F}^{j-L d/2}}
\label{eq:FeynmanINtegralFeynmanParam}
\end{align}
with $j=\sum_{i=1}^n j_i$ and where $\mathcal{U}$ and $\mathcal{F}$ are polynomials in the Feynman parameters $x_i$ with coefficients depending on the kinematic variables.
In the literature, these polynomials are called \textit{1. and 2. Symanzik polynomial}, respectively.
Both are homogeneous in the Feynman parameters, $\mathcal{U}$ of degree $L$ and $\mathcal{F}$ of degree $L+1$.
They can be constructed directly from the Feynman graph $\Gamma$ that is associated with the integral $I$ as
\begin{align}
\mathcal{U}&=\sum_{T\in\mathcal{T}_1}\prod_{e\notin T}x_e\\
\mathcal{F}&=-\mathcal{V}+\mathcal{U}\sum_{e=1}^n m_e^2 x_e\quad\text{with}\quad\mathcal{V}=\sum_{T\in\mathcal{T}_2}\left(q^T\right)^2 \prod_{e\notin T}x_e\,,
\end{align}
where $\mathcal{T}_1$ is the set of \textit{trees of $\Gamma$}, $\mathcal{T}_2$ the set of \textit{2-trees of $\Gamma$}, $e$ denotes edges of $\Gamma$ and $q^T$ is the momentum flowing into one part of a 2-tree.
A tree originates from an $L$-loop graph by deleting $L$ edges, in a way that all loops are cut.
2-trees originate from trees by deleting an additional line, resulting in two separate trees.

\begin{figure}[htp!]
\begin{center}
\includegraphics[width=0.8\textwidth]{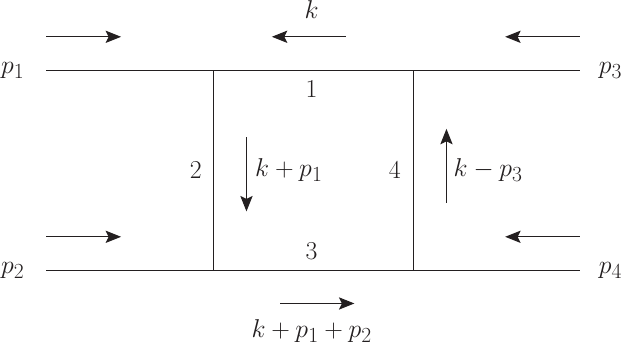}
\end{center}
\caption[Feynman diagram of the one-loop box integral. Created with \texttt{JaxoDraw} \cite{Binosi:2003}.]{Box diagram with incoming momenta $p_1$, $p_2$, $p_3$, $p_4$ and loop momentum $k$.}
\label{fig:boxdiagram}
\end{figure}

As an example for the construction, relevant to publications \ref{pub:2} and \ref{pub:3}, we consider the massless one-loop box integral
\begin{align}
I_\square\left(p_1,p_2,p_3;d\right)=\int\frac{\dx^d k}{\iu\pi^{d/2}}\,\frac{1}{k^2 (k+p_1)^2(k+p_1+p_2)^2(k-p_3)^2}\,.
\end{align}
The relevant kinematical invariants are the Mandelstam variables $s=(p_1+p_2)^2$ and $t=(p_1+p_3)^2$.
Comparing the box diagram, depicted in Figure \ref{fig:boxdiagram}, with the general loop integral formula eq.\,\eqref{eq:FeynmanINtegralFeynmanParam} we have in this case $L=1$, $j_i=1$, and $n=4$.
By deleting one line in the box we obtain the trees
$\mathcal{T}^1=\lbrace$
\begin{scalebox}{0.15}{
\begin{tikzpicture}\draw (0,1) -- (0,-1) -- (2,-1) -- (2,1);
\end{tikzpicture}}
\end{scalebox},
\begin{scalebox}{0.15}{
\begin{tikzpicture}\draw (0,-1) -- (2,-1) -- (2,1) -- (0,1);
\end{tikzpicture}}
\end{scalebox},
\begin{scalebox}{0.15}{
\begin{tikzpicture}\draw (2,-1) -- (2,1) -- (0,1) -- (0,-1);
\end{tikzpicture}}
\end{scalebox},
\begin{scalebox}{0.15}{
\begin{tikzpicture}\draw (2,1) -- (0,1) -- (0,-1) -- (2,-1);
\end{tikzpicture}}
\end{scalebox}$\rbrace$.
By deleting two lines we obtain the 2-trees 
$\mathcal{T}^2=\lbrace$
\begin{scalebox}{0.15}{
\begin{tikzpicture}\draw (0,1) -- (0,-1);\draw (2,-1) -- (2,1);
\end{tikzpicture}}
\end{scalebox},
\begin{scalebox}{0.15}{
\begin{tikzpicture}\draw (0,-1) -- (2,-1);\draw (2,1) -- (0,1);
\end{tikzpicture}}
\end{scalebox},
\begin{scalebox}{0.15}{
\begin{tikzpicture}\draw (2,1) -- (0,1) -- (0,-1);\filldraw[fill=black, draw=black] (2,-1) circle (0.1cm);
\end{tikzpicture}}
\end{scalebox},
\begin{scalebox}{0.15}{
\begin{tikzpicture}\draw (0,1) -- (0,-1) -- (2,-1);\filldraw[fill=black, draw=black] (2,1) circle (0.1cm);
\end{tikzpicture}}
\end{scalebox}, 
\begin{scalebox}{0.15}{
\begin{tikzpicture}\draw (0,-1) -- (2,-1) -- (2,1);\filldraw[fill=black, draw=black] (0,1) circle (0.1cm);
\end{tikzpicture}}
\end{scalebox}, 
\begin{scalebox}{0.15}{
\begin{tikzpicture}\draw (2,-1) -- (2,1) -- (0,1);\filldraw[fill=black, draw=black] (0,-1) circle (0.1cm);
\end{tikzpicture}}
\end{scalebox}$\rbrace$. The momenta flowing into one part of the 2-trees are $s$ for 
\begin{scalebox}{0.15}{
\begin{tikzpicture}\draw (0,1) -- (0,-1);\draw (2,-1) -- (2,1);
\end{tikzpicture}}
\end{scalebox},
$t$ for
\begin{scalebox}{0.15}{
\begin{tikzpicture}\draw (0,-1) -- (2,-1);\draw (2,1) -- (0,1);
\end{tikzpicture}}
\end{scalebox},
$s+t+u$ for
\begin{scalebox}{0.15}{
\begin{tikzpicture}\draw (2,1) -- (0,1) -- (0,-1);\filldraw[fill=black, draw=black] (2,-1) circle (0.1cm);
\end{tikzpicture}}
\end{scalebox}
and zero for the other three, corresponding to the three on-shell massless particles.
From the trees we read off the polynomials $\mathcal{U}$ and $\mathcal{V}$ as
\begin{align}
\mathcal{U}&=x_1+x_2+x_3+x_4,
\\
\mathcal{V}&=s x_1 x_3+t x_2 x_4+p_1^2 x_1 x_2+p_2^2 x_2 x_3+p_3^2 x_1 x_4+p_4^2 x_3 x_4.
\end{align}

Besides the Feynman representation, there are several other important parametric representations, which are useful in different situations.
For one, the \textit{Schwinger parameterization}, that derives from writing propagators as
\begin{align}
\frac{1}{D^j}=\frac{1}{\Gamma(j)}\int_0^\infty\dx x^{j-1} \ee^{-x D}\,,
\end{align}
which is given by
\begin{align}
I=\frac{1}{\prod_{i=1}^n\Gamma(j_i)}\int_0^\infty \left(\prod_{i=1}^n\dx x_i\,x_i^{j_i-1}\right)\mathcal{U}^{-\frac{d}{2}}\exp\left[-\frac{\mathcal{F}}{\mathcal{U}}\right]\,.
\label{eq:Schwinger_param1}
\end{align}
We will use the Schwinger parameterization when discussing dimensional shifts of Feynman integrals.
For another, there is the \textit{Lee-Pomeransky} representation \cite{Lee:2013hzt},
\begin{align}
I=\frac{\Gamma\!\left(\frac{d}{2}\right)}{\Gamma\!\left(\frac{(L+1)d}{2}-j\right)\prod_{i=1}^n\Gamma(j_i)}\int_0^\infty \left(\prod_{i=1}^n\dx x_i\,x_i^{j_i-1}\right)\mathcal{G}^{-\frac{d}{2}}\,
\label{eq:LeePom}
\end{align}
with the Lee-Pomeransky polynomial $\mathcal{G}=\mathcal{U}+\mathcal{F}$.
This representation is a particularly simple instance of an \textit{Euler integral}, which are integrals of the form
\begin{align}
I_\text{Euler}(\mathbf{s},\mathbf{j},\mathbf{d})=\int_0^\infty\!\left(\prod_{i=1}^n\frac{\dx t_i}{t_i}\,t_i^{j_i}\right)\prod_{k=1}^N \left[P_k(t_i,\mathbf{s})\right]^{d_k}
\end{align}
where the $P_k$ are polynomials in the integration parameters $t_i$ with coefficients that depend on the set of ``kinematic'' variables $\mathbf{s}$.
In applications in QFT, the $d_k$ depends on the dimensional regulator.
The behavior of the integral, especially its singularity structure, is determined by the polynomials $P_k$.

Euler integrals provide a bridge between Feynman integrals and algebraic geometry.
In the Lee-Pomeransky representation, there is only the single $\mathcal{G}$ polynomial that encodes the information.
This makes it particularly well-suited for formalizing heuristic methods developed in momentum space into algorithms.
An example relevant to this thesis is \textit{expansion by regions} discussed in section \ref{sec:ExpByReg}.
In publications \ref{pub:7} and \ref{pub:10}, we will discuss Euler representations for angular integrals; in particular, the representation developed in \ref{pub:10} is remarkably close in form to the Lee-Pomeransky representation.
Overall, Euler integrals provide a useful common language for loop and angular integrals.

\begin{readingblock}{\textbf{Further reading: }
Excellent books to learn about Feynman integrals and how to calculate them have been written by Vladimir Smirnov \cite{Smirnov:2012gma} and Stefan Weinzierl \cite{Weinzierl:2022eaz}.
}\end{readingblock}

After this discussion of loop integrals, we will devote the next section to the other type of integral relevant to this thesis: phase-space integrals.

\section{Phase space integrals}
\label{sec:PSint}
The general form of Lorentz invariant phase space integrals is given by
\begin{align}
\int\dx \mathrm{PS}_n\,f(p_j,k_i)=\prod_{i=1}^n\int\frac{\dx^{d-1} k_i}{(2\pi)^{d-1}2 E_{k_i}}\,(2\pi)^d\delta^d\left(P-\sum_{i=1}^n k_i\right)\,f(p_j,k_i)
\end{align}
where $P$ is the sum over incoming momenta and the energies are $E_k=\sqrt{m^2+\mathbf{k}^2}$.
The function $f(p_j,k_i)$ to be integrated, e.\,g. a squared amplitude, depends on the integration momenta $k_i$ and external vectors $p_j$.
Equivalently, one can re-write the $d-1$ dimensional integrals as $d$ dimensional ones subject to the on-shell condition
\begin{align}
\frac{\dx^{d-1} k_i}{2 E_{k_i}}=\dx^d k_i\,\underbrace{\delta(k_i^2-m^2)\,\theta(k_i^0)}_{=\delta^+(k_i^2-m^2)}\,.
\label{eq:PhaseSpaceToLoop}
\end{align}

To calculate a cross-section differential in a set of kinematic quantities $X$ the phase-space is split into two parts, one that is kept differential corresponding to the momenta of observed particles, and another that is integrated over.
For example, in the Drell-Yan process $p(p_1)+p(p_2)\longrightarrow \gamma^\star(q)+X\rightarrow \ell(l_1)\bar{\ell}(l_2)+X$ where we observe the complete kinematics of the lepton pair, we are interested in results at fixed $q$ and the lepton angular distribution in $\dx\Omega_{\ell\bar{\ell}}$.
Hence, the phase space for the contribution with $n$ additional particles with momenta $k_1,\dots,k_n$ is given by
\begin{align}
\int\dx\mathrm{PS}_{2+N}=\frac{1}{8(2\pi)^6}\int\dx^4 q\int\dx\Omega_{\ell\bar{\ell}}\int\dx\mathrm{PS}_N^{K\rightarrow k_1\dots k_N}\Theta(K^2)\,\Theta(K^0)\,,
\end{align}
where $K=p_1+p_2-q$.
Here, the $\mathrm{PS}_N^{K\rightarrow k_1\dots k_N}$ is the part that needs to be integrated with the squared matrix element.

Let us consider a process with $n$ incoming and $m$ observed outgoing particles in $d=4-2\eps$ dimensions.
If we allow for the radiation of $l$ additional particles, the phase space measure of the purely real corrections reads
\begin{align}
\int\dPS_{m+l}^{P_i\rightarrow P_f+k_1+\dots+k_l}=
\int\prod_{j=1}^m\dx\mu\!\left(p_{f,j}\right)\prod_{j=1}^l\dx\mu\!\left(k_{j}\right)(2\pi)^d\,\delta\!\left(P_i-P_f-\sum_{i=1}^l k_i\right)
\end{align}
where $P_i=\sum_{j=1}^n p_{i,j}$ and $P_f=\sum_{j=1}^m p_{f,j}$
and
\begin{align}
\dx\mu\!\left(p\right)=\frac{\dx^{d-1}p}{(2\pi)^{d-1}2p^0}=\frac{\dx^{d}p}{(2\pi)^{d-1}}\,\delta^+(p^2-m^2)\,.
\end{align}
Using the identity
\begin{align}
1=(2\pi)^{d-1}\int_0^\infty \dx K^2\,\dx\mu(K)\,\delta^d\left(K-\sum_{i=1}^l k_i\right)
\end{align}
we can split the phase space like
\begin{align}
\int\dPS_{m+l}^{P_i\rightarrow P_f+k_1+\dots+k_l}=
\frac{1}{2\pi}\int_0^\infty\dx K^2 \int\dPS_{m+1}^{P_i\rightarrow P_f+K}\int\dPS_{l}^{K\rightarrow k_1+\dots+k_l}.
\end{align}
We treat the first part exclusive, i.e. we keep the process fully differential in the $p_f$.
The integration measure of the second part does not depend on any of the observed particles, hence we want to integrate out this part completely.
For simplicity we will assume in the following that the additional radiated particles are massless, i.e. $k_i^2=0$. 

At leading order, when there is no real radiation, it is
\begin{align}
\int\dPS_0^{K\rightarrow\,\cdot}=(2\pi)^{4-2\eps}\,\delta^d(K)\,.
\end{align}
At NLO, when there is one additional particle, it is
\begin{align}
\int\dPS_1^{K\rightarrow k_1}=2\pi\,\delta^+(K^2)\,.
\end{align}
At NNLO, when there are two additional particles, it is
\begin{align}
\int\dPS_2^{K\rightarrow k_1+k_2}=\frac{(K^2)^{-\eps}}{2 (4\pi)^{2-2\eps}}\int\dx\Omega_{d-1}^K(k_1)\,,
\end{align}
where $\Omega_{d-1}^K(k_1)$ denotes the angular integral in $d-1$ (spatial) dimensions over the vector $k_1$ in the rest frame of $K$.
This is the first order where the integration of the (squared) amplitude is non-trivial.
A general 2-particle integral over a tree-level (squared) amplitude has the form
\begin{align}
&\int\dPS_2^{K\rightarrow k_1+k_2}\prod_{i=1}^n\frac{1}{((p_i-k_1)^2-m_i^2)^{j_i}}
\nonumber\\=&
\frac{(K^2)^{-\eps}}{2 (4\pi)^{2-2\eps}}
\left(\prod_{i=1}^n\frac{(-1)^{j_i}}{(K\cdot p_i-p_i^2+m_i^2)^{j_i}}\right)\,
\Omega_{j_1,\dots,j_n}(\lbrace v_{ii^\prime}\rbrace;\eps)
\end{align}
with the invariants 
\begin{align}
v_{ii^\prime}=v_i\cdot v_{i^\prime}=1-\frac{K\cdot p_i\,K\cdot p_{i^\prime}-K^2 p_i\cdot p_{i^\prime}}{(K\cdot p_i-p_i^2+m_i^2)(K\cdot p_{i^\prime}-p_{i^\prime}^2+m_{i^\prime}^2)}\,.
\end{align}
Note that in a propagator with $k_2$ we can always use $k_2=K-k_1$ to eliminate $k_2$ and numerator factors can be expressed as (combinations of) inverse propagators.
The appearing angular integrals $\Omega_{j_1,\cdots,j_n}(\lbrace v_{ii^\prime}\rbrace;\eps)$ are one of the central objects of study in this work.
A lot of the progress on angular integrals is made possible by transferring technology from loop integration over to the phase space setting.
Therefore, we will have a look at the relevant techniques -- at first in the loop integral setting, before we turn to angular integrals in the publications.
\begin{block}[type=note]
\textbf{Reverse unitarity: }There is a way to identify phase-space integrals with loop integrals by a method called \textit{reverse unitarity}, introduced by Charalampos Anastasiou and Kirill Melnikov \cite{Anastasiou:2002yz,Anastasiou:2003ru,Anastasiou:2004ru}. Using the rewriting of the phase-space measure from eq.\,\eqref{eq:PhaseSpaceToLoop} and expressing the delta function via
\begin{align}
\delta^+(k^2-m^2)=\frac{1}{2\pi \iu}\left[\frac{1}{k^2-m^2-\iu 0}-\frac{1}{k^2-m^2+\iu 0}\right]
\end{align}
the phase-space integral can be expressed as the branch cut discontinuity of a loop integral.
This approach is very successful in many cases since it allows to use the tools developed for Feynman integrals in a direct way.
The price to pay is however that one introduces additional denominators leading to an arguably more complicated class of integrals.
In this work, we will stay in the domain of angular integrals and transfer the loop methods directly such that they can be applied without resorting to reversed unitarity.
For comparison of approaches, we refer to the classic calculation of van Neerven \cite{vanNeerven:1985} that used the connection between the massless two-denominator angular integral and the box loop integral via the optical theorem similar in spirit to reverse unitarity, and for a modern example to the recent publication \cite{Liu:2024hfa} where reverse unitarity was used for integrals very close to the angular integrals discussed here.
\end{block}
\begin{readingblock}{
\textbf{Further reading: }For a thorough study of angular integrals see the foundational paper by Gabor Somogyi \cite{Somogyi:2011} and the publications \ref{pub:1}, \ref{pub:4}, \ref{pub:7}, \ref{pub:8}, and \ref{pub:10} of this thesis.
}\end{readingblock}

\section{IBP and reduction to master integrals}
\label{sec:IBP}
When calculating scattering amplitudes, one is quickly confronted with an unwieldy large number of loop integrals to calculate, especially if the number of loops exceeds one.
Higher order calculations are only feasible because it turns out that these integrals are not independent but related by algebraic identities.
The most important class of these are the so-called \textit{integration-by-parts (IBP)} relations \cite{Tkachov:1981}.
They connect different members of a family of Feynman integrals and ultimately allow for the reduction to a small set of \textit{master integrals}, which form a basis for the family.

An integral family is defined as
\begin{align}
I_{j_1,\dots,j_n}=\int\frac{\dx^d k_1}{\iu\pi^{d/2}}\dots \int\frac{\dx^d k_L}{\iu\pi^{d/2}}\frac{1}{D_1^{j_1}\dots D_n^{j_n}}
\end{align}
where the $D_a$ are propagators of the form $D_i=-q_i^2(p,k)+m^2$ with $q_i(p,k)$ depending on any combination of external vectors $p$ and loop momenta $k$.
Now, since boundary terms at infinity vanish -- at least for some value of $d$ and $j_i$, and by analytic continuation everywhere --, these integrals satisfy the relation
\begin{align}
0=\int\frac{\dx^d k_1}{\iu\pi^{d/2}}\dots \int\frac{\dx^d k_L}{\iu\pi^{d/2}}\,\frac{\partial}{\partial k_i^\mu}\frac{\xi^\mu}{D_1^{j_1}\dots D_n^{j_n}}\,,
\end{align}
with any loop momentum $k_i^\mu$ and vector $\xi^\mu$.
Differentiating the integrand results in an expression that can again be expressed in terms of propagators with shifted indices.
Considering all possibilities for these momenta results in a large number of linear relations between the integrals of the form
\begin{align}
0=\sum_i c_i\,I_{j_1+a_{1,i},\dots,j_n+a_{n,i}}\,.
\end{align}
In principle, these relations contain all the information necessary to perform the reduction to master integrals.
In practice, it is not a priori clear how to efficiently use the potentially large number of IBP relations for the efficient reduction of a set of integrals.

The breakthrough in this regard came with Laporta's algorithm \cite{Laporta:2000}, which opened up the systematic calculation of multi-loop amplitudes.
It introduces an ordering on the integrals based on the powers $j_1,\dots, j_n$, in general the integral is considered ``more complicated'' when $\sum_i\max(j_i,0)$ is larger and the more $j_i\neq 0$.
The latter condition defines so-called \textit{sectors} within the integral family.
Then integers $j_i$ are chosen up to some bound to generate IBP relations.
They are ordered such that they give a reduction rule of the most complicated integral in terms of simpler ones.
Taking enough of these, one obtains a large linear system, with which one can reduce all the integrals one has in a problem to master integrals.
These are the integrals considered the ``simplest'' by the ordering rule, i.\,e. for which no further IBP identities exist expressing them through simpler integrals.
Here, there is a freedom which integrals to choose as master integrals, algebraically corresponding to a change of basis.
For example, it is often useful to choose master integrals such that they are free of poles. 
This trick is used in the proof of renormalizability by 't Hooft to treat sub-divergences \cite{Hooft:1971}.

In simpler cases, it is also possible to symbolically convert the IBP relations into explicit recursion relations that reduce any integral to simpler integrals without specifying explicit integer powers.
For angular integrals, this will be done in publications \ref{pub:1}, \ref{pub:8}, and \ref{pub:10}, for two, three, and more denominators, respectively.

Overall, IBP reduction reduces the number of integrals to be calculated to a manageable scale.
In many practical applications, it is the IBP reduction that is the bottleneck to the calculation because of the potential large size of the linear system.
\begin{toolblock}{
There exist several public computer algebra implementations for IBP reduction, most prominently \texttt{AIR} \cite{Anastasiou:2004vj}, \texttt{FIRE} \cite{Smirnov:2008iw,Smirnov:2013dia,Smirnov:2014hma,Smirnov:2019qkx,Smirnov:2023yhb,Smirnov:2025prc}, \texttt{Reduze} \cite{Studerus:2009ye,vonManteuffel:2012np}, \texttt{LiteRed} \cite{Lee:2012cn,Lee:2013mka}, and \texttt{KIRA} \cite{Maierhofer:2017gsa,Klappert:2020nbg,Lange:2025fba}.
}\end{toolblock}

As an explicit example that we can do ``by hand'', let us consider the one-loop massive bubble family
\begin{align}
B_{j_1,j_2}(s,m^2;d)&=\int\frac{\dx^d k}{\iu\pi^{d/2}}\frac{1}{(-k^2+m^2-\iu 0)^{j_1}(-(k+p)^2+m^2-\iu 0)^{j_2}}
\nonumber\\
&\equiv\int\frac{\dx^d k}{\iu\pi^{d/2}}\frac{1}{D_1^{j_1}D_2^{j_2}}
\end{align}
with integers $j_1$, $j_2$ and a conventional normalization factor of $\iu\pi^{d/2}$.
This will be our go-to example not only for IBP reduction, but also for the other methods in this section.

Differentiating the propagators with respect to $k^\mu$ increases the denominator power as
\begin{align}
\frac{\partial}{\partial k^\mu}\frac{1}{D_1^{j_1}}=\frac{2j_1 k_\mu}{D_1^{j_1+1}}\,,\quad\frac{\partial}{\partial k^\mu}\frac{1}{D_2^{j_2}}=\frac{2j_2 (k_\mu+p_\mu)}{D_2^{j_2+1}}\,.
\end{align}
With this, the IBP relations for the bubble take the form
\begin{align}
0=\int\frac{\dx^d k}{\iu\pi^{d/2}}\frac{\partial}{\partial k^\mu}\frac{\xi^\mu}{D_1^{j_1}D_2^{j_2}}
=\int\frac{\dx^d k}{\iu\pi^{d/2}}\left(\frac{\partial \xi^\mu}{\partial k^\mu}+\frac{2j_1 k\cdot\xi}{D_1}+\frac{2 j_2(k\cdot\xi+p\cdot \xi)}{D_2}\right)\frac{1}{D_1^{j_1}D_2^{j_2}}
\label{eq:IBP_specific}
\end{align}
Now, for $\xi^\mu=k^\mu$ as well as $\xi^\mu=p^\mu$ we can express all scalar products involving $k$ by inverse denominators using $k^2=-D_1+m^2$ and $2k\cdot p=D_1-D_2-s$, where $s=p^2$.
This turns the left side of eq.\,\eqref{eq:IBP_specific} into integrals of the same bubble family with shifted denominators.
Taking $\xi^\mu=k^\mu$ we obtain the IBP relation
\begin{align}
0=&(d-2j_1-j_2)B_{j_1,j_2}-j_2 B_{j_1-1,j_2+1}\nonumber\\
&+2m^2 j_1 B_{j_1+1,j_2}+
(2m^2-s)j_2 B_{j_1,j_2+1}\,,
\label{eq:BubbleIBPExample}
\end{align}
and for $\xi^\mu=p^\mu$ 
\begin{align}
0=&(j_1-j_2)B_{j_1,j_2}-j_1 B_{j_1+1,j_2-1}+j_2 B_{j_1-1,j_2+1}\nonumber\\
&-s j_1 B_{j_1+1,j_2}+
s j_2 B_{j_1,j_2+1}\,.
\label{eq:BubbleIBPExample2}
\end{align}
In both cases, the first integrals of the first line have an index sum of $j_1+j_2$, those in the second line $j_1+j_2+1$.
Together with the symmetry relation $B_{j_2,j_1}=B_{j_1,j_2}$, these allow for a complete reduction to the master integrals.

We can now substitute explicit integers for $j_1$ and $j_2$ to generate a linear system of equations.
For example, taking $(j_1,j_2)=(1,0)$ and $(j_1,j_2)=(1,1)$, we have the equations
\begin{align}
0&=(d-2)\,B_{1,0}+2 m^2\,B_{2,0}\,,
\label{eq:BubbleIBPExampleExplicitA}\\
0&=(d-3)\,B_{1,1}-B_{2,0}+(4m^2-s)\,B_{2,1}\,.
\label{eq:BubbleIBPExampleExplicitB}
\end{align}
which can be solved for a reduction of $B_{2,0}$ and $B_{2,1}$ in terms of $B_{1,0}$ and $B_{1,1}$.
Explicitly,
\begin{align}
B_{2,0}&=\frac{2-d}{2 m^2}\,B_{1,0}\,,\\
B_{2,1}&=\frac{2-d}{2 m^2(4 m^2-s)}\,B_{1,0}+\frac{3-d}{4 m^2-s}\,B_{1,1}\,.
\end{align}
Taking the linear system of IBP relations from eqs.\,\eqref{eq:BubbleIBPExampleExplicitA} and \eqref{eq:BubbleIBPExampleExplicitB} large enough, we can reduce all integrals $B_{j_1,j_2}$ to $B_{1,0}$ and $B_{1,1}$ with algebraic coefficients $c_{10}$, $c_{11}$ determined by the IBP reduction,
\begin{align}
B_{j_1,j_2}=c_{10} B_{1,0}+ c_{11} B_{1,1}\,.
\end{align}
So overall, after solving the IBP reduction, we only need to calculate the integrals $B_{1,0}$ and $B_{1,1}$ explicitly to pin down the entire family $B_{j_1,j_2}$.
In the next two sections, we will discuss two methods to calculate master integrals, first approximately in a systematic expansion, then in exact kinematics, and continue with this example.

\begin{readingblock}{\textbf{Further reading: }
For a conceptual discussion of the principles based around the same example presented here, see the textbook \cite{Badger:2023eqz}.
For a recent review of actual implementations, see \cite{Smirnov:2025dfy}.
}\end{readingblock}

\section{Expansion by regions}
\label{sec:ExpByReg}
Feynman integrals with many scales are difficult to calculate.
However, in many cases, the scales are of very different size or we are interested in specific kinematic limits.
For example, external momenta can be much larger than an internal mass $|p^2|\geq m^2$, e.g. in the hard scattering of an electron or a light quark. Similarly, in cases of heavy particles, such as top quarks, we might be interested in a situation where $|p^2|\leq m^2$.
In both cases, we get much of the information from a result that is expanded in powers of the small parameter that is the fraction of both scales.
The idea is to perform this expansion on the integrand level such that the number of scales in the Feynman integral is reduced, thus facilitating its computation.
This is complicated by the fact that the loop momentum may take any magnitude.

The systematic framework to construct such an expansion in dimensional regularization is the method of  \textit{expansion by regions}, developed by Martin Beneke and Vladimir Smirnov in 1997 \cite{Beneke:1997zp}.
The solution lies in identifying the regions of loop momenta that contribute to the overall integral. 
Each region corresponds to a specific scaling of the loop momentum in terms of the expansion parameter.
Then, in each region, a Taylor expansion of the integrand can be performed.
Importantly, by virtue of dimensional regularization, each region integral can be performed over the full loop momentum space.
This only adds scaleless integrals that give a vanishing contribution.

The modern formulation identifies the regions on the level of parametric Euler integrals and this is the way it is used in publication \ref{pub:7}.
A brief introduction into the algebraic formulation of the algorithm can be found there.
Here, to get an intuition behind the method we will follow a heuristic approach in momentum space and continue with the example of the massive bubble integral 
\begin{align}
B_{1,1}(s,m^2;d)=\int\frac{\dx^d k}{\iu \pi^{d/2}}\frac{1}{(k^2-m^2)((k+p)^2-m^2)}
\end{align}
in the small-mass limit $|s|\geq m^2$, where as before $p^2=s$, and we consider the space-like case $s<0$ to not be bothered with branch cuts.
In this case, we can define the power counting parameter $\lambda$ as
\begin{align}
\lambda=-\frac{m^2}{s}\,.
\end{align} 
Here, there are two types of regions that contribute to the full integral:
\begin{itemize}
\item \textbf{Hard region:} This gives the contribution where both propagators are far off-shell and captures the UV behavior.
The loop momentum behaves as $k^\mu\sim p^\mu$, hence all components are large compared to $m$ and $k^2\gg m^2$.
\item \textbf{Soft regions:} There are two possibilities for one of the propagators to go on-shell. The first propagator can become zero when the components of the loop momenta behave like $k^\mu\sim m$, the second propagator can become zero when $k^\mu+p^\mu\sim m$ that is $k^\mu=-p^\mu+\mathcal{O}(\sqrt{\lambda})$.
\end{itemize}
After identifying the relevant regions, the next step is to use the scaling of the loop momentum to expand the integrand.

\noindent{}\textbf{Hard region.} 
In the hard region it is $m^2\ll k^2,(k+p)^2$. Expanding both propagators in powers of $m^2$ we have
\begin{align}
\frac{1}{k^2-m^2}=\frac{1}{k^2}\sum_{n=0}^\infty \left(\frac{m^2}{k^2}\right)^n\,\text{ and }
\frac{1}{(k+p)^2-m^2}=\frac{1}{(k+p)^2}\sum_{n=0}^\infty \left(\frac{m^2}{(k+p)^2}\right)^n\,.
\end{align}
Using these under the integral sign, we get for the hard region up to next-to-leading power (NLP)
\begin{align}
B_{1,1}^\text{hard}=\int\frac{\dx^d k}{\iu \pi^{d/2}}\left[
\frac{1}{k^2(k+p)^2}+\frac{m^2}{[k^2]^2(k+p)^2}+\frac{m^2}{k^2[(k+p)^2]^2}
\right]+\mathcal{O}(\lambda^2)\,.
\end{align}
Respecting the sign convention of the last section and taking the symmetry between the propagators into account, the hard region contributes as
\begin{align}
B_{1,1}^\text{hard}=B_{1,1}(s,m^2=0)-2m^2 B_{2,1}(s,m^2=0)+\mathcal{O}(\lambda^2)\,.
\end{align}
Note that all the integrals on the right only depend on a single scale and are thus much simpler to evaluate than the original two-scale integrals.
Using the massless results
\begin{align}
B_{2,1}(s,0)=\frac{d-3}{s}B_{1,1}(s,0)\text{ and }B_{1,1}(s,0)=\frac{\Gamma\!\left(2-\frac{d}{2}\right)\Gamma^2\!\left(\frac{d}{2}-1\right)}{\Gamma(d-2)}\,(-s)^{d/2-2}\,.
\end{align}
the hard region contributes in $d=4-2\eps$ with
\begin{align}
B_{1,1}^\text{hard}=\frac{1}{\Gamma(1-\eps)}\left[\frac{1}{\eps}+2-\log(-s)+\lambda\left(\frac{2}{\eps}-2\log(-s)\right)+\mathcal{O}(\eps,\lambda^2)\right].
\end{align}
\noindent{}\textbf{Soft regions.}
In the first soft region, it is $k^\mu\sim m$.
Therefore 
\begin{align}
(k+p)^2-m^2=k^2+2k\cdot p+p^2-m^2=p^2+\mathcal{O}(\sqrt{\lambda})
\end{align}
and thus
\begin{align}
B_{1,1}^\text{soft,1}=\int\frac{\dx^d k}{\iu \pi^{d/2}}\frac{1}{(k^2-m^2) p^2}+\mathcal{O}(\lambda^2)=-\frac{1}{p^2}\,B_{1,0}(m^2)+\mathcal{O}(\lambda^2)\,,
\end{align}
where the minus sign comes from the convention for the propagators in $B_{j_1,j_2}$ and the error is indeed $\mathcal{O}(\lambda^2)$ since the leading term is $\mathcal{O}(\lambda)$ and odd powers vanish upon integration by anti-symmetry.
Analogously, in the second soft region where $k^\mu+p^\mu \sim m$, it is $(k^2-m^2)=p^2+\mathcal{O}(\sqrt{\lambda})$ and thus
\begin{align}
B_{1,1}^\text{soft,2}=\int\frac{\dx^d k}{\iu \pi^{d/2}}\frac{1}{p^2((k+p)^2-m^2)}+\mathcal{O}(\lambda^2)=-\frac{1}{p^2}\,B_{1,0}(m^2)+\mathcal{O}(\lambda^2)\,,
\end{align}
the same as the first soft region, as was expected by the symmetry of the bubble integral.
As in the hard region, the remaining integral is single-scale,
explicitly we know from eq.\,\eqref{eq:Tadpole} that $B_{1,0}(m^2)=\Gamma(1-d/2)(m^2)^{d/2-1}$.
Plugging this in, the soft contributions in $d=4-2\eps$ read
\begin{align}
B_{1,1}^\text{soft,1/2}=\frac{\lambda}{\Gamma(1-\eps)}\left(-\frac{1}{\eps}-1+\log\lambda+\log(-s)\right)+\mathcal{O}(\eps,\lambda^2)\,.
\end{align}

Combining all regions gives the result of the full integral as an expansion in $\lambda$. 
Thus, in our example we have up to NLP
\begin{align}
&B_{1,1}(s,m^2;d=4-2\eps)=B_{1,1}^\text{hard}+B_{1,1}^\text{soft,1}+B_{1,1}^\text{soft,2}\nonumber\\
&=\frac{1}{\Gamma(1-\eps)}\left[\frac{1}{\eps}+2-\log(-s)+2\lambda(\log\lambda-1)+\mathcal{O}(\eps,\lambda^2)\right]\,.
\label{eq:ExpByRegionsBubbleRes}
\end{align}
Note the cancellation of spurious $\lambda/\eps$ poles between hard and soft regions.

In the next section, we will look at a method to calculate Feynman integrals without a kinematic expansion.
We will apply it to the very same example and in the end we will see that the expansion by regions result indeed agrees with the exact result  up to NLP.

\begin{toolblock}{
The method of expansion by regions is automated in several public codes, including \texttt{asy.m} \cite{Pak:2010pt,Jantzen:2012mw}, \texttt{FIESTA} \cite{FIESTA3:2014,FIESTA4:2016}, and \texttt{pySecDec} \cite{Heinrich:2022}.
}
\end{toolblock}

\begin{readingblock}{\textbf{Further reading: }
For a recent review on the method and its foundations, see \cite{Semenova:2019}.
To learn how you get from integral expansions to an effective field theory, see the introductory book to \textit{soft-collinear effective theory (SCET)} \cite{Becher:2014oda}. 
}\end{readingblock}
\section{Differential equations}
\label{sec:DiffEq}
One of the most powerful and widely used methods to calculate Feynman master integrals is the method of differential equations \cite{Kotikov:1991,Gehrmann:1999as}, introduced in its systematic form by Johannes Henn \cite{Henn:2013pwa}\footnote{In this section we will in part follow notes from one of his lectures -- which was based on the material of \cite{Henn:2014qga,Badger:2023eqz} -- taken at the 2023 QCD Masterclass.}.
Conceptually, the idea is to differentiate the integrals, algebraically relate the derivative to the original integrals, and solve the ensuing system of differential equations.
This provides an algorithmic approach to ``integration by differentiation''.

To showcase the principle, instead of directly looking at a loop integral, it is illustrative to first look at the family of special functions
\begin{align}
\mathbf{f}(x;\eps)=\left(\begin{array}{c}\eps^2 \dilog(1-x)\\
\eps\log(x)\\
1
\end{array}\right).
\end{align}
Recalling that $\log x=\int_1^x \frac{\dx t}{t}$ 
and $\dilog (x)=-\int_0^x \frac{\dx t}{t}\log(1-t)$,
we find by differentiating $\mathbf{f}$ with respect to $x$
\begin{align}
\partial_x\mathbf{f}(x;\eps)=\left(\begin{array}{c}\eps^2 \frac{\log x}{1-x}\\
\frac{\eps}{x}\\
0
\end{array}\right)
=\eps\underbrace{\left[\frac{A_0}{x}+\frac{A_1}{x-1}\right]}_{\equiv A(x)}\cdot\mathbf{f}(x;\eps)
\label{eq:DiffEqExample}
\end{align}
with
\begin{align}
A_0=\left(\begin{array}{ccc} 0&0&0\\0&0&1\\0&0&0\end{array}\right)\quad\text{and}\quad
A_1=\left(\begin{array}{ccc} 0&-1&0\\0&0&0\\0&0&0\end{array}\right).
\end{align}

In differential form with $\dx=\partial_x \dx x$ this reads
\begin{align}
\dx\mathbf{f}(x;\eps)=\eps\left[A_0\dx\!\log x+A_1\dx\!\log(1-x)\right]\cdot\mathbf{f}(x;\eps)\,,
\label{eq:DiffEqExampleDlog}
\end{align}
which is of the typical form of differential equations encountered for loop integrals.
``Forgetting'' our explicit definition of $\mathbf{f}(x;\eps)$ from which we derived it, only from the differential equation we can reconstruct $\mathbf{f}(x;\eps)$ by integrating the $\dx\!\log$-forms.
Directly, we can read off its singularities at $x=0$, $x=1$, and by change of variables $y=1/x$ at $x=\infty$.
This already greatly restricts the possible function space of the solutions.

Note that the $\eps$-dependence of eq.\,\eqref{eq:DiffEqExampleDlog} is very simple, only an overall proportionality factor of $\eps$.
This makes it such that we can solve iteratively by making the ansatz
\begin{align}
\mathbf{f}(x;\eps)=\sum_{k=0}^\infty\eps^k f^{(k)}(x)\,.
\end{align}
Plugging this into eq.\,\eqref{eq:DiffEqExample},
\begin{align}
\sum_{k=0}^\infty \eps^k\partial_x \mathbf{f}^{(k)}(x)=\sum_{k=0}^\infty \eps^{k+1}A(x)\cdot \mathbf{f}^{(k)}(x)\,,
\end{align}
the orders decouple and we have
\begin{align}
\partial_x\mathbf{f}^{(0)}(x)=0\,,\quad
\partial_x\mathbf{f}^{(1)}(x)=A(x)\cdot \mathbf{f}^{(0)}(x)\,,\quad
\partial_x\mathbf{f}^{(2)}(x)=A(x)\cdot \mathbf{f}^{(1)}(x)\,,\dots
\label{eq:OrderDecoupling}
\end{align}
Now taking the boundary value $\mathbf{f}(x=1;\eps)=(0,0,1)$, it is
\begin{align}
\mathbf{f}^{(0)}(x)=\left(
\begin{array}{c}
0\\0\\1
\end{array}\right)
\end{align}
and iteratively integrating yields
\begin{align}
\mathbf{f}^{(1)}(x)=\left(\begin{array}{c}0\\ \log x\\ 0\end{array}\right)\quad\text{and}\quad
\mathbf{f}^{(2)}(x)=\left(\begin{array}{c}\dilog(1-x) \\ 0 \\ 0\end{array}\right).
\end{align}
In this particular example, the higher orders all vanish exactly since $\partial_x \mathbf{f}^{(3)}(x)=A(x)\cdot\mathbf{f}^{(2)}(x)=0$ and thus $\mathbf{f}^{(3)}(x)=0$.
Hence, we recover the original function from the differential equation.

In many cases, viewing master integrals in terms of their associated differential equations is very helpful.
For a family of Feynman integrals $\mathbf{f}(x;\eps)$, the general structure of differential equations is
\begin{align}
\partial_x \mathbf{f}(x;\eps)=A(x;\eps)\cdot \mathbf{f}(x;\eps)
\label{eq:DiffEqGen}
\end{align}
where $x$ is a kinematic invariant, $\eps$ the dimensional regulator and the matrix $ A(x;\eps)$ has only simple poles, meaning near a singularity at $x_0$ the matrix has the form
\begin{align}
A(x;\eps)=\frac{A_0(\eps)}{x-x_0}+\mathcal{O}(1)\,.
\end{align}
This property is called \textit{Fuchsian}.

An example for a non-Fuchsian differential equation would be
\begin{align}
\partial_x f(x)=\frac{a}{x^2}\,f(x)
\end{align}
with solution $f(x)=\ee^{-a/x}$, which has an essential singularity at $x=0$.
The Fuchsian property ensures that such behavior does not appear in Feynman integrals.

Here, another important aspect is the ``gauge dependence'' of the differential equation eq.\,\eqref{eq:DiffEqGen}.
Considering a transformation 
\begin{align}
\mathbf{f}(x;\eps)=T(x)\cdot \mathbf{g}(x;\eps)
\end{align}
with any invertible matrix $T(x)$ we get a new differential equation\footnote{Which looks just like a gauge transformation with $T=\ee^{\iu g \omega_a(x)t^a}$ and $A=-\iu g A_\mu^a t^a$ (compare e.\,g. \cite{Collins:2011zzd}).}
\begin{align}
\partial_x \mathbf{g}(x;\eps)=B(x;\eps)\cdot \mathbf{g}(x;\eps)\quad\text{with}\quad
B=T^{-1}A T-T^{-1}\partial_x T\,.
\end{align}
Now, taking for example
\begin{align}
T=\left(\begin{array}{ccc}
1+x & 0 &0\\
0 & -x & 0\\
0 & 0 & 1
\end{array}\right)
\text{ it is }
B=\left(\begin{array}{ccc}
-\frac{1}{1+x} & \frac{\eps x}{(x-1)(1+x)} &0\\
0 & -\frac{1}{x} & -\frac{\eps}{x^2}\\
0 & 0 & 0
\end{array}\right)
\end{align}
This new matrix $B$ is no longer proportional to $\eps$, has a double pole at $x=0$, and a spurious singularity at $x=-1$.
Hence, in practice, we will need to find suitable ``gauge transformations'' to bring the system into canonical form, since only then the properties of the differential equation become transparent.

The formal solution of eq.\,\eqref{eq:DiffEqGen} is a path ordered exponential\footnote{Which has the same form as a Wilson line in QCD with $A=-\iu g A_\mu^a t^a$ (compare again e.\,g. \cite{Collins:2011zzd})}
\begin{align}
\mathbf{f}(x;\eps)=\mathcal{P}\exp\left[
\int_C\dx x^\prime\, A(x^\prime;\eps)
\right]\cdot\mathbf{f}(x_0,\eps)\,.
\label{eq:PExpSolution}
\end{align}
Eq.\,\eqref{eq:PExpSolution} is a formal expansion in terms of iterated integrals that becomes useful when $A$ is in canonical form, since then $A\sim\eps$, which we can use as an expansion parameter.
Here, $\eps$ plays the role the QCD coupling plays for Wilson lines.
So for $A(x;\eps)=\eps A(x)$ it is
\begin{align}
\mathbf{f}(x;\eps)&=\mathcal{P}\exp\left[
\eps\int_C\dx x^\prime\, A(x^\prime)
\right]\cdot\mathbf{f}(x_0;\eps)
\nonumber\\
&=\left[1+\eps\int_C\dx x^\prime\, A(x^\prime)+\frac{\eps^2}{2}\int_C\dx x^\prime\int_C\dx x^{\prime\prime}\, \mathcal{P}\left[A(x^\prime)A(x^{\prime\prime})\right]+\dots \right]\cdot\mathbf{f}(x_0;\eps)
\end{align} 
where the $\eps^k$ term is a $k$-fold iterated integral and
we can truncate the exponential in a systematic expansion.
If $A(x)$ consists only of $\dx\!\log$-forms the result is in the space of multiple polylogarithms.

We now continue the example of the massive bubble integral from the last section where we set $d=2-2\eps$ for reasons that will become clear later.
How to connect the result back to $d=4-2\eps$ will be a matter of section \,\ref{sec:DimShift}.
We left off with the master integrals $B_{0,1}$ and $B_{1.1}$, which we now want to calculate using differential equations.
The integrals depend on the invariants $m^2$ and $s\equiv p^2$.
The derivative of the propagators with respect to $m^2$ can be calculated directly, for $\partial_s$ we can make the ansatz $\partial_s=\beta p\cdot \partial_p$, since $p$ is the only vector the integral depends on.
From $\partial_s p^2\overset{!}{=}1$ it follows
\begin{equation}
\partial_s=\frac{1}{2 s}p\cdot \partial_p\,.
\end{equation}
Straightforwardly, we find
\begin{align}
\partial_{m^2}\left(\begin{array}{c}
B_{0,1}\\ B_{1,1}
\end{array}\right)=
\left(\begin{array}{c}
-B_{0,2}\\ -2B_{2,1}
\end{array}\right)\,.
\label{eq:BubbleMDeriv}
\end{align}
Now, we can use the IBP relations from eq.\,\eqref{eq:BubbleIBPExampleExplicitA} and \eqref{eq:BubbleIBPExampleExplicitB} to express the right side in terms of the basis integrals $B_{0,1}$ and $B_{1,1}$.
This results in
\begin{align}
\partial_{m^2}\underbrace{\left(\begin{array}{c}
B_{0,1}\\ B_{1,1}
\end{array}\right)}_{=\mathbf{b}}
=\left[\underbrace{\left(\begin{array}{cc}
0&0\\ 0 & -\frac{2}{4m^2-s}
\end{array}\right)}_{=A_{m^2,0}}+\eps \underbrace{\left(\begin{array}{cc}
-\frac{1}{m^2}&0\\ -\frac{2}{m^2(4m^2-s)} & -\frac{4}{4m^2-s}
\end{array}\right)}_{=A_{m^2,1}}
\right]\left(\begin{array}{c}
B_{0,1}\\ B_{1,1}
\end{array}\right)
\end{align}
Doing the same for $\partial_s$, we end up with the system
\begin{align}
\partial_{m^2} \mathbf{b}=A_{m^2}\cdot \mathbf{b}\\
\partial_s\mathbf{b}=A_{s}\cdot \mathbf{b}
\end{align}
where $A_{m^2}=A_{m^2,0}+\eps A_{m^2,1}$ and $A_{s}=A_{s,0}+\eps A_{s,1}$ with the explicit matrices
\begin{align}
A_{s,0}=\left(\begin{array}{cc}
0&0\\0 & \frac{s-2m^2}{s(4m^2-s)}
\end{array}\right)\quad\text{and}\quad
A_{s,1}=\left(\begin{array}{cc}
0&0\\\frac{2}{s(4m^2-s)} & \frac{1}{4m^2-s}
\end{array}\right).
\end{align}

We note that the differential equations are not yet in canonical form and we require a basis transformation
\begin{align}
\mathbf{f}=T\cdot\mathbf{b}
\end{align}
to eliminate the constant term such that the system becomes 
\begin{align}
\partial_{m^2} \mathbf{f}&=\eps \tilde{A}_{m^2,1}\cdot \mathbf{f}\\
\partial_s \mathbf{f}&=\eps \tilde{A}_{s,1}\cdot \mathbf{f}\,.
\label{eq:DiffEqExampleTrafo}
\end{align}
Here the $\tilde{A}$ are transformed matrices that do not depend on $\eps$.
To achieve this, the relevant equation is the ``gauge transformation''
\begin{align}
\eps \tilde{A}_{x,1}=T A_x T^{-1}-T\partial_x T^{-1}
\end{align}
for $x=m^2, s$. Using
\begin{equation}
0=\partial_x\underbrace{(T T^{-1})}_{=1}=(\partial_x T)T^{-1}+T(\partial_x T^{-1})
\end{equation}
it is
\begin{align}
\eps \tilde{A}_{x,1}=(T A_{x,0}+\eps T A_{x,1}+\partial_x T)\cdot T^{-1}\,.
\end{align}
Assuming $T$ is independent of $\eps$, we can ``integrate out'' the constant term by demanding
\begin{align}
\partial_x T=-T A_{x,0}
\end{align}
which is solved by
\begin{align}
T=\left(\begin{array}{cc}
1&0\\0 & \sqrt{-s(4m^2-s)}
\end{array}\right).
\end{align}
Hence, our new basis reads
\begin{align}
\mathbf{f}=\left(\begin{array}{c}
B_{0,1}\\
\sqrt{-s(4m^2-s)}B_{1,1}
\end{array}\right)
\label{eq:DiffEqBasisChange}
\end{align}
and the transformed matrices are $\tilde{A}_{x,1}=T A_{x,1} T^{-1}$.
Combining eq.\,\eqref{eq:DiffEqExampleTrafo} into a total differential, it is
\begin{align}
\dx \mathbf{f}&=\dx m^2\partial_{m^2}\mathbf{f}+\dx s\partial_{s}\mathbf{f}
\nonumber\\
&=\eps\underbrace{\left[\tilde{A}_{m^2,1}\dx m^2+\tilde{A}_{s,1}\dx s\right]}_{\overset{!}{=}\dx\mathcal{A}}\cdot \mathbf{f}\,.
\end{align}
This means we need to find $\mathcal{A}$ such that
\begin{align}
\partial_{m^2}\mathcal{A}=\tilde{A}_{m^2,1}\quad\text{and}\quad
\partial_{s}\mathcal{A}=\tilde{A}_{s,1}\,.
\end{align}
Now this is the only place in the calculation where we need to perform an explicit integration.
We find
\begin{align}
\mathcal{A}=\left(\begin{array}{cc}
-\log m^2 & 0\\
-2\log\left(\frac{\sqrt{1-4m^2/s}-1}{\sqrt{1-4m^2/s}+1}\right) & -\log(4m^2-s)
\end{array}
\right)
\end{align}
satisfying the canonical differential equation
\begin{align}
\dx\mathbf{f}=\eps(\dx\mathcal{A})\cdot\mathbf{f}\,.
\end{align}
From $\mathcal{A}$, we can read off the \textit{symbol alphabet}, i.e. the singularities of the differential equation, which determines the function class of the solution.
Here, the alphabet reads
\begin{align}
\left\lbrace
m^2,4m^2-s,\frac{\sqrt{1-4m^2/s}-1}{\sqrt{1-4m^2/s}+1}
\right\rbrace\,.
\end{align}

We have already learned above that the general solution to the differential equation is given by an iterated integral.
For this, we require a boundary condition.
Here, a particularly simple point is $s=0$, $m^2=1$ since the second component of $\mathbf{f}$ vanishes because the square root in eq.\,\eqref{eq:DiffEqBasisChange} is zero.
So we only need to know the simpler integral $B_{10}$ at this point.
This is an elementary integration and from eq.\,\eqref{eq:Tadpole} we already know the result.
Adjusting the normalization factors, it is
\begin{align}
B_{0,1}=\Gamma(\eps)\left(m^2\right)^{-\eps}\,
\label{eq:TadpoleExplicit}
\end{align}
so
\begin{align}
\mathbf{f}(s=0,m^2=1,d=2-2\eps)=\left(\begin{array}{c}\Gamma(\eps)\\ 0\end{array}\right)\,.
\end{align}
To finally perform the integration explicitly in terms of multiple polylogarithms, we need to perform a last simplification:
We need to change variables such that the alphabet becomes rational, i.e. free of square roots.
In this case, rationalization can be achieved\footnote{See \cite{Besier:2018} to learn how.} by introducing the new variable $x$ as
\begin{align}
s=-m^2\frac{(1-x)^2}{x}\,,
\end{align}
where we assume $0<x<1$ to be away from branch cuts of the square roots.
In the new variable, the alphabet becomes $\lbrace m^2, x, 1+x\rbrace$.

Making a final basis change that divides out the overall mass dimension and the $\eps$-dependence of the tadpole
\begin{align}
\tilde{\mathbf{f}}(x;\eps)=\frac{1}{(m^2)^{-\eps}\Gamma(\eps)}\left(
\begin{array}{c}
B_{0,1}\\
\sqrt{(-s)(4m^2-s)}B_{1,1}
\end{array}
\right)
\end{align}
leaves us with only a dependence on the dimensionless variable $x$.
The new basis satisfies the canonical differential equation
\begin{align}
\dx\tilde{\mathbf{f}}(x;\eps)=\eps
\underbrace{\left(\begin{array}{cc}
0 & 0\\
-2\dx\!\log x & \dx\!\log\left(\frac{x}{(1+x)^2}\right)
\end{array}
\right)}_{=\dx\tilde{\mathcal{A}}}\tilde{\mathbf{f}}(x;\eps)
\end{align}
with alphabet $\lbrace x,1+x\rbrace$.
This tells us already that the solution is a special subset of harmonic polylogarithms.
We recall that the solution to the differential equation in canonical form is given order by order by eq.\,\eqref{eq:OrderDecoupling}.
\begin{toolblock}{  
Public codes for the transformation of differential equations into canonical form are \texttt{CANONICA} \cite{Meyer:2017joq} and \texttt{Fuchsia} \cite{Gituliar:2017vzm}.
}
\end{toolblock}

Now we bear witness the power of iterated integrals, making the rest algorithmic:
Starting from
\begin{align}
\tilde{\mathbf{f}}^{(0)}(x)=\left(\begin{array}{c}
1\\0
\end{array}\right)
\end{align}
by the boundary condition, the first non-trivial order is
\begin{align}
\dx \tilde{\mathbf{f}}^{(1)}(x)=\dx\tilde{\mathcal{A}}\cdot\tilde{\mathbf{f}}^{(0)}(x)=\left(\begin{array}{c}
0\\-2\,\dx\!\log x
\end{array}\right),
\end{align}
thus
\begin{equation}
\tilde{\mathbf{f}}^{(1)}(x)=\left(\begin{array}{c}
0\\-2 \goncharov(0;x)
\end{array}\right)
\end{equation}
where we wrote the logarithm in Goncharov form $\log x=G(0;x)$ for convenient further integration.
In the next order in $\eps$ we have
\begin{align}
\dx \tilde{\mathbf{f}}^{(2)}(x)=\dx\tilde{\mathcal{A}}\cdot\tilde{\mathbf{f}}^{(1)}(x)=\left(\begin{array}{c}
0\\-2 \goncharov(0;x)\dx\!\log \frac{x}{(1+x)^2}
\end{array}\right),
\end{align}
so upon integration $f^{(2)}_1=0$ and using the iterative definition of Goncharov polylogarithms
\begin{align}
f^{(2)}_2&=-2\int_1^x\dx\!\log\!\left(\frac{x}{(1+x)^2}\right)\,\goncharov(0;x)
\nonumber\\
&
=-2\int_1^x\left[\frac{\dx x}{x}-2\frac{\dx x}{x+1}\right]\goncharov(0;x)
\nonumber\\
&
=-2[\goncharov(0,0;x)-G(0,0;1)]+4[\goncharov(-1,0;x)-\goncharov(-1,0;1)]
\end{align}
where the second term in each bracket originates from the boundary value at $x=1$.
Since the involved functions are of weight $2$, we can express all Goncharov polylogarithms in terms of logarithms, dilogarithms, and zeta values.
In particular
\begin{align}
\goncharov(0,0;x)=\frac{\log^2 x}{2!}\quad\text{and}\quad
\goncharov(-1,0;x)=\dilog(-x)+\log x\log(1+x)\,,
\end{align}
thus
\begin{align}
f^{(2)}_2(x)=4\dilog(-x)+4\log x\log(1+x)-\log^2 x+\frac{\pi^2}{3}\,.
\label{eq:f22def}
\end{align}

Finally putting everything together, we have the result for the massive scalar bubble integral in $d=2-2\eps$ dimensions
\begin{align}
B_{1,1}=\frac{\Gamma(1+\eps)(m^2)^{-\eps}}{\sqrt{-s(4m^2-s)}}\left[
-2\log x+\eps\,f^{(2)}_2(x)+\mathcal{O}\left(\eps^2\right)
\right]\,,
\label{eq:DiffEqB11res}
\end{align}
with
\begin{align}
x=\left(\frac{\sqrt{1-4m^2/s}-1}{\sqrt{1-4m^2/s}+1}\right)\,.
\end{align}
We will continue this example in the next section and show how to use the expansion in $d=2-2\eps$ to construct the one in physical dimensionality $d=4-2\eps$.

Of course, for this toy example, other, less sophisticated methods would also have been possible.
The great strength of the differential equation technique is that it scales well with the complexity of the problem when taking into account more variables, more loops, and a larger number of master integrals.
Cutting-edge calculations using the differential equation technique have been pushed to astonishing heights:
four loops for single-scale three-point functions \cite{Lee:2023dtc}, three loops for $2\rightarrow 2$ kinematics with over a hundred master integrals \cite{Canko:2021xmn}, and two loop five-point functions with well over a hundred letters \cite{Abreu:2021smk,Henn:2024ngj}.

In a more modest setting, we will transfer the differential equation approach to phase-space integrals.
In particular, we will calculate the three and four denominator angular integrals in publication \ref{pub:8} and publication \ref{pub:10}, respectively.
In publication \ref{pub:8} we will discuss in detail what adaptations need to be made in this new setting of angular integrals.
One of the relevant tricks that will make our lives easier is discussed in the next section.

\begin{readingblock}{\textbf{Further reading: }
For a conceptual discussion of the principles based on the same example presented here, see again the textbook \cite{Badger:2023eqz}, co-authored by Henn, or also his lecture notes \cite{Henn:2014qga} with similar content.
}\end{readingblock}

\section{Dimensional shifts}
\label{sec:DimShift}
As we have seen in section \ref{sec:DimReg}, Feynman integrals naturally become analytic functions of the space-time dimension $d$.
It turns out that there is more to this analytic behavior than only ``analytic continuation'' between dimensions:
Oleg Tarasov found algebraic identities between Feynman integrals that connect the $d$ dimensional integral with those in $d\pm 2$. 
These are called \textit{dimensional shift relations} \cite{Tarasov:1996br}. 

The starting point in his derivations is writing the general loop integral
\begin{align}
I(d)=\int\left(\prod_{l=1}^L\frac{\dx^d k_l}{\iu\pi^{d/2}}\right)\frac{1}{D_1^{j_1}\dots D_n^{j_n}}
\end{align}
in Schwinger parameterization, compare eq.\,\eqref{eq:Schwinger_param1},
\begin{align}
I(d)=\int_0^\infty \left(\prod_{i=1}^n\frac{\dx x_i\, x_i^{j_i-1}}{\Gamma(j_i)}\right)\frac{\exp\left[\frac{\mathcal{V}}{\mathcal{U}}-\sum_{i=1}^n x_i m_i^2\right]}{\mathcal{U}^{d/2}}\,,
\label{eq:Schwinger_param}
\end{align}
with the Symanzik polynomials $\mathcal{U}(x_i)$ and  $\mathcal{V}(x_i,s_{ij})$.
Their explicit form follows from the associated Feynman graph and was discussed in sec.\,\ref{sec:loop}, for the following argument it is only necessary to recall that $\mathcal{U}$ is homogeneous of degree $L$ in the parameters $x_i$ with constant coefficients and $\mathcal{V}$ is homogeneous of degree $L+1$ with coefficients depending on the momenta that flow through the graph but crucially, both are independent of the masses.
Hence, the only mass dependence sits in $\exp[-\sum_{i=1}^nx_i m_i^2]$.
We can use that to construct the dimensional shift by observing that, replacing the arguments in the polynomial $\mathcal{U}(x_i)$ by the respective mass derivatives $\partial_{m_i^2}$,
\begin{align}
\mathcal{U}(\partial_{m_i^2}) \exp\left[-\sum_{i=1}^nx_i m_i^2\right]=(-1)^L \mathcal{U}(x_i)\exp\left[-\sum_{i=1}^nx_i m_i^2\right]
\end{align}
where we used the homogeneity of $\mathcal{U}$ to extract the factor of $-1$ and recover the original polynomial as a pre-factor.
Thus, acting with $\mathcal{U}(\partial_{m_i^2})$ on both sides of eq.\,\eqref{eq:Schwinger_param}, Tarasov found
\begin{align}
\mathcal{U}(\partial_{m_i^2})I(d)=(-1)^L I(d-2)\,.
\end{align}
From what we have seen in section \ref{sec:DiffEq}, the left hand side can also be expressed in terms of integrals with shifted indices, staying in $d$ dimensions.
Subsequently, both sides of the equation can be reduced to master integrals by using the IBP relations in their respective dimension.
That leads to an identity of the schematic form
\begin{align}
I_k(d-2)=\sum_j C_{kj} I_j(d)\,,
\end{align}
where the $I_j$ are master integrals and $C_{kj}$ is a rational function depending on the momenta, masses, and dimension.
This identity can also be inverted to express an integral in $d$ dimensions in terms of $d+2$ dimensional ones, i.\,e. of the form
\begin{align}
I_k(d+2)=\sum_j \tilde{C}_{kj} I_j(d)\,.
\end{align}

These shifts $d\rightarrow d\pm2$ are not only of conceptual interest but also of practical relevance.
Structurally, they are very similar to the IBP relations only that they act on the dimensionality rather than the exponents.
They provide an independent set of identities that can be used to investigate Feynman integrals.
Importantly, shifting the dimension can bring an integral that diverges at $\eps=0$ to a dimension where it converges.
In this case, the pole becomes explicit algebraically.

In section \ref{sec:DiffEq}, we looked at the differential equation of the one-loop bubble integral and calculated the $\eps$-expansion in $d=2-2\eps$.
This choice was made such that the master integral was of uniform transcendental weight.
However, in practice we may be interested in the physical case of $d=4-2\eps$ dimensions.
The dimensional shift relations provide a way to directly re-use the expansion in the lower dimension.

To see the workings of the method explicitly, we start from our example integral $B_{11}(s,m^2;d)$ in Feynman parameterization, which in this case is even simpler than using the general Schwinger parameterization approach, where we would need to assume different masses first.
From eq.\,\eqref{eq:FeynmanParamB11} we recall
\begin{align}
B_{1,1}(s,m^2;d)=\Gamma\!\left(2-\frac{d}{2}\right)\int_0^1\dx x\left[m^2-x(1-x)s\right]^{d/2-2}\,.
\end{align}
Now, acting with $\partial_{m^2}$, we get
\begin{align}
&\partial_{m^2}B_{1,1}(s,m^2;d)=\Gamma\!\left(2-\frac{d}{2}\right)\int_0^1\dx x \left(\frac{d}{2}-2\right)\left[m^2-x(1-x)s\right]^{d/2-3}\,\nonumber\\
&=-\Gamma\!\left(2-\frac{d-2}{2}\right)\int_0^1\dx x\left[m^2-x(1-x)s\right]^{(d-2)/2-2}=-B_{1,1}(s,m^2;d-2)\,.
\end{align}
From section \ref{sec:DiffEq}, we know that we can alternatively express $\partial_{m^2}B_{11}$ as a shift on the exponents that can subsequently be reduced to master integrals using the IBPs,
\begin{align}
\partial_{m^2}B_{1,1}(d)\overset{\eqref{eq:BubbleMDeriv}}{=}-2 B_{2,1}(d)\overset{\eqref{eq:IBP_specific}}{=}\frac{d-2}{m^2(4m^2-s)}B_{1,0}(d)+\frac{2(d-3)}{4m^2-s}B_{1,1}(d)\,.
\end{align}

Combining the two expressions for $\partial_{m^2}B_{1,1}(d)$, we have the dimensional shift relation
\begin{align}
B_{1,1}(d-2)=\frac{2-d}{m^2(4m^2-s)}B_{1,0}(d)+\frac{2(3-d)}{4m^2-s}B_{1,1}(d)\,.
\label{eq:BubbleDimShiftMinus}
\end{align}
To make the connection $d\rightarrow d+2$, we need to invert this relation.
Recalling from eq.\,\eqref{eq:Tadpole} we readily find\footnote{Of course, we could also use Tarasov's method to establish this relation for $B_{1,0}$ without resorting to the closed form result.}
\begin{align}
B_{1,0}(d+2)=-\frac{2m^2}{d}B_{1,0}(d).
\label{eq:TadpoleDimShiftPlus}
\end{align}
Shifting $d\rightarrow d+2$ in eq.\,\eqref{eq:BubbleDimShiftMinus}, plugging in eq.\,\eqref{eq:TadpoleDimShiftPlus}, and solving for $B_{1,1}(d+2)$, we obtain the second dimensional shift relation
\begin{align}
B_{1,1}(d+2)=\frac{1}{d-1}B_{1,0}(d)+\frac{s-4m^2}{2(d-1)}B_{1,1}(d)\,.
\end{align}

Now we can set $d=2-2\eps$, plug in the $\eps$-expansions we calculated in section \ref{sec:DiffEq} in this dimensionality on the right and obtain the result for the expansion in $d=4-2\eps$ as a result.
Explicitly, we find
\begin{align}
B_{1,1}(d=4-2\eps)=\frac{(m^2)^{-\eps}}{\Gamma(1-\eps)}\left[\frac{1}{\eps}+2+\sqrt{1-\frac{4m^2}{s}}\log\!\left(\frac{\sqrt{1-\frac{4m^2}{s}}-1}{\sqrt{1-\frac{4m^2}{s}}+1}\right)+\mathcal{O}(\eps)\right].
\end{align}
If we were now to expand this result in $\lambda=-m^2/s$, we would recover exactly the result found by expansion by regions in eq.\,\eqref{eq:ExpByRegionsBubbleRes}.
We will revisit this example for a last time in the next section and obtain the result in $d=2-2\eps$ with an independent method.

Dimensional shift identities for angular integrals are used in publication \ref{pub:8} to simplify the calculation of master integrals by differential equations and are developed in a more general direction in \ref{pub:10}, where they are used to reduce the scales of master integrals by breaking them into smaller pieces.

\begin{readingblock}{\textbf{Further reading: }
A good introduction to dimensional shifts is given by Tarasov's original paper \cite{Tarasov:1996br}.
For a method to use dimensional shifts to analytically calculate loop integrals, see \cite{Lee:2009dh}.
}
\end{readingblock}

In the last technical section before we get to the publications, we discuss another independent approach to integrals in pQCD, suited for both numerical and analytical treatment.
\section{Mellin transform and Mellin-Barnes integrals}
\label{sec:Mellin}
The \textit{Mellin transform} is, in contrast to other integral
transforms which arose in physics, a product of pure mathematics that only later found application in high energy physics.
Already Bernhard Riemann used it in his famous 1859 paper, where he formulated the Riemann hypothesis about the location of the zeros of the Zeta function \cite{Riemann1859}.
The transform is named after the Finnish mathematician Hjalmar Mellin\footnote{Later in his life, Mellin was an outspoken critic of Einstein's theory of relativity and devoted his private life to strengthening Finnish national identity \cite{paris2001asymptotics}.}, who conducted the first systematic studies \cite{Mellin:1897}.

The Mellin transform of a function $f(x)$ is formally defined as
\begin{align}
\mathcal{M}[f](N)=\int_0^\infty\dx x\,x^{N-1} f(x)\,.
\end{align}
When $f(x)\sim x^{-a+\eps}$ for $x\rightarrow 0$ and $f(x)\sim x^{-b-\eps}$ for $x\rightarrow \infty$ for an $\eps>0$ and $a<b$, then this defines an analytic function of $N$ in the strip $a<\mathrm{Re}(N)<b$.
In particle physics, we are mainly concerned with functions with support only on $0\leq x\leq 1$, so there will be an (implicit) $\Theta(1-x)$ cutting of the integral at $x=1$, hence often the Mellin transform is defined directly as 
\begin{align}
\mathcal{M}[f](N)=\int_0^1\dx x\,x^{N-1} f(x)\,.
\end{align}
Its main benefit in the context of perturbative QCD comes from the fact that it transforms Mellin-convolution integrals,
\begin{align}
f\otimes g(x)=\int_x^1\frac{\dx \xi}{\xi}\,f(\xi)\,g\!\left(\frac{x}{\xi}\right)\,,
\end{align}
the typical structure encountered in factorization theorems and evolution equations, into ordinary products,
\begin{align}
\mathcal{M}[f\otimes g](N)=\mathcal{M}[f](N)\,\mathcal{M}[g](N)\,.
\end{align}
In practice, this is especially convenient in the context of PDF evolution and fitting parton distribution functions.

The inverse transform is given by
\begin{align}
\mathcal{M}^{-1}[f](x)=\int_{-\iu\infty}^{\iu\infty}\frac{\dx N}{2\pi \iu}\,x^{-N}\,f(N)
\end{align}
where the integration contour needs to be chosen such that all poles are to its left.

To gain some intuition about the mapping from $x$ to Mellin-$N$ space, let us consider the simple example
\begin{align}
f(x)=x^a (1-x)^b\,\theta(1-x)
\label{eq:MellinExampleX}
\end{align}
which we may think of either as a simple model for a PDF or a coefficient function.
In this case, the Mellin transform is simply a Beta function
\begin{align}
\mathcal{M}[f](N)=\int_0^1\dx x\, x^{N+a-1}(1-x)^b=\frac{\Gamma(N+a)\Gamma(1+b)}{\Gamma(N+1+a+b)}\,.
\label{eq:MellinExample}
\end{align}

\begin{figure}
\centering
\includegraphics[width=0.75\textwidth]{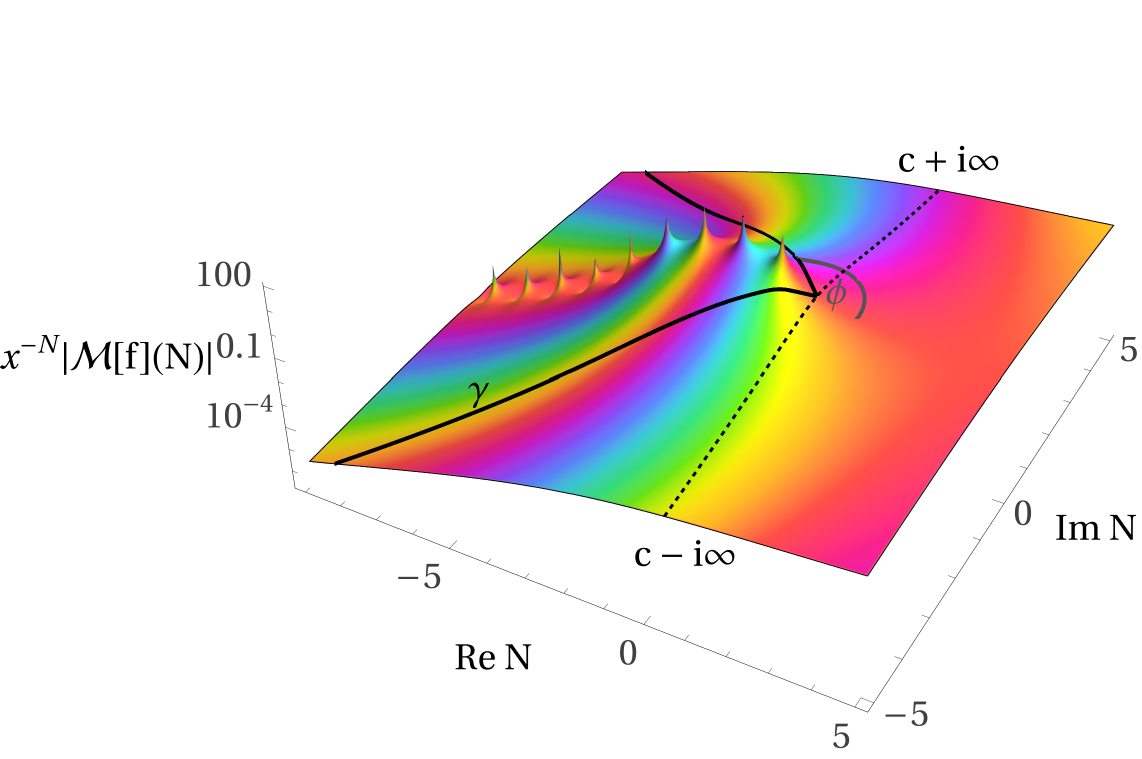}
\caption[Graphical illustration of Mellin back-transform. Created with \texttt{Mathematica} \cite{Mathematica13}.]{The function $x^{-N}\mathcal{M}[f](N)$ with $f$ from eq.\,\eqref{eq:MellinExampleX} in Mellin-space for $x=0.5$, $a=-0.5$, and $b=3.2$ as it appears in the kernel of the inverse Mellin transform. The profile gives the absolute value of the function on a logarithmic scale, the complex phase is color-coded. 
The original inversion contour is the dashed straight line $c-\iu\infty$ to $c+\iu\infty$, the bold contour $\gamma$, tilted by the angle $\phi$, enhances numerical convergence by using the exponential fall-off of $x^{-N}$ for large negative real parts of $N$. The poles appear as spikes on the negative real axis.
}
\label{fig:MellinExample}
\end{figure}

The question we want to answer is how the $x$-space behavior is encoded in Mellin-space.
For this, we need to think about $\mathcal{M}[f](N)$ as a function in the complex $N$-plane.
We observe that eq.\,\eqref{eq:MellinExample} has poles at
\begin{equation}
N=-a,-a-1,-a-2,\dots
\label{eq:MellinExamplePolePositions}
\end{equation}
from $\Gamma(N+a)$.

In figure \ref{fig:MellinExample}, where $x^{-N}\mathcal{M}[f](N)$ -- the kernel for the Mellin-inversion -- is displayed, they appear as distinct spikes in the absolute value.
The original integration contour for the inversion (dashed line) runs from $c-\iu\infty$ to $c+\iu\infty$ where $c$ needs to be right of the rightmost pole, i.e. here $c>-a$.
By analyticity, the value for the integral does not depend on the exact choice of $c$, for better numerical convergence it is however advisable to choose $c$ not too close to the pole and tilt the integration contour to the path $\gamma$ (thick black line) parameterized as $\gamma(t)=c+|t|\ee^{\mathrm{sign}(t)\iu\phi}$, which makes use of the exponential fall-off of $x^{-N}$ for $\mathrm{Re[N]}\rightarrow-\infty$.
Along the path $\gamma$, the integral can be calculated as
\begin{align}
\mathcal{M}^{-1}[f](x)=\frac{1}{\pi}\int_0^\infty\dx t\,\mathrm{Im}\left[
\ee^{\iu\phi}x^{-c-t\ee^{\iu\phi}}f(c+t\ee^{\iu\phi})
\right].
\end{align}

Analytically, we see from the pole positions of eq.\,\eqref{eq:MellinExamplePolePositions} that there is a correspondence of the right-most pole at $N=-a$ and the small-$x$ behavior of $f(x)\sim x^{a}$.
Indeed, considering only the pure pole and taking its Mellin-inverse, we find by the residue theorem
\begin{align}
\mathcal{M}^{-1}\left[\frac{1}{N+a}\right]=\int_{-\iu\infty}^{\iu\infty}\frac{\dx N}{2\pi\iu}\frac{x^{-N}}{N+a}=\mathrm{Res}\left[\frac{x^{-N}}{N+a}\right]_{N=-a}\!\!\!\!\!\!\Theta(1-x)=x^a\,\Theta(1-x)\,.
\end{align}
Note that for $0<x<1$ the factor $x^{-N}$ decays exponentially for $\mathrm{Re}N\rightarrow-\infty$, hence we needed to close the contour to the left.
This confirms that we can read off the leading small-$x$ behavior as
\begin{align}
f(x)\overset{x\rightarrow 0}{\sim} x^{-N_\text{rightmost pole}}\,.
\end{align}
More generally, each pole in $N$-space corresponds to a contribution $x^{-N_\text{pole}}$.

Next up, we want to investigate how the $x\rightarrow 1$ behavior\footnote{In application, $x\rightarrow 1$ may for example correspond to the threshold-limit of a cross-section.} is encoded in Mellin-space.
For this, we look at the asymptotic behavior for large $|N|$ of eq.\,\eqref{eq:MellinExample}.
By Stirling's approximation of the Gamma function (see eq.\,\eqref{eq:GammaApprox}), it is, away from the negative real axis,
\begin{align}
\mathcal{M}[f](N)=\frac{\Gamma(N+a)\Gamma(1+b)}{\Gamma(N+1+a+b)}\sim N^{-b-1}\Gamma(1+b)+\mathcal{O}\!\left(N^{-b-2}\right).
\end{align}
Here, the $x\rightarrow 1$ behavior shows up as $N^{-b-1}$.
Considering the Mellin-inverse of the pure asymptotic term, we have
\begin{align}
\mathcal{M}^{-1}[N^{-b-1}]
=\int_{-\iu\infty}^{\iu\infty}\frac{\dx N}{2\pi\iu} N^{-b-1}x^{-N}
=(\mp\log x)^{b}\int_{-\iu\infty}^{\iu\infty}\frac{\dx N}{2\pi\iu}N^{-b-1}\ee^{\pm N}\,,
\end{align}
where the sign is chosen such that $\mp\log x$ is positive, i.e., the upper for $0<x<1$, else the lower.  
We see that $N^{-b-1}$ has a branch cut along the negative real axis.
For $x>1$, we have $\ee^{-N}$, which decays for $\mathrm{Re}N\rightarrow\infty$.
Closing the contour to the right encircles no poles or branch cuts hence the integral vanishes.
For $0<x<1$, we have $\ee^{N}$, which decays for $\mathrm{Re}N\rightarrow-\infty$.
Closing the contour to the left, we have to encircle the cut along the negative real axis.
Using
\begin{align}
N^{-b-1}=\left\lbrace\begin{array}{cc}
|N|^{-b-1}\ee^{-\iu\pi(1+b)} & \text{above the cut}\\
|N|^{-b-1}\ee^{\iu\pi(1+b)} & \text{below the cut}\\
\end{array}\right.,
\end{align}
we find
\begin{align}
\mathcal{M}^{-1}[N^{-b-1}]
&=(-\log x)^b\,\Theta(1-x) \int_\text{cut}\frac{\dx N}{2\pi\iu} N^{-b-1}\ee^N\nonumber\\
&=-\frac{\sin(\pi b)}{\pi}(-\log x)^b\,\Theta(1-x)\underbrace{\int_0^\infty\dx|N|\,|N|^{-b-1}\ee^{-|N|}}_{=\Gamma(-b)}\nonumber\\
&=\frac{(-\log x)^b\Theta(1-x)}{\Gamma(1+b)}\,,
\end{align}
where we used the Euler reflection formula in the last step.
Thus, expanding about $x=1$, 
\begin{align}
\mathcal{M}^{-1}[N^{-b-1}]=\frac{(1-x)^b\Theta(1-x)}{\Gamma(1+b)}+\mathcal{O}\!\left((1-x)^{b+1}\right)
\end{align}
Therefore, the threshold behavior of $x$ manifests itself in Mellin-space in the large-$N$ asymptotics. 
\begin{align}
\text{If }\mathcal{M}[f](N)\overset{N\rightarrow\infty}{\sim} N^p\,\Gamma(-p), \text{ then }f(x)\overset{x\rightarrow 1}{\sim} (1-x)^{-1-p}\,.
\end{align}
This is used especially in threshold resummation, which captures the most singular parts of the cross-section near $x=1$.
Mellin-space techniques are most powerful for single-scale quantities, but can be extended to multi-variable cases, such as SIDIS, using multiple Mellin transforms \cite{Abele:2021,Abele:2022,Goyal:2025}.

Closely related to Mellin space representations are \textit{Mellin-Barnes integrals}.
They allow for a flexible representation and, in many cases, explicit calculation of parametric integrals.
The key identity is the representation
\begin{align}
(x+y)^\lambda=\frac{1}{\Gamma(-\lambda)}\int_{-\iu\infty}^{\iu\infty}\frac{\dx z}{2\pi\iu}\Gamma(-z)\Gamma(-\lambda+z)\,x^{\lambda-z}\,y^z\,,
\label{eq:MBBinomi}
\end{align}
which factorizes $(x+y)^\lambda$ and is essentially a direct generalization of the binomial theorem to non-natural number exponents $\lambda$.
The integration contour needs to be such that it separates the ``right'' poles of $\Gamma(-z)$ from the ``left'' poles of $\Gamma(-\lambda+z)$.
If $\lambda<0$, this can be achieved by a straight contour from $-\iu\infty+c$ to $\iu\infty+c$ where $\lambda<c<0$, else the contour needs to appropriately circle around the poles.
This is always possible if the poles do not overlap, which happens when $\lambda$ is a non-negative integer.

A general Mellin-Barnes integral in the context of this thesis is of the form
\begin{align}
\int_{-\iu\infty}^{\iu\infty}\frac{\dx z_1}{2\pi\iu}\dots\int_{-\iu\infty}^{\iu\infty}\frac{\dx z_n}{2\pi\iu}\,\frac{\prod_{i=1}^N\Gamma\!\left(a_i+\sum_{j=1}^n\lambda_{ij} z_j\right)}{\prod_{i=1}^M\Gamma\!\left(b_i+\sum_{j=1}^n\mu_{ij} z_j\right)}\,x_1^{z_1}\dots x_n^{z_n}
\label{eq:MBIntegralGeneral}
\end{align}
with constants $a_i$, $b_i$, $\lambda_{ij}$, and $\mu_{ij}$ and where the integration contour is such that it separates left ($\lambda_{ij}>0$) and right poles ($\lambda_{ij}<0$) for each $\Gamma$-function $\Gamma(a_i+\dots)$ in each integration variable $z_j$.
Typical Mellin-Barnes integrals arising from Feynman and phase-space integrals are \textit{balanced}, meaning that for each $z_j$ it holds
\begin{align}
\sum_{i=1}^N \lambda_{ij}-\sum_{i=1}^M\mu_{ij}=0\,,
\end{align}
so the ``sum of $z_j$ entries in Gamma functions'', where occurrences in the denominator are counted negative, vanishes.
In many cases, this property allows to flexibly convert the Mellin-Barnes integral to parametric integrals via the Beta integral representation of the (multi-variable) Beta function.
Switching between MB and parametric representations will be key in the analytic calculation of angular integrals in $d$ dimensions in publication \ref{pub:1}.

Multifold MB-integrals are a direct generalization of multi-variable hypergeometric functions and eq.\,\eqref{eq:MBIntegralGeneral} is also known as the multi-variable $\mathrm{H}$-function of $n$ variables \cite{hai1995convergence,mathai2009h,Somogyi:2011}.
However, this becomes so general that it can only be borderline considered a ``special function''\footnote{Even though it has a name.}.
In even more general Mellin-Barnes integrals we can also admit more general analytic functions $f(x_1,\dots,x_n;z_1,\dots,z_n)$ instead of the monomials and derivatives of gamma functions.

\begin{toolblock}{
\textbf{Tricks for the analytic evaluation of Mellin-Barnes (MB) integrals: }
To factorize terms in order to calculate parametric integrals or, vice-versa, to evaluate factorized MB integrals, we can use
\begin{align}
&(x_1+x_2+\dots+x_n)^{\lambda}=\frac{1}{\Gamma(-\lambda)}\int_{-\iu\infty}^{\iu\infty}\frac{\dx z_1\dots\dx z_{n-1}}{(2\pi\iu)^{n-1}}
\,\Gamma(-z_1)\dots\Gamma(-z_{n-1})\nonumber\\
&\quad\times\Gamma(-\lambda+z_1+\dots+z_{n-1})\,x_n^{\lambda-z_1-\dots-z_{n-1}}x_1^{z_1}\dots x_{n-1}^{z_{n-1}}\,.
\end{align}
Very important and versatile MB representations are those of the Gauss hypergeometric function $\ghy$, which admits two different forms,
\begin{align}
&\ghy(a,b,c;x)=\frac{\Gamma(c)}{\Gamma(a)\Gamma(b)}\int_{-\iu\infty}^{\iu\infty}\frac{\dx z}{2\pi\iu}\frac{\Gamma(a+z)\Gamma(b+z)\Gamma(-z)}{\Gamma(c+z)}(-x)^z
\label{eq:MB2F1a}\\
&=\frac{\Gamma(c)}{\Gamma(a)\Gamma(b)\Gamma(c-a)\Gamma(c-b)}\int_{-\iu\infty}^{\iu\infty}\frac{\dx z}{2\pi\iu}\,\Gamma(a+z)\Gamma(b+z)\Gamma(c-a-b-z)\nonumber\\
&\phantom{\frac{\Gamma(c)}{\Gamma(a)\Gamma(b)\Gamma(c-a)\Gamma(c-b)}}\qquad\times\Gamma(-z)(1-x)^z\,.
\end{align}
This can be used to transform MB integrals into each other.\\
Explicit integration of many MB integrals, which are free of $x$-variables, is possible using Barnes' lemmas. 
The first one is
\begin{align}
&\MBint{z}\Gamma(\alpha+z)\Gamma(\beta+z)\Gamma(\gamma-z)\Gamma(\delta-z)\nonumber\\
&=\frac{\Gamma(\alpha+\gamma)\Gamma(\alpha+\delta)\Gamma(\beta+\gamma)\Gamma(\beta+\delta)}{\Gamma(\alpha+\beta+\gamma+\delta)}
\end{align}
and the second one 
\begin{align}
&\MBint{z}\frac{\Gamma(\alpha_1+z)\Gamma(\alpha_2+z)\Gamma(\alpha_3+z)\Gamma(1-\beta_1-z)\Gamma(-z)}{\Gamma(\beta_2+z)}
\nonumber\\
&=\frac{\Gamma(\alpha_1)\Gamma(\alpha_2)\Gamma(\alpha_3)\Gamma(\alpha_1+1-\beta_1)\Gamma(\alpha_2+1-\beta_1)\Gamma(\alpha_3+1-\beta_1)}{\Gamma(\beta_2-\alpha_1)\Gamma(\beta_2-\alpha_2)\Gamma(\beta_2-\alpha_3)},
\end{align}
which holds provided that the Saalschütz condition $\beta_2=\alpha_1+\alpha_2+\alpha_3+1-\beta_1$ is met.
}\end{toolblock}

We want to finish this methods chapter by returning to our example integral $B_{1,1}$ for a final time.
Here, we will use the presented Mellin-Barnes machinery to evaluate the Feynman parameter representation in closed form in $d$ dimensions in terms of hypergeometric functions.
This offers an alternative route to the $\eps$-expansion.

To convert the Feynman representation recalled from eq.\,\eqref{eq:FeynmanParamB11},
\begin{align}
B_{1,1}=\Gamma\!\left(2-\frac{d}{2}\right)\int_0^1\dx x\,\left[m^2-x(1-x) s\right]^{\frac{d}{2}-2}\,,
\end{align}
into a MB integral, we use eq.\,\eqref{eq:MBBinomi} to factorize the bracket $[\dots]^{d/2-2}$.
Upon interchanging parametric and MB integration, this leads to
\begin{align}
B_{1,1}=\left(m^2\right)^{\frac{d}{2}-2}\MBint{z}\,\Gamma\!\left(2-\frac{d}{2}+z\right)\Gamma(-z)\left(-\frac{s}{m^2}\right)^z \int_0^1\dx x\, x^z (1-x)^z\,.
\end{align}
The remaining integral over the Feynman parameter $x$ can now be evaluated in terms of Gamma functions by eqs.\eqref{eq:BetaFctInt1} and \eqref{eq:BetaAsGamma} as
\begin{align}
\int_0^1\dx x\, x^z (1-x)^z=\frac{\Gamma^2(1+z)}{\Gamma(2+2z)}\overset{\eqref{eq:LegendreDuplication}}{=}\frac{\sqrt{\pi} 2^{-2z-1}\Gamma(1+z)}{\Gamma\!\left(\frac{3}{2}+z\right)}\,.
\end{align}
The resulting MB representation matches eq.\,\eqref{eq:MB2F1a} of the Gauss hypergeometric function.
Therefore,
\begin{align}
B_{1,1}=\left(m^2\right)^{\frac{d}{2}-2}\frac{\sqrt{\pi}}{2}\MBint{z}\,\frac{\Gamma(1+z)\Gamma\!\left(2-\frac{d}{2}+z\right)\Gamma(-z)}{\Gamma\!\left(\frac{3}{2}+z\right)}\,\left(-\frac{s}{4m^2}\right)^z\nonumber\\
=\Gamma\!\left(2-\frac{d}{2}\right)\left(m^2\right)^{\frac{d}{2}-2}\ghy\!\left(1,2-\frac{d}{2},\frac{3}{2};\frac{s}{4m^2}\right).
\end{align}
This is a closed form analytic result in $d$-dimensions in terms of a special function.
To get from here to the $\eps$-expansion, we can use the established knowledge about $\ghy$.
To compare with the result from section \ref{sec:DiffEq}, which we found by the differential equation technique, we consider $d=2-2\eps$ and for simplicity let us only check the $\mathcal{O}(\eps^0)$ term.
Then, we can use the reduction identity
\begin{align}
\ghy\!\left(1,1,\frac{3}{2};x\right)\overset{\eqref{eq:PfaffEuler3}}{=}\frac{\ghy\!\left(\frac{1}{2},\frac{1}{2},\frac{3}{2};x\right)}{\sqrt{1-x}}\overset{\href{https://dlmf.nist.gov/15.4}{\text{DLMF}\,15.4.5}}{=}\frac{\log\left(\sqrt{-x}+\sqrt{1-x}\right)}{\sqrt{-x}\,\sqrt{1-x}}
\end{align}
to directly get to
\begin{align}
B_{1,1}(d=2-2\eps)=\frac{4\Gamma(1+\eps)(m^2)^{-\eps}}{\sqrt{-s}\,\sqrt{4m^2-s}}\left[\log\!\left(\sqrt{-\frac{s}{4m^2}}+\sqrt{1-\frac{s}{4m^2}}\right)+\mathcal{O}(\eps)\right]\,.
\end{align}
This indeed matches the earlier result from eq.\,\eqref{eq:DiffEqB11res} upon recognizing the algebraic identity
\begin{align}
\left(\sqrt{-x}+\sqrt{1-x}\right)^2=\frac{\sqrt{1-\frac{1}{x}}+1}{\sqrt{1-\frac{1}{x}}-1}\quad(\text{for }x<0)\,,
\end{align}
which is admittedly easier to find when one knows what one is looking for.

This concludes the study of this example.
On the way towards the result, we got to know the most important techniques for loop integrals that will be transferred to the setting of angular integrals in the next chapter.
\begin{readingblock}{\textbf{Further reading: }
A modern review on Mellin-Barnes techniques can be found in \cite{Dubovyk:2022obc}, co-authored by Gabor Somogyi. 
Hence, unsurprisingly, it also features angular integrals.
}\end{readingblock}

\end{fancychapter2}
\cleardoublepage
\begin{fancychapter2}{Collected publications}{To the arXiv!}{Why did you do this?}{Yuri Kovchegov}
\label{ch:Publications}

Now that we have sufficiently discussed the basics of perturbative QCD and its analytical methods, we are in the position to have a look at the main body of this doctoral work: the publications.
There are ten of them, so it is in order to give a brief overview before looking at each one individually.
Overall, they can be grouped into four thematic categories,
phase-space integrals (see \ref{pub:1}, \ref{pub:4}, \ref{pub:7}, \ref{pub:8}, \ref{pub:10}), loop integrals (see \ref{pub:1}, \ref{pub:2}, \ref{pub:3}), collider processes (\ref{pub:5}, \ref{pub:9}), and parton distributions (see \ref{pub:6});
Fig.\,\ref{fig:Paper_categories} provides a graphical overview.

\begin{figure}[h]  % from the float package
  \centering
  \begin{minipage}{0.6\linewidth}
     \includegraphics[width=\linewidth]{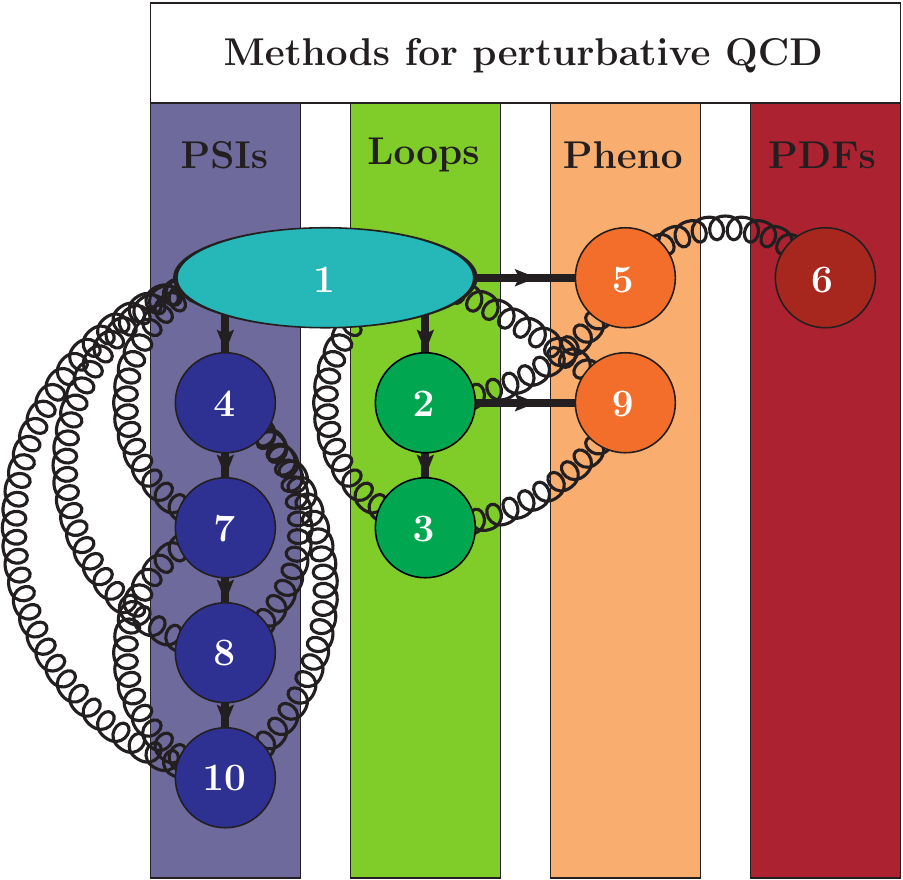}
  \end{minipage}\hfill
  \begin{minipage}{0.35\linewidth}
     \caption[Graphical overview of the publications and their connections. Created with \texttt{JaxoDraw} \cite{Binosi:2003}.]{Graphical overview of the publications, each represented by a numbered circle or ellipse, their thematic sub-field, and their connections. Arrows indicate a publication builds on a previous one, curly lines indicate thematic connections. Numbers refer to chronological order, corresponding to the section numbering in this chapter. The categories are phase-space integrals (PSIs), loop integrals (Loops), collider processes (Pheno), and parton distributions (PDFs).}
\label{fig:Paper_categories}
  \end{minipage}
\end{figure}

The largest block is concerned with phase-space integrals, specifically angular integrals.
In a series of publications, we significantly improved the state-of-the-art for this class of integrals, known for over 40 years in the pQCD literature, by finding a number of new structural connections, allowing for a higher number of denominators and scales, and a more careful treatment of kinematic limits.
While angular integrals have been used a lot for the analytic phase-space integration of $2\rightarrow 1+X$ processes at NNLO, the new results bring analytic phase-space integration for $2\rightarrow n+X$ processes within reach.

A second line of research was concerned with loop integrals, with a main focus on the branch cut structure of one-loop box integrals.
These are important for several important collider processes, such as electron-positron hadroproduction, SIDIS, and Drell-Yan.

The loop results found direct application in the third category of publications concerned with collider phenomenology for the Drell-Yan process and SIDIS.
Here, we studied the transverse momentum spectrum for Drell-Yan with a novel expansion technique and simplified analytic NNLO results for SIDIS using single-valued polylogarithms, respectively.

The fourth and last sub-field of publications is about parton distributions. 
Specifically we introduced a new method for solving evolution equations semi-analytically and conducted a proof-of-principle study on the DGLAP equation.

In the remainder of this chapter we will go through all of the publications in chronological order.
Each paper comes with its own section that starts with an overview block that summarizes the most important aspects.
This block is structured as follows:
\vspace{1cm}
\begin{center}
\Large{\textbf{Title of the publication}}
\end{center}
\begin{summaryblock}
{
\textbf{Why?\,: }Gives a short motivation why the topic is interesting.
\vspace{0.1cm}
\\
\textbf{What?\,: } Briefly states what was studied in the publication.
}
{
\textbf{Foundations: }The topics the reader should be familiar with to understand the details of the paper. These were discussed in chapters \ref{ch:Intro}, \ref{ch:SpecialFunctions}, and \ref{ch:Methods}.
\vspace{0.1cm}
\\
\textbf{Novel methods: }Methodical advances that made the work possible.
}
{
\textbf{Results: }Here comes a list of results of the publication.
\\
\textbf{Implications: } States what we can learn from the results and what can be done with them in future research.
}
\end{summaryblock}
\vspace{0.2cm}
After that block, we give some additional context about the publication.
This will explain how the research came about, highlight key aspects, and also what the connections are between the different projects to embed them into the general narrative.
The idea is that without reading the publications in detail, the casual reader is able to grasp the main points of this thesis by only reading the prefaces to the manuscripts.
The actual publications are then reserved for anyone interested in the technical details -- for example when implementing the presented analytic methods in pQCD practice.

\FloatBarrier
\cleardoublepage
\section[New ideas for loop and angular integrals]{Publication 1: New ideas for handling of loop and angular integrals in D-dimensions in QCD}
\label{pub:1}
\begin{summaryblock}
{
\textbf{Why?\,: }Analytic higher order calculations in pQCD require loop and phase-space integrals in dimensional regularization.
While a lot is known in this regard from studies in the past 50 years we could contribute a few new insights helpful in practice.
\vspace{0.1cm}
\\
\textbf{What?\,: } We calculated tensor one-loop integrals and two-denominator angular integrals.
}
{
\textbf{Foundations: }Dimensional regularization, Feynman parameterization, Mellin-Barnes integrals, hypergeometric functions, polylogarithms.
\vspace{0.1cm}
\\
\textbf{Novel methods: }Transferred several loop methods to angular integrals with two denominators; introduced the \textit{two-point splitting lemma} that allows for mass-reduction by partial fractioning.
}
{
\textbf{Results:}
\begin{itemize}
\item[(i)] New orthogonal Passarino-Veltman algorithm for tensor reduction of one-loop bubble, triangle, and box integrals.
\item[(ii)] Systematic partial fraction algorithm for angular integrals.
\item[(iii)] Hypergeometric form of double-massive two-denominator angular integral in $d$ dimensions.
\item[(iv)] All-order $\eps$-expansion for all angular integrals with up to two denominators, going beyond $d=4$ for the double-massive case after 40 years.
\end{itemize}
\textbf{Implications:}
The results simplify the analytic phase-space calculation for a large number of pQCD processes that use angular integrals, e.\,g. DY or SIDIS differential in transverse momentum, especially when additional poles require higher orders in $\eps$.
}
\end{summaryblock}

\vspace{0.2cm}
This is the first publication that was done during my doctoral work -- in collaboration with Valery Lyubovitskij, scientific staff in Tübingen, and Alexey Zhevlakov, post-doctoral researcher in Dubna, Russia -- and laid the foundation for many later developments.
It consists of two main parts that are mostly independent:
The first part (Section 2 and Appendices A and B) is about loop integrals and discusses a modification of the Passarino-Veltman technique to use an orthogonal basis, an idea of my co-author Valery Lyubovitskij.
Using a covariant basis constructed from external momenta  in the kinematics relevant to the Drell-Yan process, it discusses the massless tensor bubble-, triangle-, and box-integrals.

The second part (Section 3 and Appendices C to G) discusses angular integrals with up to two denominators.
This work builds upon the studies of van Neerven and especially Somogyi; the reader interested in the details of the historical development of angular integrals is referred to the review in the second half of the publication's introduction.
Using Mellin-Barnes techniques pioneered by Somogyi, we establish closed-form analytic results in terms of hypergeometric functions for up to two denominators and two masses, where the double-massive case is a genuinely new result.
Furthermore, this paper discusses for the first time structural relations between angular integrals in a systematic way, where before mainly long lists of explicit results were given.
This includes differential equations (though not yet used to calculate the integrals) and IBP relations, derived using an explicit coordinate representation.
Starting from the hypergeometric results, the angular master integrals are expanded to all orders in $\eps$.
This constitutes the second\footnote{Before our paper was completed, reference \cite{Isidori:2020} appeared where the $\mathcal{O}(\eps)$ of the integral $I_{1,1}^{(2)}$ was needed for QED corrections in exclusive $\overline{B}\to \overline{K}{\mathrm{\ell}}^{+}{\mathrm{\ell}}^{-}$ decay. To obtain the result, which is presented in a somewhat more complicated form compared to ours, private communication with Somogyi is cited. An even earlier account for the $\mathcal{O}(\eps)$ result can be found in \cite{Wunder:2020}.} published result for the double-massive integrals beyond $d=4$, the first major progress in the analytic calculation of this specific integral forty years after the first calculation of the integral in four dimensions in 1981 \cite{Schellekens:1981}.
The central insight that allowed for this achievement is a clever use of partial fractioning to reduce the double-massive to the single-massive case.
This idea is systematically developed in Appendix D of the publication.

The publication sets the foundation for the subsequent study of angular integrals starting from \ref{pub:4} where the two-denominator angular integral is revisited in kinematic limits. Later publications on higher-denominator angular integrals \ref{pub:7}, \ref{pub:8}, \ref{pub:10} all draw from the experience of this work.
The loop part of the paper found application in the Drell-Yan calculation of \ref{pub:5} and ties to publications \ref{pub:2} and \ref{pub:3} where the branch cut structure of the box integral was studied.

The results of this work have been presented at the \textit{CFNS Workshop: Precision QCD predictions for ep Physics at the EIC} in Stony Brook\footnote{On Long Island, close to where the EIC is built.} in August 2022.

\includepdf[pages=-, pagecommand={}, offset=3mm 0mm]{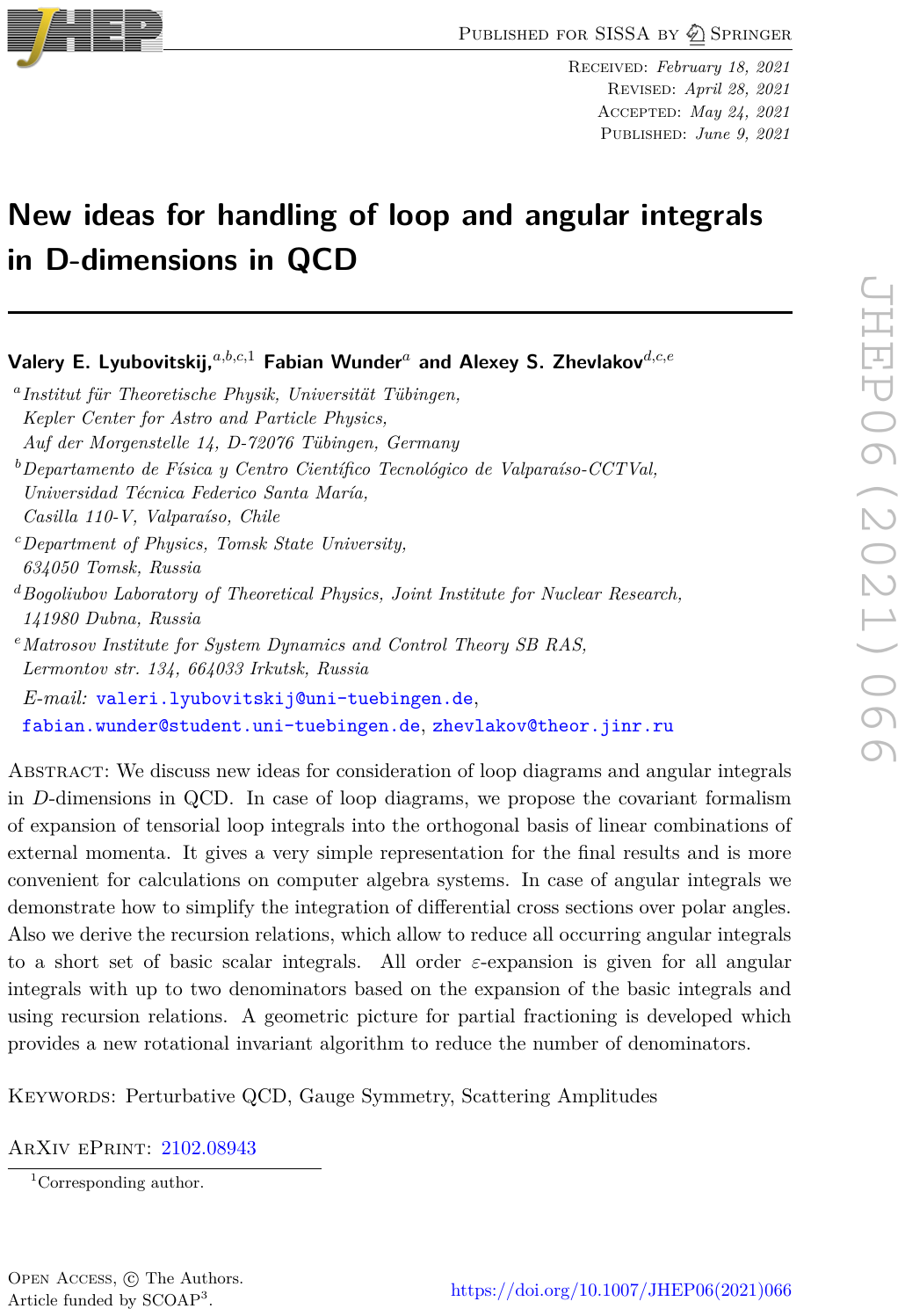}
\cleardoublepage
\section[The massless single off-shell scalar box integral]{Publication 2: The massless single off-shell scalar box integral --- branch cut structure and all-order epsilon expansion}
\label{pub:2}

\begin{summaryblock}
{
\textbf{Why?\,:} NNLO calculations in pQCD require careful handling of imaginary parts of loop amplitudes.
For the one-loop box these were not known to higher orders in the dimensional regularization parameter $\eps$.
\vspace{0.1cm}
\\
\textbf{What?\,:} We calculated the scalar box integral with one external off-shell particle and studied its branch cut structure to all orders in dimensional regularization.
}
{
\textbf{Foundations: }
Dimensional regularization, Feynman parameterization, hypergeometric functions, polylogarithms, branch cuts. 
\vspace{0.1cm}
\\
\textbf{Novel methods: }Used additional regulator for spurious branch cuts; introduced new class of single-valued polylogarithms to remove spurious branch cuts.
}
{
\textbf{Results:}
\begin{itemize}
\item[(i)] Hypergeometric representation of the single off-shell box integral valid in all kinematic regions.
\item[(ii)] All-order $\eps$-expansion of the single off-shell box integral, free of spurious branch cuts, with explicit imaginary parts in all kinematic regions.
\end{itemize}
\textbf{Implications: }
This solves the problem of extracting imaginary parts for the one-loop corrections of DY, SIDIS, and $e^+e^-$ to all orders in $\eps$.
}
\end{summaryblock}

\vspace{0.2cm}
This second publication is the first work initiated independently with fellow doctoral student Juliane Haug and evolved from the question of how to consistently treat the branch cut structure of one-loop integrals at higher orders in $\eps$ that are necessary for certain NNLO calculations, for example the real-virtual corrections to the SIDIS coefficient functions.

The paper presents a comprehensive derivation of the all-order $\eps$-expansion of the massless one-loop box integrals with one external particle off-shell, the kinematics relevant for $e^+ e^-$-annihilation, SIDIS\footnote{For the specific case of SIDIS kinematics, the branch cut structure of the box integral had been independently studied in \cite{Gehrmann:2022cih} shortly before.}, and the Drell-Yan process.
In the calculation, the branch cut structure is carefully tracked by keeping the Feynman $+\iu 0$ prescription for propagators  throughout.
One key step occurs from eq.\,(2.25) to (2.26) where an additional regulator $+\iu \tilde{0}$ is introduced to also track spurious branch cuts that develop when expressing the box integral in terms of a sum of three hypergeometric functions in eq.\,(2.36).
From there, the $\eps$-expansion is performed in a way that makes the cancellation of the spurious branch cut between the different hypergeometric functions explicit.
This naturally leads to an expression in terms of \textit{single-valued} polylogarithms (SVP).
The specific version of SVPs used here was not discussed in the literature before  and is introduced in Appendix C.
The central result of the work is eq.\,(3.27), in combination with the expression for the single-valued version of the hypergeometric function $\mathfrak{F}$ from eq.\,(3.15). 

The result of the publication was subsequently generalized to two diagonally opposite off-shell particles in publication \ref{pub:3}.
Most importantly, this paper set the foundation for the simplification of the SIDIS NNLO structure function in \ref{pub:9}, where the one-loop box integral plays a central role.
It was also useful for the study in \ref{pub:5} even though there, known results from the older literature would have sufficed since higher orders in $\eps$ did not contribute.

The results of this work have been presented at the FOR2926 research group meeting in February 2023 and the CFNS-CTEQ School in Stony Brook in 2023 by Juliane Haug, at the QCD Masterclass 2023 in Saint-Jacut-de-la-Mer, France by myself, and at the 2023 Amplitudes school at CERN in a joint presentation.  
\includepdf[pages=-, pagecommand={}, offset=3mm 0mm]{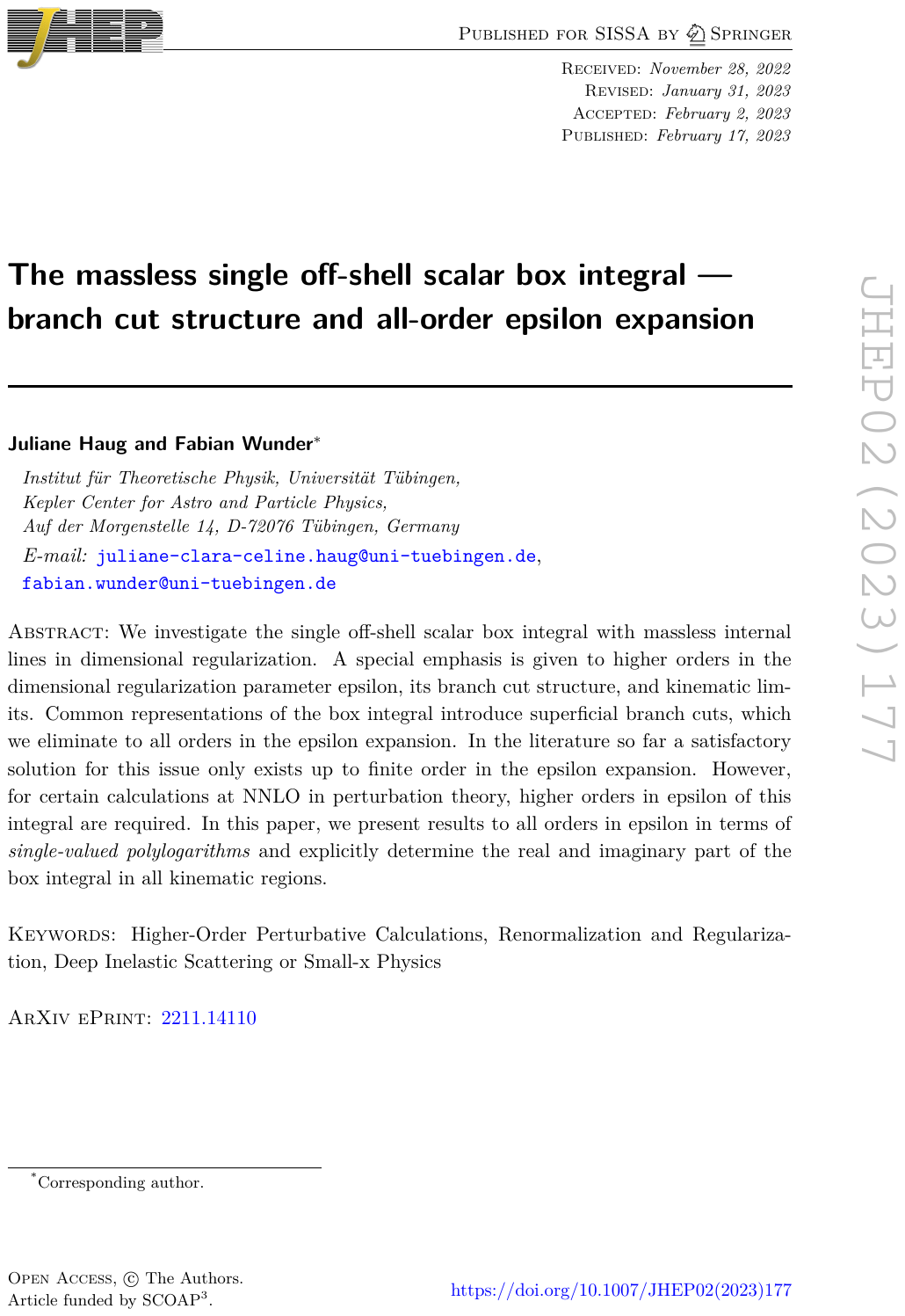}
\cleardoublepage
\section[The massless non-adjacent double off-shell scalar box integral]{Publication 3: The massless non-adjacent double off-shell scalar box integral --- branch cut structure and all-order epsilon expansion}
\label{pub:3}
\begin{summaryblock}
{
\textbf{Why?\,: }After finishing publication \ref{pub:2}, the natural question arose whether the calculation method generalizes to other loop integrals. The most direct generalization is taking another particle off-shell.
\vspace{0.1cm}
\\
\textbf{What?\,:} We calculated the scalar box integral with two external off-shell particles at diagonally opposite corners and studied its branch cut structure to all orders in dimensional regularization.
}
{
\textbf{Foundations: }
Methods from publication \ref{pub:2}.
\vspace{0.1cm}
\\
\textbf{Novel methods: } Symmetric treatment of Mandelstam variables and virtualities for the box integral.
}
{
\textbf{Results:}
\begin{itemize}
\item[(i)] Hypergeometric representation of the non-adjacent double-off-shell box integral valid in all kinematic regions.
\item[(ii)] All-order $\eps$-expansion with explicit imaginary parts in all regions free of spurious branch cuts.
\end{itemize}
\textbf{Implications: }The generalizations of the results from publication \ref{pub:2} reveal a symmetric structure between the virtualities of the off-shell particles and the Mandelstam variables describing the kinematics.
A further generalization of the method to the adjacent double, triple, and quadruple off-shell cases as well as the inclusion of masses would open up the treatment of a wide class of processes. This may require a more general class of single-valued polylogarithms.
Also, it would be interesting to study to what extend single-valued polylogarithms and their generalizations can be used to remove spurious branch cuts for multi-loop integrals.
}
\end{summaryblock}

\vspace{0.2cm}
This is a short follow-up paper on publication \ref{pub:2} that takes one more particle off-shell and discusses the case of two diagonally opposite off-shell particles at the one-loop massless box.
The work was motivated by a referee remark asking to what extend the calculation of \ref{pub:2} is generalizable.

From an ansatz with four off-shell particles, one realizes that for the calculation to work in a way close to \ref{pub:2}, two non-adjacent particles need to be taken on-shell with $p^2=0$.
Keeping the other two off-shell however allows for a calculation in terms of hypergeometric functions in the style of \ref{pub:2}, again by introducing additional regulators for spurious branch cuts.
The new result of eq.\,(2.14) in terms of a sum of four hypergeometric functions, that is a direct generalization of eq.\,(2.36) from \ref{pub:2}, reveals a symmetry between Mandelstam variables and off-shell momenta that is opaque in the special case of eq.\,(2.36).
The all-order $\eps$-expansion, where all spurious branch cuts cancel, is again expressed in terms of single-valued polylogarithms.

After publication of the preprint, we were in communication with Oleg Tarasov and learned about the possibility of calculating the box integral with an even higher number of scales in terms of hypergeometric functions by functional reduction \cite{Tarasov:2015wcd,Tarasov:2019mqy,Tarasov:2022clb,Tarasov:2022pwt}.
This idea is close to the use of the mass reduction for angular integrals.

The results were presented together with \ref{pub:2} at the FOR2926 research group meeting in February 2023 by Juliane Haug, at the QCD Masterclass 2023 in Saint-Jacut-de-la-Mer, France by myself, and at the 2023 Amplitudes school at CERN in a joint presentation.  

\includepdf[pages=-, pagecommand={}, offset=3mm 0mm]{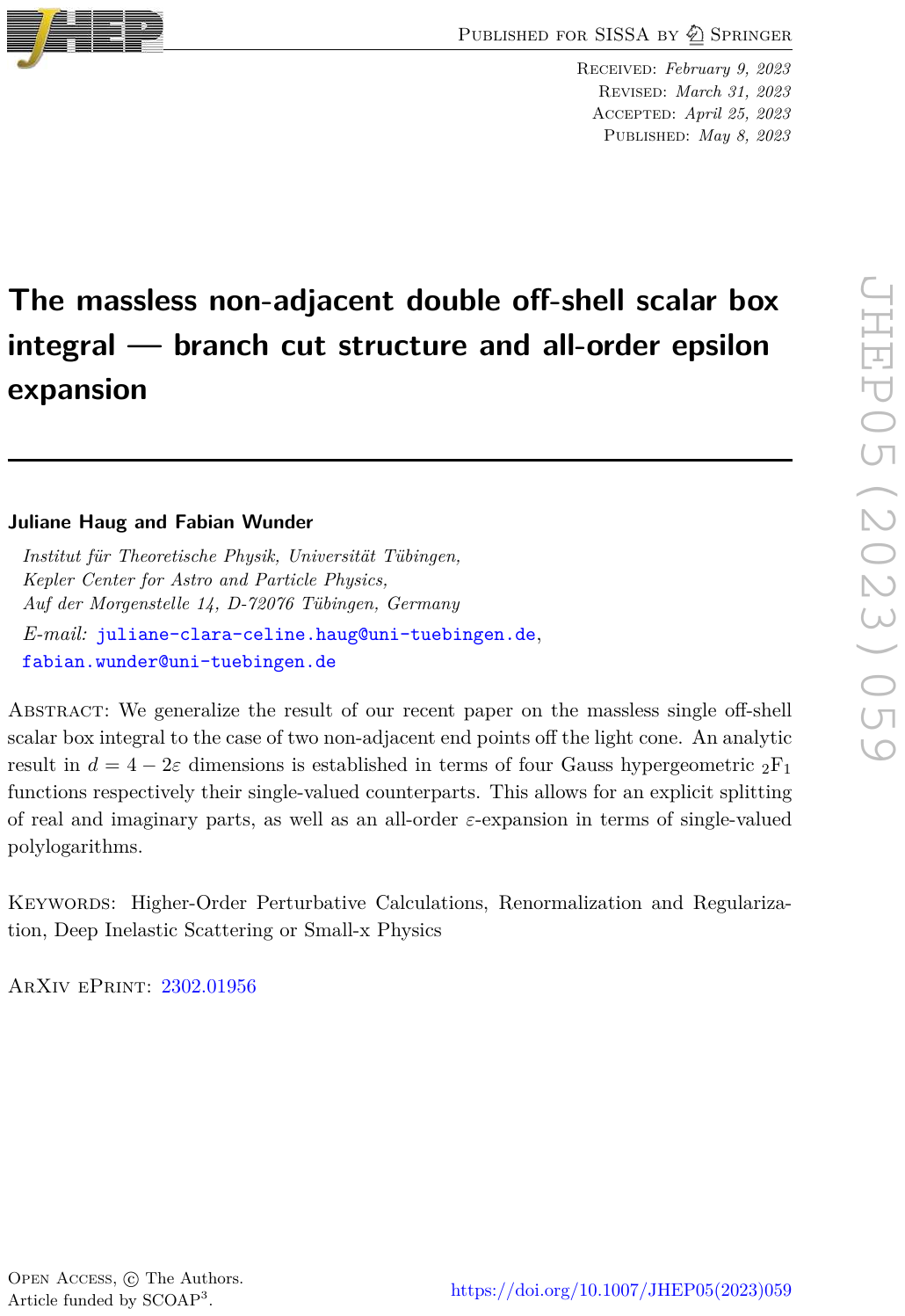}
\cleardoublepage
\section[Asymptotic behavior of angular integrals in the massless limit]{Publication 4: Asymptotic behavior of angular integrals in the massless limit}
\label{pub:4}
\begin{summaryblock}
{
\textbf{Why?\,: }
In pQCD processes such as quarkonium production at NLO, angular integrals appear that need to be integrated over the denominator ``mass''.
Existing $\eps$-expansions were insufficient since expanding in $\eps$ does in general not commute with taking the massless limit.
\vspace{0.1cm}
\\
\textbf{What?\,: } 
We calculated $\eps$-expansions of the two-denominator angular integral in a form where the massless limit is explicit, called ``asymptotic'' here.
}
{
\textbf{Foundations: }
Dimensional regularization, polylogarithms, angular integrals from publication \ref{pub:1}.
\vspace{0.1cm} 
\\
\textbf{Novel methods: } Systematic application of the mass-splitting lemma from \ref{pub:1}.
}
{
\textbf{Results:}
\begin{itemize}
\item[(i)] Asymptotic $\eps$-expansions for the massive one-denominator angular integral.
\item[(ii)] Asymptotic $\eps$-expansions for the single-massive two-denominator angular integral.
\item[(iii)] Asymptotic $\eps$-expansions for the double-massive two-denominator angular integral.
\end{itemize}
\textbf{Implications: }The new form of the $\eps$-expansion of the double-massive two-denominator angular integral allows for the analytic phase-space integration in quarkonium production at NLO.
}
\end{summaryblock}

\vspace{0.2cm}
This single-authored paper revisits the two-denominator integral treated in publication \ref{pub:1} and focuses on the behavior of the $\eps$-expansion in the massless limit.
This can be problematic, since in general $\eps$-expansion and kinematic limits do not commute, and therefore a result expanded in $\eps$ cannot always be plugged into an integral that includes the kinematic limit at the end-point.

This study was motivated by discussions with Mathias Butenschön at the FOR2926 meeting in Hamburg in November 2023, who presented ongoing work towards quarkonium production at NNLO \cite{Butenschön:20future}, which required a double-massive angular integral integrated from the massless endpoint.
A problem appears already at NLO, where the master integrals were known in the form of $\eps$-expansions, but not in a form that could be used directly.\footnote{In fact, initially, the $\eps$-expansions were plugged in directly, since this worked for almost all master integrals. The problem surfaced because of inconsistencies of the poles that could be checked against results using phase-space slicing \cite{Harris:2001sx}.
Overall, it took half a year to get to the origin of the problem.}
Hence, the goal was to extract an $\eps$-expansion in a form that commutes with the massless limit.\footnote{Key results for this publication were worked out on the train ride home from Hamburg back to Tübingen.}

The main tool for the extraction of the behavior of the integral in the massless limit is the \textit{two-point splitting lemma} (see eq.\,(5) in the publication or eq.\,\eqref{eq:twopointsplit} in the main text) that has been introduced in publication \ref{pub:1}.
Using this in a graphical way allows to see that all two-denominator angular integrals can be re-expressed by integrals that either do not change the number of masses in the limit under consideration or the two vectors of the integral coincide in the limit.
For the latter case, the $\eps$-expansion of the single-massive two-denominator integral in the form of eq.\,(11) is well-behaved, since the logarithmic structure of the limit is grasped by the un-expanded factor $(v_{11}/v_{12}^2)^\eps$.
Together, this allows to find $\eps$-expansions that are well-behaved in the massless limit for all angular integrals with up to two denominators.

The discussion of the asymptotic behavior of angular integrals was continued in \ref{pub:7} in collaboration with Vladimir Smirnov.
I presented this work at the FOR2926 meeting in Regensburg in July 2024 and at the 2025 SCET Workshop in Tucson, Arizona.
\cleardoublepage
\includepdf[pages=-, pagecommand={}, scale=0.95, offset=6mm 0mm]{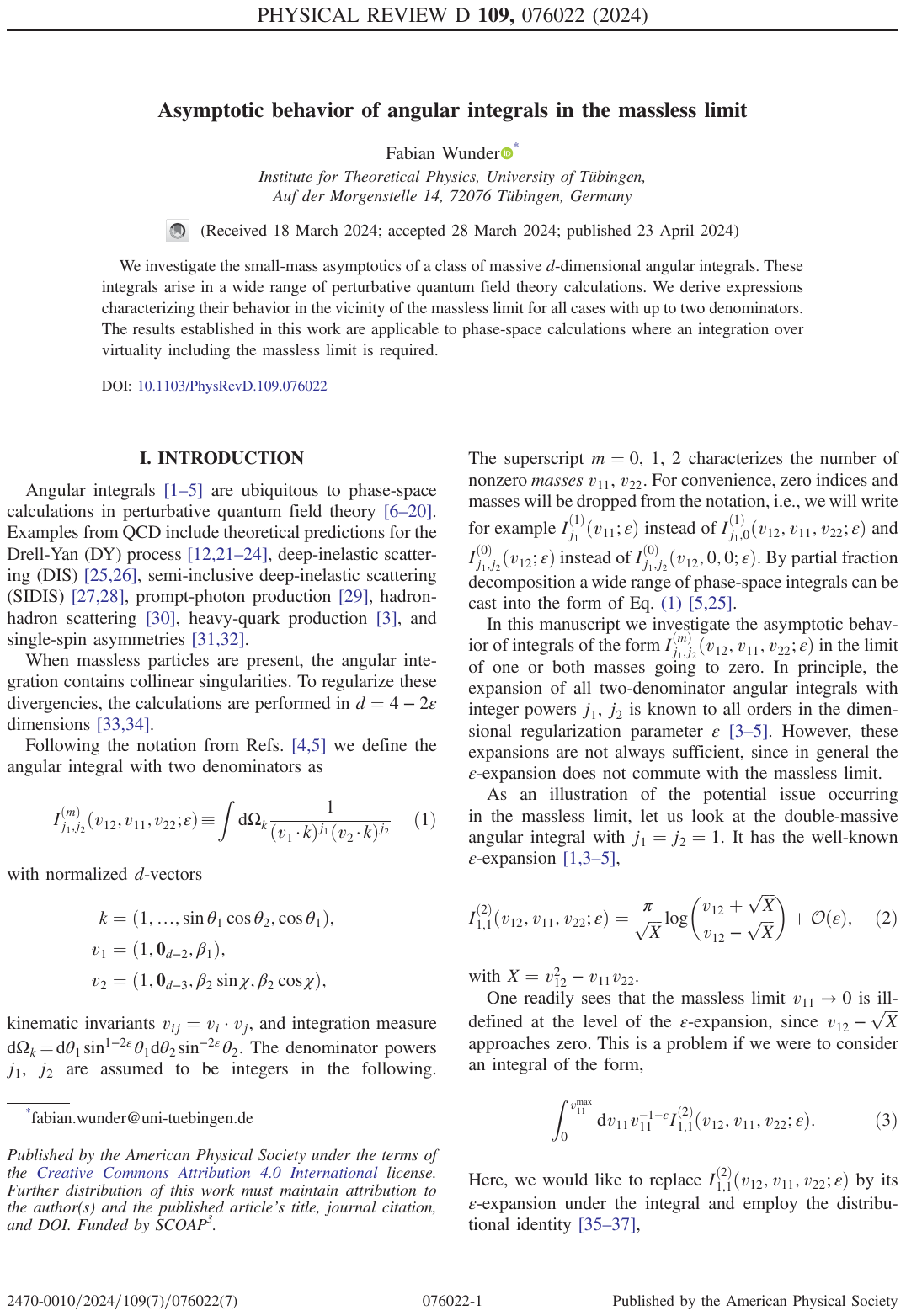}
\cleardoublepage
\section[Perturbative $T$-odd asymmetries in the Drell-Yan process revisited]{Publication 5: Perturbative $\mathbf{T}$-odd asymmetries in the Drell-Yan process revisited}
\label{pub:5}
\begin{summaryblock}
{
\textbf{Why?\,: }The Drell-Yan process is central to measurements at proton-proton colliders such as the LHC. For large transverse momentum $\Qt$, it is described in collinear factorization, for small transverse momentum, TMD physics sets in.
A better understanding of the interplay of both regimes is needed to understand the full transverse momentum spectrum.
\vspace{0.1cm}
\\
\textbf{What?\,: } We calculated a specific subset of helicity structure functions, called $T$-odd, that can be perturbatively generated by QCD radiation in collinear factorization, but also by TMDs. To connect both, we expand our results for small $\Qt$.
}
{
\textbf{Foundations: }Factorization, Drell-Yan process, $T$-odd observables, one-loop integrals.
\vspace{0.1cm}
\\
\textbf{Novel methods: }Systematic expansion of collinearly factorized results for small $\Qt$.
}
{
\textbf{Results:}
\begin{itemize}
\item[(i)] one-loop results for the three $T$-odd helicity structure functions in collinear factorization for $\gamma/Z$ and $W^\pm$ channels.
\item[(ii)] Next-to-next-to-leading power expansion in transverse momentum. 
\item[(iii)] Non-zero values for $T$-odd contributions from perturbative corrections, in qualitative agreement with ATLAS data for $Z$ production. A quantitative comparison remained inconclusive due to large experimental uncertainties and missing higher order corrections.
\end{itemize}
\textbf{Implications: }
The DY-$\Qt$ spectrum for $T$-odd asymmetries requires further investigation.
}
\end{summaryblock}

This publication -- done in collaboration with Valery Lyubovitskij, my advisor Werner Vogelsang, and Alexey Zhevlakov -- is concerned with $T$-odd effects\footnote{What is meant by $T$-odd effects is explained in the box following the summary.} in the Drell-Yan process differential in transverse momentum $\Qt$ for both $\gamma/Z$  and $W^\pm$ induced lepton pair production.
In the angular distribution $\dx N/\dx\Omega$, $T$-odd structure functions have characteristic angular modulations proportional to $\sin \phi$ and $\sin 2\phi$, which allow for extraction from data (see eqs.\,(2), (3) and FIG.\,1.).
In collinear factorization, the leading $T$-odd effect is generated by the imaginary part of the interference between one-loop diagrams -- here the results from publication \ref{pub:2} could be applied -- and the Born level amplitude.
We calculated the contributions from both channels that contribute at this order,  $q\bar{q}$-annihilation and the $qg$-Compton-like channel for all three $T$-odd structure functions (eqs. (20) to (34)).
The collinear factorization approach is formally valid at $\Lambda_\text{QCD}\ll Q\sim\Qt$.
For $\Qt\ll Q$, the appropriate framework is TMD factorization.
To match our calculation to the TMD region smoothly, we formally expand our result about $\Qt=0$.
In contrast to the earlier literature, this is done beyond leading power in a way that generalizes to arbitrary power corrections.\footnote{For reference, the most relevant identities for this expansion are summarized in a tool-block before the re-print.}
FIG. 4 shows the importance of NLP contributions -- counted in powers of $\Qt^2/Q^2$ -- already at $\Qt/Q\sim 0.2$.
For the only case where data is available, $Z$-exchange near $Q=M_\text{Z}$, we confronted our predictions with data from the ATLAS experiment at the LHC.
FIGs. 8 and 9 show agreement regarding non-vanishing $T$-odd effects, but beyond that, match poorly.
A key problem lies in the smallness of the $T$-odd effects on the percent level.
Nevertheless, the mismatch indicates the necessity of further investigations of the Drell-Yan angular distribution, in particular for the $T$-odd structure functions, both from collinear and TMD factorization approaches.
From the TMD side, studies of NLP effects -- counted in powers of $\Qt/Q$ -- appeared in the meantime \cite{Piloneta:2024aac,Arroyo-Castro:2025slx}.

Following up on this paper, my co-authors Valery Lyubovitskij and Alexey Zhevlakov continued the project and included the $T$-even structure functions and the Drell-Yan forward-backward asymmetry in a series of articles \cite{Lyubovitskij:2024jlb,Lyubovitskij:2025oig,Anikin:2025vqy}.
The appearance of derivatives of PDFs by the $\Qt$-expansion formalism, starting at NLP, contributed to motivating the study of PDF evolution in a way that analytically controls the $x$-space behavior of PDFs in publication \ref{pub:6}.

I presented this work at the QCD evolution workshop in May 2024 in Pavia, Italy.

\begin{block}[type=note]
\textbf{What is a $T$(ime reversal)-odd observable?}\\
To understand what a $T$-odd observable is, we recall some basic facts about amplitudes in QFT:
We start from the unitarity of the $S$-matrix, i.e. $1\!\!1=\hat{S}\,\hat{S}^\dagger$.
Based on the $S$-matrix, the transition matrix $\hat{T}$ is given by $\hat{S}=1\!\!1+i \hat{T}$ and 
\begin{equation}
\langle f| \hat{T}| i\rangle =(2\pi)^4 \delta^4 (P_f-P_i)
 \mathcal{M}_{fi}
\end{equation}
defines the matrix element $\mathcal{M}_{fi}$ for a transition from the initial state $i$ to the final state $f$.
From the unitarity of the $S$-matrix it follows that the matrix element satisfies the generalized \textit{optical theorem} in the form
\begin{equation}
\mathcal{M}_{fi}-\mathcal{M}_{if}^*\!=\!i\overbrace{\sum_X \!\mathcal{M}_{X f}^* \mathcal{M}_{Xi}}^{\mathcal{A}_{fi}}\,.
\end{equation}
Also, it is symmetric under time-reversal of momenta and spin with simultaneous interchange of initial and final states, $i\leftrightarrow f$, turning particles into anti-particles (denoted by bars), i.\,e., $|\mathcal{M}_{fi}|^2=|\mathcal{M}_{\bar{i}\bar{f}}|^2$.

Now, taking the absolute square of the optical theorem, we have
\begin{equation}
|\mathcal{M}_{fi}|^2=|\mathcal{M}_{if}|^2-2 \,\mathrm{Im}(\mathcal{M}_{if} \mathcal{A}_{fi})+|\mathcal{A}_{fi}|^2.
\end{equation}
Subtracting $|\mathcal{M}_{\bar{f}\bar{i}}|^2$ on both sides, we arrive at the defining equation for a $T$-odd effect, meaning the change-of-sign of an observable under \textit{naive} time reversal, i.e. reversal of momenta and spin without interchange of initial and final state,
\begin{equation}
\underbrace{\textcolor{red}{|\mathcal{M}_{fi}|^2-|\mathcal{M}_{\bar{f}\bar{i}}|^2}}_{T\text{-odd effect}}=\underbrace{|\mathcal{M}_{if}|^2-|\mathcal{M}_{\bar{f}\bar{i}}|^2}_{\text{vanishes by }T\text{-invariance}}-\underbrace{\textcolor{red}{2\,\mathrm{Im}(\mathcal{M}_{if} \mathcal{A}_{fi})}}_{\sim \mathcal{M}^3}+\underbrace{|\mathcal{A}_{fi}|^2}_{\sim \mathcal{M}^4}.
\end{equation}
This equation shows that the sign flip of the expression on the left dominantly comes from the interference term $\textcolor{red}{2\,\mathrm{Im}(\mathcal{M}_{if} \mathcal{A}_{fi})}$.
$T$-odd effects can occur in theories invariant under true time reversal, including a wide range of QCD scattering phenomena.
In Drell-Yan, they appear as angular asymmetries in the angle between the lepton and hadron planes ($\sim \sin\phi$ and $\sin 2\phi$, which has been studied e.\,g. in \cite{Hagiwara:1984hi,Mirkes:1992,Yokoya:2007xe,Benic:2024fvk}).
\end{block}
\vspace{1.2cm}
\begin{toolblock}{\textbf{Transverse momentum expansion beyond leading power: }\\
For the expansion of the coefficient functions 
\begin{align}
&W(x_1,x_2,\rho^2) =
\frac{1}{x_1 x_2} \, \sum\limits_{a,b}
\, \int\limits_{x_1}^1 \, dz_1
\, \int\limits_{x_2}^1 \, dz_2
 \, f_{a/H_1}\Big(\frac{x_1}{z_1}\Big) 
\, f_{b/H_2}\Big(\frac{x_2}{z_2}\Big)\nonumber\\
&\qquad\times\underbrace{\delta\left((1-z_1)(1-z_2)
-\frac{\rho^2}{1+\rho^2}z_1 z_2\right)}_\text{from two-particle phase space} \tilde{w}^{ab}(z_1,z_2,\rho^2)
 \,\quad
 \label{eq:DYStructure functions}
\end{align}
\begin{wrapfigure}[16]{r}{0.4\textwidth}
\vspace{-0.5cm}
\includegraphics[width=0.4\textwidth]{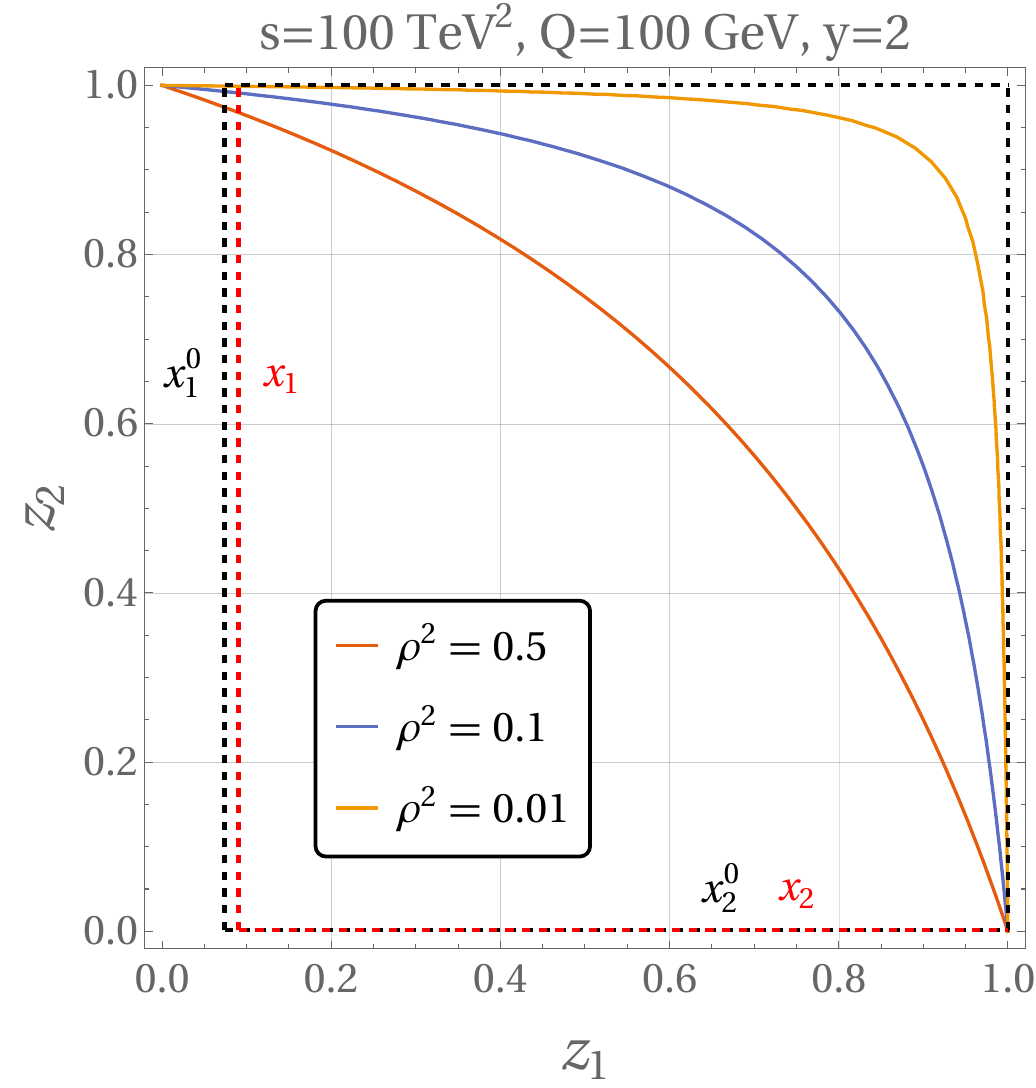}
\caption[Integration domain of $T$-odd coefficient function. Created with \texttt{Mathematica} \cite{Mathematica13}.]{Integration domain of eq.\,\eqref{eq:DYStructure functions}. The solid lines show the delta function condition for different values of the transverse momentum.}
\end{wrapfigure}
in powers of $\rho^2=\frac{\Qt^2}{Q^2}$, there are two sources of transverse momentum dependence.
First, from the dependence of the integrand on  $\rho^2$.
Second, from the implicit dependence of the lower integration bounds $x_{1,2}$ on the transverse momentum through
\begin{align}
x_{1,2}   &= e^{\pm y} \sqrt{\frac{Q^2(1+\rho^2)}{s}}
\end{align}
with rapidity $y$, which is to be expanded about $x_{1,2}^0=e^{\pm y} \sqrt{\frac{Q^2}{s}}$.\\
\textbf{Integrand expansion. }To expand the integrand beyond leading power, we generalized the expansion of the phase-space delta function -- which was well-known to leading power \cite{Boer:2006eq} -- to, in principle, arbitrary powers by making systematic subtractions in the integrand and isolating the distributional contributions until the remainder is sufficiently suppressed by powers of $\rho^2$.
Explicitly, to $\mathcal{O}(\rho^4)$, the expansion of the delta function reads
\begin{align}
\hspace{-0.2cm}&\delta\!\left((1-z_1)(1-z_2)-\frac{z_1 z_2 \rho^2}{1+\rho^2}\right)
\nonumber\\
\hspace{-0.2cm}&=
\frac{\delta(1-z_2)}{(1-z_1)_+}
+\frac{\delta(1-z_1)}{(1-z_2)_+}
-\log\rho ^2\,\delta(1-z_1) \delta(1-z_2)
   \nonumber\\
\hspace{-0.2cm}&+\rho^2 \left[\frac{\delta
   ^{(1)}(1-z_2)}{(1-z_1){}^2_{\text{+,}1}}
   -\frac{\delta
   ^{(1)}(1-z_2)}{(1-z_1)_+}
   +
\frac{\delta
   ^{(1)}(1-z_1)}{(1-z_2){}^2_{\text{+,}1}}
   -\frac{\delta
   ^{(1)}(1-z_1)}{(1-z_2)_+}
   \right.
   \nonumber\\
\hspace{-0.2cm}&
\left.   
\quad
   +\frac{\delta(1-z_2)}{(1-z_1)_+}
   -\frac{\delta(1-z_2)}{(1-z_1){}^2_{\text{+,}1}}
   +\frac{\delta(1-z_1)}{(1-z_2)_+}
   -\frac{\delta(1-z_1)}{(1-z_2){}^2_{\text{+,}1}}
   +\delta(1-z_1)\delta(1-z_2) 
   \right.
   \nonumber\\
\hspace{-0.2cm}&
\left.   \quad 
   +\log\rho^2 \left(-\delta^{(1)}(1-z_1) \delta^{(1)}(1-z_2)
   +\delta ^{(1)}(1-z_1) \delta(1-z_2)
      \right.\right.
   \nonumber\\
\hspace{-0.2cm}&
\left. \left.\quad
   +\delta(1-z_1) \delta ^{(1)}(1-z_2)-\delta(1-z_1)\delta
   (1-z_2)\right)\vphantom{\left(\frac{X}{X}\right)}\right]
   +\mathcal{O}(\rho^4\log\rho^2)\,.
\end{align}
Here, the $\delta^{(k)}(1-z)$ mean derivatives of the delta function defined by
\begin{align}
\int_0^1\dx z\,\delta^{(k)}(1-z)f(z)=(-1)^{k}\left.\frac{\dx^k f(z)}{\dx z^k}\right|_{z=1}
\end{align}
and there appear generalized plus distributions (here only for $l=0$) 
\begin{align}
&\int_0^1\dx z\,\left[\frac{\log^{l}(1-z)}{(1-z)^{m+1}}\right]_{+;m}\,f(z)
=\int_0^1\dx z\,\frac{\log^{l}(1-z)}{(1-z)^{m+1}}\left[f(z)-\mathcal{T}^m_1 f(z)\right],
\end{align}
where $\mathcal{T}^m_1 f(z)$ means the $m$-th order Taylor polynomial of $f$ about $z=1$,
\begin{align*}
\mathcal{T}^m_1 f(z)=\sum_{j=0}^m\frac{(-1)^j}{j!}(1-z)^j \,\left.\frac{\dx^j f(z)}{\dx z^j}\right|_{z=1}\,.
\end{align*}
\textbf{Boundary expansion. }
After expansion of the integrand, we receive convolutions of the form 
\begin{equation}
I(x)=\int_x^1 \frac{\dx z}{z}\,w(z)\,q\!\left(\frac{x}{z}\right).
\end{equation}
We can expand these about $x_0\equiv x(\rho^2=0)$ using
\begin{align}
I(x)=\sum_{n=0}^\infty\frac{(x-x_0)^n}{n!}\int_{x_0}^{1^+}\frac{\dx z}{z}\,\left(\frac{\dx^n}{\dx z^n}\,\left[w(z)\Theta(1-z)\right]\right)\,\left(\frac{z}{x_0}\right)^n q\!\left(\frac{x_0}{z}\right),
\end{align}
where $1^+$ means the upper bound is to be taken slightly above one\footnote{This is implicitly understood whenever $\delta(1-z)$ distributions are involved to ensure that they give a well-defined contribution. Here, the $1^+$ is made explicit to stress that upon partial integration $\theta(1-z)$ terms do not receive a contribution from the upper boundary.} and derivatives of plus distributions are calculated as
\begin{align}
\frac{\dx }{\dx z}\left[\frac{\log^{l}(1-z)}{(1-z)^{n+1}}\right]_{+,n}\!\!=&\,(n+1)\left[\frac{\log^{l}(1-z)}{(1-z)^{n+2}}\right]_{+,n+1}
\nonumber
\\&-l\left[\frac{\log^{l-1}(1-z)}{(1-z)^{n+2}}\right]_{+,n+1}
-\delta_{l0}\sum_{k=0}^{n+1}\frac{1}{k!}\,\delta^{(k)}(1-z)\,.
\end{align}
Note that by definition all plus distributions are understood to have an implicit $\Theta(1-z)$, i.e. 
\begin{equation}
\left[\frac{\log^{l}(1-z)}{(1-z)^{n+1}}\right]_{+,n}\equiv \left[\frac{\log^{l}(1-z)}{(1-z)^{n+1}}\right]_{+,n}\Theta(1-z),
\end{equation}
which is required for consistency with partial integration -- they do not make a contribution at the upper integration boundary at $1^+$.
}\end{toolblock}
\includepdf[pages=-, pagecommand={}, scale=0.95, offset=6mm 0mm]{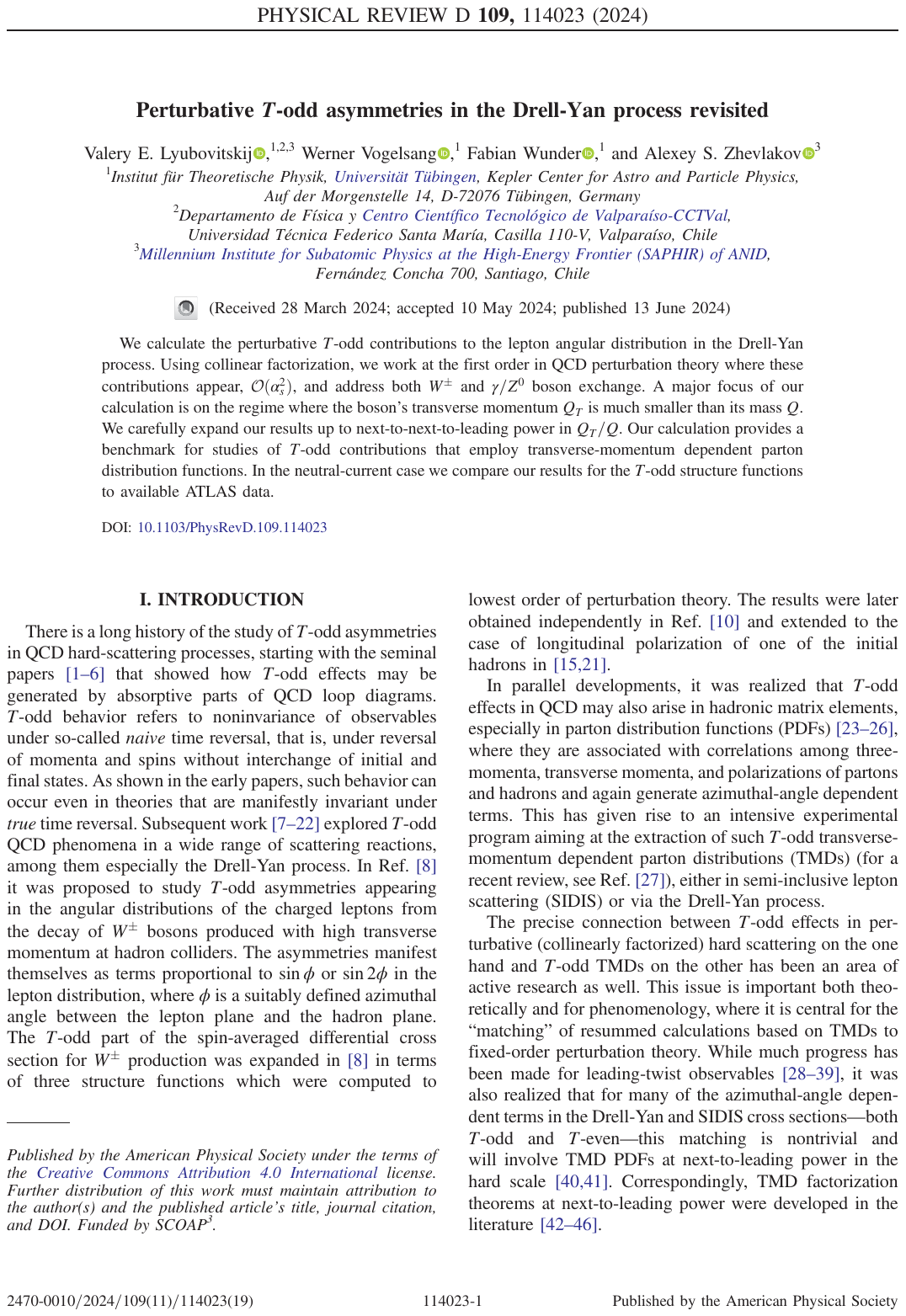}
\cleardoublepage
\section[A semi-analytical $x$-space solution for parton evolution]{Publication 6: A semi-analytical $\mathbf{x}$-space solution for parton evolution — Application to non-singlet and singlet DGLAP equation}
\label{pub:6}
\begin{summaryblock}
{
\textbf{Why?\,: }Parton distributions depend on the scale at which they are probed; this dependence is governed by evolution equations. Numerically efficient solutions of these equations are essential for phenomenology and especially the extraction of parton distributions from global fits.
Existing methods either use discretization in the momentum fraction $x$ or Mellin transform.
\vspace{0.1cm}
\\
\textbf{What?\,: } 
We developed a novel semi-analytic method to solve parton evolution which captures part of the analytic $x$-space behavior and is applicable to cases where Mellin transformation is not possible. As a proof-of-principle we applied it to the evolution of PDFs at LO.
}
{
\textbf{Foundations: }
Parton distribution functions, DGLAP equation, Magnus expansion (reviewed in the publication).
\vspace{0.1cm}
\\
\textbf{Novel methods: }  Analytical basis in $x$-space to transform the integro-differential evolution equation into a system of ordinary differential equations that is solved numerically via a matrix exponential (\texttt{POMPOM} method); $x$-space ansatz for DGLAP with polynomials and $\log x$.
}
{
\textbf{Results:}
\begin{itemize}
\item[(i)] The proposed semi-analytic method for solving the DGLAP equation in the non-singlet and singlet case has been implemented in the code \texttt{POMPOM} in both \texttt{Mathematica} and \texttt{Python}.
\item[(ii)] Results show good agreement with benchmarks, also internal consistency checks through sum-rule violation and convergence checks show good accuracy across a wide $x$ range, $10^{-7}<x<1$.
\item[(iii)] For fixed number of basis functions, a clever choice of basis functions can give orders-of-magnitude improvements in the numerics.
\end{itemize}
\textbf{Implications: }
The new method to solve evolution equations proved itself promising. Hence, it is worth considering to implement it for higher order DGLAP evolution, to generalize it to other evolution equations, and to investigate further optimization potential through systematic choice-of-basis.
}
\end{summaryblock}

\vspace{0.2cm}
In this publication, done in collaboration with my fellow doctoral students Juliane Haug and Oliver Schüle, we introduced a novel semi-analytical approach to solving parton evolution equations and demonstrated it on the DGLAP equation, which governs the scale dependence of parton distribution functions.
The main idea is to solve this equation in a way that keeps the $x$-space behavior in terms of scale-independent basis functions and shifts the scale dependence to numerical coefficients.
\begin{toolblock}{\textbf{The \texttt{POMPOM}-method for solving evolution equations:}\\
We start from a general integro-differential equation with an integral operator $\mathbf{P}\otimes$ acting on a set of parton distributions $\mathbf{f}$ in $\vec{x}$-space
			\begin{equation}
				\frac{\dx}{\dx \mu} \mathbf{f}(\mu,\vec{x}) \,=\, \left( \mathbf{P}\otimes \mathbf{f} \right) (\mu,\vec{x})\,.
				\label{eq:General_integro-differential_equation}
			\end{equation}
To solve it, we make an ansatz with a suitable set of spanning functions $\mathbf{f}_m(\vec{x})$
			\begin{align}
				\mathbf{f}(\mu,\vec{x}) \,=\, \sum_m\, a_m(\mu)\, \mathbf{f}_m(\vec{x}).
			\end{align}
This transforms eq.\,\eqref{eq:General_integro-differential_equation} into an infinite-dimensional system of ordinary differential equations for the coefficients,
			\begin{align}
				\frac{\dx}{\dx\mu} a_m(\mu) \,=\, \mathcal{P}_{mn}(\mu)\, a_n(\mu) \,.
				%\label{eq:General_ODE_for_coefficients}
			\end{align}
Next, we truncate this system in a controlled way -- such that we keep the numerically important contributions -- into a finite system and solve it numerically via a matrix exponential.
Once the evolution matrix $\mathcal{P}$ is calculated, it can be applied to any initial condition.
}\end{toolblock}

The paper reviews the theoretical prerequisites about evolution equations in sections 1 and 2, the semi-analytic method to decouple integro-differential evolution equations into a system of ordinary differential equations is introduced in section 3.1.
In the remainder of section 3, it is used on the most elementary example case, the leading order DGLAP equation.
To assess the numerical performance of the method, especially the numerical error from truncating to a finite system, we compared relative and absolute deviations from benchmark evolutions showing overall good agreement in a broad range of $x$. 
For $\mathcal{O}(200)$ basis functions, sum rules are violated by less than $10^{-6}$ and truncation effects are shown to systematically improve with increasing number of basis functions.

The proof-of-principle for the numerical viability of the \texttt{POMPOM} method motivates building it into a full evolution tool for the DGLAP equation by incorporating higher orders and applying it to similar evolution equations where existing methods are less well developed.
In both cases, an investigation of how to systematically improve the choice of basis functions and cut-off would be valuable.

Along with the publication, public \texttt{Mathematica} and \texttt{Python} implementations were provided.
This work has been presented at the QCD Evolution Workshop in May 2024 in Pavia, the QCD Masterclass 2024 in Saint-Jacut-de-la-Mer, and the FOR2926 meeting in Hamburg in February 2025 by Juliane Haug and in the group seminar of Barbara Jäger in Tübingen in June 2024 by myself.

\includepdf[pages=-, pagecommand={}, offset=3mm 0mm]{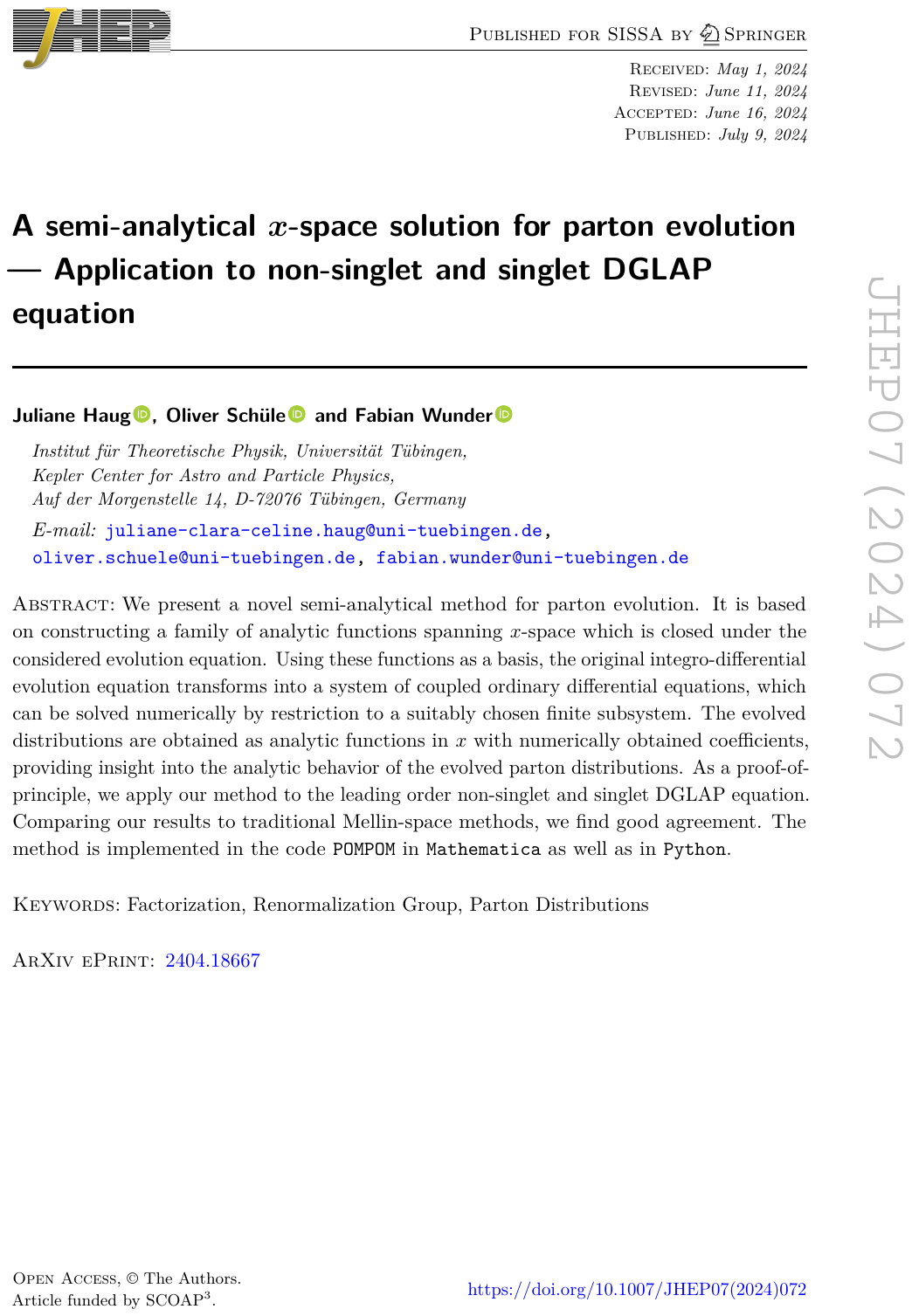}
\begin{block}[type=note]
\textbf{Erratum: }After publication, we noted a few typos in the integrals listed in Appendix A that are corrected in the latest arXiv version \cite{Haug:2024asl} but not in the journal version reprinted here. The corrected appendix reads:\\
\vspace{0.5cm}\\
\begin{footnotesize}
\textbf{A\quad Master integrals}\\

In this appendix, we list the results for the master integrals appearing in eq.\,(3.5).
First, we have
	\begin{align}
		&I_1^{m,n} \,\equiv\, x^n \int_x^1 \dx\xi\, \frac{\left(\xi^{-n} - 1\right)}{1-\xi} \frac{\ln^m(x/\xi)}{m!} &
		\\
		&=\, \left\{
		\begin{array}{ll}
			-\sum_{k=0}^{-n-1} \left[ \sum_{j=0}^m \frac{1}{(k+1)^{m+1-j}}\; \frac{\ln^j(x)\,x^n}{j!}  \,-\, \frac{x^{k+n+1}}{(k+1)^{m+1}} \right] & \text{if } n \,\leq -2 \,,
			\\
			-\frac{1}{x}\sum_{k=0}^m \frac{\ln^k(x)}{k!} \,+\, 1 & \text{if } n \,= -1 \,,
			\\
			0 & \text{if } n \,=\, 0 \,,
			\\
			x^n \left[ (-1)^m \sum_{k=1}^{n-1} \frac{x^{-k}}{k^{m+1}} \,-\, \sum_{k=0}^m (-1)^{k}\, H_{n-1,k+1}\, \frac{\ln^{m-k}(x)}{(m-k)!} \,-\, \frac{\ln^{m+1}(x)}{(m+1)!}\right]  & \text{if } n \,\geq\, 1 \,.
		\end{array}
		\right. 
	\end{align}
The second master integral is given by
	\begin{align}
		I_2^{m,n} \,&\equiv\, x^n \int_x^1 \frac{\dx\xi}{(1-\xi)}\, \frac{\ln^m(x/\xi) \,-\, \ln^m(x)}{m!}
		\\
		\,&=\, x^n \left[ \sum_{k=0}^{m-1} \zeta_{m-k+1} \frac{\ln^k(x)}{k!} \,+\, \frac{\ln^m(x)}{m!}\sum_{k=1}^{\infty} \frac{x^k}{k} \,-\, \sum_{k=1 }^{\infty} \frac{x^k}{k^{m+1}} \right] .
	\end{align}
For the singlet DGLAP equation, we also need the third master integral
	\begin{align}
		I_ 3^{m,n} \,&\equiv\,  x^n \int_x^1 \dx\xi\, \frac{\ln^m(x/\xi)\, \xi^{-n}}{m!}
		\\
		&=\, \left\{
		\begin{array}{ll}
			\sum_{j=0}^m \frac{1}{(1-n)^{m+1-j}} \frac{\ln^j(x)\, x^n}{j!} \,-\, \frac{x}{(1-n)^{m+1}} &\quad \text{if } n \,\neq\, 1 \,,
			\\
			- \frac{\ln^{m+1}(x)\, x}{(m+1)!} &\quad \text{if } n \,=\, 1 \,.
		\end{array} 
		\right.
	\end{align}
Here, $H_{n,m} \equiv \sum_{k=1}^{n}\frac{1}{k^m}$ are the generalized harmonic numbers, while $\zeta_m \equiv \sum_{k=1}^{\infty}\frac{1}{k^m}$ are the integer values of the Riemann zeta function.
\end{footnotesize}
\end{block}
\cleardoublepage
\section[Expansion by regions meets angular integrals]{Publication 7: Expansion by regions meets angular integrals}
\label{pub:7}
\begin{summaryblock}
{
\textbf{Why?\,: }
Expansion by regions is a method foundational to the study of multi-scale Feynman integrals.
Transferring it to angular integrals allows for a systematic extension of the study from publication \ref{pub:4} beyond two denominators. 
\vspace{0.1cm}
\\
\textbf{What?\,: }By finding a suitable integral representation, we could use the method of \textit{expansion by regions} to study the small mass asymptotics of multi-denominator angular integrals.
}
{
\textbf{Foundations: } Dimensional regularization, Feynman parameterization, Euler integrals, expansion by regions, angular integrals.
\vspace{0.1cm}
\\
\textbf{Novel methods: }Used expansion by regions, which was developed for loop integrals, on angular integrals.
}
{
\textbf{Results:}
\begin{itemize}
\item[(i)] First Euler integral representation for angular integrals, making it accessible to existing expansion by regions codes.
\item[(ii)] Small-mass results at leading power for three- and four-denominator angular integrals to order $\eps^0$ and $\eps^{-1}$, respectively.
\item[(iii)] Based on the structure for up to four denominators, conjectured the collinear pole of angular integrals for the general case of $n$ denominators and $m$ masses.
\end{itemize}
\textbf{Implications:}
Euler integrals proved themselves as a useful unifying language between loop and angular integrals which made expansion by regions applicable with the usual loop integral tools.
Hence, searching for Euler integral representations for other types of phase-space integrals would be a promising direction. 
Also, deepening the investigation of the contributing regions in momentum space might lead to a better understanding of the overall collinear structure of amplitudes. The most general poles from angular integrals seem to have a universal simple structure. Furthermore, giving proofs for the conjectured results for a general number of denominators might be a worthwhile endeavor.
}
\end{summaryblock}
\vspace{0.2cm}

In this publication -- done in collaboration with Vladimir Smirnov -- we used the method of expansion by regions\footnote{Co-invented by my co-author \cite{Beneke:1997zp}.} to study angular integrals.
This extends the previous work in \ref{pub:4} to the multi-denominator case in the limit of small masses.
It is the first systematic study of the $\eps$-expansion of multi-denominator angular integrals.

Expansion by regions was originally developed for loop integrals and is formalized in a geometric way in terms of parametric Euler integrals.
To make it applicable to angular integrals, we established parametric integrals of the same form.
This allowed us to use existing tools to identify the relevant regions \cite{Pak:2010pt}.
Calculating the two-, three- and four-denominator integrals we found that there is always a single region associated with the fully massless integral and an additional region for each mass.
In the case of two denominators, the regions could be identified with the parts of the algebraic decomposition found in publication \ref{pub:4}.
The regions associated with massive vectors take a universal form shown in eq.\,(3.42) which we proved up to $n=4$ and conjecture to hold true for any $n$.
Since the fully massive integral is free of collinear poles, this allows to deduce the general structure of collinear $1/\eps$ poles in eq.\,(3.46).

Going beyond the limit of small masses, we continued the study of multi-propagator angular integrals in publications \ref{pub:8} and \ref{pub:10}.

I presented this work at the FOR2926 meeting in Regensburg in July 2024 and at the SCET Workshop in March 2025 in Tucson, Arizona.

\includepdf[pages=-, pagecommand={}, offset=3mm 0mm]{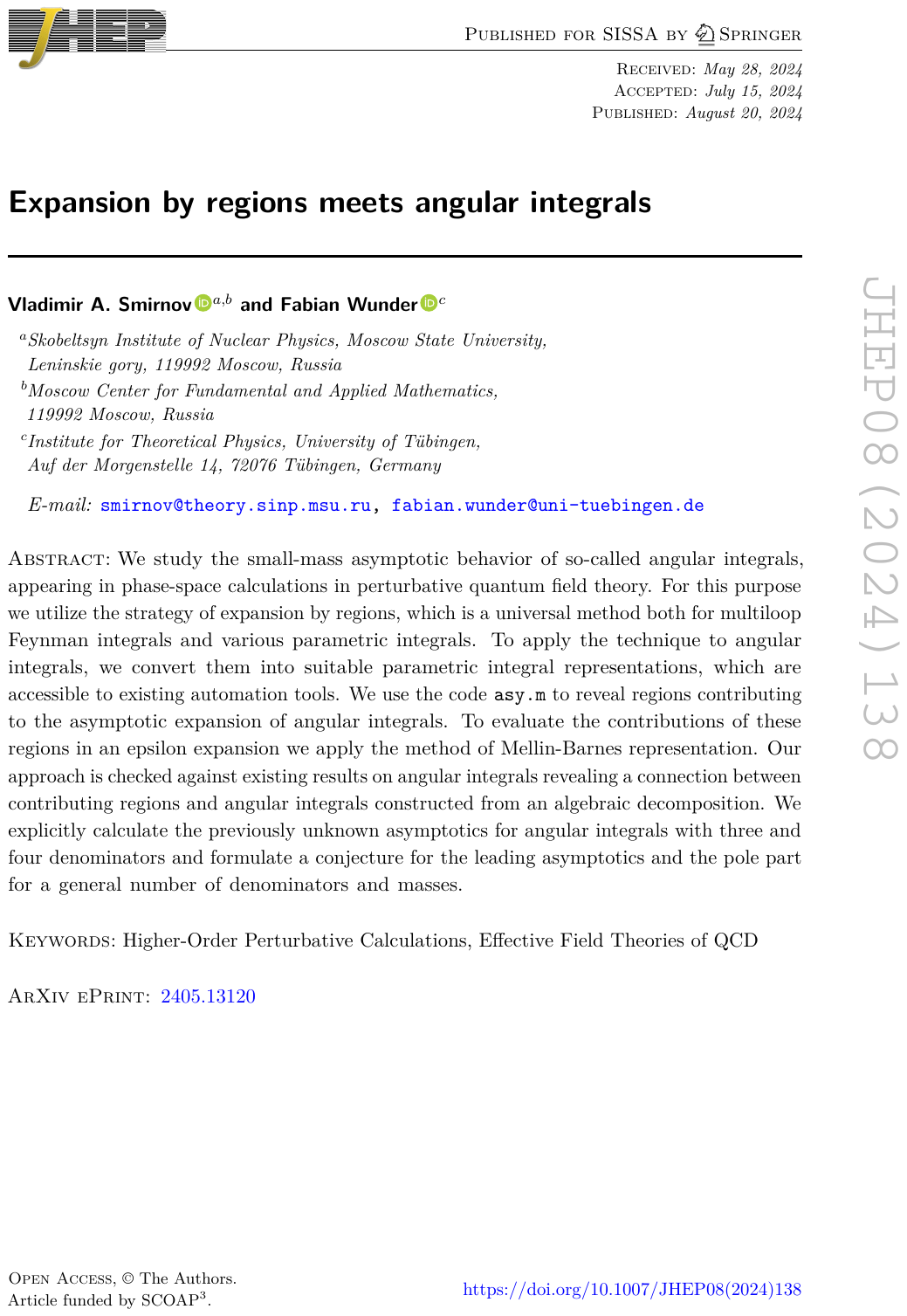}
\cleardoublepage
\section[Angular integrals with three denominators]{Publication 8: Angular integrals with three denominators via IBP, mass reduction, dimensional shift, and differential equations}
\label{pub:8}
\begin{summaryblock}
{
\textbf{Why?\,:} For processes with a larger number of observed particles, angular integrals with more than two denominators are required for analytic phase space integration.
For these, analytic results were virtually unknown in the literature, hence these phase-spaces have been treated mostly numerically.
Continuing the work of publication \ref{pub:7}, where a leading power expansion was applied, we generalize to general kinematics in the three denominator case. 
\vspace{0.1cm}
\\
\textbf{What?\,: } We calculate the angular integral with three denominators for any number of masses using methods from multi-loop integration.
}
{
\textbf{Foundations: }
Dimensional regularization, IBP, two-point splitting lemma, differential equations, polylogarithms.
\vspace{0.1cm}
\\
\textbf{Novel methods: } Parameterization independent IBPs and differential equations on the level of angular integrals; clean mass-reduction by Gram-determinant scaling; dimensional shift to simplify differential equation; Clausen functions for angular integrals.
}
{
\textbf{Results:}
\begin{itemize}
\item[(i)] IBP relations for angular integrals with three denominators that can be combined into explicit recursion relations, not needing the Laporta algorithm.
\item[(ii)] Clean form of the mass reduction expressed in terms of Gram determinants.
\item[(iii)] Dimensional shift relation for angular integrals with three denominators.
\item[(iv)] $\eps$-expansion of angular integral with three denominators to order $\eps$ for any number of masses, also implemented in \texttt{Mathematica}.
\item[(v)] Geometric interpretation of the massless angular master integral with three denominators in six dimensions.
\end{itemize}
\textbf{Implications: }
This work constitutes a major step towards analytically integrating the phase-space of $2\rightarrow n+X$ processes at NNLO for $n>1$. Whenever the external vectors of the angular integral correspond to external particles that can be treated as 4-dimensional, there are a maximum of three linearly independent spatial vectors. Hence, by partial fractioning, the three-denominator angular integral is sufficient for any number of observed particles.
Using the three-denominator integral in practice, there can however be additional soft singularities that require extracting specific terms to higher order in $\eps$. 
This is addressed in publication \ref{pub:10}.
}
\end{summaryblock}

\vspace{0.2cm}
In this publication -- done together with Juliane Haug -- we continued the study of angular integrals.
For the first time, we calculated the $\eps$-expansion of an angular integral with more than two denominators in general kinematics.
Before, only a result up to $\mathcal{O}(\eps^0)$ for the massless three-denominator integral and results based on the small-mass approximation from publication \ref{pub:7} were known.
The case of three denominators is of particular interest, since it opens the path towards analytic calculation of phase spaces for more exclusive processes, beyond the $2\rightarrow 1+X$ kinematics.

The progress in this work is made possible by systematic application of loop techniques to angular integrals with three denominators.
While there was already usage of several of these techniques in publication \ref{pub:1}, here they were applied in a more mature way closely matching the differential equation technique by Henn \cite{Henn:2013pwa} and translating it to a phase-space setting.
For a graphical summary of the detailed set-up, see Figure 1.
In sections 2 to 7 of the paper, all relevant steps are explained in a pedagogical way.

The paper contains several new results.
Notably, the IBP relations are combined into explicit recursion relations that allow for a reduction to master integrals without needing Laporta's algorithm.
The mass reduction that allows to only consider the massless and single-massive case, while known since publication \ref{pub:1}, is refined such that all pre-factors take the form of Gram determinants.
Of great help for establishing the $\eps$-expansion is the dimensional shift identity $d\rightarrow d+2$. 
This allows to extract the pole and finite part of the three-denominator integral ``for free'' in terms of known lower-denominator integrals.
Only the remaining part needs to be solved by differential equations.
Here, the most challenging part is the appearance of square-roots that require rationalization.
Upon integration of the differential equation, the genuine three-denominator part gives rise to a combination of Clausen functions (reviewed in Appendix H).
In the massless case, these can be interpreted in an intriguing geometric way, combining quantities from euclidean, spherical, and hyperbolic geometry in a single formula (see eq.\,(8.7) and Figure 3).
While broader relevance of this finding remains unclear, it constitutes a personal highlight of this thesis.

Parallel to this work, a group from Regensburg, led by Taushif Ahmed, independently calculated the three-denominator angular integral based on a Mellin-Barnes approach \cite{Ahmed:2024pxr}.
Prior to the publication we cross-checked results numerically and coordinated uploads of the preprints.

Later, the presented differential equation approach was shown to straightforwardly generalize to include four denominators in publication \ref{pub:10}.
I presented this work at the FOR2926 meeting in Hamburg in February 2025 and at the SCET Workshop in March 2025 in Tucson, Arizona.

\includepdf[pages=-, pagecommand={}, offset=3mm 0mm]{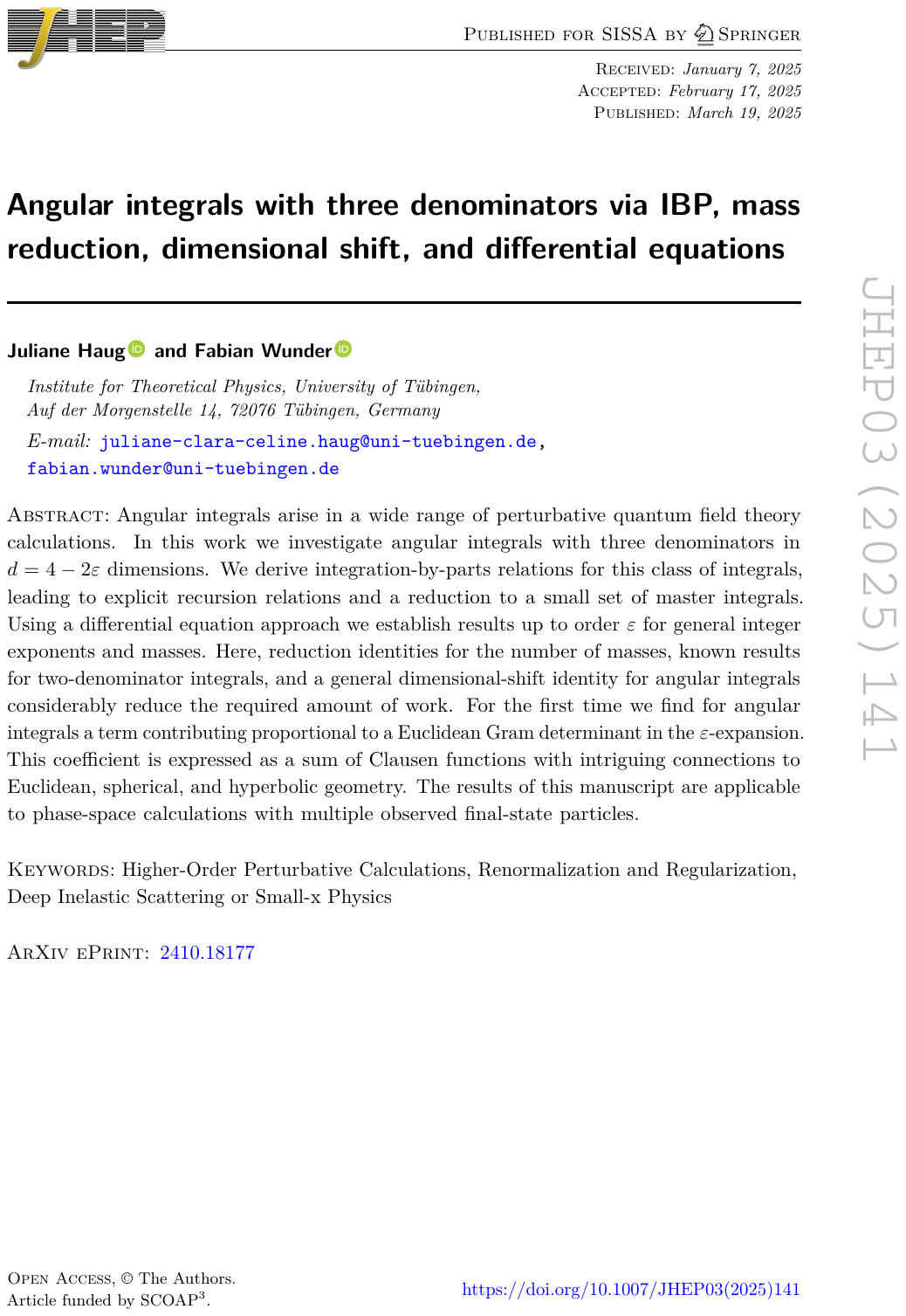}
\cleardoublepage
\section[SIDIS with single-valued polylogarithms]{Publication 9: Single-valued representation of unpolarized and polarized semi-inclusive deep inelastic scattering at next-to-next-to-leading order}
\label{pub:9}
\begin{summaryblock}
{
\textbf{Why?\,: }SIDIS is a key process at the EIC and essential to polarized PDF determination. For good theoretical control, the inclusion of higher order corrections in QCD for the partonic coefficient functions is necessary.
Recently, NNLO results for the coefficients have become available that contain case distinctions.
\vspace{0.1cm}
\\
\textbf{What?\,: } We removed the case distinctions in the coefficient functions, which originate from spurious branch cuts in loop integrals, by applying the results from publication \ref{pub:2}.
}
{
\textbf{Foundations: }
SIDIS, factorization, loop integrals, branch cuts.
\vspace{0.1cm}
\\
\textbf{Novel methods: }
Used box master integral free of spurious branch cuts from publication \ref{pub:2}; applied single-valued polylogarithms to an observable in pQCD.
}
{
\textbf{Results:}
\begin{itemize}
\item[(i)] Established results free of case distinctions for the unpolarized and polarized NNLO SIDIS coefficients, 30\% to 60\% shorter than the original ones.
\item[(ii)] Explored the use of the compactified structure functions for analytic Mellin transform and showcased that integer moments of the parts containing  single-valued polylogarithms are analytically calculable.
\item[(iii)] Developed the \texttt{C++} special functions library \texttt{BEAVER} that reduces the computation time for the SIDIS coefficient functions by a factor of about 5.
\end{itemize}
\textbf{Implications:}
For the upcoming EIC, it is desirable to have an efficient, preferably public, numerical codebase for important processes. This work is a step towards such an efficient numerical implementation for SIDIS based on more compact analytical results.
Directly, \texttt{BEAVER} offers a significant speed-up when using existing implementations, for example for the Mellin-grid generation in polarized PDF fits.
A challenging but highly useful --  especially for PDF fits -- follow-up would be the calculation of the full analytic Mellin transform.
}
\end{summaryblock}

\vspace{0.2cm}
This publication -- again done in collaboration with Juliane Haug -- deals with the compactification of the recently published NNLO results for SIDIS, calculated by two independent groups \cite{Bonino:2024unpol,Bonino:2024pol,Goyal:2023unpol,Goyal:2024pol,Goyal:2024emo}.
These coefficients contain case distinctions in the kinematic $(x,z)$-plane of the SIDIS variables that originate from the one-loop box integral.

Based on the experience with the box integral from publication \ref{pub:2}, we knew that the box integral admits a representation free of case distinctions when expressed through single-valued polylogarithms (SVPs).
This led my co-author, who was working with the unpolarized NNLO coefficients, to check what happens when the parts containing case distinctions are re-expressed through SVPs. Indeed, the case distinctions vanished, which facilitated further analytic calculation with them.
While attending a workshop\footnote{CFNS-INT Joint Program: Precision QCD with the Electron Ion Collider.} at the Institute for Nuclear Theory (INT) at the University of Washington in Seattle, kindly organized by my advisor Werner Vogelsang, we discussed these findings with Thomas Gehrmann and Sven Moch, each part of one of the groups who published the original SIDIS NNLO results, which led us to decide to introduce SVPs also to the polarized case and present the findings in a short publication.\footnote{This was done while at the INT, which we thank for their hospitality.}

The introduction of SVPs considerably reduces the length of the NNLO coefficients by about 30\% to 60\% and constitutes the first use of SVPs in the simplification of an observable level quantity beyond the Regge limit\footnote{High energy limit at fixed momentum transfer, i.\,e. in Mandelstam variables $s\rightarrow\infty$ for fixed $t$.}.
Use cases for the more compact results include more efficient numerical implementations as well as an analytic Mellin transform of the result, both relevant for polarized PDF fits.
These directions are discussed in Appendices A and C, respectively.
When working on the numerical implementation of the coefficient functions, it turned out that at present, the limiting factor is the evaluation time of special functions rather than the algebraic expression size.
To overcome this bottleneck, we developed the special functions library \texttt{BEAVER} (Better Evaluate A Very Efficient Rational) that is presented in Appendix B -- useful beyond the scope of SIDIS since the major speed-up results from efficient implementations of the logarithm and the dilogarithm\footnote{What might Kepler have said if someone would have told him that someday we can calculate a logarithm in five nanoseconds?}.
In preliminary tests, it led to a speed-up of about a factor of five across all SIDIS NNLO coefficients. 

I presented the results of this work at the FOR2926 workshop in Tübingen in July 2025.

\includepdf[pages=-, pagecommand={}, scale=0.95, offset=6mm 0mm]{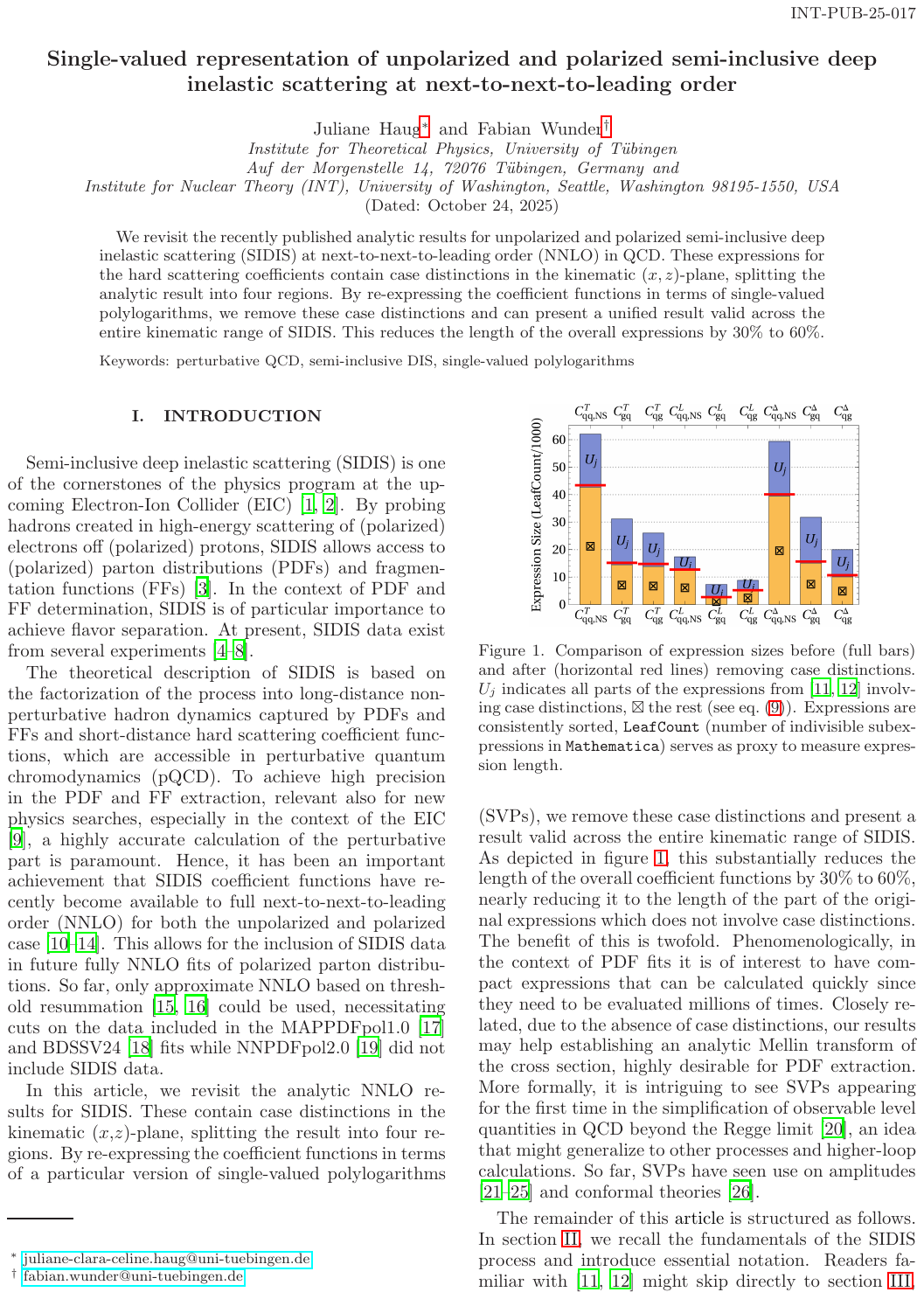}
\cleardoublepage
\section[ On multi-propagator angular integrals]{Publication 10: On multi-propagator angular integrals}
\label{pub:10}
\begin{summaryblock}
{
\textbf{Why?\,: }
To amend the body of knowledge that can be used for analytic phase-space integration and to investigate to what extend loop methods translate into this new setting, we continued the study of angular integrals.
\vspace{0.1cm}
\\
\textbf{What?\,: } We generalized our previous work on angular integrals to a higher number of denominators looking for general patterns between them.
}
{
\textbf{Foundations: }Dimensional regularization, IBP, dimensional recurrence, methods from publications \ref{pub:7} and \ref{pub:8}.
\vspace{0.1cm}
\\
\textbf{Novel methods: }
Classification of coefficients in terms of Gram determinants;
introduced the concept of \textit{branch integrals} which allow for a systematic simplification of angular integrals by scale reduction.
}
{
\textbf{Results:}
\begin{itemize}
\item[(i)] Simplified Euler representation for angular integrals that looks like a Lee-Pomeransky representation
\item[(ii)] General Laporta-free recursion relations for $n$ denominators.
\item[(iii)] Dimensional shift relation for $n$ denominators.
\item[(iv)] Results to order $\eps^0$ for the four denominator integral with arbitrary masses.
\item[(v)] All-order $\eps$-expansion of the massless three-denominator angular integral including a resummation of soft logarithms.
\item[(vi)] Novel idea of decomposing master integrals into \textit{branch integrals} reduces the number of scales from $(n+1)n/2$ to only $n$.
\end{itemize}
\textbf{Implications: }
With this work, the theory of angular integrals is now in a rather satisfactory state.
With the inclusion of soft-logarithm resummation, the three-denominator results are expected to be ready-to-use for higher-multiplicity processes.
Generalizing to other phase-space integrals, the natural next target is the inclusive $\dx \mathrm{PS}_3$ with general propagators; this would correspond to the two-loop level.
Also, it might be interesting to translate the novel concept of branch integrals back to loop integrals and see whether they offer new insights there.
}
\end{summaryblock}

\vspace{0.2cm}
This final publication of the doctoral work continues and concludes, at least for now, the study of angular integrals.
The project combined two of my former collaborators on the subject, Juliane Haug and Vladimir Smirnov, in a study of multi-propagator angular integrals with a focus on structural properties.

In 2024, based on the integral representation provided in publication \ref{pub:7}, Giulio Salvatori calculated the single-massive four-denominator angular integral as an example for a novel subtraction scheme \cite{Salvatori:2024nva}.\footnote{In the first version, there was a mistake in the integral result that got corrected after correspondence.}
Using the mass-splitting identities, we used these to infer the results for the double- and triple-massive cases, which resulted in rather complicated expressions -- that was before we realized the clean Gram-determinant formulation of mass-splitting in publication \ref{pub:8}.
After finishing the project on the three-denominator integral, we used the same differential equation method to re-derive Salvatori's result in a more compact form and extend it to the massless case.
The calculation turned out to be surprisingly straightforward, even the integrals that occur when integrating the differential equation were identical.
Together with the better understanding of the mass-reduction, this led to surprisingly elegant results, again in terms of Clausen functions.

While studying the IBP reduction and dimensional shift formulas, several patterns occurred for the emerging coefficients and all of them could be expressed in terms of Gram determinants and objects derived from them.
This strongly suggests that these patterns hold for a general number of denominators -- including an explicit recursion that allows to reduce the $n$-denominator integral to master integrals. 
Of conceptual interest is also the novel Euler representation, improving on the one given in \ref{pub:7} that very closely matches the Lee-Pomeransky representation for loop integrals, compare eq.\,\eqref{eq:LeePom}.
One of the most interesting novel methods of this paper comes in section 6.
Here, the method of dimensional recurrence is used to reduce the scales of master integrals by splitting those into newly introduced \textit{branch integrals}\footnote{They metaphorically speaking allow to successively grow the whole master integral ``tree'' from simple ``root'' integrals (compare Figure 2).} that depend on a smaller number of scales.
This allows for a reduction from $(n+1) n/2$ to $n$ scales\footnote{In the paper, it erroneously says $n+1$ scales.
When looking at eq.\,(6.4) this clearly depends on $n-1$ Gram type variables since the product starts at $i=2$. Hence, at the bottom of page twelve it should correctly say $n-1$ Gram type variables plus a mass in the massive case. So, overall, a maximum of $n$ scales.}.

Some time after publication of the preprint, results for the four denominator master integrals, expressed in considerably more lengthy form\footnote{Comparing the \texttt{LeafCount} in \texttt{Mathematica}, the results from \cite{Ahmed:2025yrx} are over 20 times longer for the massless case (0.6 MB expression size), for the single-massive case even more than 300 times longer (17.7 MB expression size).}, also appeared by the Regensburg group around Taushif Ahmed \cite{Ahmed:2025yrx}.
Their independent calculation was again based on a Mellin-Barnes approach.

I presented this work at the FOR2926 workshop in Regensburg in November 2025.

\includepdf[pages=-, pagecommand={}, scale=0.95, offset=3mm 0mm]{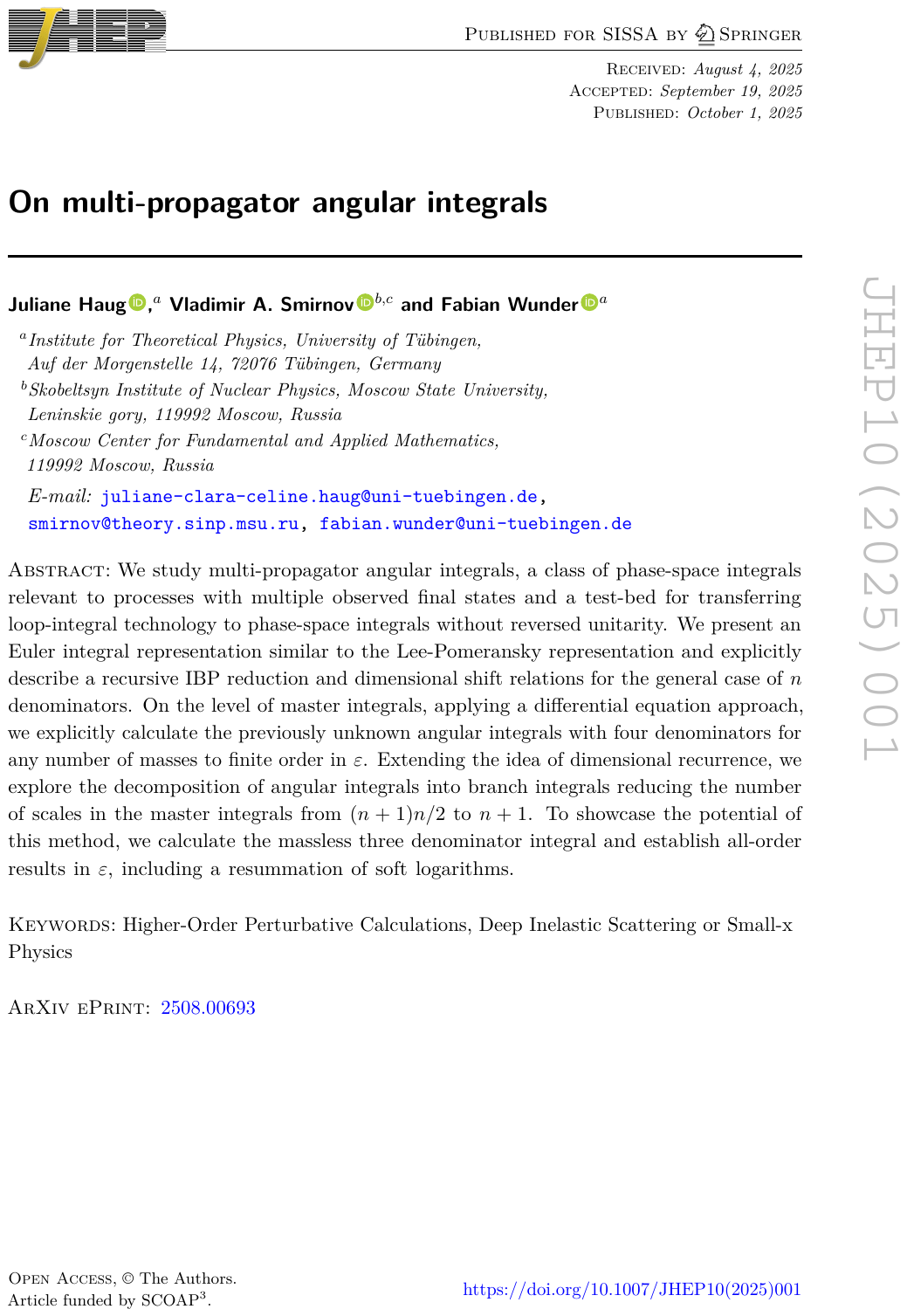}
\end{fancychapter2}
\cleardoublepage
\begin{fancychapter2}{What we learned and where it leads}{Final remarks}{The aim remains: to understand the world.}{John Bell}
\label{ch:Conclusion}

\section{Recap of findings}
This thesis has developed and applied a set of analytic methods for perturbative QCD with the objective of improving theoretical control over ingredients that enter precision collider phenomenology.
The individual publications addressed distinct problems -- ranging from the structure of phase-space integrals to the analytic properties of one-loop integrals, kinematic expansions of observables, and the scale evolution of parton densities -- that share a common purpose: to advance the analytic toolset underlying calculations of scattering processes and the extraction of the proton structure from experimental data.

A central part of this work concerned angular integrals, where several techniques traditionally associated with Feynman integrals were transferred to a setting in which they previously had not been exploited to their full potential.
This allowed us to put our knowledge of angular integrals on a solid foundation, bridging the gap between classic developments of the 1980s and the contemporary toolbox of multi-loop techniques.
This program led to new closed-form results, in particular a number of all-order expansions in $\eps$, clarified the analytic structure of these integrals, and demonstrated the utility of modern loop integral methods in a phase-space setting.
These developments are not only of technical value but also provide building blocks for the real radiation part of higher-order corrections to phenomenologically relevant processes.

A second theme involved the analytic structure of one-loop integrals and the use of single-valued polylogarithms to eliminate spurious branch cuts.
This led to considerably more compact expressions for the NNLO coefficient functions in SIDIS.
Besides improving structural transparency, such compact representations are advantageous both for further analytical treatment, for example in the calculation of Mellin transforms, as well as for fast numerical implementation for phenomenology and global analysis of parton densities.
For SIDIS, both are particularly relevant in view of the precision program at the EIC.

Furthermore, this thesis discussed a systematic expansion in powers of the transverse momentum to analyze the kinematic dependence of hard-scattering coefficients in the Drell-Yan process, and applied it to the $T$-odd helicity structure functions.
This expansion helps to clarify the transition between the fixed-order collinear regime and the domain of TMD factorization.
In addition, a semi-analytical ansatz for the scale evolution of parton distributions was explored, demonstrating that analytical information and numerical methods can be combined to solve integro-differential evolution equations -- as showcased on the example of the DGLAP equation.

Taken together, these results provide a modest contribution to the analytical methods for perturbative QCD.
Beyond the specific applications discussed in this thesis, the techniques developed here are expected to find use in the theoretical preparation for the EIC and other precision facilities.

\section{Directions for future work}
Going forward, there are several directions for future work to build on the results discussed here.
First, the new results for angular integrals should be exploited in phenomenology for new analytic higher order results.
Some of them already found application, for example in the study of mass effects in parton evolution \cite{Assi:2023}, in Higgs+jet production \cite{Pal:2023}, for prompt photons \cite{Rein:2024}, for N$^3$LO soft functions within subtraction schemes \cite{Baranowski:2024ysi}, for transverse nucleon single-spin asymmetries in hadron and jet production \cite{Rein:2025qhe}, in threshold resummation of four-top quark production \cite{vanBeekveld:2025ghw}, and the study of super-leading logarithms in top quark pair production \cite{Banerjee:2025kkq}.
Yet a substantial part of the novel machinery for angular integrals -- especially for those with three or more denominators -- has  so far not been used in phenomenology.
Identifying phenomenologically interesting, cleanly probeable, semi-inclusive processes in a $2\rightarrow 2+X$ kinematic, where irreducible three-denominator angular integrals arise -- either directly or in subtraction terms --, would be a natural next step.
Such processes with higher multiplicities are interesting in view of the upcoming EIC that will allow for high statistics also in multi-differential measurements.

Another, technically quite demanding, direction is going to higher orders in the real-radiation phase space.
Allowing for the additional radiation of a third particle that is integrated over, the angular integration measure needs to be replaced by the three-particle phase space $\int\dPS_3^{K\rightarrow k_1+k_2+k_3}$.
Splitting this phase space into two coupled two-particle phase spaces, the corresponding N$^3$LO phase space measure can be parameterized as
\begin{align}
\int\dPS_3^{K\rightarrow k_1+k_2+k_3}=\frac{(K^2)^{1-2\eps}}{2 (4\pi)^{5-4\eps}}\int_0^1\dx\xi\,\xi^{-\eps}(1-\xi)^{1-2\eps}\int\dx\Omega_{d-1}^K(k_1)\int\dx\Omega_{d-1}^{K_{23}}(k_2)\,,
\label{eq: PS3 parametrization}
\end{align}
where $K_{23}=k_2+k_3$ and $\xi\equiv K_{23}^2/K^2$.
It is a challenge for future work to establish a similar body of knowledge for this class of integrals with general denominators in the integrand.
If successful, it would make the next perturbative order analytically tractable for the same set of processes where the traditional angular integrals of van Neerven were indispensable.
A natural starting point would be establishing suitable Euler representations.

Also beyond angular integrals, there remain interesting open questions that follow from this work.
Regarding the study of branch cuts in loop integrals, it will be intriguing to see whether single-valued polylogarithms and generalizations thereof can help to simplify the analytic structure beyond one loop.
In application, the simplified SIDIS results provide a natural stepping stone towards efficient, preferably public, numerical implementations in preparation for the EIC.
Concerning tool building, also the \texttt{POMPOM} algorithm deserves further attention to mature from a proof-of-concept into a tool that can be employed in state-of-the-art analyses.
Furthermore, as the study of the transverse momentum spectrum for $T$-odd structure functions showcased, there remains much to be understood about the interplay of different kinematic regimes in differential observables, especially when the predicted effects are small.

Ultimately, the unifying goal of these future research directions is preparation for the precision data offered by the EIC.
This necessitates higher order calculations in perturbation theory for multi-differential processes and precise frameworks for global analysis of hadron structure.
These frontiers will continue to drive the demand for advancing the analytic methods that are at our disposal.
Finally, the theoretical and experimental effort will -- once the EIC is taking high-statistics data -- hopefully culminate in new discoveries about the structure of matter, adding yet another layer to answering the question: \textit{What is the world made of?} 
\flushbottom
\newpage
\thispagestyle{empty}
\cleardoublepage
\makeatletter
\let\ps@plain\ps@fancy
\makeatother
\appendix
\thispagestyle{empty}
\begingroup
\vspace*{\fill}
\begin{center}
    {\scalebox{2}{\bfseries\Huge Appendix}}
    \addcontentsline{toc}{chapter}{Appendix}
\end{center}
\vspace*{\fill}
\endgroup
\cleardoublepage
\pagestyle{fancy}
\section*{List of abbreviations}
\addcontentsline{toc}{section}{List of abbreviations}
\markright{List of abbreviations}
\begin{table}[h!]
\centering
\renewcommand{\arraystretch}{1.15}
\begin{tabularx}{0.95\textwidth}{lX}
\hline\hline
\textbf{Abbreviation} & \textbf{Meaning} \\ \hline
\texttt{BEAVER}   & Special function library for SIDIS \\
BSM      & Beyond Standard Model\\
DGLAP    & Dokshitzer--Gribov--Lipatov--Altarelli--Parisi (equation) \\
DIS      & Deep inelastic scattering \\
DY       & Drell--Yan process \\
EFT      & Effective field theory \\
EIC      & Electron--Ion Collider \\
$e^\mp$       & Electron/Positron \\
$\eps$   & Dimensional regularization parameter\\
FF       & Fragmentation function \\
$g$      & Gluon \\
$\gamma^{(\ast)}$ & (virtual) photon\\
IBP      & Integration-by-parts (identities) \\
LHC      & Large Hadron Collider \\
LO       & Leading order \\
MB       & Mellin--Barnes (representation) \\
NLO      & Next-to-leading order \\
NNLO     & Next-to-next-to-leading order \\
N$^{3}$LO & Next-to-next-to-next-to-leading order \\
(N)LP      & (Next-to-)leading power \\
pQCD     & Perturbative quantum chromodynamics \\
PDF      & Parton distribution function \\
\texttt{POMPOM}   & Semi-analytic $x$-space DGLAP evolution method \\
$\dx\mathrm{PS}_n$      & $n$-particle phase-space \\
$q$        & Quark \\
$\bar{q}$ & Antiquark \\
QCD      & Quantum chromodynamics \\
QFT      & Quantum field theory \\
$Q^{2}$  & Hard scale (momentum transfer) \\
$Q_T$    & Transverse momentum \\
SCET     & Soft-collinear effective theory \\
SIDIS    & Semi-inclusive deep inelastic scattering \\
SM       & Standard Model \\
SVP      & Single-valued polylogarithm \\
TMD      & Transverse-momentum dependent (functions/factorization) \\
\hline\hline
\end{tabularx}
\end{table}
\clearpage
\addcontentsline{toc}{section}{List of Figures}
\listoffigures
\addcontentsline{toc}{section}{List of Tables}
\listoftables
\addcontentsline{toc}{section}{Bibliography}
\bibliographystyle{JHEP}
\bibliography{Literature_Phd.bib}
\clearpage
\thispagestyle{empty}
\cleardoublepage
\section*{Erklärung über die Verwendung generativer KI}
\addcontentsline{toc}{section}{Erklärung über die Verwendung generativer KI}
Generative KI-Systeme wurden gemäß den ,,Richtlinien zum Umgang mit generativer KI in der Lehre und Forschung'' der Universität Tübingen vom 30.04.2024 ausschließlich unterstützend verwendet. Insbesondere wurde ChatGPT 5.1 zur sprachlichen Verbesserung und zur formalen Qualitätssicherung herangezogen. Wesentliche inhaltliche Beiträge, Analysen und Schlussfolgerungen entstammen vollständig dem Autor; alle KI-Ausgaben wurden kritisch geprüft.
\thispagestyle{empty}
\cleardoublepage
\thispagestyle{empty}
\section*{Eigenständigkeitserklärung}
\addcontentsline{toc}{section}{Eigenständigkeitserklärung}
Ich erkläre hiermit, dass ich die zur Promotion eingereichte Arbeit selbständig verfast, nur die angegebenen
Quellen und Hilfsmittel (auch KI Tools) benutzt und wörtlich oder inhaltlich übernommene Stellen als solche
gekennzeichnet habe. Ich erkläre, dass die Leitlinien zur Sicherung guter wissenschaftlicher Praxis der Universität
Tübingen (Beschluss des Senats vom 11.2.2021) beachtet wurden. Ich versichere an Eides statt, dass diese
Angaben wahr sind und dass ich nichts verschwiegen habe. Mir ist bekannt, dass die falsche Abgabe einer
Versicherung an Eides statt mit Freiheitsstrafe bis zu drei Jahren oder mit Geldstrafe bestraft wird.\\
\vspace{1.5cm}
\\
\begin{tabularx}{0.9\textwidth}[b]{p{5cm} X p{5cm}} \cline{1-1} \cline{3-3}
Name & & Ort, Datum
\end{tabularx}
\end{fancychapter2}
\end{document}